# Quantitative Partition Models and Benford’s Law

## ABSTRACT

Benford’s Law predicts that the first significant digit on the leftmost side of numbers in real-life data is proportioned between all possible 1 to 9 digits approximately as in LOG(1 + 1/digit), so that low digits occur much more frequently than high digits in the first place. The two essential prerequisites for data configuration with regards to compliance with Benford’s Law are high order of magnitude and positive skewness with a tail falling to the right of the histogram, so that quantitative configuration is such that the small is numerous and the big is rare. In this article various quantitative partition models are examined in terms of the quantitative and digital behavior of the resultant set of parts. The universal feature found across all partition models is having many small parts but only very few big parts, while Benford’s Law is valid only in some particular partition cases and under certain constraints. Hence another suggested vista of Benford’s Law is viewing it as a particular subset of the broader positive skewness phenomenon in quantitative partitioning. Significantly, such a vista is true in all other causes and explanations of Benford’s Law where the small consistently outnumbers the big also in partial structures of the model or well before full convergence to Benford is achieved - endowing the principle universality in a sense. In conclusion, either the active act of partitioning or the passive consideration of a large quantity as the composition of smaller parts can be considered as another independent explanation for the widespread empirical observation of Benford’s Law in the physical sciences.

Alex Ely Kossovsky
akossovsky@gmail.com

# PART 1: BENFORD'S LAW

# PART 2: QUANTITATIVE PARTITION MODELS

# PART 1:

# BENFORD'S LAW

## [I] The First Digit on the Left Side of Numbers

It has been discovered that the first digit on the left-most side of numbers in real-life data sets is most commonly of low value such as {1, 2, 3} and rarely of high value such as {7, 8, 9}.
As an example serving as a brief and informal empirical test, a small sample of 40 values relating to geological data on time between earthquakes is randomly chosen from the data set on all global earthquake occurrences in 2012 – in units of seconds. Figure A depicts this small sample of 40 numbers. Figure B emphasizes in bold and black color the 1st digits of these 40 numbers.

| | | | |
|---|---|---|---|
| 285.29 | 185.35 | 2579.80 | 27.11 |
| 5330.22 | 1504.49 | 1764.41 | 574.46 |
| 1722.16 | 815.06 | 3686.84 | 1501.61 |
| 494.17 | 362.48 | 1388.13 | 1817.27 |
| 3516.80 | 5049.66 | 2414.06 | 387.78 |
| 4385.23 | 2443.98 | 2204.12 | 1224.42 |
| 1965.46 | 3.61 | 1347.30 | 271.23 |
| 3247.99 | 753.80 | 1781.45 | 593.59 |
| 1482.64 | 1165.04 | 4647.39 | 1219.19 |
| 251.12 | 7345.52 | 1368.79 | 4112.13 |

**Figure A**: Sample of 40 Time Intervals between Earthquakes

| | | | |
|---|---|---|---|
| **2**85.29 | **1**85.35 | **2**579.80 | **2**7.11 |
| **5**330.22 | **1**504.49 | **1**764.41 | **5**74.46 |
| **1**722.16 | **8**15.06 | **3**686.84 | **1**501.61 |
| **4**94.17 | **3**62.48 | **1**388.13 | **1**817.27 |
| **3**516.80 | **5**049.66 | **2**414.06 | **3**87.78 |
| **4**385.23 | **2**443.98 | **2**204.12 | **1**224.42 |
| **1**965.46 | **3**.61 | **1**347.30 | **2**71.23 |
| **3**247.99 | **7**53.80 | **1**781.45 | **5**93.59 |
| **1**482.64 | **1**165.04 | **4**647.39 | **1**219.19 |
| **2**51.12 | **7**345.52 | **1**368.79 | **4**112.13 |

**Figure B**: The First Digits of the Earthquake Sample

Clearly, for this very small sample, low digits occur by far more frequently on the first position than do high digits. A summary of the digital configuration of the sample is given as follows:

| | |
|---|---|
| Digit Index: | { 1, 2, 3, 4, 5, 6, 7, 8, 9 } |
| Digits Count totaling 40 values: | { 15, 8, 6, 4, 4, 0, 2, 1, 0 } |
| Proportions of Digits with '%' sign omitted: | {38, 20, 15, 10, 10, 0, 5, 3, 0 } |

Assuming (correctly) that these 40 values were collected in a truly random fashion from the large data set of all 19,452 earthquakes occurrences in 2012; without any bias or attempt to influence first digits occurrences; and that this pattern is generally found in many other data sets, one then may conclude with the phrase "not all digits are created equal", or rather "not all first digits are created equal", even though this seems to be contrary to intuition and against all common sense.

The focus here is actually on the first meaningful digit – counting from the left side of numbers, excluding any possible encounters of zero digits which only signify ignored exponents in the relevant set of powers of ten of our number system. Therefore, the complete definition of the **First Leading Digit** is the first non-zero digit of any given number on its left-most side. This digit is the first significant one in the number as focus moves from the left-most position towards the right, encountering the first non-zero digit signifying some quantity; hence it is also called the **First Significant Digit**. For 2365 the first leading digit is 2. For 0.00913 the first leading digit is 9 and the zeros are discarded; hence even though strictly-speaking the first digit on the left-most side of 0.00913 is 0, yet, the first significant digit is 9. For the lone integer 8 the leading digit is simply 8. For negative numbers the negative sign is discarded, hence for -715.9 the leading digit is 7. Here are some more illustrative examples:

**6**,719,525 → digit 6
0.0000**7**61 → digit 7
-0.**2**81264 → digit 2
**8**75 → digit 8
**3** → digit 3
-**5** → digit 5

For a data set where all the values are greater than or equal to 1, such as in the sample of the earthquaqe data, the first digit on the left-most side of numbers is also the First Leading Digit and the First Significant Digit, and necesarily one of the nine digits {1, 2, 3, 4, 5, 6, 7, 8, 9}; while digit 0 never occurs first on the left-most side.

## [II] Benford's Law and the Predominance of Low Digits

Benford's Law states that:

$$\text{Probability[First Leading Digit is d]} = LOG_{10}(1 + 1/d)$$

$LOG_{10}(1 + 1/1) = LOG(2.00) = 0.301$
$LOG_{10}(1 + 1/2) = LOG(1.50) = 0.176$
$LOG_{10}(1 + 1/3) = LOG(1.33) = 0.125$
$LOG_{10}(1 + 1/4) = LOG(1.25) = 0.097$
$LOG_{10}(1 + 1/5) = LOG(1.20) = 0.079$
$LOG_{10}(1 + 1/6) = LOG(1.17) = 0.067$
$LOG_{10}(1 + 1/7) = LOG(1.14) = 0.058$
$LOG_{10}(1 + 1/8) = LOG(1.13) = 0.051$
$LOG_{10}(1 + 1/9) = LOG(1.11) = 0.046$
---------
1.000

Figure C depicts the distribution. Figure D visually depicts Benford's Law as a bar chart. This set of nine proportions of Benford's Law is sometimes referred to in the literature as **'The Logarithmic Distribution'**. Remarkably, Benford's Law is confirmed in almost all real-life data sets with high order of magnitude, such as in data relating to physics, chemistry, astronomy, economics, finance, accounting, geology, biology, engineering, governmental census data, and many others.

| Digit | Probability |
|---|---|
| 1 | 30.1% |
| 2 | 17.6% |
| 3 | 12.5% |
| 4 | 9.7% |
| 5 | 7.9% |
| 6 | 6.7% |
| 7 | 5.8% |
| 8 | 5.1% |
| 9 | 4.6% |

**Figure C**: Benford's Law for First Digits

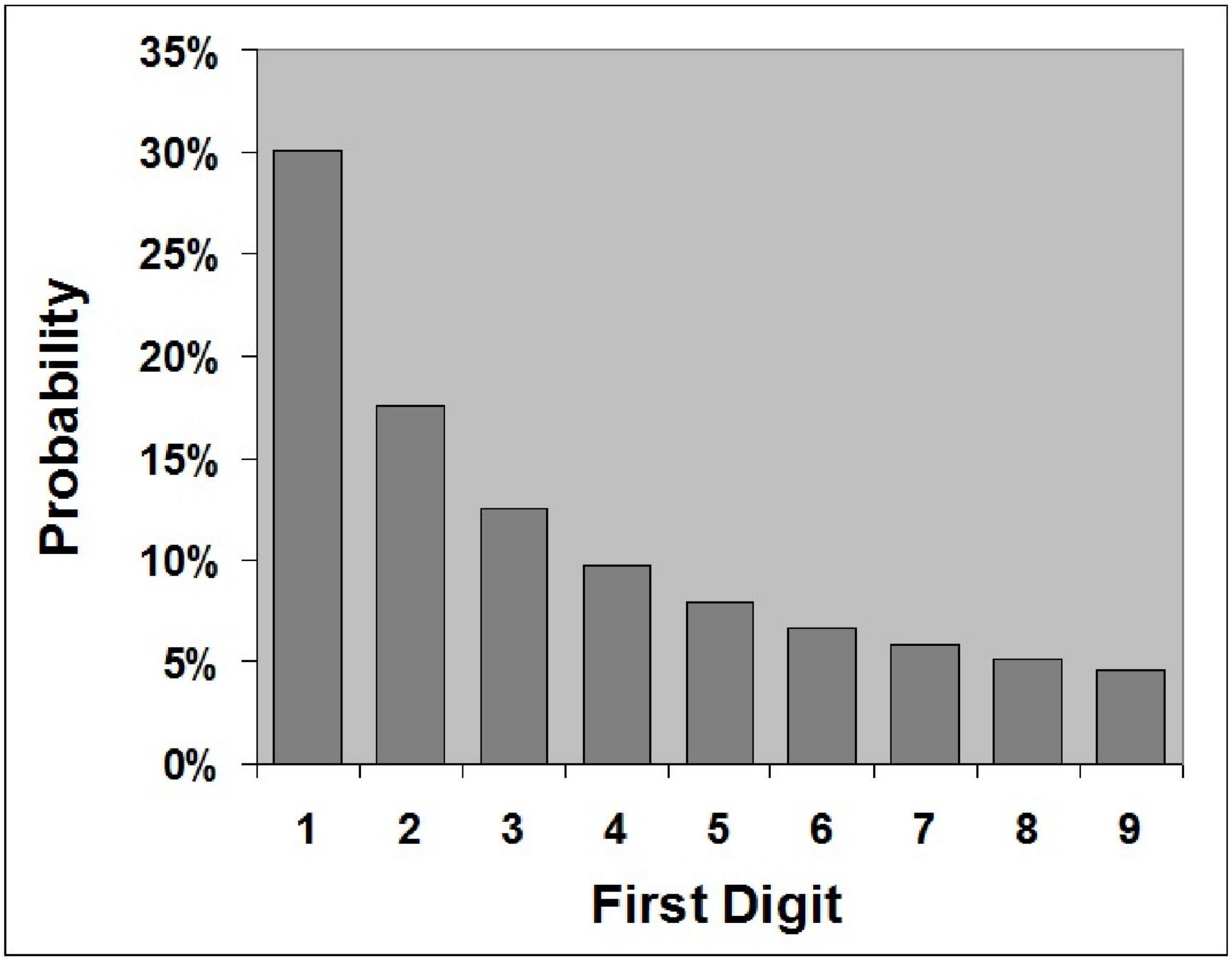

**Figure D**: Benford's Law – Probability of First Leading Digit Occurrences as a Bar Chart

## [III] Sum of Squares Deviation Measure (SSD)

It is necessary to establish a standard measure of 'distance' from the Benford digital configuration for any given data set. Such a numerical measure could perhaps tell us about the conformance or divergence from the Benford digital configuration of the data set under consideration. This is accomplished with what is called **Sum Squares Deviations (SSD)** defined as the sum of the squares of the 'errors' between the Benford expectations and the actual/observed values (in percent format – as opposed to fractional**/**proportional format):

$$SSD = \sum_{1}^{9} \left( \text{Observed \% of d} - 100 * \text{LOG}(1 + \frac{1}{d}) \right)^{\mathbf{2}}$$

For example, for observed 1st digits proportions of {31.1, 18.2, 13.3, 9.4, 7.2, 6.3, 5.9, 4.5, 4.1} with '%' sign omitted, SSD measure of distance from the logarithmic is calculated as:

**SSD** = $(31.1 - \mathbf{30.1})^2 + (18.2 - \mathbf{17.6})^2 + (13.3 - \mathbf{12.5})^2 + (9.4 - \mathbf{9.7})^2 +$
$+ (7.2 - \mathbf{7.9})^2 + (6.3 - \mathbf{6.7})^2 + (5.9 - \mathbf{5.8})^2 + (4.5 - \mathbf{5.1})^2 + (4.1 - \mathbf{4.6})^2 = \mathbf{3.4}$

SSD generally should be below 25; a data set with SSD over 100 is considered to deviate too much from Benford; and a reading below 2 is considered to be ideally Benford.

## [IV] Physical Order of Magnitude of Data (POM)

Rules regarding expectations of compliance with Benford's Law rely heavily on measures of order of magnitude (i.e. variability) of data.

Physical order of magnitude of a given data set is a measure that expresses the extent of its variability. It is defined as the ratio of the maximum value to the minimum value. The data set is assumed to contain only positive numbers greater than zero.

$$\text{Physical Order of Magnitude (POM)} = \text{Maximum}/\text{Minimum}$$

The classic definition of order of magnitude involves also the application of the logarithm to the ratio maximum/minimum, transforming it into a smaller and more manageable number.

$$\text{Order of Magnitude (OOM)} = \text{LOG}_{10}(\text{Maximum}/\text{Minimum})$$
$$\text{Order of Magnitude (OOM)} = \text{LOG}_{10}(\text{Maximum}) - \text{LOG}_{10}(\text{Minimum})$$

Since such logarithmic transformation has a monotonic one-to-one relationship with max/min, it does not provide for any new insight or information, but could rather be looked upon in a sense as simply the use of an alternative scale, still measuring the same thing. For this reason the complexity of logarithm can be avoided altogether by referring only to the simple POM measure.

The more profound reason for using **POM** instead of **OOM** is its feature as a universal measure of variability, totally independent of societal number system in use, as well as being independent on the arbitrary choice of base 10, derived from the chanced or random occurrence of us having 10 fingers. This is the motivation behind the use of the term 'physical', expressing real and physical measure of variability, divorced from any numerical inventions, and especially so when data relates to the natural world such as in scientific figures and physical information.

OOM is perhaps more appropriate for a single isolated number, where it is re-defined as simply $\text{LOG}_{10}(\text{Number})$ without any reference to maximum, minimum, or any ratio. If we can assume that that number is an integral whole number without any [trailing] fractional part, then an alternative meaning of this OOM definition is simply expressing how many digits approximately are necessary to write the number. Surely in general, the bigger the [integral] number the more digits it takes to write it! For example, $\text{LOG}_{10}(8{,}200{,}135) = 6.9$, which is about 7, and that's exactly how many digits the number involves. As another example, $\text{LOG}_{10}(10{,}000{,}000) = 7.0$ which is exactly one digit less than the number of digits involved in writing the number, namely 8 digits.

In the extreme case where all the numbers in the data set are identical, having the value R say, variability is then nonexistent, and POM = maximum/minimum = R/R = 1.

NOTE: LOG(X) or log(x) notation in this article would always refer to our decimal base 10 number system, hence the more detailed notation of $\text{LOG}_{10}(X)$ is often (but not always) avoided.

## [V]  A Robust Measure of Physical Order of Magnitude (CPOM)

It is perhaps unfortunate that the literature in statistics does not seem to contain any robust definition of order of magnitude. Such a measure should prove steady and consistent for all types of data sets, strongly resisting outliers, preventing them from overly influencing the numerical measure of data variability.

In order to accomplish exactly that, and also to preserve the advantage of avoiding dependencies on arbitrary societal number systems and particular bases, the basic (independent) structure of POM shall be used, but with the added modification of simply eliminating any possible outliers on the left for small values and on the right for big values. This is accomplished by narrowing the focus exclusively onto the core 80% part of the data. This brutal purge eliminates any malicious and misleading outliers as well as any innocent and proper data points which happened to stray just a little bit away from the core part of the data. The measure shall be called Core Physical Order of Magnitude and it is defined as follows:

Core Physical Order of Magnitude (CPOM) = $Q_{90\%} / Q_{10\%}$

The definition simply reformulates POM by substituting the 10th percentile (in symbols $Q_{10\%}$) for the minimum, and by substituting the 90th percentile (in symbols $Q_{90\%}$) for the maximum.

The 10th percentile is the value below which about 10% of the data points may be found. The 50th percentile is the median, below which about half of all the data points may be found. The 90th percentile is the value below which about 90% of the data points may be found.

In general, the rejection of outliers appearing in a given data set may be justified, or it may actually be misguided. For example, if over 50,000 students at a large university are surveyed with regards to height, and the top value is say 7.25 meter, then this outlier is certainly some kind of an error in recording and should be excluded from further analysis. If the top value is say 2.37 meter, then this 2.37 outlier is actually an integral part of the data set, and especially so if the well-known tall student is ordered to appear at the administration office, rudely interrupting his exciting basketball game at the court, and another measurement is taken, confirming his 2.37 meter height as well as his existence. This is not simply a matter of mere semantics, and there is a compelling argument not to classify this 2.37 value as an outlier, although in reality it depends on the context.

## [VI] Two Essential Requirements for Benford Behavior

One of the two essential prerequisites or conditions for data configuration with regards to compliance with Benford's Law is that the value of the order of magnitude of the data set should be approximately over 3; in other words, that $LOG_{10}$(Maximum/Minimum) > 3, and that therefore (Maximum/Minimum) > $10^3$. This in turn implies that the threshold POM value (separating compliance from non-compliance) is about 1000, namely that POM > 1000 constitutes the condition for compliance.

The above prerequisite for compliance totally ignores the thorny issue of outliers and edges, and in that sense it is too simplistic and even completely erroneous for some data sets. Hence, using the CPOM qualification is essential in judging whether or not a given data set is expected or not expected to comply with Benford's Law. The proper qualification for expectance of compliance with the law in the approximate - obtained via extensive empirical studies - is then as follows:

Core Physical Order of Magnitude $= Q_{90\%} / Q_{10\%} > 100$

Actually, even lower CPOM values such as 50 and 30 are expected to yield Benford, but falling below 30 does not bode well for getting anywhere near the logarithmic distribution.

Skewness of data where the histogram comes with a prominent tail falling to the right is the second essential criterion necessary for Benford behavior. Indeed, most real-life physical data sets are generally skewed in the aggregate, so that overall their histograms have tails falling on the right, and consequently the quantitative configuration is such that the small is numerous and the big is rare, while low first digits decisively outnumber high first digits.

The asymmetrical, Exponential, Lognormal, k/x [and many other distributions] are typical examples of such quantitatively skewed configuration, and therefore they are approximately, nearly, or exactly Benford - respectively. The symmetrical Uniform, Normal, Triangular, Circular-like, and other such distributions are inherently non-Benford, or rather anti-Benford, as they lack skewness and do not exhibit any bias or preference towards the small and the low.

Symmetrical distributions are always non-Benford, no matter what values are assigned to their parameters. By definition they lack that asymmetrical tail falling to the right, and such lack of skewness precludes Benford behavior regardless of the value of their order of magnitude. Order of magnitude simply does not play any role whatsoever in Benford behavior for symmetrical distributions. For example, first digits of the Normal($10^{35}$, $10^8$) or the Uniform(1, $10^{27}$) are not Benford at all, and this is so in spite of their extremely large orders of magnitude. In summary: Benford behavior in extreme generality can be found with the confluence of sufficiently large order of magnitude together with skewness of data - having a histogram falling to the right. The combination of skewness and large order of magnitude is not a guarantee of Benford behavior, but it is a strong indication of likely Benford behavior under the right conditions. Moderate [overall] quantitative skewness with a tail falling too gently to the right implies that digits are not as skewed as in the Benford configuration. Extreme [overall] quantitative skewness with a tail falling sharply to the right implies that digits are severely skewed, even more so than they are in the Benford configuration.

Bowley Skewness for example, defined as [(Q3 – Q2) – (Q2 – Q1)] / [Q3 – Q1] is an intuitive measure of skewness but its numerical value fluctuates greatly across data sets. Calculated Bowley Skewness values for numerous logarithmic data sets and distributions do not yield any consistent result, except that all values come out above 0.3 and below 1.0, and which is consistent with the fact that all logarithmic data sets are positively skewed in the aggregate. In sharp contrast, non-logarithmic data sets generally come out with decisively lower Bowley Skewness values below 0.25 and above 0. In contrast to Bowley's unstable value for logarithmic data sets, Benford's Law is a very consistent and almost exact measure of skewness, with very little fluctuations across logarithmic data sets.

## [VII] Data Skewness is More Prevalent than Benford's Law

All data sets obeying Benford's Law (i.e. logarithmic data) are structured in such a way that in the aggregate there are more small quantities than big quantities. In other words, that in the aggregate the histogram is falling to the right, except perhaps in the beginning on the very left for low values where it temporarily rises for a very small portion of overall data, as well as in few and minors reversals along the way where it rises briefly. This quantitative configuration is called 'positive skewness' in mathematical statistics.

The expression or motto for this quantitative phenomenon regarding size configuration is coined as '**small is beautiful**'. The term 'beautiful' in this context is not meant literally, but rather metaphorically, as it signifies the connotation associated with the adjectives numerous, plentiful, frequent, and most common. The terms 'small' and 'big' refer to relative quantities within the framework of any given data set under consideration, and never to any absolute quantities or some imaginary fixed and universal benchmark values applicable to all existing data sets.

The small is beautiful phenomenon has by far much wider scope and it is much more prevalent in the physical world and in the realm of abstract mathematics than the more particular Benford quantitative configuration. This statement does not imply that Benford's Law is not prevalent in scientific, physical, and numerous other data types, on the contrary, it is highly prevalent. The statement only implies that in almost all the counter examples and exceptions to Benford's Law, the small is beautiful phenomenon is still valid, albeit with different quantitative configurations than that of the Benford one (and which are typically milder, but at times even skewer).

The assertion is derived from concrete experience with real-life numerical examples and from general research in Benford's Law. While this discussion may sound vague, in fact it is rather a very essential overview in the entire quantitative phenomenon of Benford's Law and of real-life data analysis in general. For those statisticians and data analysts who have worked on data sets and the Benford phenomenon for many years, including doing theoretical research, this generic statement seems natural, fundamental, and quite necessary.

All this can be stated more succinctly in three ways:

I) The Benford's Law configuration is a subset of the small is beautiful phenomena.
II) The small is beautiful phenomenon is even more prevalent than Benford's Law.
III) A significant portion of non-Benford real-life data is quantitatively structured in the spirit of the small is beautiful phenomenon.

## [VIII] The Random and the Deterministic Flavors in Benford's Law

Not all logarithmic data sets are created equal, but rather they come with two distinct flavors, the random flavor and the deterministic flavor. The essential distinction between these two flavors is the way digits behave throughout the entire range of the data locally, on smaller sub-intervals.

If a given data set with a range say between 3 and 2789 is nearly perfectly Benford, could we then conclude that small parts of the data are also Benford, just because they have been cut out from a whole Benford configuration? For example, is the sub-set of the data belonging to the sub-range between 3 and 1652 also Benford? Is the sub-set of the data belonging to the sub-range between 10 and 100 also Benford? Does the whole endow its Benford property to its parts?

Globally, from the minimum on the very left part of the range, all the way to the maximum on the very right part of the range, digits proportions are as in LOG(1 + 1/d) overall, as predicated by Benford's Law. Yet, for random-flavored data, local mini digit distributions on smaller sub-intervals show a remarkably consistent pattern of differentiation, as digits develop from near digital equality on the left for low values, to approximately the Benford digital configuration around the middle, and finally to extreme digital inequality on the far right for big values, where low digits overwhelm high digits, and where digit 1 typically usurps leadership by earning 40% or even more than 50% proportion in some cases. This pattern in random-flavored data is coined as "**Digital Development Pattern**".

In order to be able to observe Digital Development Pattern, it is necessary to partition the relevant section of the x-axis into sub-intervals standing between integral powers of ten, such as 0.01, 0.1, 1, 10, 100, and so forth. This partition is the most natural one in the context of Benford's Law since these points signify the beginnings and the ends of all the first digit cycles.

The vast majority of real-life data is of the random flavor. A tiny minority - such as deterministic exponential growth series and data relating to k/x distribution - come with very consistent local digit distributions throughout the entire range of data, namely that of the Benford digital configuration, which is found equally on the left, in the center, and on the right, without any development or changes whatsoever.

The coining of the terms '**deterministic'** and '**random'** is usually appropriate in most cases, but these terms should not be taken literally, because the distinction here is actually not about randomness in data versus predictable events and deterministic generation of resultant numbers, but rather about localized digital behavior within the entire data range. The choice of these two terms is due to the fact that almost all random data come with such differentiated local digital behavior, while the particular case of deterministic and predictable exponential growth series comes with the consistent Benford behavior throughout its entire range. But these two terms would seem awkward when random data has (very rarely) the consistent Benford behavior throughout its entire range [*such as in the case of exponential growth series with variable and random growth rate, as well as in the case of k/x statistical distribution*], or when deterministic data comes with digital development. Perhaps future authors would coin the alternative terms of the 'consistent flavor' and the 'developmental flavor' in Benford's Law.

The k/x distribution is the only density that perfectly obeys Benford's Law for a range standing between two adjacent IPOT points, such as (1, 10), (10, 100), or (100, 1000), and so forth. For such adjacent IPOT ranges, there exists no other distribution that perfectly obeys Benford's Law (with all higher orders considered) except k/x distribution! On such particular intervals the k/x distribution is unique!

It should be noted that k/x is also perfectly Benford whenever it is defined between any two points A and B such that log difference LOG(B) – LOG(A) is an integer greater than 1, such as say the interval (1.22835, 12283.5) where log difference is the integral value of 4, but k/x is not unique on such wider interval, and there are in principle infinitely many other distributions that are perfectly Benford as well.

Certainly the case of k/x distribution is quite exceptional in the field of Benford's Law, yet its highly-consistent Benfordian feature (totally lacking Digital Development Pattern) renders it quite irrelevant to practically all types of real-life random data! 'Paradoxically', k/x has loyally served us in deciphering GLORQ, which is the parent law of Benford!

Such is the seductive power of k/x distribution in the context of Benford's Law that some misguided authors and overly enthusiastic students of Benford's Law start their article or essay by basing it on some assumption or feature regarding the k/x distribution and then proceed to draw far reaching conclusions, mistakenly extrapolating the odd case of k/x to all real-life random data. Such regrettable trend has led to several erroneous conclusions, published in respectable journals, and officially certified by expert mathematicians as true. This author has taken on the dissenting role of an agitator as well as a prophet of doom, preaching the virtue of separating the random from the deterministic and of becoming aware of this crucial distinction in the field, and predicting the encountering of contradictions between the empirical and the theoretical in all such misguided pseudo-mathematical endeavors.

# PART 2:

# QUANTITATIVE PARTITION MODELS

## [1] Partitions Almost Always Lead to Quantitative Skewness

The mathematical field of Integer Partition investigates the ways an integral quantity can be expressed as the sum of (smaller) integers. It deals with very simple and straightforward questions such as: "In how many ways and exactly how can the quantity 7 be broken into integral parts?" The answer to this question is the exhaustive list of all possible integral partitions of 7 as shown in detail below:

7
6 + 1
5 + 2
5 + 1 + 1
4 + 3
4 + 2 + 1
4 + 1 + 1 + 1
3 + 3 + 1
3 + 2 + 2
3 + 2 + 1 + 1
3 + 1 + 1 + 1 + 1
2 + 2 + 2 + 1
2 + 2 + 1 + 1 + 1
2 + 1 + 1 + 1 + 1 + 1
1 + 1 + 1 + 1 + 1 + 1 + 1

If the focus is on the quantity 5 instead of the quantity 7, then the question posed is: "In how many ways and exactly how can the quantity 5 be broken into integral parts?" The answer to this question is the exhaustive list of all possible partitions of 5, and this is shown below together with some philosophical comments about sizes:

| | |
|---|---|
| 5 | ← few big parts |
| 4 + 1 | ← few big parts |
| 3 + 2 | |
| 3 + 1 + 1 | |
| 2 + 2 + 1 | |
| 2 + 1 + 1 + 1 | ← many small parts |
| 1 + 1 + 1 + 1 + 1 | ← many small parts |

These two examples demonstrate a very profound, universal, and yet extremely simple principle regarding how a conserved quantity can be partitioned into parts, namely the observation that: **'One big quantity is composed of numerous small quantities'**, or equivalently: '**Numerous small quantities are needed to merge into one big quantity**".

Partitioning a fixed quantity into parts can be done roughly-speaking in two extreme styles, either via a breakup into many small parts, or via a breakup into few big parts. For example, two extreme styles of integer partition of 7 are {1, 2, 1, 1, 1, 1} with many small parts, and {3, 4} with fewer but relatively bigger parts. A more moderate style perhaps would be to have a mixture of all kinds of sizes, consisting of many small ones, some medium ones, and few big ones. The above conceptual outline is one of the chief causes why so often real-life data sets are skewed quantitatively, having numerous small values, but only very few big values.

The principle merits additional visual presentations. Figure 1 demonstrates all possible integer partitions of 5, where the small clearly outnumbers the big in the entire scheme. Naturally, 3 could be designated as the middle-size quantity here. Consequently, 1 & 2 are designated as the small ones, while 4 & 5 are designated as the big ones. According to such classification of sizes, there are only 2 big quantities, but as many as 16 small quantities. In any case one should not lose sight of the main aspect seen in Figure 1 which shows a mixture of all sorts of sizes, but with a strong bias towards the small, discriminating against the big. This is of course true in all Integer Partitions, and not only for integer 5. Figure 2 which organizes all the parts of Figure 1 nicely according to size, clearly demonstrates the above principle, and if modified as a proper histogram then it can be said to be positively skewed with a tail falling on the right.

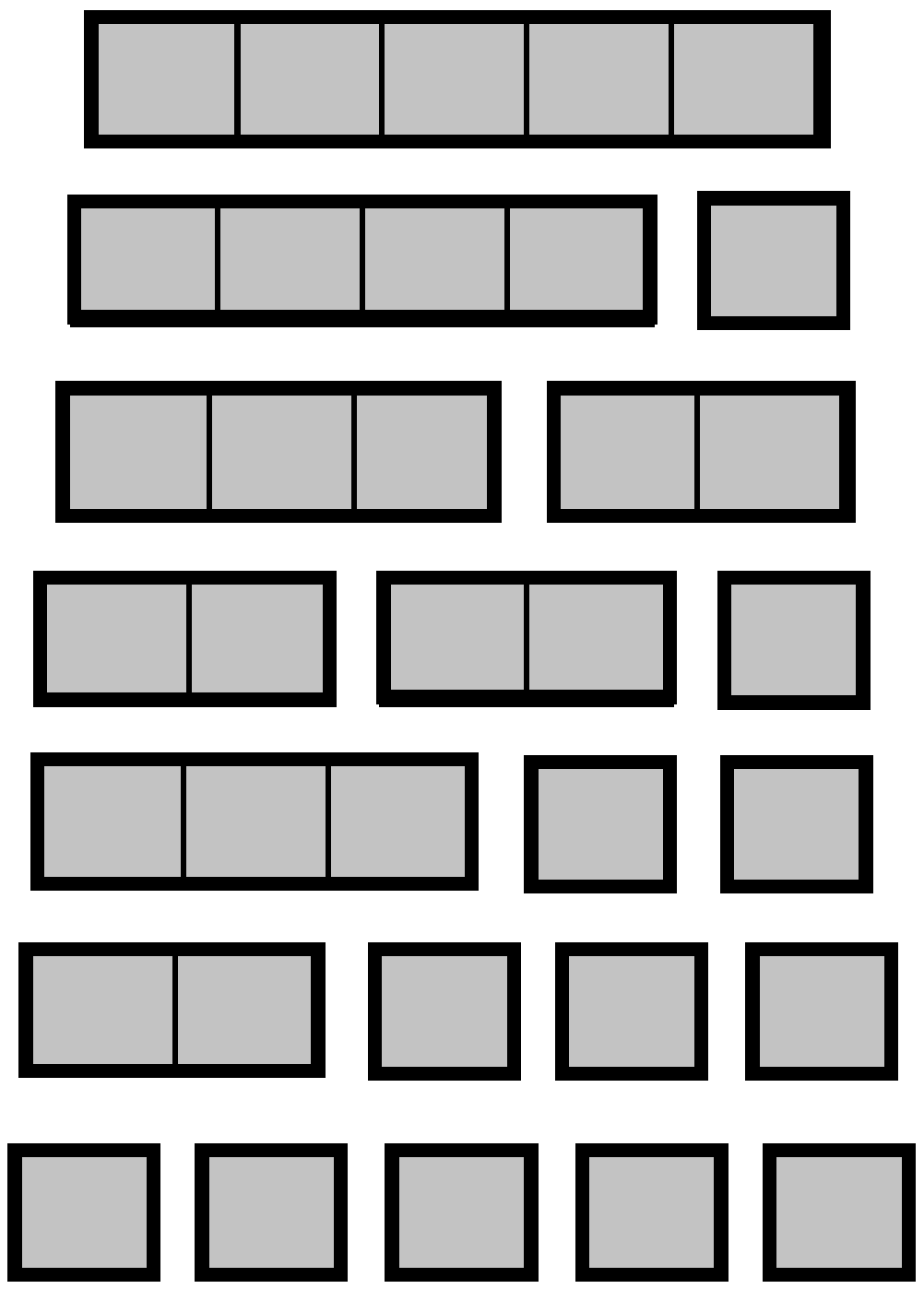

**Figure 1**: The Small Decisively Outnumbers the Big in Integer Partitions of 5

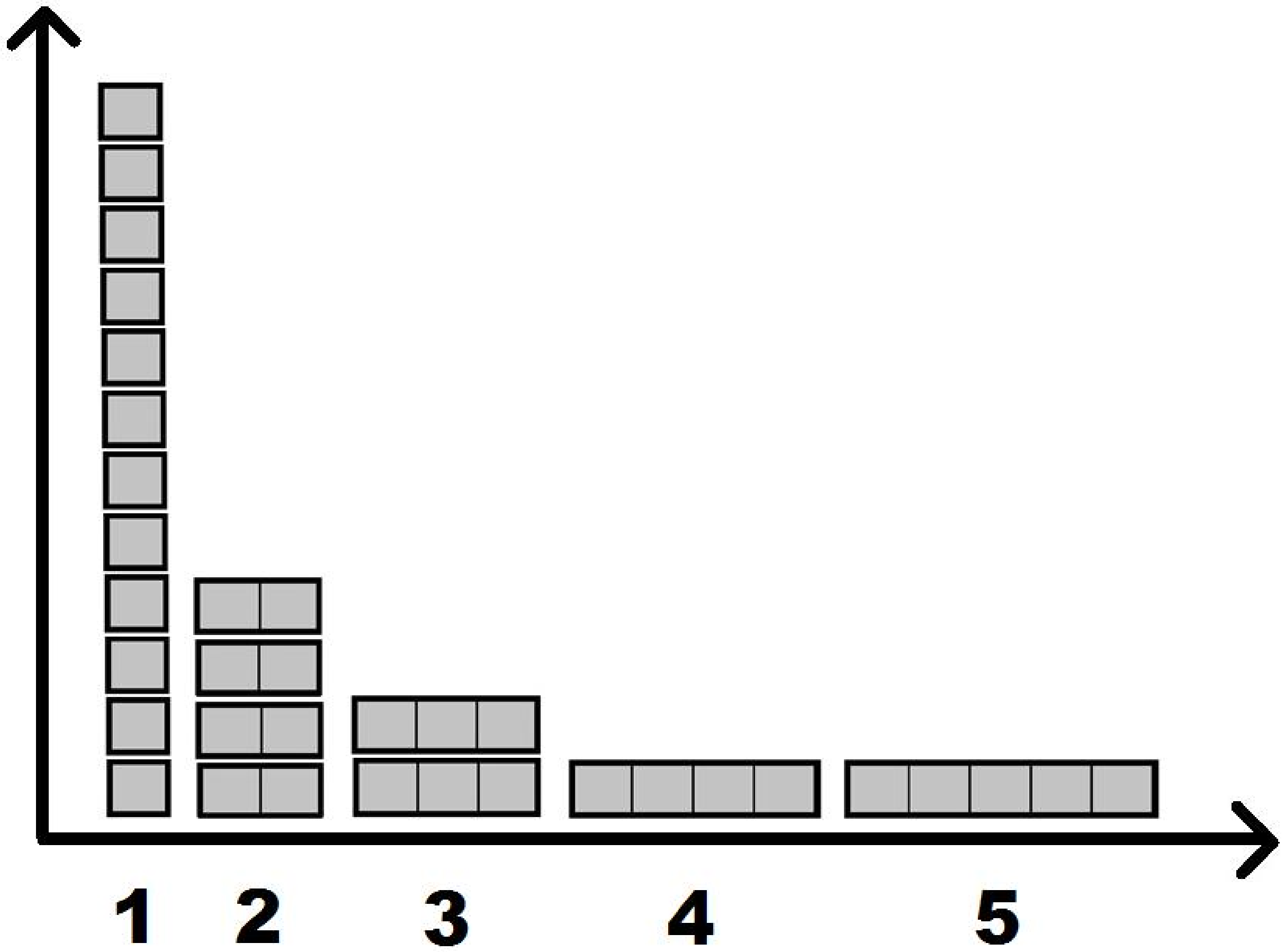

**Figure 2**: The Organization of All Possible Integer Partitions of 5 – A Histogram of Sorts

The abundance of small integers and the rarity of big integers is an intrinsic feature of Integer Partitions, and this is certainly not restricted to the examples of Integer Partition of 7 and Integer Partition of 5. As an additional example, Integer Partition of 13 is similarly analyzed. There are 101 distinct possible partitions for integer 13; as compared with only 15 possible partitions for integer 7, and only 7 possible partitions for integer 5. Only five partition examples out of the complete set of 101 possibilities are shown below as follows:

13 = 7 + 4 + 1 + 1
13 = 4 + 3 + 3 + 2 + 1
13 = 10 + 1 + 1 + 1
13 = 5 + 5 + 3
13 = 7 + 6

Figure 3 depicts the histogram of <u>all</u> the integers from 1 to 13 within <u>all</u> 101 possible partitions. There are 556 integers in total residing within these 101 partitions of 13. The histogram is consistently and monotonically falling to the right, except at the very end for integer 12 with frequency 1 and for integer 13 with frequency 1. Since integer 12 and integer 13 occur exactly once in the entire scheme, namely as 13 = 13 and as 13 = 12 + 1, the histogram is actually flat and horizontal there for the tiny part on the right-most part of the histogram.

All histograms in Integer Partitions of whatsoever integer N are flat at the end of the tail on the right-most part for the largest two integers. This is so since N = N and N = (N – 1) + 1.

Yet the histogram never retreats and it never rises, and thus in the case of Integer Partition it can be said in general that the small is consistently more numerous than the big (with the irrelevant and insignificant exception of the largest two integers).

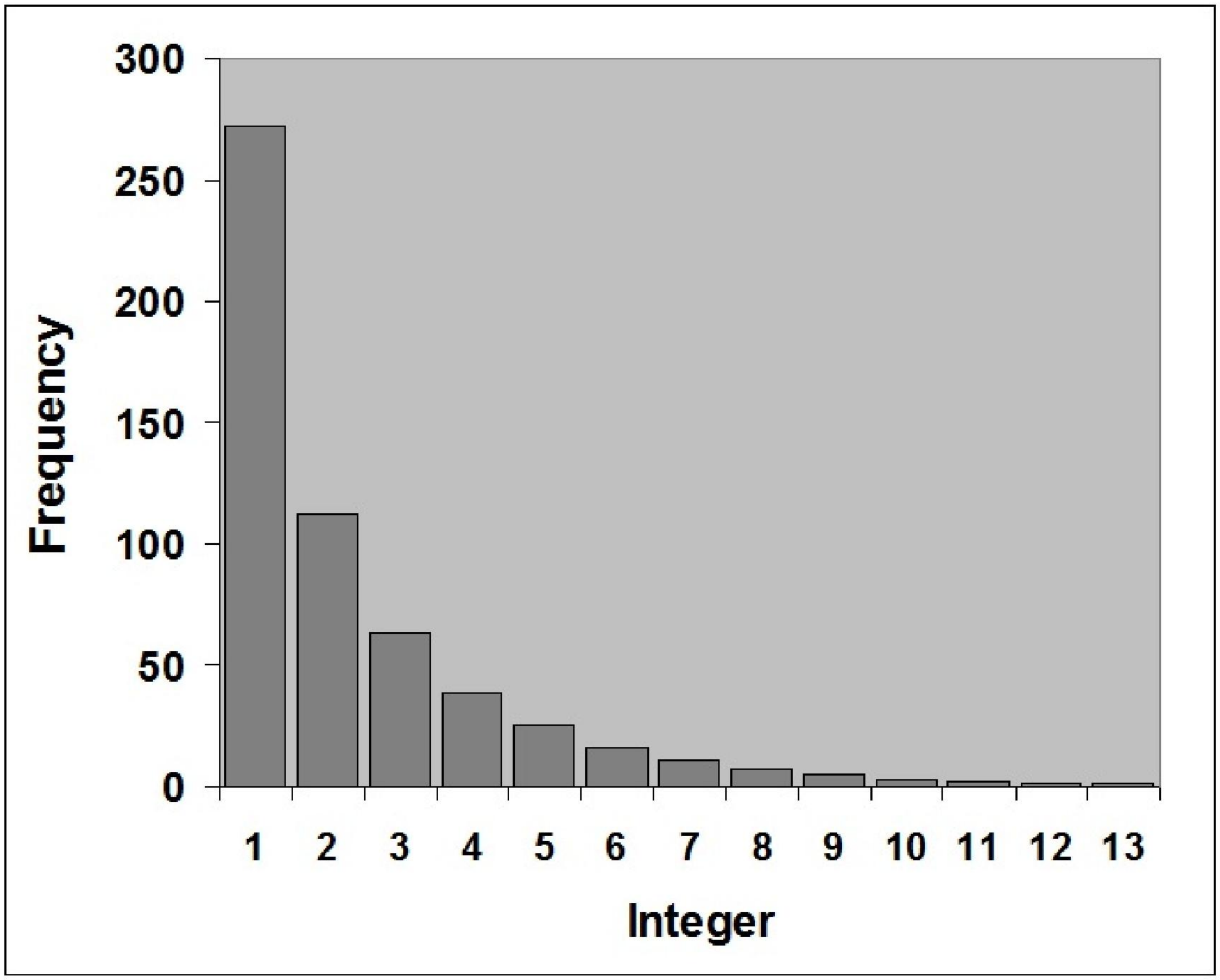

**Figure 3**: Histogram of All Possible Integer Partitions of 13 – Small is Numerous

If we ask a friend to partition a given integral quantity into smaller integral parts only once, in any way he or she sees fit, then no grand conceptual principles can be applied. The friend could favor the big, or could favor the small. If possible, the friend might even partition the quantity into completely even pieces perhaps, where all the parts are of the same quantity, resulting in one size only. Clearly, no quantitative prediction can be made whatsoever for a single partition. Yet, if we ask the friend to randomly repeat partitioning many times over, or if he or she has the time and the patience to perform all possible partitions and to present them as one vast data set, then the skewed configuration where the small is numerous is inevitable! Such resultant blind bias toward the small is observed as long as he or she does not favor any particular sizes, treating all sizes equally and fairly. This generic bias towards the small is almost a universal principle in most other partition models, including those where fractional parts are allowed.

In Figure 4 the entire area in the shape of an oval representing the original value is partitioned in two distinct ways according to size. The first partition in the left panel divides the oval-shape area into big parts, and therefore the low number of only 5 parts is obtained. The second partition in the right panel divides the [same] oval-shape area into much smaller parts, and therefore the high number of 13 parts is obtained. Staring at Figure 4 reinforces the inevitability of the consequence of skewnes where the small is numerous in all partitions of a conserved quantity into parts having a variety of sizes. This territorial example broadens the quantitative application of Integer Partitions into fractional partitions as well.

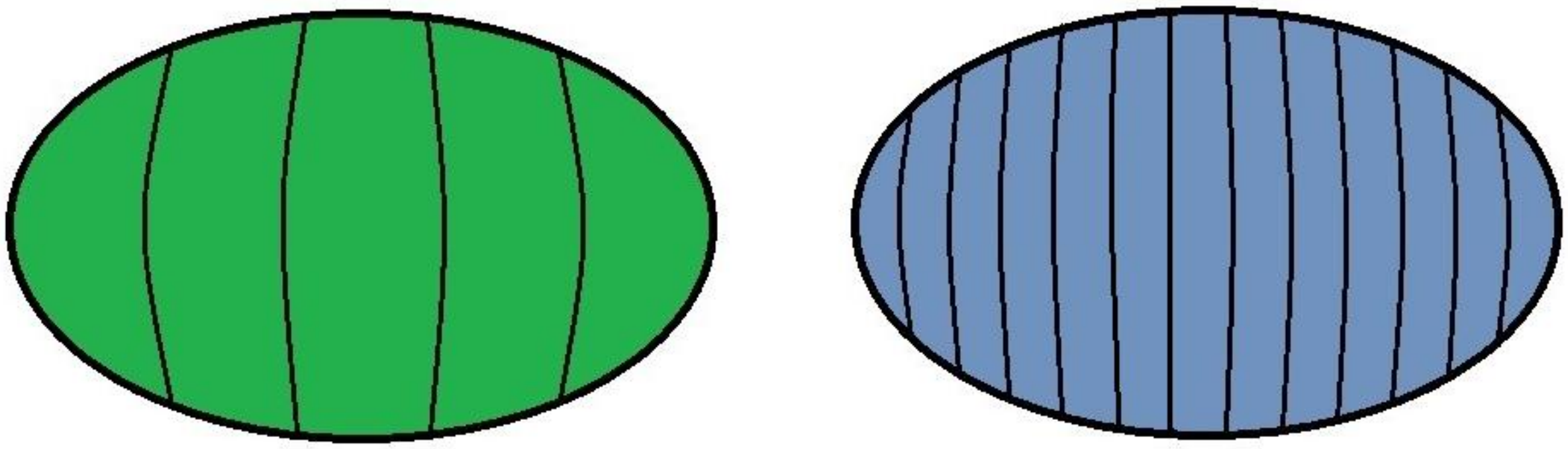

**Figure 4**: A Given Quantity is Partitioned into Either Few Big Parts or Many Small Parts

Figure 5 provides another visual and intuitive demonstration that a given random partition containing the big, the small, and the medium; and where all sizes mix together randomly, many more small pieces should be found than big pieces. Figure 5 focuses on the area as the quantitative variable, but the lesson learnt from it is generic and applicable to any other types of quantities. Figure 5 depicts one possible random partition in the natural world where approximately 1/3 of the entire oval area consists of big parts (around the left side); approximately 1/3 of the entire oval area consists of small parts (around the center); approximately 1/3 of the entire oval area consists of medium parts (around the right side), namely endowing equal portions of overall quantity fairly to each size without any bias. Surely in nature there exists no such order and grouping by size around the left-center-right sections or along any other dimensions. In nature, the big, the small, and the medium are all mixed in chaotically. But the order for the sizes along the left-center-right sections shown in Figure 5 is made for pedagogical purposes, to reinforce visually for the reader the profound quantitative consequences affecting sizes in typical partition models. The example given via Figure 5 broadens the quantitative application of Integer Partitions into fractional partitions as well.

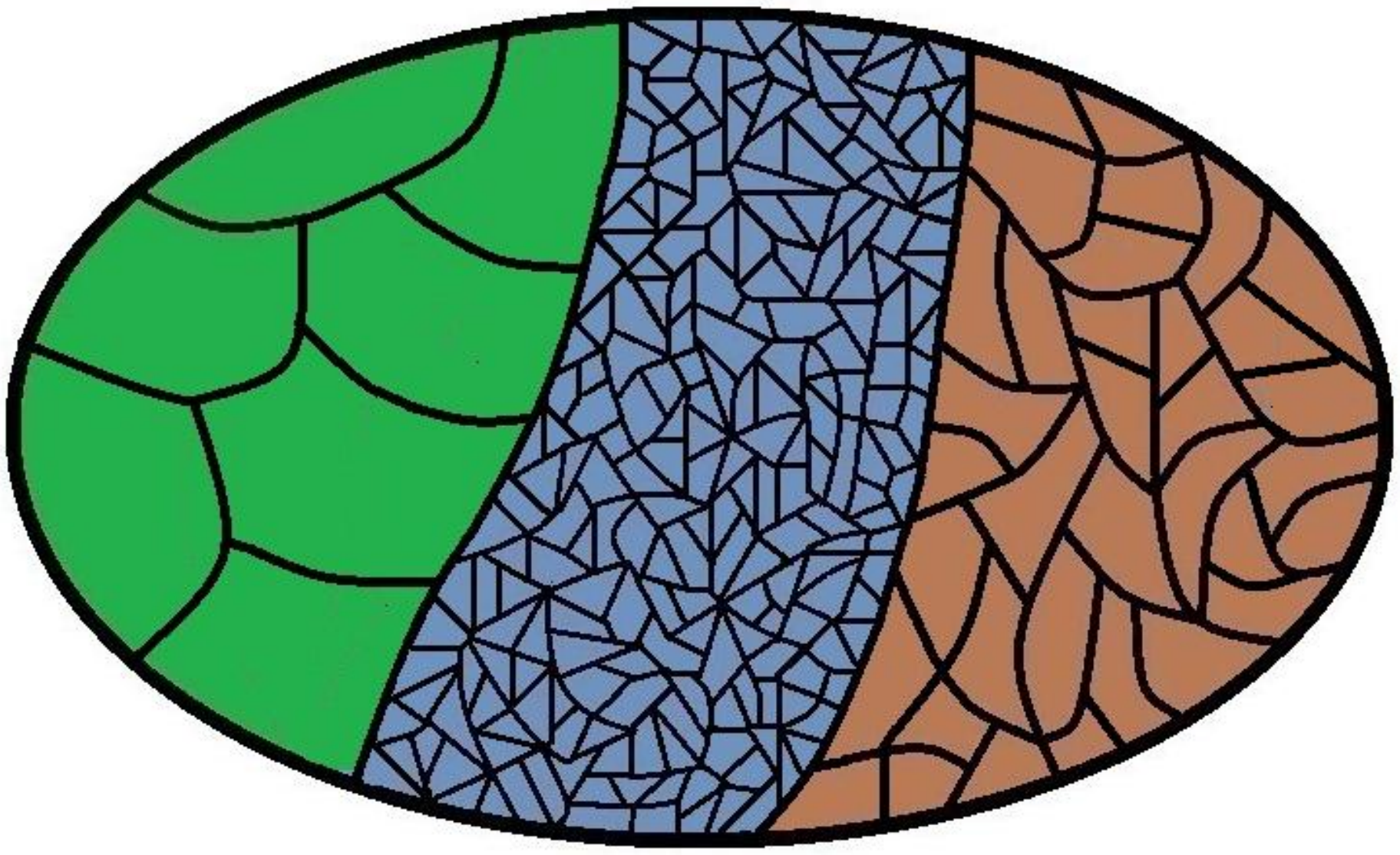

**Figure 5**: An Equitable Mix of Small, Medium, and Big Yielding 'Small is Beautiful'

Hence, the data of Figure 5 is structured in such a way that adding all small values yields approximately the same sum as adding all big values (or medium). Yet, in spite of such equitable allocation of quantitative portions to the 3 sizes, the small outnumbers the big. Vaguely counting the number of enclosed areas for each size leads to the decisive conclusion that the small is by far more numerous than the median, and that the median is definitely more numerous than the big. Careful count of the enclosed areas in the oval shape above shows that there are **220** small parts in the middle, **35** medium parts on the right, and only **7** big parts on the left.

It should be emphasized that in fact, real-life data sets frequently deviate from the 33.3% fair allocation of overall quantitative portion for the 3 sizes. Also evidently, a great deal depends on the exact definition of what should constitute small, big, and medium. Since real-life data sets almost never occur nicely with exactly and merely 3 sizes, but rather mostly with numerous distinct values, it is necessary to arbitrarily group all values into 3 camps according to size – assuming one wishes to stick with the small, big, and medium categories.

For example, for the Integer Partition of 5 discussed earlier, the large data set of all possible partitions is {5, 4, 1, 3, 2, 3, 1, 1, 2, 2, 1, 2, 1, 1, 1, 1, 1, 1, 1, 1}. Overall quantity is (5 + 4 + 1+ 3 + 2 + 3 + 1 + 1 + 2 + 2 + 1 + 2 + 1 + 1 + 1 + 1 + 1 + 1 + 1 + 1) = 35. Considering 1 and 2 as small and 4 and 5 as big, the portion of big is (5 + 4)/(35) = (9)/(35) = **25.7%**; the portion of small is (1 + 2 + 1 + 1 + 2 + 2 + 1 + 2 + 1 + 1 + 1 + 1 + 1 + 1 + 1 + 1)/(35) = (20)/(35) = **57.1%**; and the portion of medium is (3 + 3)/(35) = (6)/(35) = **17.1%**. Such state of quantitative affairs is extremely in favor of the small, much more so than the 33.3% equal allocation for all 3 sizes. But surely, the definitions of small, big and medium here are arbitrary.

Considering small as {1}, big as {5}, and medium as {2, 3, 4}, also leads to the small is beautiful conclusion, although of slightly different intensity in beauty. Such re-definition of sizes also yields different allocation of portions from the 35 overall quantity. The portion of big is (5)/(35) = **14.3%**; the portion of small is (1 + 1 + 1 + 1 + 1 + 1 + 1 + 1 + 1 + 1 + 1 + 1)/(35) = (12)/(35) = **34.3%**; and the portion of medium is ( 4 + 3 + 2 + 3 + 2 + 2 + 2)/(35) = (18)/(35) = **51.4%**.

Figure 6 depicts another quantitative configuration that could also occur in the physical world. Here the small constitutes approximately 80% of the entire area (i.e. entire quantity), while the big constitutes only approximately 10% of the area, and the medium only about 10% as well. Portions such as 40%, 50% or even 60% are much more typical for the small in the physical world, while the 80% depicted here is an exaggeration that does not really happen often. Here the small is by far more numerous than the big, over and above the configuration of Figure 5.

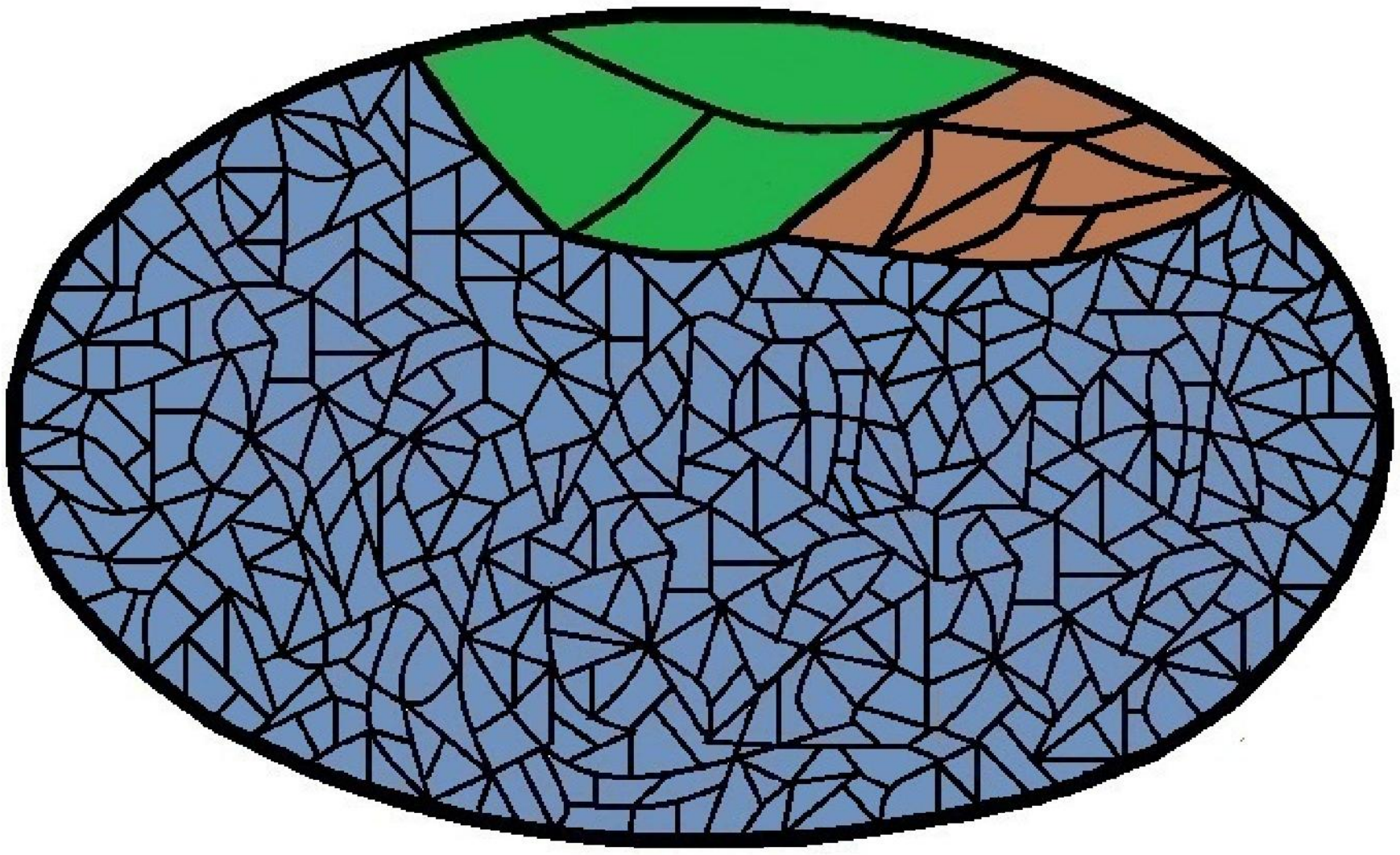

**Figure 6**: Uneven Mix with Too Many Small Parts Yielding 'Small is Exceedingly Beautiful'

Figure 7 depicts an unnatural quantitative configuration that rarely occurs in the physical world. Here the big constitutes approximately 85% of the entire area (i.e. entire quantity), while the small constitutes only approximately 5% of the area, and the medium only about 10%. Here the big managed to be more numerous than the small as well as more numerous than the medium by strongly dominating overall area at the expense of the small and the medium.

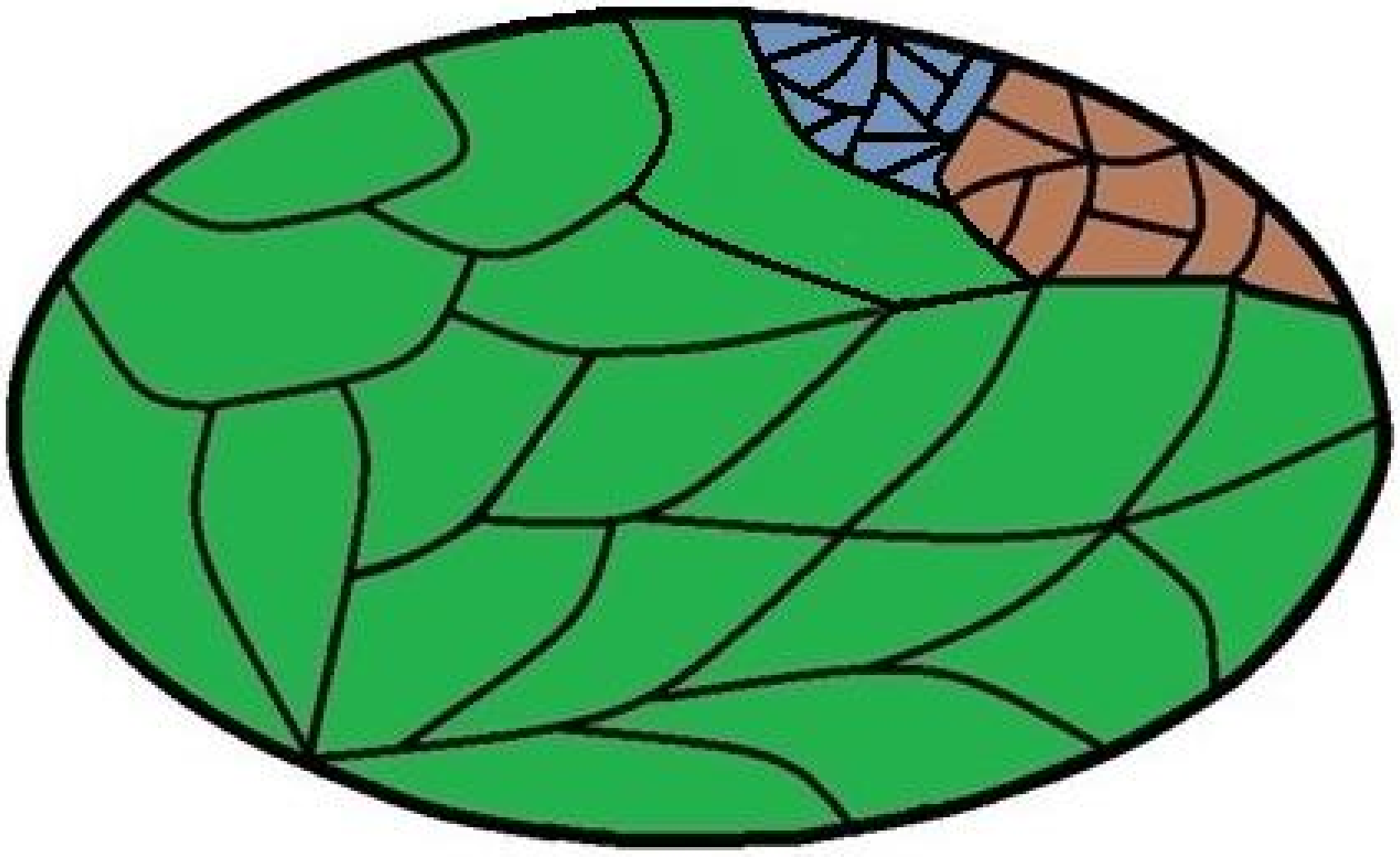

**Figure 7**: Uneven Mix with Too Many Big Parts Yielding Unnatural and Rare Configuration

Apart from any static mathematical arguments, one could actually postulate the equitable mix of all sizes via heuristic argument involving the existence of random consolidation and fragmentation dynamic forces acting upon the quantities. These forces are thought of as physical, chemical, geological, or biological forces, as opposed to any abstract mathematical model of quantitative consolidations and fragmentations. The mechanism driving the system in the direction towards size-equilibrium is the differentiated intensity that these consolidations and fragmentations tendencies occur which depends on the current size configuration of the pieces. The bigger the pieces the stronger is the tendency towards fragmentation. The smaller the pieces the stronger is the tendency towards consolidation. Hence a situation with too many big quantities as in Figure 7 is unstable, as nature tends to break the many existing big pieces into smaller ones, much more so than it tends to consolidate the very few existing small pieces into bigger ones, thereby gradually changing the composition of sizes in favor of the small. In a similar fashion, a situation with too many small quantities as in Figure 6 is unstable, as nature tends to consolidate the many existing small pieces into bigger ones, much more so than it tends to break up the very few existing big pieces into smaller ones, thereby gradually changing the composition of sizes in favor of the big. The equitable and unbiased mix of Figure 5 where all 3 sizes share about 1/3 of the entire territory (quantity) is stable. Perhaps such equitable mix of all sizes is the long term quantitative configuration to which many physical systems tend.

A hypothetical example can be given in the political arena of ancient times. A huge empire is hard to control. Communication and transportation are not well-developed, and so it's difficult for the ruling center in the capital to obtain information about occurrences on the peripheries, and even harder to transport soldiers and material there. Local commanders appointed to rule those far-flung regions are tempted to claim independence and take full control themselves. Hence as a general principle, the larger the empire the greater are the chances and the forces acting upon it towards dissolution and fragmentation. On the other hand, for very small principalities or municipalities ruled by brutal warlords, scheming princes, petty dictators, and such, the opposite forces of conquest, mergers, and consolidations are normally very strong. Here the fear of being absorbed by the well-known proximate neighboring rulers who could easily move armies and material through these short distances is overwhelming and constant. The smaller the political territory the fewer the defenses it can master and the more insecure it feels, and thus the greater the tendency towards conspiracies, attacks, and absorptions. All this leads to plans of preemptive attacks by all sides in order to avoid surprises and conquest, and at times simply due to greed and the desire to rule and exploit larger territory. Such tense state of affairs inevitably ends up with just one successful strongman absorbing his neighboring rivals and ruling all the proximate territory, thus achieving as a consequence much greater political stability. Long term tendencies work against having many big empires or too many small principalities. Thus a happy balance on the political map between the small and the big is achieved, ensuring that this configuration of the territorial sizes is steady and durable.

Surely, not many entities in the world are derived from actual or physical partition processes. Yet, if we substitute the words '**Composition**', '**Constitution**', or '**Consolidation**' for '**Partition'**, and think of the entity represented by the data set under consideration as something composed of many parts, then the partitioning vista and its consequences could be appropriate and applicable for some real-life cases. The three oval-shape entities of Figures 5, 6, and 7 and the conclusions from the related analysis are valid whether it is believed that the entities are actively going to be physically partitioned along the lines inside, or that they are passively composed of all these inner parts – and these two points of view are equivalent as far as resultant quantitative configuration is concern. For example, atomic analysis of the typical chocolate consumed by people shows that it is composed mostly of very small hydrogen atoms as well as of medium-size atoms such as carbon, nitrogen, and oxygen, but that it contains only very few and rare big atoms such as zinc, iron, and calcium. A piece of chocolate is certainly not some continuous chunk of primordial matter initially, only to be 'partitioned' later and sorted into its constituents atoms, yet, allowing such fictitious description of the piece of chocolate leads to the same quantitative and size configuration! Clearly, allowing the interpretation of partition as composition enlarges the scope of the analysis.

In other words, applications of partition models in the natural sciences do not have to explicitly assume actual or physical fragmentations of the whole into parts for the results outlined in this chapter to hold. Indeed, a natural entity at times can be thought of as being composed of much smaller parts, or that it exists as the consolidation of numerous separate and smaller parts held together by some physical force or because of any other reason, leading to the same conclusion regarding quantitative and size configuration as for actual partition.

In truth, the exact 3 categories for sizes, namely Small, Medium, and Big shown in Figures 5, 6, and 7 rarely occur as such in nature; they are artificial and arbitrary drawing for the sake of demonstrating the principles of involved regarding partitions. Typically, Mother Nature cannot be as exact, orderly, and regimented when randomly partitioning an overall quantity into parts by sticking only to 3 sizes throughout the entire process. Yet, the neat arrangements of partitions into parts with only 3 possible sizes in the above figures are presented for pedagogical purposes, in order to gain broad insight about the different possibilities of partitions, and in order to inform on the fact regarding which generic configuration is more common and which is relatively rare in real-life data and partitions. There exist two extreme poles above and below such exact and neat partitioning into 3 sizes as in the above figures, as in the following two opposing scenarios:

**(I)** A partition into identical parts, utilizing only one size for all the parts, as shown approximately in Figure 8 (difficulties in artistically drawing the exact same size for all areas led to slight variations). The small is beautiful phenomenon cannot be manifested here a priori since there is only one size involved. Certainly, Mother Nature would almost never partition in such even and controlled manner, except in some exceedingly rare occasions and cases.

**(II)** A partition into totally distinct parts, utilizing as many sizes as there are parts, as shown in Figure 9. This is the most common scenario in refine random partition processes having the flexibility of producing parts of any real, fractional, or integral values whatsoever. True randomness in typical partition processes ensures that Mother Nature would produce her parts in a thoroughly chaotic manner, and without any repetition of any single quantity/size whatsoever.

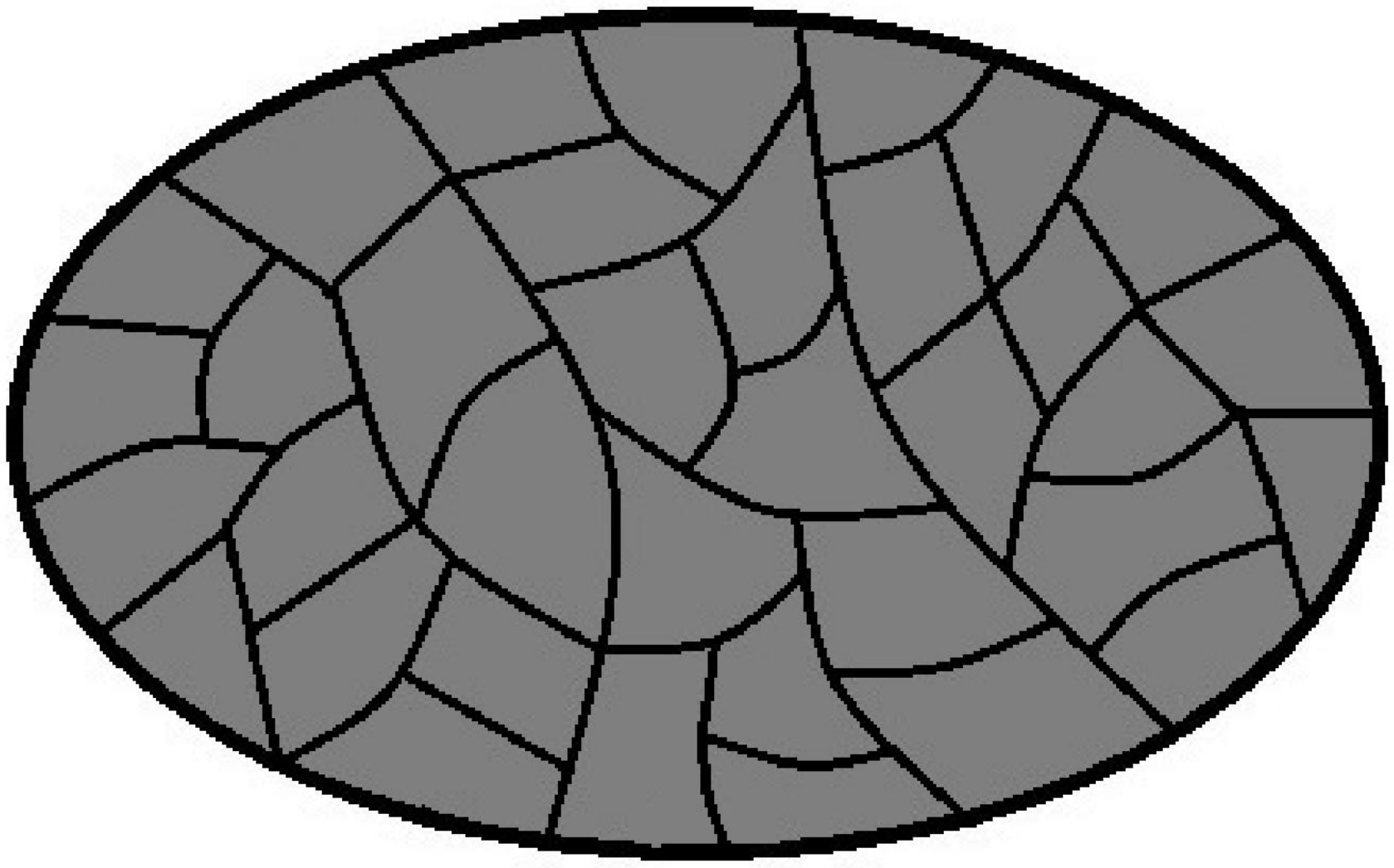

**Figure 8**: Partitioning a Quantity into Equal Parts Utilizing One Size Only

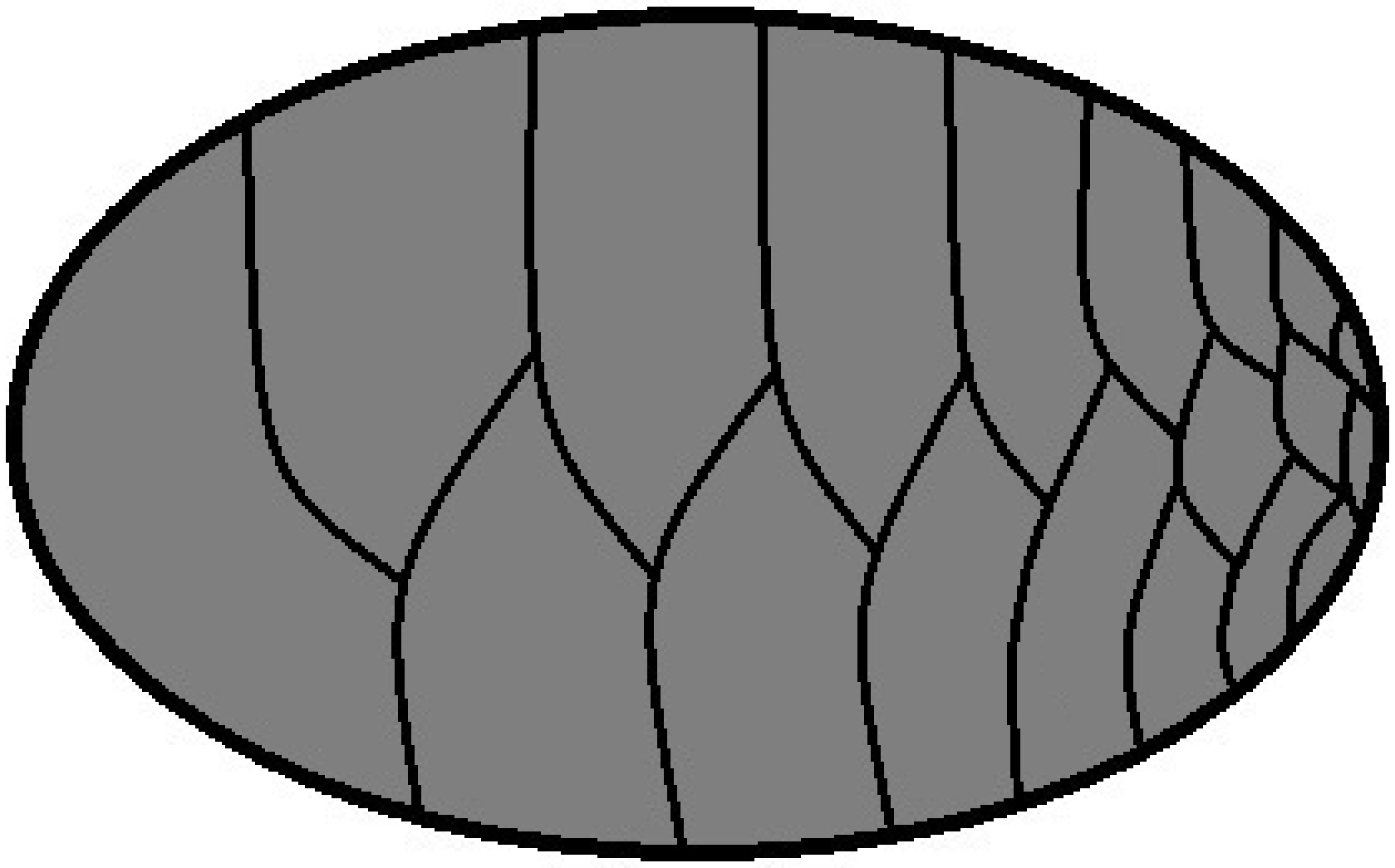

**Figure 9**: Partitioning a Quantity into Totally Distinct Parts – Small May be Beautiful

Yet, guaranteeing that the resultant set of parts contains distinct parts without any repetition of values does not immediately imply the small is beautiful phenomenon (i.e. positive skewness), rather a careful examination of the resultant set of parts must be performed. Surely the fanatical and stubborn statistician can superficially manufacture the phenomenon by generously defining Small on a wide range endowing it many parts, while restricting the definition of Big to a narrower range and thus depriving it of the ability to earn many parts. And surely, another statistician with opposite tendencies, a contrarian, might place the threshold or cutoff boundary points between Small, Medium, and Big in such a way as to greatly benefit Big, letting it earn the majority of parts. Yet, proper definition of sizes requires that we assign equal sub-interval length to all sizes; and that the entire range is divided fairly into equal territorial segments for the various sizes. Each oval-shape type of partition as in Figure 9 with totally distinct parts requires careful definition of sizes according to the entire range, as well as subsequent careful count of sizes according to these size definitions – in order to determine whether or not the small is beautiful phenomenon manifests itself there.

The next two chapters regarding refine random partition processes (allowing the parts to be any real numbers without restricting them to integral values) illustrate the main driving force behind the manifestation of the small is beautiful phenomenon for their resultant sets of parts. These two random partition processes will be computer-simulated, and their resultant sets of parts will be shown to exhibit the small is beautiful phenomenon.

## [2] Random Dependent Partition is Always Skewed

The generic small is beautiful principle found in Integer Partitions is applicable to fractional partitions as well, and this fact lends the principle much wider scope and nearly universal applicability in almost all partition schemes. One particular partition scheme in this context is of a well-structured arrangement of repeatedly partitioning a single quantity randomly into many parts, and this is coined as '**Random Dependent Partition**'. This process is best exemplified by randomly breaking a big rock in multiple stages into much smaller pieces, and this example of the generic idea is coined as '**Random Rock Breaking'**. Surely the description of a rock only serves as a vivid example, and the generic idea here is the repeated break up of a single quantity into smaller and smaller values, culminating in the final set of much smaller values.

In spite of the liberal use of the adjective 'random', the process of Random Dependent Partition actually follows a strict partition procedure with exact and carefully executed stages. In the first stage the rock is broken into 2 pieces using a random pair of percentage values, such as 23% and 77% for example. Then the second stage starts with the orderly breaking of each of the 2 pieces in a random fashion using two new random pairs of percentage values, resulting in 4 pieces altogether. In the third stage, each of the 4 pieces is broken into two pieces using four new random pairs of percentage values, resulting in 8 pieces altogether, and so forth.

An essential feature leading rapidly to the small is beautiful phenomenon here is the random manner by which each piece is broken into two smaller pieces, namely that the pair of percentage values are always chosen randomly anew at each stage and for each piece. It is helpful to envision an especially-manufactured roulette in a respectable and honest casino. The roulette wheel contains 99 pockets of numbers and one thrown ball which randomly falls into one of the pockets. These 99 values are marked as {1%, 2%, 3%, … , 97%, 98%, 99%} so that all possible (integral) percentage values can be obtained by chance. For example, if the ball lands inside the 25% pocket, then the 75% - 25% pair of percentages are employed to break the next piece of rock. Figure 10 depicts the physical arrangement of such roulette in one imaginary casino.

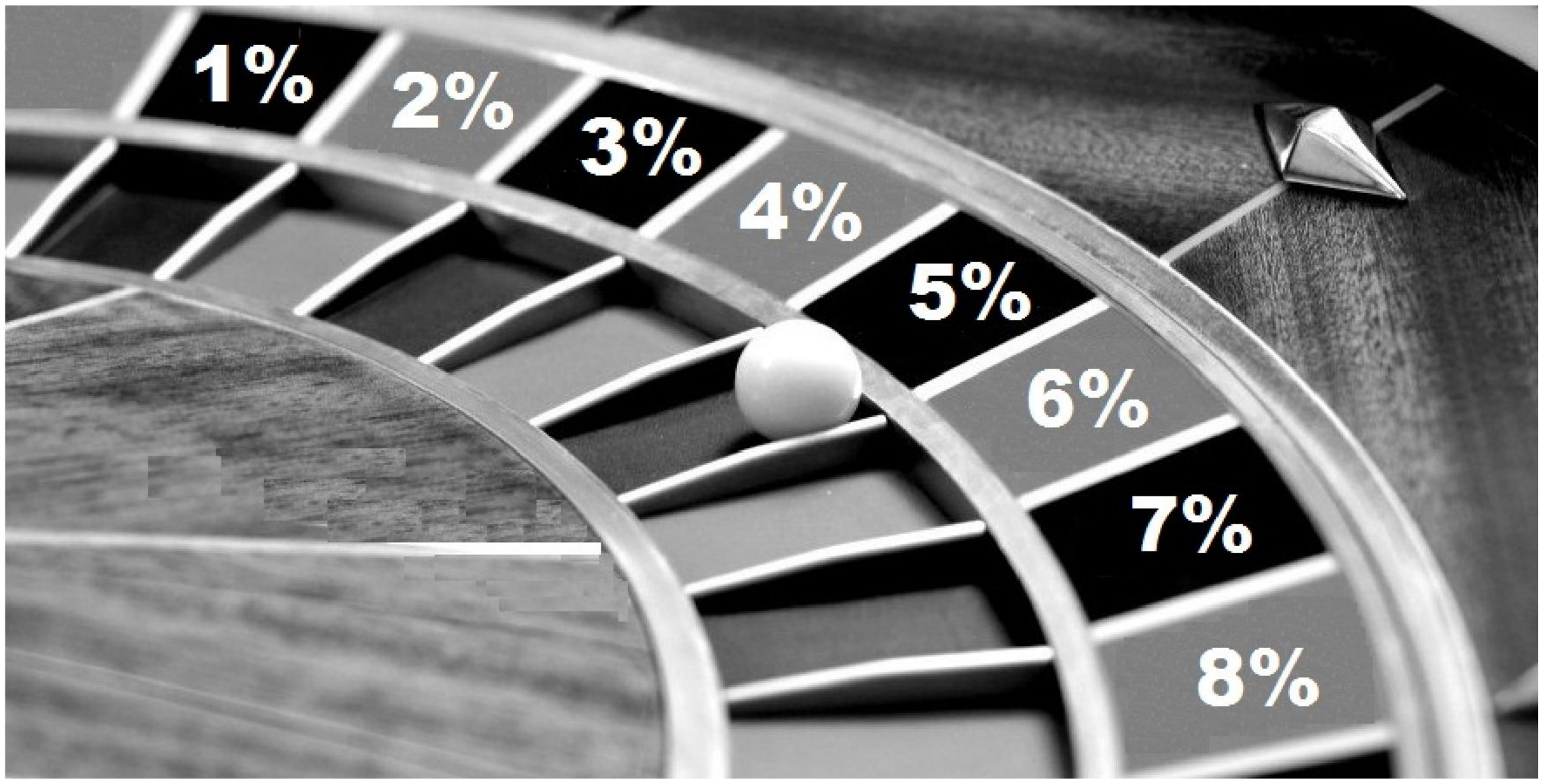

**Figure 10**: Roulette with 99 Pockets Determining the Percent Breakup in Rock Partition

Let us illustrate Random Dependent Partition with one concrete example of a six-stage process.

A **500** kilogram rock is broken in the first stage via the random 18% - 82% percentage pair, yielding two new pieces [500]**x**[18/100] = **90** and [500]**x**[82/100] = **410**.

In the second stage, firstly, the 90 piece is broken via the random 71% - 29% percentage pair, yielding two new pieces [90.0]**x**[71/100] = **63.9** and [90.0]**x**[29/100] = **26.1**.
Secondly, the 410 piece is broken via the random 86% - 14% percentage pair, yielding two new pieces [410]**x**[86/100] = **352.6** and [410]**x**[14/100] = **57.4**.

In the third stage, the four random pairs of percentage values for the 4 pieces above about to be broken are respectively: 72% - 28%, 29% - 71%, 38% - 62%, 52% - 48%.
This leads to 8 new pieces {**46.0, 17.9, 7.6, 18.5, 134.0, 218.6, 29.8, 27.6**}.

Figure 11 depict the process from the original 500-kilogram rock to the end of the third stage.

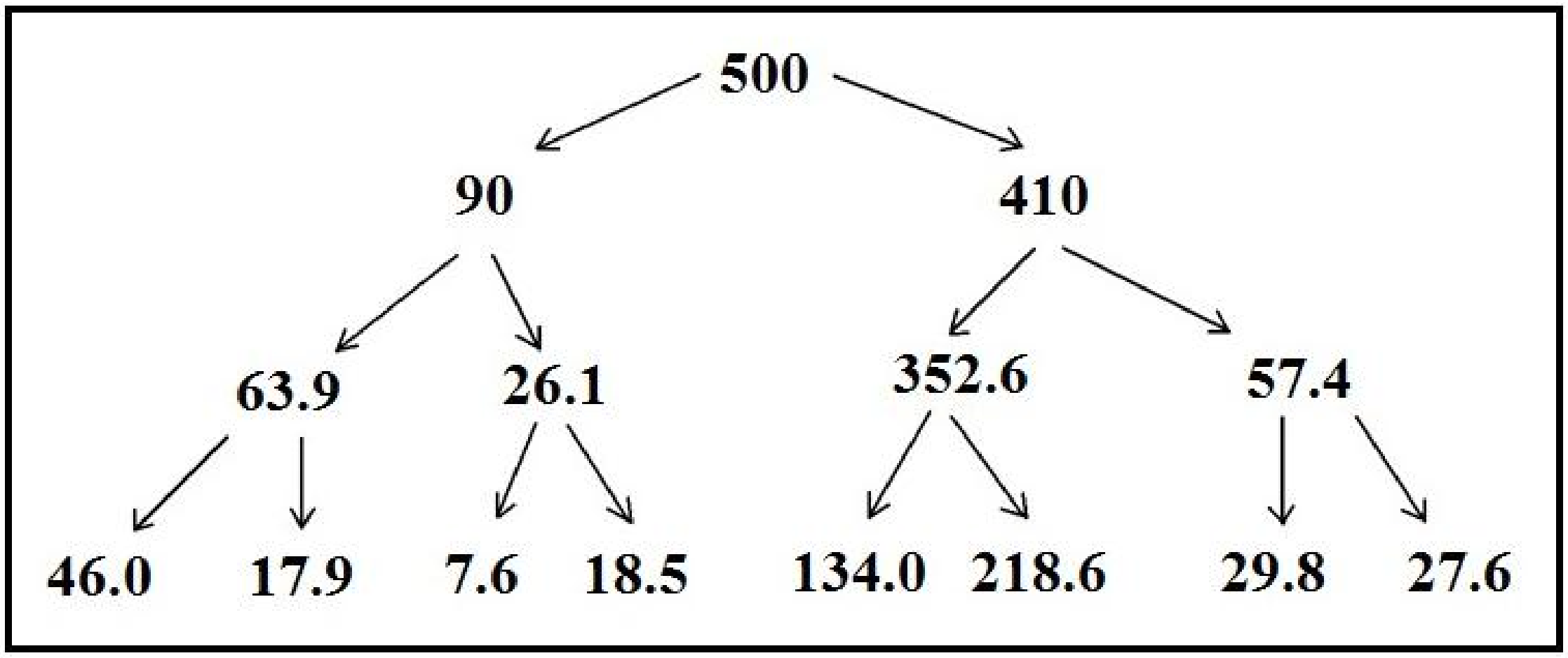

**Figure 11**: Random 500-Kilogram Rock Breaking – From Original Piece to the Third Stage

In the fourth stage, these 8 pieces are broken in a similar fashion, leading to 16 new pieces [*percentage values are not shown for the sake of brevity*]:

| | | | | | | | | | | | |
|---|---|---|---|---|---|---|---|---|---|---|---|
| 8.3 | 37.7 | 2.5 | 15.4 | 5.1 | 2.5 | 15.8 | 2.8 | 61.6 | 72.4 | 54.7 | 164.0 |
| 17.6 | 12.2 | 2.2 | 25.3 | | | | | | | | |

In the fifth stage, these 16 pieces are broken in a similar fashion, leading to 32 new pieces:

| | | | | | | | | | | | |
|---|---|---|---|---|---|---|---|---|---|---|---|
| 3.8 | 4.5 | 17.7 | 20.0 | 0.3 | 2.2 | 14.9 | 0.5 | 0.2 | 4.9 | 1.2 | 1.3 |
| 11.2 | 4.6 | 2.4 | 0.3 | 30.2 | 31.4 | 61.5 | 10.9 | 45.9 | 8.7 | 113.1 | 50.8 |
| 4.2 | 13.4 | 5.1 | 7.1 | 0.5 | 1.7 | 22.6 | 2.8 | | | | |

In the sixth stage, these 32 pieces are broken in a similar fashion, leading to 64 new pieces:

| | | | | | | | | | | | |
|---|---|---|---|---|---|---|---|---|---|---|---|
| 1.87 | 1.94 | 3.00 | 1.48 | 15.25 | 2.48 | 4.00 | 16.00 | 0.27 | 0.06 | 1.00 | 1.18 |
| 2.69 | 12.24 | 0.14 | 0.32 | 0.07 | 0.08 | 1.52 | 3.39 | 0.23 | 0.99 | 1.07 | 0.20 |
| 10.18 | 1.01 | 2.65 | 1.92 | 2.10 | 0.34 | 0.24 | 0.10 | 12.68 | 17.52 | 18.55 | 12.89 |
| 19.68 | 41.82 | 6.62 | 4.23 | 35.35 | 10.56 | 3.76 | 4.98 | 66.75 | 46.38 | 9.15 | 41.68 |
| 2.96 | 1.27 | 10.44 | 2.94 | 2.36 | 2.78 | 6.81 | 0.28 | 0.48 | 0.00 | 1.63 | 0.09 |
| 13.08 | 9.48 | 0.86 | 1.92 | | | | | | | | |

The final set of the 64 pieces after the sixth stage, sorted low to high, is as follows:

| | | | | | | | | | | | |
|---|---|---|---|---|---|---|---|---|---|---|---|
| 0.005 | 0.06 | 0.07 | 0.08 | 0.09 | 0.10 | 0.14 | 0.20 | 0.23 | 0.24 | 0.27 | 0.28 |
| 0.32 | 0.34 | 0.48 | 0.86 | 0.99 | 1.00 | 1.01 | 1.07 | 1.18 | 1.27 | 1.48 | 1.52 |
| 1.63 | 1.87 | 1.92 | 1.92 | 1.94 | 2.10 | 2.36 | 2.48 | 2.65 | 2.69 | 2.78 | 2.94 |
| 2.96 | 3.00 | 3.39 | 3.76 | 4.00 | 4.23 | 4.98 | 6.62 | 6.81 | 9.15 | 9.48 | 10.18 |
| 10.44 | 10.56 | 12.24 | 12.68 | 12.89 | 13.08 | 15.25 | 16.00 | 17.52 | 18.55 | 19.68 | 35.35 |
| 41.68 | 41.82 | 46.38 | 66.75 | | | | | | | | |

A cursory look at the above set of numbers roughly confirms the small is beautiful feature of the pieces, and this observation is decisively confirmed by the detailed histogram of Figure 12. It should be noted that the first three bins in the histogram are of 1-kilogram width, while the last four bins are of 5-kilogram width. There are 5 pieces weighing over 23 kilogram which are not shown in the histogram as they are very sparsely and thinly spread far out on the horizontal axis, rendering the big there even rarer.

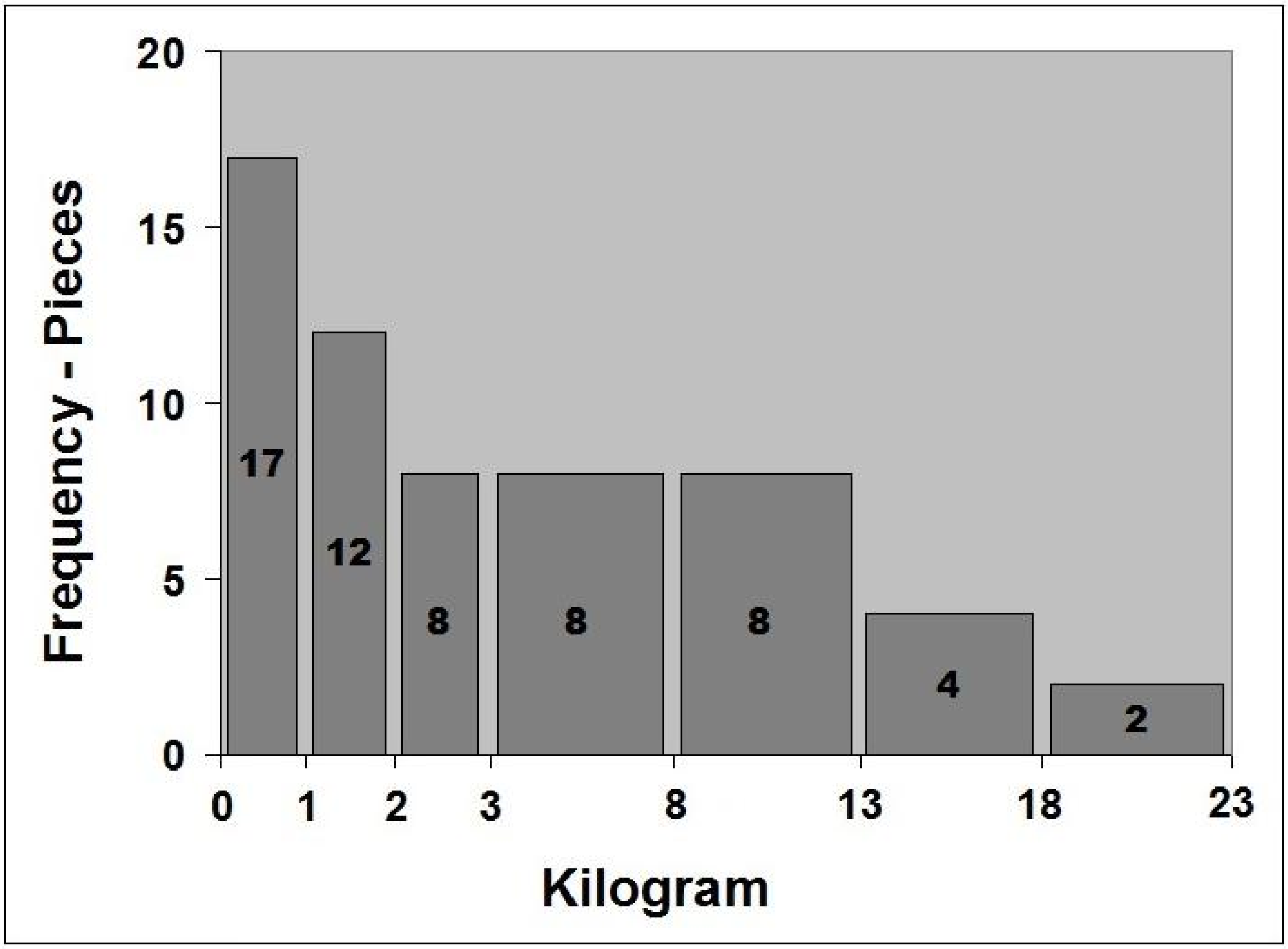

**Figure 12**: Histogram of Pieces After the 6th Stage – Random 500-Kilogram Rock Breaking

Skeptical or cynical readers might suspect the author for possible manipulations of the roulette simulations so as to arrive at many small pieces but only few big pieces. Yet the above results could be easily re-produced and verified with the aid of a standard personal computer. MS-Excel for example provides random number generator called RAND() which yields random values uniformly and evenly distributed between 0 and 1, such as 0.964125, and 0.176387, and so forth. These random fractions could be interpreted here as the random percentage values. Indeed, this more general usage of fractions from the **real number line on (0, 1)** - without restricting random percentages to integral values - is what constitutes part of the setup and the real definition of Random Dependent Partition.

There are two fundamental differences between the small is beautiful model of Integer Partition as outlined in the previous chapter, and this model of Random Dependent Partition. Firstly, Integer Partition model strictly and exclusively considers only integral values, while Random Dependent Partition liberally includes any fractional and real ones. Secondly, in Integer Partition the model incorporates or aggregates all possible partition scenarios into one vast data set, while in Random Dependent Partition there is no need to aggregate distinct partition scenarios, and instead only a single trajectory or scenario of the breakup of the original rock is considered.

The general dependency of the weights of the pieces at each stage upon the weights of the pieces in the previous stage should be noted carefully. Surely, this dependency on the previous stage also involves dependency on random elements, namely the random percentage values which determine how to break the pieces of the previous stage. Hence the process concocts new values at each stage out of the old values of the previous stage, mixing in random elements as well.

Let us construct a pie chart for the resultant set of 64 rock pieces in order to demonstrate the quantitative breakdown by size as in the oval-like Figures 5, 6, and 7. Defining the 3 sizes fairly, each on a third of the entire range (66.75 – 0.005)/3 = 22.25, would endow Small such a huge advantage over the other sizes, that the phenomenon would be very strongly manifested. But since the biggest 66.75 piece is suspected of being an outlier and anomaly, including this anomalous value in the definitions of all the sizes may not be appropriate. In any case, in order to illustrate the persistency and the tenacity of the small is beautiful phenomenon here, it shall be shown to hold true even under a different [yet quite reasonable] set of definitions of sizes having a slight bias against the Small and in favor of the Big, as follows:

**Small:** Less than 10 kilogram.
**Medium:** From 10 to 20 kilogram.
**Big**: Over 20 kilogram.

This size criterion divides the entire set of 64 pieces into 3 classes:

| | | | | | | | | | | | |
|---|---|---|---|---|---|---|---|---|---|---|---|
| 0.005 | 0.06 | 0.07 | 0.08 | 0.09 | 0.10 | 0.14 | 0.20 | 0.23 | 0.24 | 0.27 | 0.28 |
| 0.32 | 0.34 | 0.48 | 0.86 | 0.99 | 1.00 | 1.01 | 1.07 | 1.18 | 1.27 | 1.48 | 1.52 |
| 1.63 | 1.87 | 1.92 | 1.92 | 1.94 | 2.10 | 2.36 | 2.48 | 2.65 | 2.69 | 2.78 | 2.94 |
| 2.96 | 3.00 | 3.39 | 3.76 | 4.00 | 4.23 | 4.98 | 6.62 | 6.81 | 9.15 | 9.48 | |
| | | | | | | | | | | | |
| 10.18 | 10.44 | 10.56 | 12.24 | 12.68 | 12.89 | 13.08 | 15.25 | 16.00 | 17.52 | 18.55 | 19.68 |
| | | | | | | | | | | | |
| 35.35 | 41.68 | 41.82 | 46.38 | 66.75 | | | | | | | |

Calculations of total weight by size for each of the 3 size classes, as well as the proportion of each size-total within the entire system-weight of 500 kilograms, yields the following results:

**Small:** **47** pieces weighing a total of 98.9 kilograms, or 98.9/500 = **20%** of overall weight.
**Medium: 12** pieces weighing a total of 169.1 kilograms, or 169.1/500 = **34%** of overall weight.
**Big**: **5** pieces weighing a total of 232.0 kilograms, or 232.0/500 = **46%** of overall weight.

Figure 13 depicts the quantitative portion of each piece, sorted low to high in the clockwise direction, and where relative area signifies relative quantity. In addition, the color for Small is blue; the color for Medium is brown; and the color for Big is green. Small is bitter about having only 20% portion of overall weight, and it accuses Big of stealing a portion of its weight, while Medium is not suspected of any theft. This accusation is based on Small's misguided notion that it has a divine right to earn **33.3%** of total quantity in all systems, cases, and situations whatsoever. Big retorts that it didn't do anything of the sort; that this is all the fault of the definer who has created unreasonable criteria for sizes; and that Small should have been assigned the wider range from 0 to 15 kilograms at the expense of Medium, instead of the narrow range of 0 to 10 kilograms. Big's excuse immediately creates animosity between Small and Medium and triggers fierce competition between them in attempts to establish preferential criterion for sizes. Nonetheless, Small is actually quite content, being by far the most numerous size within the set of 64 pieces, having 47 pieces of its own. Big turned out to be the least popular and quite rare, having only 5 pieces out of a total of 64.

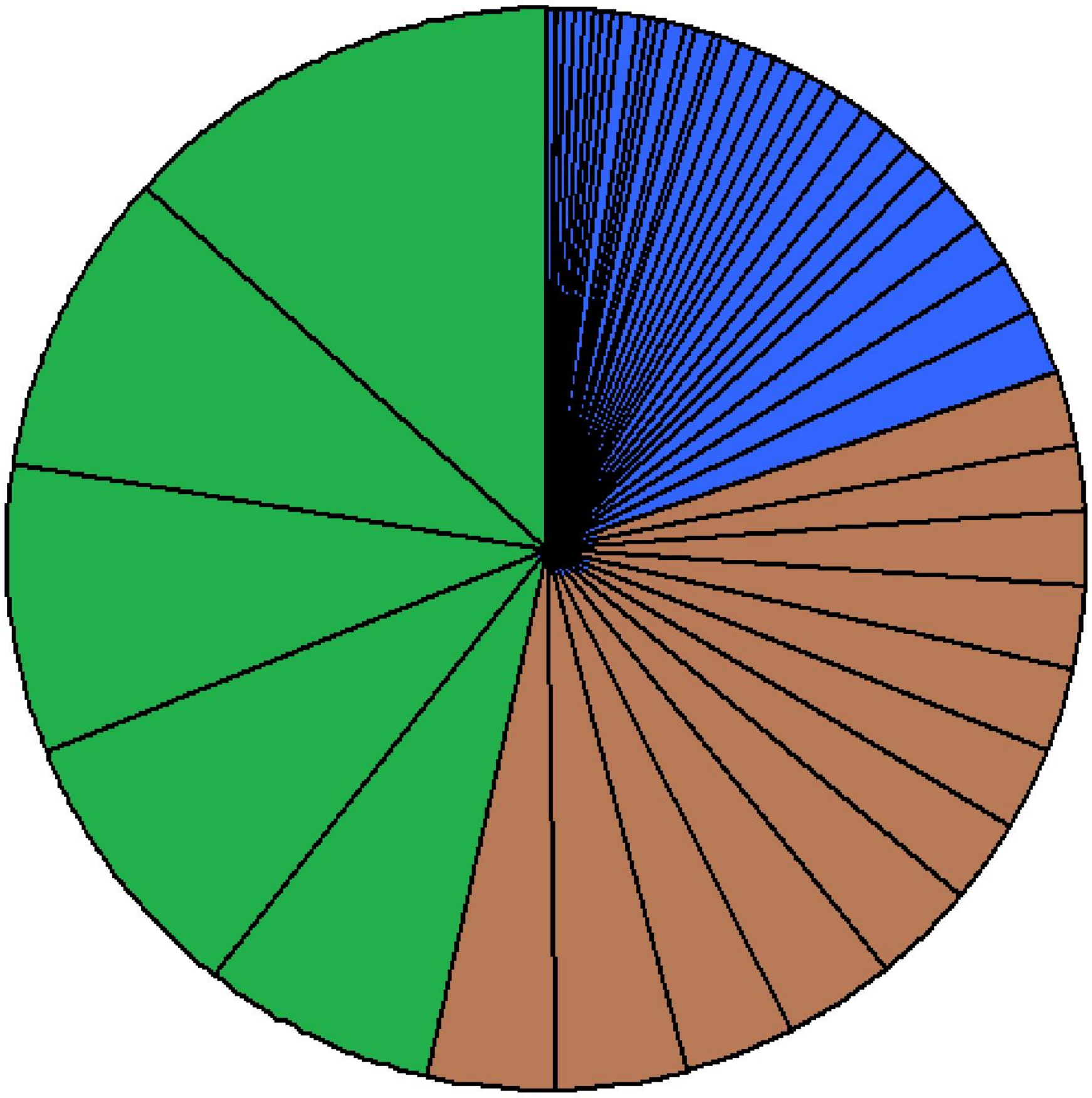

**Figure 13**: Nearly Equitable Mix of Small, Medium, and Big – Random Dependent Partition

## [3] Random Real Partition is Always Skewed

A random and spontaneous partitioning process lacking any dependent stages leads to the small is beautiful phenomenon just as well. Here the phenomenon is encountered without any constraints, free of strict sequential procedures, encompassing integral as well as fractional values. Instead of breaking the original quantity in sequential stages, a comprehensive plan of how and where to cut or break is contemplated beforehand, and then at an opportune and appropriate grand moment the partition is fully executed, cutting or breaking simultaneously in all the planned places and locations. The process is coined as '**Random Real Partition**'. This process is best exemplified by the cutting of one-dimensional long pipe at random locations along its length, and this example of the generic idea is coined as '**Random Pipe Breaking'**.

Let us provide a concrete numerical example. A 15-meter long metal pipe is to be randomly partitioned into 30 parts. This is accomplished by obtaining 29 independent random points along the pipe - prior to the moment of the actual cutting - to serve as marks indicating where the pipe should be partitioned. These marks constitute the grand plan of the entire partition process, and they are generated via independent simulations from the continuous **Uniform(0, 15)**.

Actually, distances from the left edge of the pipe are obtained via 29 computer simulations using the random number generator uniformly distributed between 0 and 1. These 29 simulated numbers then provide random percentage values to determine the positions of the marks along the 15-meter length pipe, as for example (0.28)*(15) = 4.2, or (0.71)*(15) = 10.7, and so forth.

Following the simulation of each such random position, a mark on the physical pipe is made accordingly with a black marker or by gently scratching the location with a saw to indicate where to do the actual cutting later when the grand moment of execution arrives. It is only at the end of the long sequence of 29 simulations and markings that actual cutting and sawing at these marked locations take place, breaking the long pipe into 30 parts of (usually) totally distinct lengths. Figure 14 depicts the actual 29 marks along the pipe obtained via computer simulations.

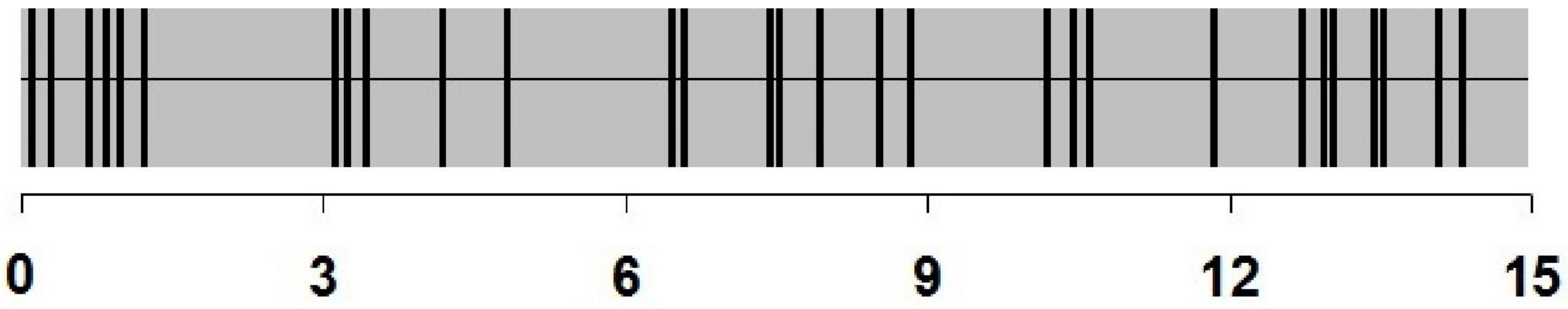

**Figure 14**: The 29 Marks Randomly Scratched along the 15-Meter Pipe Prior to Partition

The 29 random values generating the marks on the pipe, order from low to high; including the left edge of 0.000 as extra, and the right edge of 15.000 as extra, are outlined as follows:

0.000 0.140 0.301 0.710 0.868 1.003 1.243 3.163 3.273 3.453 4.181 4.871
6.500 6.606 7.453 7.550 7.962 8.555 8.847 10.207 10.464 11.652 11.882 12.750
12.947 13.060 13.452 13.564 14.123 14.376 15.000

Calculating distances between the marks (i.e. differences) by subtracting from each value greater than zero its adjacent value on the left, we obtain the set of the lengths of the parts of the pipe:

| | | | | | | | | | | | |
|---|---|---|---|---|---|---|---|---|---|---|---|
| 0.140 | 0.161 | 0.409 | 0.158 | 0.135 | 0.239 | 1.921 | 0.109 | 0.180 | 0.728 | 0.689 | 1.629 |
| 0.106 | 0.846 | 0.097 | 0.412 | 0.593 | 0.292 | 1.361 | 0.257 | 1.188 | 0.230 | 0.868 | 0.197 |
| 0.113 | 0.392 | 0.112 | 0.559 | 0.253 | 0.624 | | | | | | |

Ordering these distances low to high, we obtain the final ordered set of the lengths of the parts:

| | | | | | | | | | | | |
|---|---|---|---|---|---|---|---|---|---|---|---|
| 0.097 | 0.106 | 0.109 | 0.112 | 0.113 | 0.135 | 0.140 | 0.158 | 0.161 | 0.180 | 0.197 | 0.230 |
| 0.239 | 0.253 | 0.257 | 0.292 | 0.392 | 0.409 | 0.412 | 0.559 | 0.593 | 0.624 | 0.689 | 0.728 |
| 0.846 | 0.868 | 1.188 | 1.361 | 1.629 | 1.921 | | | | | | |

Clearly, the small is beautiful principle is evidently valid for this set of 30 values. Had this pipe been cut deterministically and fairly into 30 equal parts, then each part would have been 15/30 = 0.5 meter long, hence the value of 0.5 serves as the benchmark for the 'truly middle size'. In contrast, for this random partition here, the majority of the parts are shorter than 0.5 meter; fewer are longer than 0.5 meter; and all this is surely in the spirit of the small is beautiful principle.

Figure 15 depicts the histogram of these 30 parts after partition, confirming visually the manifestation of the small is beautiful phenomenon for this random partition process.

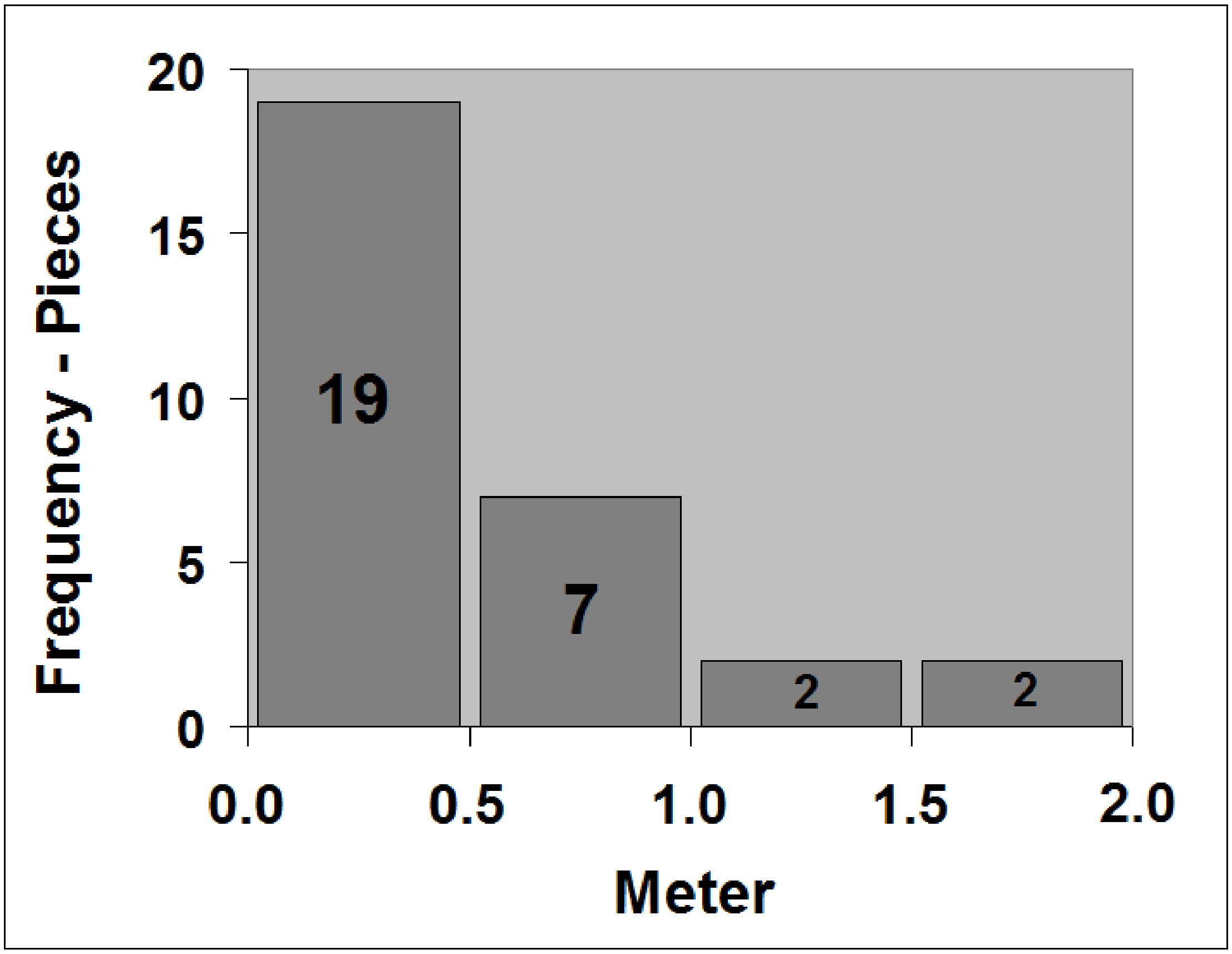

**Figure 15**: Histogram of Resultant 30 Parts in Random Real Partition of 15-Meter Pipe

Let us construct a pie chart for the resultant set of 30 pipe parts in order to demonstrate the quantitative breakdown by size as in the oval-like Figures 5, 6, and 7. Defining the 3 sizes fairly, each on a third of the entire range (1.921 – 0.097)**/**3 = 0.61, would endow Small decisive advantage over the other sizes, and the phenomenon would be strongly manifested. But since the biggest 1.921 part may perhaps appear as an outlier, including this value in the definitions of all the sizes might not be prudent. In any case, in order to illustrate the persistency and the tenacity of the small is beautiful phenomenon here, it shall be shown to hold true even under a different [yet quite reasonable] set of definitions of sizes having a slight bias against the Small and in favor of the Big, as follows:

**Small:** Less than 0.5 meter.
**Medium:** From 0.5 to 1.0 meter.
**Big**: Over 1.0 meter.

This size criterion divides the entire set of 30 parts into 3 classes:

0.097 0.106 0.109 0.112 0.113 0.135 0.140 0.158 0.161 0.180 0.197 0.230
0.239 0.253 0.257 0.292 0.392 0.409 0.412

0.559 0.593 0.624 0.689 0.728 0.846 0.868

1.188 1.361 1.629 1.921

Calculations of total length by size for each of the 3 size classes, as well as the proportion of each size-total within the entire system-length of 15 meters, yields the following results:

**Small:** **19** parts with a total length of 3.993 meters, or 3.993**/**15 = **27%** of overall length.
**Medium:** **7** parts with a total length of 4.909 meters, or 4.909**/**15 = **33%** of overall length.
**Big**: **4** parts with a total length of 6.098 meters, or 6.098**/**15 = **41%** of overall length.

Figure 16 depicts the quantitative portion of each part, sorted low to high in the clockwise direction, and where relative area signifies relative quantity. In addition, the color for Small is blue; the color for Medium is brown; and the color for Big is green. Here all 3 sizes earn close to the **33.3%** ideal, equitable, and expected proportions of overall length.

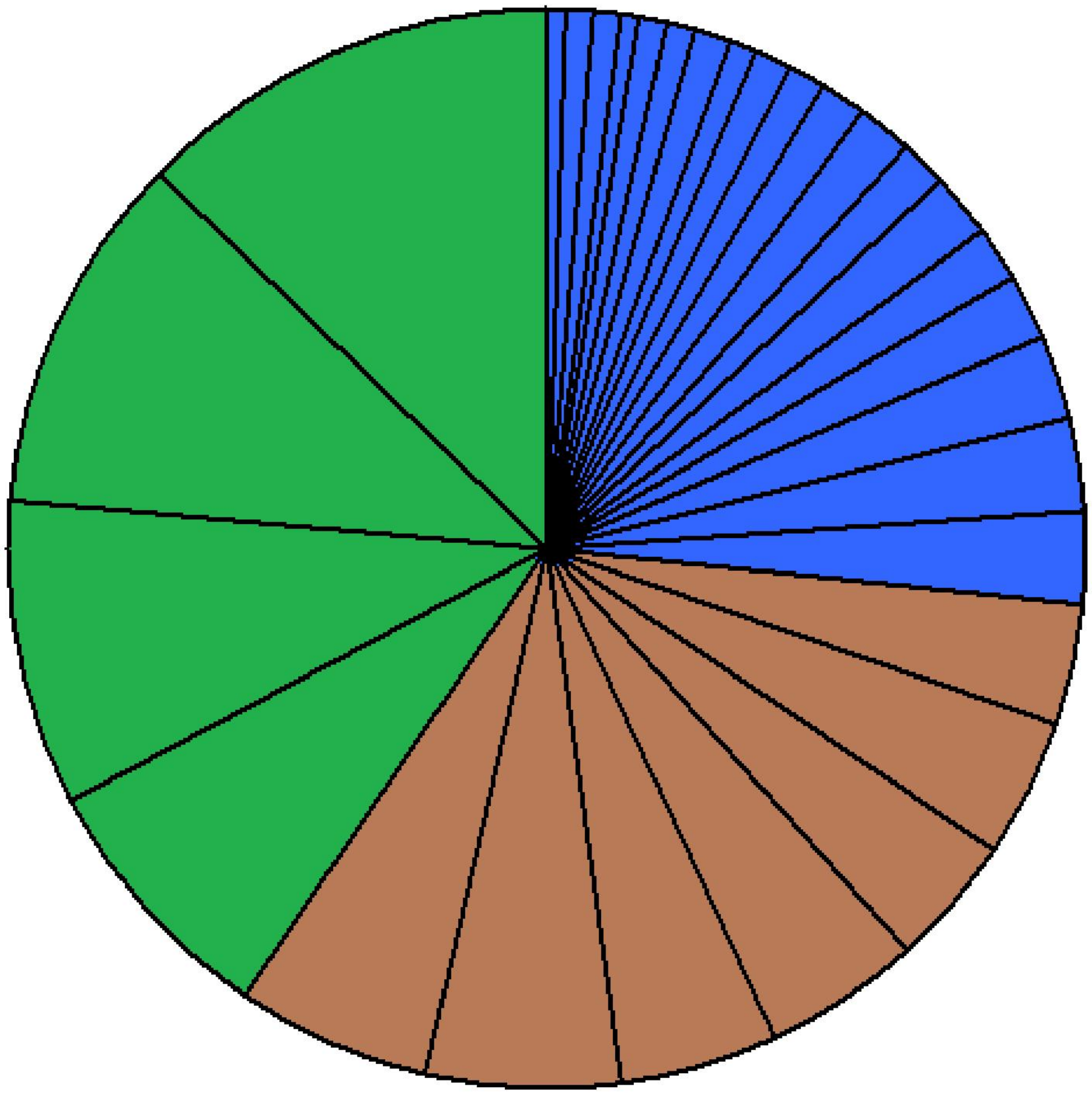

**Figure 16**:  Equitable Mix of Small, Medium, and Big – Random Real Partition

The process of Random Real Partition involves random and totally **independent** numerical choices for the partition of a given quantity. Typical examples of the process involve the cutting of one-dimensional long **pipe** at random locations along its length – giving each section and each part of the pipe equal probability in being marked by using the continuous Uniform. Such pipe examples fit naturally with the defined process since it is very easy and straightforward to mark those locations along the one-dimensional pipe beforehand as planned partition. In contrast, **rock** is three-dimensional, and marking surfaces and areas or presenting diagrams and maps for cutting beforehand is extremely cumbersome and complex, and totally impractical.

The term '**real**' in the phrase 'Random Real Partition' refers to the fact that there is no restriction on having only integral values as was the case for Integer Partition, and that the parts or the pieces can attain any fractional, rational, real, and irrational value. In addition, the positions of the marks along the one-dimensional pipe as in Figure 14 invoke the concept of the **real line**, or the **real number line** in mathematics, which is the line whose points are the set R of all real numbers (also viewed as a geometric or Euclidean space).

## [4] Numerous & Distinct Parts in Partitions are Necessary Conditions

In order to arrive at the resultant small is beautiful quantitative configuration in random partitions, it is necessary at a minimum that the process thoroughly breaks the original quantity into a set of numerous and mostly distinct smaller parts. Some numerical examples follow:

Breaking {15} into {5, 5, 5} or into {1, 2, 3, 3, 3, 3} involve repeated values such as 5 or 3, and thus these partition processes are not conducive to the small is beautiful phenomenon. In addition, these partitions yield very few parts, and this is also not conducive to the phenomenon.

Breaking {15} into {1, 2, 3, 4, 5} is a bit more conducive to the phenomenon since the parts are of distinct quantities without any repetition of values, yet since this process partitions 15 into only 5 smaller parts, it is rendered insufficient to lead to the phenomenon.

Breaking {15} into the set of 30 parts as in the example of pipe breaking of the previous chapter yields the following set of parts (sorted from low to high):

{0.097, 0.106, 0.109, 0.112, 0.113, 0.135, 0.140, 0.158, 0.161, 0.180, 0.197, 0.230, 0.239, 0.253, 0.257, 0.292, 0.392, 0.409, 0.412, 0.559, 0.593, 0.624, 0.689, 0.728, 0.846, 0.868, 1.188, 1.361, 1.629, 1.921}

This pipe partition process is quite conducive to the phenomenon for two reasons. Firstly, the process produces numerous parts, namely thirty. Secondly, the resultant set of parts is of totally distinct quantities, without any repetition of values whatsoever. In other words, when viewed as an ordered set from the low of 0.097 to the high of 1.921, there is constantly some 'growth' in the quantity of the parts; they are steadily increasing; guaranteeing that the entire set of parts are of totally distinct values. As shall be discussed later, the third requirement is that the (percent-wise) growth is steady or varied, but without any overall downward trend.

Certainly, Random Dependent Partition (Rock) with only 1, 2, or 3 stages say, producing only 2, 4, or 8 pieces, does not result in the small is beautiful quantitative configuration, due to scarcity in the number of resultant pieces. In the same vein, Random Real Partition (Pipe) with only 2 to 8 parts say, does not result in the small is beautiful quantitative configuration, due to scarcity in the number of resultant parts. For these two partition processes to arrive at the small is beautiful quantitative configuration it is necessary that enough randomness has been executed throughout the system, and that the system has produced plenty of parts.

Breaking {15} into the set of 30 parts, with 15 parts of value 0.40, 10 parts of value 0.50, and only 5 parts of value 0.80, is a partition process which does indeed produce numerous parts.

{0.40, 0.40, 0.40, 0.40, 0.40, 0.40, 0.40, 0.40, 0.40, 0.40, 0.40, 0.40, 0.40, 0.40, 0.40, 0.50, 0.50, 0.50, 0.50, 0.50, 0.50, 0.50, 0.50, 0.50, 0.50, 0.80, 0.80, 0.80, 0.80, 0.80}

The resultant set of parts is typically not of distinct values, since there are frequent repetitions here, nonetheless, this partition process has been carefully and deliberately calibrated to produce more small parts than big parts, hence it manifests the small is beautiful phenomenon. Surely, it must be acknowledged that this meticulous partition process was not random in any way, but rather deliberately produced and planned.

Breaking {15} into the set of 30 parts, with 10 parts of value 0.20, 10 parts of value 0.50, and 10 parts of value 0.80, is a partition process which does indeed produce numerous parts.

{0.20, 0.20, 0.20, 0.20, 0.20, 0.20, 0.20, 0.20, 0.20, 0.20, 0.50, 0.50, 0.50, 0.50, 0.50, 0.50, 0.50, 0.50, 0.50, 0.50, 0.80, 0.80, 0.80, 0.80, 0.80, 0.80, 0.80, 0.80, 0.80, 0.80}

Yet, the resultant set of parts is typically not of distinct values, since there are frequent repetitions of values. This partition process is definitely not in the spirit of the phenomenon, since all 3 sizes are of equal [10 times] frequency. Surely, it must be acknowledged that also this meticulous partition process was not random in any way, but rather deliberately produced and planned. We might view the difficulty of arriving at the small is beautiful phenomenon for this partition process as arising from the fact that there is very little '**growth**' for the parts, and that the '**growth factors**' $X_{N+1}/X_N$ are almost always 1.0; except on the occasion of growth from 0.20 to 0.50 with its growth factor of 2.5; and on the other occasion of growth from 0.50 to 0.80 with its growth factor of 1.6.

It is useful to adopt this particular growth vista for the resultant set of parts (ordered from low to high), artificially viewing the set as exponential growth series. Conceptually of course, there exists no growth in partition processes in any sense whatsoever, and there exists no time dimension here within which quantities 'grow', yet, such a vista is helpful in the description of the conditions necessary for arriving at the small is beautiful phenomenon.

The focus in such growth analysis is on the set of individual factors leading from one value [part] to its next adjacent value [part] on the right, namely on $F = X_{N+1}/X_N$.

The set of 30 parts of the partitioned 15-meter pipe of the previous chapter can be viewed as exponential growth for 29 periods from the smallest 0.097 part to the biggest 1.921 part. The set of 29 growth factors of the ordered set of parts is as follows:

{1.091, 1.028, 1.023, 1.007, 1.200, 1.035, 1.131, 1.017, 1.122, 1.091, 1.168, 1.040, 1.056, 1.018, 1.134, 1.345, 1.043, 1.006, 1.359, 1.061, 1.052, 1.104, 1.057, 1.162, 1.025, 1.369, 1.146, 1.198, 1.179}

Average factor is 1.1126, although for a more precise effective/overall factor we solve for F in the expression First*$(F)^{29}$ = Last, namely 0.097*$(F)^{29}$ = 1.921, leading to F = 1.1084.
The crucial feature in the set of these growth factors leading to the small is beautiful phenomenon is that there exists no overall downward trend, and that the growth factors fluctuate randomly up and down around their average value.

As another example, the final set of the 64 pieces of the broken 500-kilogram rock after the sixth stage in chapter 2 is viewed as exponential growth for 63 periods from the smallest 0.005 piece to the biggest 66.75 piece.

{0.005, 0.06, 0.07, 0.08, 0.09, 0.10, 0.14, 0.20, 0.23, 0.24, 0.27, 0.28, 0.32, 0.34, 0.48, 0.86, 0.99, 1.00, 1.01, 1.07, 1.18, 1.27, 1.48, 1.52, 1.63, 1.87, 1.92, 1.92, 1.94, 2.10, 2.36, 2.48, 2.65, 2.69, 2.78, 2.94, 2.96, 3.00, 3.39, 3.76, 4.00, 4.23, 4.98, 6.62, 6.81, 9.15, 9.48, 10.18, 10.44, 10.56, 12.24, 12.68, 12.89, 13.08, 15.25, 16.00, 17.52, 18.55, 19.68, 35.35, 41.68, 41.82, 46.38, 66.75}

The set of 63 growth factors of this ordered set of pieces is as follows:

{ **12**, 1.167, 1.143, 1.125, 1.111, 1.400, 1.429, 1.150, 1.043, 1.125, 1.037, 1.143, 1.063, 1.412, 1.792, 1.151, 1.010, 1.010, 1.059, 1.103, 1.076, 1.165, 1.027, 1.072, 1.147, 1.027, **1.000**, 1.010, 1.082, 1.124, 1.051, 1.069, 1.015, 1.033, 1.058, 1.007, 1.014, 1.130, 1.109, 1.064, 1.058, 1.177, 1.329, 1.029, 1.344, 1.036, 1.074, 1.026, 1.011, 1.159, 1.036, 1.017, 1.015, 1.166, 1.049, 1.095, 1.059, 1.061, 1.796, 1.179, 1.003, 1.109, 1.439}

There is only one repetition in the set of pieces, namely 1.92 and 1.92, leading to only one factor of 1.000. The first extraordinarily large factor of 12 (from piece 0.005 to piece 0.06) is viewed as an outlier and anomaly, thus the smallest 0.005 pieces is excluded from further analysis, and only the remaining 62 'growth periods' are considered beginning from the 2nd-smallest 0.06 piece.

Excluding factor 12, the average factor is 1.1293, although for a more precise effective/overall factor we solve for F in the expression First*$(F)^{62}$ = Last, namely 0.06*$(F)^{62}$ = 66.75, so that F = 1.1198. The crucial feature in the set of these growth factors leading to the small is beautiful phenomenon is that there exists no overall downward trend, and that the growth factors fluctuate randomly up and down around their average value.

Since rock breaking and pipe breaking are performed randomly, and since these partition processes are not limited to integral values but are rather performed on the basis of all possible real numbers, there should normally be no repetition, and the chances of finding many/most/all parts having identical values are exceedingly small, and formally the probability for repeated real values is zero. Given that computer simulations are thorough and refine, then not even a single repetition of parts should be found in these random partition processes. The same can be said about the probability of finding all the parts totally distinct but well-structured and being neatly arranged having the same exact fixed growth factor for all – and which would border on the extraordinary or magical. Randomness ensures that almost nothing is steady and equal in the final resultant quantitative configuration of the set of parts after partition. Even a split concentration of growth factors between two dominant groups, with say one group of a variety of factors between 1.05 to 1.10, and another group of a variety of factors between 1.20 and 1.25, and nothing in between 1.10 and 1.20, would constitute extraordinary and extremely rare occurrence, and this is certainly not expected.

Without arguing mathematically and theoretically a priori against the possibility that random partition processes could result in a clear upward or downward overall trend in the growth factors, what is found empirically and repeatedly is that these growth factors are almost always found to be spread randomly over a range clustered around the value of some average growth factor, showing some gentle downward trend on the left for small pieces, and some gentle upward trend on the right for big pieces, mimicking U-shape histogram of sorts, albeit in a zigzag manner.

In the context of the small is beautiful analysis, the utilization of a single fixed growth factor for the entire partition process shall now be attempted, as it differs little from the U-shape structure of factors. In other words, the assumption is taken that the substitution for all these fluctuating growth factors by a single fixed and deterministic growth factor constitutes a somewhat similar model for random partition processes. Let us then attempt to represent Random Real Partition as in the example of Random 15-meter Pipe Breaking of the previous chapter in terms of a particular deterministic growth model with a constant $\mathbf{F}_{\text{FIXED}}$ growth factor standing as the average factor for the entire process, and where 30 parts shall be created. In addition, the value of the smallest part is arbitrarily set to 0.045. The value of $\mathbf{F}_{\text{FIXED}}$ shall be determined by calibrating it in such a way as to yield the total quantity of 15 for the entire deterministic growth process of 29 expansions. Such calibration yields $\mathbf{F}_{\text{FIXED}} = 1.1365$.

By deriving each consecutive part from its adjacent smaller part on the left as 13.65% growth, we ensure that all the parts are of distinct values, without a single repetition. By fixing the value of all the growth rates at 13.65%, we eliminate the mild upward and downward trends in the set of growth factors, and create a model where they are neither rising nor falling. By choosing many parts, namely 30 values in total, we ensure that there are ‘numerous’ parts in the system. It is conjectured that these three achievements are enough to guarantee the manifestation of the small is beautiful phenomenon. Indeed, this conjecture can be easily proven mathematically from the expressions of the growing quantity themselves, namely B, BF, BFF, BFFF, BFFFF, and so forth.

Let us examine the resultant parts for this abstract 13.65% growth model from 0.045 base value:

{0.045 0.051 0.058 0.066 0.075 0.085 0.097 0.110 0.125 0.142 0.162 0.184
0.209 0.237 0.270 0.307 0.349 0.396 0.450 0.512 0.582 0.661 0.751 0.854
0.970 1.103 1.253 1.424 1.619 1.840}

Each value here is derived from its adjacent smaller value on the left via multiplication by the factor 1.1365, namely by growing at 13.65%.

The entire range of {0.045, 1.840} is divided fairly into 3 equal sections for the classification of Small, Medium, and Big. Each section is of the width (1.840 - 0.045)**/**(3) = (1.795)**/**(3) = 0.598. It follows that Small is from (0.045) to (0.045 + 0.598), and that Medium is from (0.045 + 0.598) to (0.045 + 0.598 + 0.598).

Hence, the 3 classes for size are as follows:

**Small:** From 0.045 to 0.643.
**Medium:** From 0.643 to 1.241.
**Big**: From 1.241 to 1.840.

Utilizing this fair classification to calculate sub-totals per size, we obtain:

Small Total = 0.045 + 0.051 + 0.058 + 0.066 + 0.075 + 0.085 + 0.097 + 0.110 + 0.125 +
0.142 + 0.162 + 0.184 + 0.209 + 0.237 + 0.270 + 0.307 + 0.349 + 0.396 +
0.450 + 0.512 + 0.582 = **4.512**

Medium Total = 0.661 + 0.751 + 0.854 + 0.970 + 1.103 = **4.339**

Big Total = 1.253 + 1.424 + 1.619 + 1.840 = **6.136**

System Total = 4.512 + 4.339 + 6.136 = **15**

This entire set of 30 quantities can be conceptualized as a deterministic partition process of the original quantity of 15 into 30 smaller parts, instead of viewing it as exponential growth series.

According to above fair classification:

**Small:** **21** parts with a total value of 4.512, or 4.512**/**15 = **30.1%** of system total.
**Medium:** **5** parts with a total value of 4.339, or 4.339**/**15 = **29.0%** of system total.
**Big**: **4** parts with a total value of 6.136, or 6.136**/**15 = **40.9%** of system total.

Figure 17 depicts the quantitative portion of each part, sorted low to high in the clockwise direction, and where relative area signifies relative quantity. In addition, the color for Small is blue; the color for Medium is brown; and the color for Big is green. Here, all 3 sizes earn in the approximate the **33.3%** ideal, equitable, and expected proportions of overall quantity.

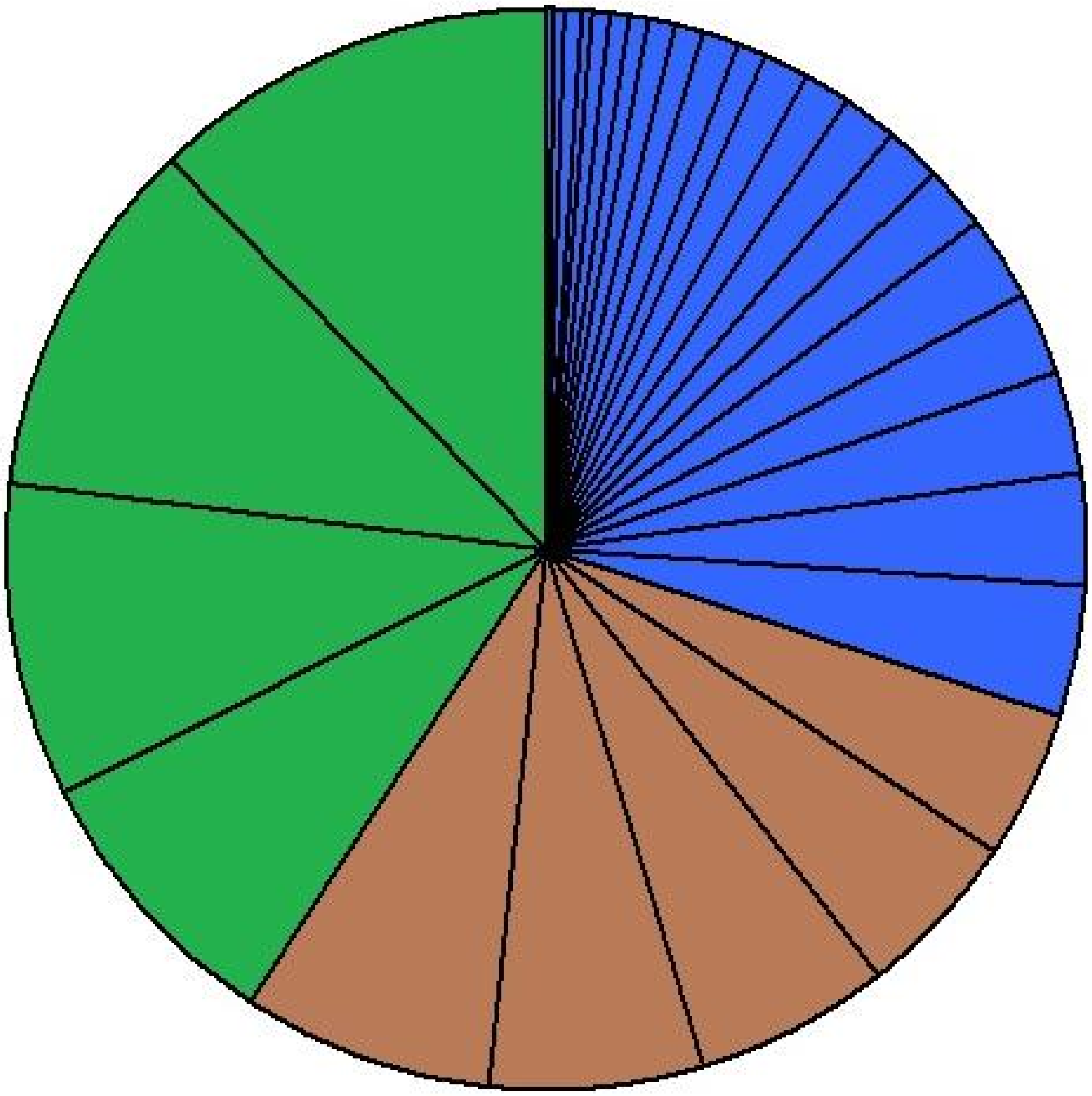

**Figure 17**:  Equitable Mix of Small, Medium, and Big in Generic Presentation of Partitions

Hence the attempt to represent a random partition process as simply a set of totally distinct parts with a constant growth factor appears reasonable, while the small is beautiful phenomenon is manifested in both; namely in the partition of the 15-meter pipe with its randomly fluctuating growth factors, as well as in the abstract model of constant quantitative growth.

In order to drive the point that the small is beautiful phenomenon is not much disturbed due to the switch from the original random model to the deterministic model, and that both models yield approximately the same quantitative structure, a comparison chart is added. Figure 18 depicts the bar chart of the two sets of parts for comparison, clearly demonstrating that both sets have an almost identical quantitative configuration, and that the small is beautiful phenomenon is manifested in both cases. In a sense, the deterministic model can be thought of simply as applying '**Data Smoothing'** on the random model. Data Smoothing is the use of an algorithm to remove noise from a data set, allowing important patterns to stand out.

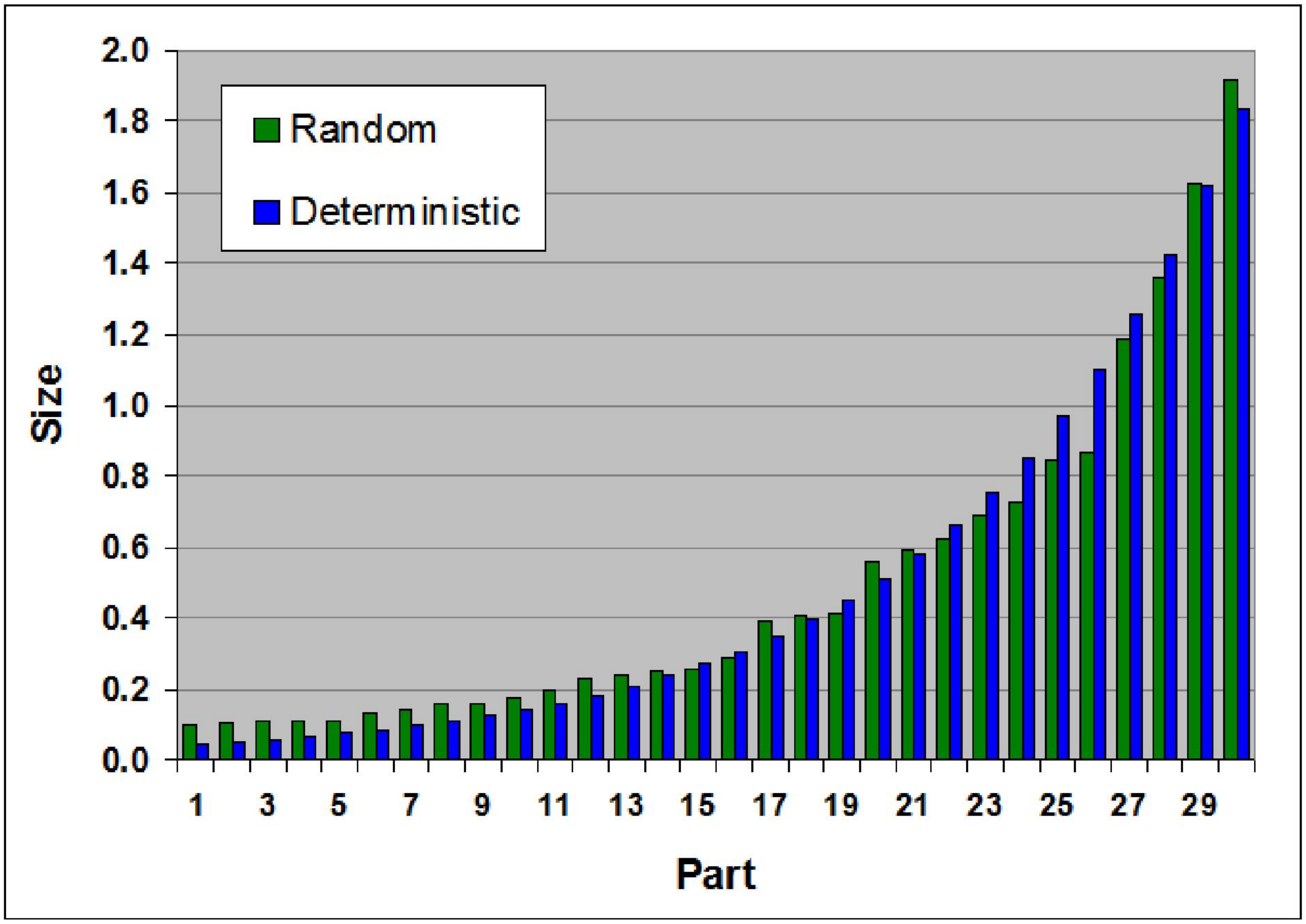

**Figure 18**: Correspondence in Overall Quantitative Configuration - Random & Deterministic

It should be emphasized that the two conditions of this chapter for partitions, namely having numerous and mostly distinct parts, are not entirely sufficient for the manifestation of the small is beautiful phenomenon. The necessary third condition for the manifestation of the phenomenon is that the values of the parts are increasing in a multiplicative manner such as $X_{N+1} = F^*X_N$, where F is a constant, or where F varies randomly but without any overall downward trend. The phenomenon is not manifested for a model where the parts are increasing in an additive manner such as $X_{N+1} = X_N + C$, where C is a constant, or where C varies randomly but without any overall downward or downward trend. An additive manner of growth does not yield the small is beautiful phenomenon but rather an equitable and fair size configuration where all sizes are equally numerous. Indeed, an additive manner of growth yields the Uniform Distribution which is size-neutral, and where the small is beautiful phenomenon does not manifest itself.

As a numerical example of the (deterministic) additive partition model, the initial base value (the smallest part) is arbitrarily set to 0.70, and the constantly added value is 0.22. In addition, it is determined that 32 parts shall be created, so that 31 additions of the value 0.22 are executed.

{0.70, 0.92, 1.14, 1.36, 1.58, 1.80, 2.02, 2.24, 2.46, 2.68, 2.90,
3.12, 3.34, 3.56, 3.78, 4.00, 4.22, 4.44, 4.66, 4.88, 5.10, 5.32,
5.54, 5.76, 5.98, 6.20, 6.42, 6.64, 6.86, 7.08, 7.30, 7.52}

The entire range of {0.70, 7.52} is divided fairly into 3 equal sections for the classification of Small, Medium, and Big. Each section is of the width (7.52 - 0.70)**/**(3) = (6.82)**/**(3) = 2.27. Small is from (0.70) to (0.70 + 2.27). Medium is from (0.70 + 2.27) to (0.70 + 2.27 + 2.27). Hence, the 3 classes for size are as follows:

**Small:** From 0.70 to 2.97.
**Medium:** From 2.97 to 5.25.
**Big**: From 5.25 to 7.52.

Utilizing this classification to calculate sub-totals per size, we obtain:

Small Total = 0.70 + 0.92 + 1.14 + 1.36 + 1.58 + 1.80 + 2.02 + 2.24 + 2.46 + 2.68 + 2.90
= **19.8**

Medium Total = 3.12 + 3.34 + 3.56 + 3.78 + 4.00 + 4.22 + 4.44 + 4.66 + 4.88 + 5.10
= **41.1**

Big Total = 5.32 + 5.54 + 5.76 + 5.98 + 6.20 + 6.42 + 6.64 + 6.86 + 7.08 + 7.30 + 7.52
= **70.6**

System Total = 19.8 + 41.1 + 70.6 = **131.5**

This entire set of 32 quantities can be conceptualized as a deterministic partition process of the original quantity of 131.5 into 32 smaller parts, instead of viewing it as an additive growth.

According to above fair classification, all sizes occur with nearly the same frequency:

**Small:** **11** parts with a total value of 19.8, or 19.8**/**131.5 = **15.1%** of system total.
**Medium:** **10** parts with a total value of 41.1, or 41.1**/**131.5 = **31.3%** of system total.
**Big**: **11** parts with a total value of 70.6, or 70.6**/**131.5 = **53.7%** of system total.

Figure 19 depicts the quantitative portion of each part, sorted low to high in the clockwise direction, and where relative area signifies relative quantity. In addition, the color for Small is blue; the color for Medium is brown; and the color for Big is green. Here the 3 sizes are not even close to the **33.3%** ideal division of overall quantity, rather, the larger the size, naturally the more proportion it earns.

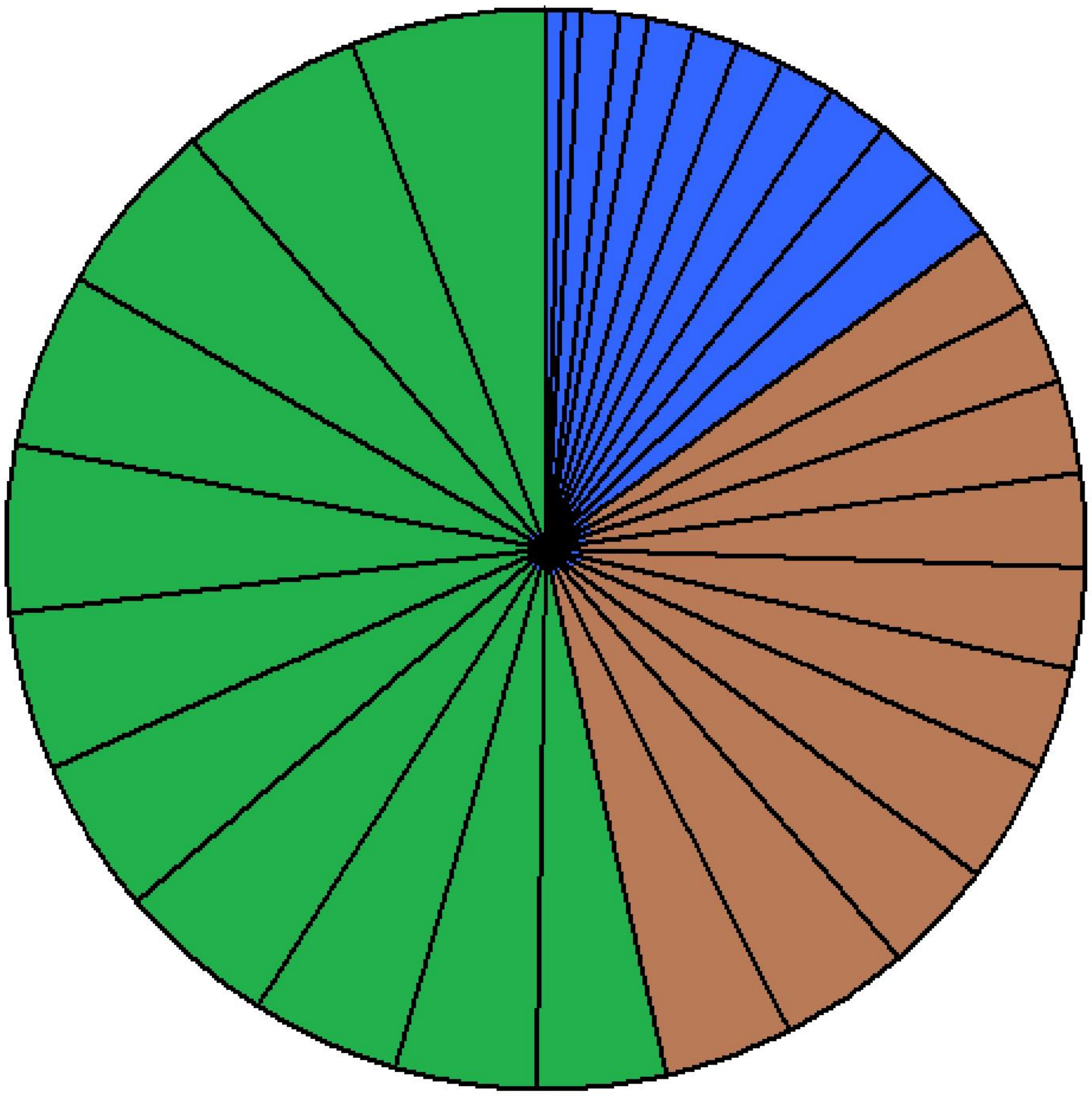

**Figure 19**: Equality in Frequency - Inequality in Quantity - Additive Presentation of Partitions

An alternative perspective for the failure of the (deterministic) additive partition model to manifest the small is beautiful phenomenon is that the set of growth factors (in a multiplicative sense) shows a consistent downward trend. The set of growth factors here are as follows:

{1.314, 1.239, 1.193, 1.162, 1.139, 1.122, 1.109, 1.098, 1.089, 1.082, 1.076, 1.071, 1.066, 1.062, 1.058, 1.055, 1.052, 1.050, 1.047, 1.045, 1.043, 1.041, 1.040, 1.038, 1.037, 1.035, 1.034, 1.033, 1.032 1.031, 1.030}

Clearly, these growth factors of the additive model are monotonically and consistently decreasing in value, showing a distinctive downward trend from the smallest part to the biggest part, thus precluding the small is beautiful phenomenon from manifesting itself.

The opposite pole of this case, namely a partition where the growth factors are with an upward trend, yields the small is beautiful phenomenon even more decisively and forcefully so. Such sets of quantities with constantly increasing growth rates, growing ever faster, are sometimes referred to as ‘super exponential growth’.

## [5] Conclusion: Random Partitions and Skewness

As shall be mentioned later in this article, the significant digits configuration of the resultant set of parts in Random Dependent Partition is derived in a mathematically rigorous way, and this digital result relates to the small is beautiful quantitative configuration found for the set of parts.

In Mathematical Statistics it is shown that distances between random markings along the length of one spacial dimension are distributed as in the Exponential Distribution. Hence, Random Real Partition is intimately connected with the Exponential Distribution. For example, L-meter pipe is cut randomly into N parts via (N – 1) random markings between 0 and L applying the Uniform Distribution. This implies the rate of (N - 1)/(L) markings per unit distance, and pointing to the Exponential Distributions with (N - 1)/(L) Lambda parameter value.

The Exponential Distribution is strictly positively skewed, monotonically falling to the right, and having a strong preference for the small. This result implies that Random Real Partition exhibits the small is beautiful phenomenon as well.

Surely there are in principle many random partition processes other than Random Dependent Partition and Random Real Partition. These two processes are perhaps the most straightforward or natural ways to break a quantity into parts in a random fashion. The concluding general statement regarding the manifestation of the small is beautiful phenomenon in random partition processes is based on very consistent and broad empirical testing and evidences, as well as sound conceptual underpinning. In part it also rests on some rigorous mathematical results.

The general statement is based on three assumptions:

(I) Given that partition is performed on the real number basis and not exclusively on integers.
(II) Given that partition is truly random.
(III) Given that partition thoroughly breaks the original quantity into numerous refined parts.

Then, the small is beautiful quantitative configuration is found in the resultant set of parts; all the parts are of distinct values; with all growth factors greater than 1 and fluctuating randomly around an average factor value, showing some gentle downward trend on the left for small pieces, and some gentle upward trend on the right for big pieces, mimicking U-shape histogram of sorts (namely exhibiting a multiplicative manner of growth rather than an additive one).

## [6] Quantitative Partitions Models Lead to Skewness and Often to Benford

Random Real Partition and Random Dependent Partition shall now be revisited and further analyzed, and various other quantitative partition models shall be examined in terms of the quantitative and digital behavior of the resultant set of parts. Almost all the models are of the random type, and only a few are of the deterministic type. The [nearly] universal feature found across [almost] all partition models is having many small parts but only few big parts, namely quantitative skewness, while Benford's Law is valid only in some particular partition cases and under certain constraints. Hence the phenomenon of Benford's Law is viewed as a particular subset of the broader small is beautiful phenomenon in quantitative partitioning.

Significantly, such a vista is true in the context of other causes and explanations of Benford's Law where the small consistently outnumbers the big in [almost all] partial structures of the model under consideration, or well before full convergence to Benford is achieved, while Benford is often found only in full model structures and only after the complete convergence of the model. These results endow the vista much broader scope.

Further in this article it will be shown how either the active process of partitioning or the passive consideration of a large quantity as the composition of smaller parts can be considered as another independent explanation for the widespread empirical observation of Benford's Law, as well as another cause of the more universal observation of the small is beautiful phenomenon.

The following is brief summary of the partition models discussed in this article:

**Complete Equipartition**: The set of all possible Integer Partitions of N.
**Refined Equipartition**: The limited set of all Integer Partitions of N restricted to $n << N$.
**Irregular Equipartition**: The set of Integer Partitions of N restricted in an arbitrary manner.
**Partition as a Set of Marks on the x-axis**: The lengths of the intervals between the random marks established on the x-axis constitute the parts in this random partition process.
**Singular Balanced Partition**: The balanced partitioning of quantity X into many identical parts yielding only one size exclusively.
**Random Real Partition as in Random Pipe Breaking**: Generalized beyond the standard use of the Uniform Distribution in generating the marks on the x-axis to include also other skewed and symmetrical distributions.
**Random Dependent Partition as in Random Rock Breaking**: Generalized via the use of a deterministic fixed ratio/percentage of breakup.
**Chaotic Rock Breaking**: A variation on Random Dependent Partition via the introduction of an extra measure of randomness in the partition process.
**Random Minimum Breaking**: A variation on Random Dependent Partition where the minimum is repeatedly broken via random ratio/percentage of breakup.
**Random Maximum Breaking**: A variation on Random Dependent Partition where the maximum is repeatedly broken via random ratio/percentage of breakup.
**Balls Distributed inside Boxes**: The partitioning of a singular set of numerous discrete balls via their placement into fewer discrete boxes.

## [7] Refined Equipartition Parable

A group of 33 well-trained spies are about to be sent to the enemy state for long periods of military sites observations, sabotage activities, political meddling and subversion. The secret service commander thinks that it is best to separate them into independent cells of no more than 5 spies each, in case one spy is uncovered, so that not all would be lost as a result in such adverse instances. In addition, the work is more suitable for small cells, as well as for an individual spy operating totally alone. Not having any specific plans of spying assignments beforehand, the commander first simply divides these 33 future spies in an arbitrary or random manner before their departure. Later, once they are all settled in their new environments, each cell would be assigned specific work according to its size and abilities. Some possible partitions are:

{5, 5, 5, 5, 5, 5, 2, 1}
{2, 4, 5, 5, 1, 3, 1, 1, 1, 5, 5}
{1, 1, 1, 1, 1, 1, 1, 1, 1, 1, 2, 2, 2, 2, 2, 3, 5, 5}
{4, 5, 4, 5, 4, 5, 2, 2, 2}
{3, 3, 3, 3, 3, 3, 3, 3, 3, 3, 3}
etc. etc.

The bored statistician at the secret service headquarters who hadn't gotten any work assignment in several days has decided to entertain himself by fictitiously inventing the abstract large data set of all such possible partitions (arbitrarily aggregated as a singular data set) with the intention of thoroughly analyzing it quantitatively. The aggregated data set that he has in mind is:

{5, 5, 5, 5, 5, 5, 2, 1, 2, 4, 5, 5, 1, 3, 1, 1, 1, 5, 5, 1, 1, 1, 1, 1, 1, 1, 1, 1, 1, 2, 2, 2, 2, 2, 3, 5, 5, 4, 5, 4, 5, 4, 5, 2, 2, 2, 3, 3, 3, 3, 3, 3, 3, 3, 3, 3, 3, etc. etc.}

He has calculated that there are exactly 918 possible partitions here, and that the aggregated data set contains 14,608 numbers in total. Is his analysis practical in any way? To what use could such a project be applied? Interestingly, by studying the relative occurrences of {1, 2, 3, 4, 5} in such abstract data set he discovered that their frequencies are inversely proportional to their size, so that a lone spy occurs 6,905 times, a cell consisting of a pair of spies occurs 3,228 times, a trio spy cell occurs 2,017 times, a cell of 4 spies occurs 1,408 times, and a cell of 5 spies occurs only 1,050 times. Moreover, he was able to reasonably fit these 5 values into an algebraic expression which he mathematically derived in the abstract and in general for all such types of partitions.

Could such highly abstract study be put into some concrete flesh-and-blood scenario so that real-life applications are made? Yes indeed! Imagine the headquarters of the American or Soviet intelligence departments during the height of the Cold War employing tens of thousands of spies. In one comprehensive spying program involving thousands of such 33-spy groups, each group is launched by a different officer, who randomly and independently partitions his 33-group into smaller cells of no more than 5 spies each. Each officer gives each possible partition equal chance (i.e. weight) of being applied to the 33 brave men and women about to depart on their patriotic mission. The statistical desk at the secret service headquarters is interested in the overall relative frequencies of the sizes of the cells occurring worldwide. Hence it was calculated that:

**47%** [ *6,905* / *14,608* ] of all the cells worldwide are of **1** spy.
**22%** [ *3,228* / *14,608* ] of all the cells worldwide are of **2** spies.
**14%** [ *2,017* / *14,608* ] of all the cells worldwide are of **3** spies.
**10%** [ *1,408* / *14,608* ] of all the cells worldwide are of **4** spies.
**7%** [ *1,050* / *14,608* ] of all the cells worldwide are of **5** spies.

Certainly, the small is beautiful principle manifests itself decisively here, even though we have restricted the Integral Partition of 33 exclusively to small integers that are less than 6.

In a statistical sense, even for a single partition decision by a commanding officer responsible for only one such 33-spy group about to depart, one can inquire about the probability of any particular cell size, assuming the officer does not have any bias or preference for any particular size and equally chooses one partition among all the possibilities.

## [8] Complete, Refined, and Irregular Equipartition Models

Three Equipartition models are discussed in this article, each based on some variation of Integer Partition. These three models are named Complete Equipartition, Refined Equipartition, and Irregular Equipartition.

**Complete Equipartition** is simply the entire set of all possible Integer Partitions of N, where all partitions are aggregated into one vast data set. This model allows for any integer from 1 to N to become a part of the N whole, without any limits or constraints. Indeed we insist that all the integers from 1 to N, small or big, must participate in the partition, and not only once, but as many times as possible. Complete Equipartition is called '**complete**' since it incorporates all possible breakups, completely, without leaving out any type of partition whatsoever. Complete Equipartition breaks up quantity N into very small integral parts such as 1 and 2; or into very big integral parts such as N - 2 and N – 1; and including the improper partition where N is actually not being broken up at all and instead it is being left intact as {N}. Since all possible partitions are given equal weights within the model, and are simply aggregated into that vast data set without any preferential adjustments, the prefix '**equi**' is included in the term 'Equipartition'.

For example, the Complete Equipartition of N = 27 is that vast data set containing among others {2, 5, 10, 10}, {1, 1, 25}, {1, 1, 1, 1, 1, 1, 1, 20}, {27}, {1, 26}, {1, 1, 25}, and so forth. This is written in full as {2, 5, 10, 10, 1, 1, 25, 1, 1, 1, 1, 1, 1, 1, 20, 27, 1, 26, 1, 1, 25, and so forth}. The entire data set of Complete Equipartition of 27 is way too long and vast to describe here.

In chapter 1, three examples of Complete Equipartition are discussed, namely those with N = 5, N = 7, and N = 13. Figures 1 and 2 clearly demonstrate the skewness of Complete Equipartition of 5. Figure 3 depicts the histogram of Complete Equipartition of 13, and that histogram is skewed as well, as it is consistently and monotonically falling to the right, except at the very end. At the high end of each Complete Equipartition, the partitions {1, N - 1} and {N} are the unique partitions involving integers N and N – 1, hence integers N and N – 1 are with equal frequency of 1, and so the histogram is flat on that small and insignificant portion.

Complete Equipartition is inherently skewed, and this is so regardless of the value of integer N. Yet, in spite of its innate skewness, Complete Equipartition is not Benford, regardless of the value of N, and this is so due to the extreme skewness associated with Complete Equipartition.

Complete Equipartition is even skewer than the Benford skewed configuration. For example, Complete Equipartition of N = 9 yields the vector of the count of the integers 1 to 9 occurrences {67, 26, 15, 8, 5, 3, 2, 1, 1}, and this translates into the vector of percents of occurrences of the integers - or rather 'digits' - of {52.3%, 20.3%, 11.7%, 6.3%, 3.9%, 2.3%, 1.6%, 0.8%, 0.8%}. Here integer or 'digit' 1 has the very high frequency of 52.3%, while in Benford's Law digit 1 has the relatively milder frequency of 30.1%.

**Refined Equipartition** refers to Integer Partitions of N where only {1, 2, 3, … , n} are allowed in the partitions, and where n is much smaller than N. Since only partitions involving integers that are much smaller than N are allowed, this ensures that N is being broken into much smaller and '**refined**' parts, resulting in a significant fragmentation of the original quantity N. It should be noted that <u>all</u> integers from 1 to n are allowed to participate in the partition and are included, without any gaps or exclusions. Indeed we insist that all the integers from 1 to n must participate in the partition, and not only once, but as many times as possible.

The Equipartition Parable of a group of 33 spies - being separated into independent cells of no more than 5 spies each - is an example of Refined Equipartition. Here N = 33 and n = 5.

An article by Don Lemons in 1986 inspired and led Steven Miller to publish another related article in 2015 presenting a mathematically rigorous proof that Refined Equipartition endows the same quantitative portion from the entire quantity of the entire equipartition scheme to each integer - in the limit as N approaches infinity while n is kept fixed and finite so that n << N. Rephrasing Miller's assertion: The quantitative sum from each allowed integer in Refined Equipartition is a constant, granting each integer equal portion from the entire quantity embedded within that vast data set. Conceptually, Miller argues that even though a small integer is of low value, yet it has high frequency, so that its total quantity should be equivalent to the total quantity of a big integer of high value but of low frequency. In other words, that there exist perfect cancelation and offsetting effects between size and frequency.

Calculating quantitative sum for each integer in Refined Equipartition of 33 limited to 5 of the Equipartition Parable, we obtain: {(1)*(6905), (2)*(3228), (3)*(2017), (4)*(1408), (5)*(1050)}, namely {6905, 6456, 6051, 5632, 5250}. While total quantity per integer is not truly constant here, because Miller's constraints have not been fully met, yet the five portions are all nearly of about the same level, fluctuating around their average value of 6059, with relatively little variation between them. The quantitative configuration here is very similar to that of the oval shape in Figure 5 where Big, Medium, and Small all have about the same territory (i.e. the same total quantity). Therefore Refined Equipartition of 33 limited to 5 is very much in the spirit of the small is beautiful principle. Yet it certainly cannot be Benford with only 5 integers/digits, and where OOM = Log(5/1) = 0.70. Benford's Law applies only to data with high variability.

For a more detailed example, let us examine Refined Equipartition of 13 limited to 3. Here quantity 13 is broken into smaller integral parts that are not bigger than 3. This leads to the set of the following partitions involving only 1, 2, or 3 as parts:

{3, 3, 3, 3, 1}
{3, 3, 3, 2, 2}
{3, 3, 3, 2, 1, 1}
{3, 3, 3, 1, 1, 1, 1}
{3, 3, 2, 2, 2, 1}
{3, 3, 2, 2, 1, 1, 1}
{3, 3, 2, 1, 1, 1, 1, 1}
{3, 3, 1, 1, 1, 1, 1, 1, 1}
{3, 2, 2, 2, 2, 2}
{3, 2, 2, 2, 2, 1, 1}
{3, 2, 2, 2, 1, 1, 1, 1}
{3, 2, 2, 1, 1, 1, 1, 1, 1}
{3, 2, 1, 1, 1, 1, 1, 1, 1, 1}
{3, 1, 1, 1, 1, 1, 1, 1, 1, 1, 1}
{2, 2, 2, 2, 2, 2, 1}
{2, 2, 2, 2, 2, 1, 1, 1}
{2, 2, 2, 2, 1, 1, 1, 1, 1}
{2, 2, 2, 1, 1, 1, 1, 1, 1, 1}
{2, 2, 1, 1, 1, 1, 1, 1, 1, 1, 1}
{2, 1, 1, 1, 1, 1, 1, 1, 1, 1, 1, 1}
{1, 1, 1, 1, 1, 1, 1, 1, 1, 1, 1, 1, 1}

This yields the aggregated set of all the above partitions as follows:

{3, 3, 3, 3, 1, 3, 3, 3, 2, 2, 3, 3, 3, 2, 1, 1, 3, 3, 3, 1, 1, 1, 1, 3, 3, 2, 2, 2, 1, 3, 3, 2, 2, 1, 1, 1, 3, 3, 2, 1, 1, 1, 1, 1, 3, 3, 1, 1, 1, 1, 1, 1, 1, 3, 2, 2, 2, 2, 2, 3, 2, 2, 2, 2, 1, 1, 3, 2, 2, 2, 1, 1, 1, 1, 3, 2, 2, 1, 1, 1, 1, 1, 1, 3, 2, 1, 1, 1, 1, 1, 1, 1, 1, 3, 1, 1, 1, 1, 1, 1, 1, 1, 1, 1, 2, 2, 2, 2, 2, 2, 1, 2, 2, 2, 2, 2, 1, 1, 1, 2, 2, 2, 2, 1, 1, 1, 1, 1, 2, 2, 2, 1, 1, 1, 1, 1, 1, 1, 2, 2, 1, 1, 1, 1, 1, 1, 1, 1, 1, 2, 1, 1, 1, 1, 1, 1, 1, 1, 1, 1, 1, 1, 1, 1, 1, 1, 1, 1, 1, 1, 1, 1, 1, 1, 1, 1, 1}

The ordered set is then as follows:

{1, 1, 1, 1, 1, 1, 1, 1, 1, 1, 1, 1, 1, 1, 1, 1, 1, 1, 1, 1, 1, 1, 1, 1, 1, 1, 1, 1, 1, 1, 1, 1, 1, 1, 1, 1, 1, 1, 1, 1, 1, 1, 1, 1, 1, 1, 1, 1, 1, 1, 1, 1, 1, 1, 1, 1, 1, 1, 1, 1, 1, 1, 1, 1, 1, 1, 1, 1, 1, 1, 1, 1, 1, 1, 1, 1, 1, 1, 1, 1, 1, 1, 1, 1, 1, 1, 1, 1, 1, 1, 1, 1, 1, 1, 1, 1, 1, 1, 1, 1, 1, 1, 2, 2, 2, 2, 2, 2, 2, 2, 2, 2, 2, 2, 2, 2, 2, 2, 2, 2, 2, 2, 2, 2, 2, 2, 2, 2, 2, 2, 2, 2, 2, 2, 2, 2, 2, 2, 2, 2, 2, 2, 2, 2, 2, 2, 2, 3, 3, 3, 3, 3, 3, 3, 3, 3, 3, 3, 3, 3, 3, 3, 3, 3, 3, 3, 3, 3, 3, 3, 3, 3, 3, 3}

The counts of integer 1, 2, 3 - namely their frequencies - are 102, 45, 27 respectively.
The entire data set of this Refined Equipartition contains 102 + 45 + 27 = 174 numbers in total.
Integers 1, 2, 3 are with frequency-percentages of 58.6%, 25.9%, 15.5% respectively.

While 3 << 13 is not actually true since 3 is not that much smaller than 13; and while N is still considered to be quite small with its value of 13, instead of approaching infinity or being some truly large number, nonetheless let us attempt to examine Miller's assertion in this case:

Total quantity of the entire data set of Refined Equipartition of 13 limited to 3 is obtained by simply summing up all the 174 integers, and this yields 273.

Total quantitative portion of any integer is simply (integer)*(frequency of the given integer), and here this yields:

Total quantity for integer 1: (1)*(102) = 102
Total quantity for integer 2: (2)*(45) = 90
Total quantity for integer 3: (3)*(27) = 81
----------------------------------------------------------
Total quantity of Equipartition: 273

While total quantity per integer is not truly constant here, because Miller's constraints have not been fully met, yet the three quantities are not that much different from each other, having little variation of only about 10 units above or below their average value of 91.

Finally, for properly constrained Refined Equipartition models where n << N and N → ∞, the observation that total quantity per integer is constant implies that frequency (or density) of an integer is inversely proportional to the value of the integer, and therefore frequency (or density) is of the form **k/x distribution**, known for its exact Benford behavior whenever defined range has an integral exponent difference between the max and min values [*LOG(max) – LOG(min) = Integer*], such as on integral powers of ten ranges. Hence Miller's assertion implies that Refined Equipartition models with particularly proper rangers are Benford. The following equations reinforce the above argument:

(total quantitative portion of the integer) = (integer)*(frequency of the integer)
(frequency of the integer) = (total quantitative portion of the integer)**/**(integer)
(frequency of the integer) = (constant for all the integers)**/**(integer)
Integer Frequency = Constant **/** Integer

**Irregular Equipartition** refers to Integer Partitions of N in which the parts are restricted and chosen from an arbitrary set of allowed integers. This arbitrary set does not necessarily increase nicely by one integer at a time, having possibly some gaps where some integers are skipped.

For example, for N = 25 representing the entire integral quantity to be partitioned, we allow only the arbitrary set of integers {1, 3, 6, 7, 11, 19} to be used in the partition. Integers 2 or 5 for example are not allowed to participate in the partition. Some possible partitions are:
{19, 6}, {19, 3, 3}, {11, 7, 3, 3, 1}, {6, 6, 6, 3, 3, 1}, and so forth.

Here, no mathematical results are offered for such messy and highly irregular Equipartition models, although these rare cases might emerge in some particular applications of partitions.

## [9] Partition as a Set of Marks on the x-axis

It is essential to visualize each possible partition in Integer Partition of N as a sort of a scheme that places marks along the x-axis on the interval (0, N). These marks must be placed only upon the integers, avoiding the spaces between them. Figure 20 depicts such representation for the singular partition of 33 into {1, 1, 1, 1, 1, 1, 1, 2, 2, 2, 2, 3, 3, 3, 4, 5}. This is just one possible partition within the entire set of all 918 possible partitions for this Refined Equipartition of 33 limited to 5 of the Equipartition Parable.

Initially the integral parts are represented in order from 1 to 5 for convenience, first showing seven ones, then four twos, then three threes, a four, and finally a five - as seen in the upper panel of Figure 20. Such an order helps in organizing the parts carefully onto the x-axis. Yet, equipartition does not necessitate order of the parts in any way, so this is not the best or the only way of viewing partitions in our context where the focus is on the set of all possible partitions, or equivalently on random partitioning, and the implications about occurrences of relative quantities. Hence it would be better to randomize each singular occurrence of the parts, and this is shown in the lower panel of Figure 20 where the set of parts is shown in a randomized order as the equivalent partition {2, 5, 1, 3, 1, 1, 2, 2, 1, 1, 3, 4, 3, 2, 1, 1}.

Physical order of magnitude here is POM = Max/Min = 5/1 = 5, while order of magnitude is OOM = LOG(Max/Min) = LOG(5/1) = 0.7, and which are quite low.

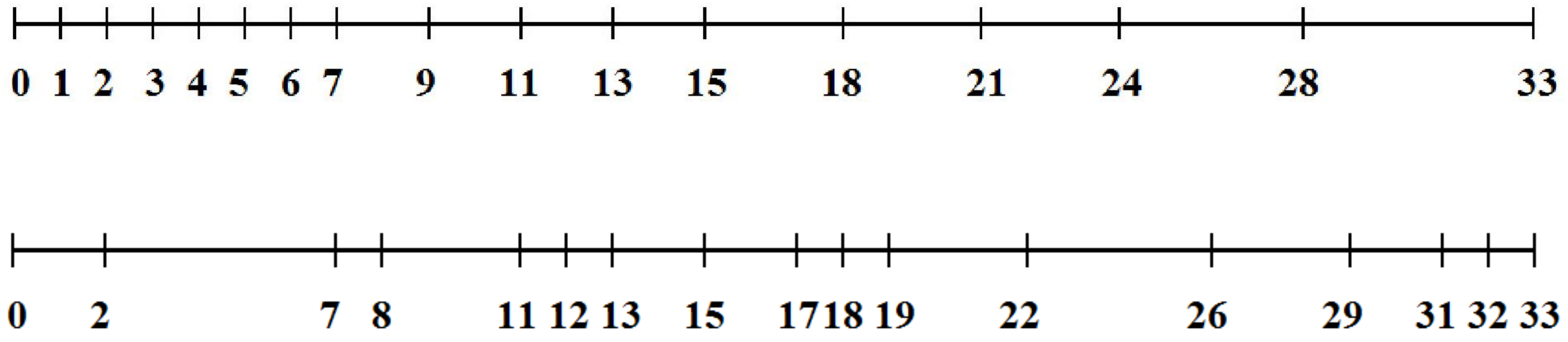

**Figure 20**: Integer Partition of 33 into {1, 1, 1, 1, 1, 1, 1, 2, 2, 2, 2, 3, 3, 3, 4, 5} along x-axis

Figure 21 depicts several other partition possibilities visualized as marks along the x-axis on the interval (0, 33). The last two partitions at the bottom are the two extremes, where either the set of the biggest possible pieces [integer 5 aided by integer 3] or the set of the smallest possible pieces [integer 1] are used to partition 33. It should be noted that it is very easy to obtain many more partitions here by simply starting out with a blank line between 0 and 33 and then randomly marking/poking marks along the integers as one sees fit. There is no need to calculate anything (the setup guarantees that it will all add up to 33 automatically) as long as we limit the gap between any two adjacent marks to the maximum length of 5.

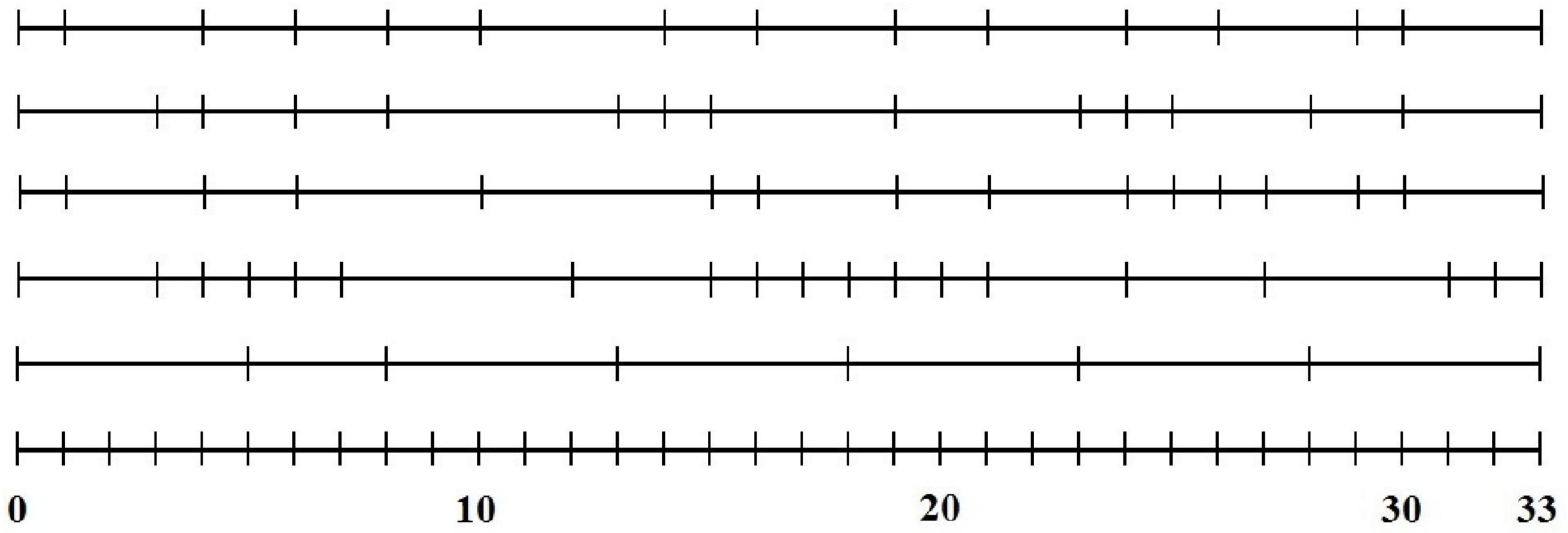

**Figure 21**: A Variety of Partitions of 33 into {1, 2, 3, 4, 5} Integral Parts along the x-axis

In each partition here of Figure 21 we are forcing the system into some significant fragmentation of the original quantity of 33 into much smaller parts. This is so due to the partial and limited compliance with the constraint n << N, although the statement 5 << 33 is not really true. In any case this partition still constitutes Refined Equipartition since it restricts the parts into smaller integers less than 6.

Even the usage of the combination of the largest parts possible near the bottom of Figure 21, namely {5, 3, 5, 5, 5, 5, 5} with six fives and one three, still results in significant fragmentation of the original 33 quantity into relatively much smaller parts.

It would take drawing 918 lines for Figure 21 in order to extend it fully until all possible partitions are visible, and for it to be properly called 'Equipartition', since by definition any such model must incorporate all possible partitions within the system.

Had we further restricted the partition of 33 to the set {2, 3, 5} say, as in Irregular Equipartition, then there would be even less variety in the ways distinct partitions can be formed, as only 3 sizes are allowed to participate in the partition, leading to fewer partitions.

Had we further restricted the partition of 33 to an even smaller set of {2, 3}, as in Irregular Equipartition, then there would be now even less variety in the ways distinct partitions can be formed, as only 2 sizes are allowed to participate in the partition, leading to even fewer partitions.

Had we further restricted the partition of 33 to the tiny set of {3} with only one integer, as in Irregular Equipartition, then there would be no variety whatsoever in the ways distinct partitions can be formed, as only 1 size is allowed now to participate in the partition, leading to one partition only. Indeed, this is the extreme case of **Singular Balanced Partition**, namely the case of a conserved quantity X partitioned into many identical parts yielding only one size exclusively. In this scenario, the parts are restricted exclusively to Q (which was 3 in the above example), assuming $Q < X$ and where Q is a perfect divisor of X as in $X/Q = I$, where I is an integer. Here there is no need to give equal weights to numerous partitions because there is only one possible partition to consider in the entire process, namely {Q, Q, Q, … I times} under the assumption that (Q + Q + Q + … I times) = X.

Singular Balanced Partitions come with two flavors. The first flavor is of a conserved integral quantity X partitioned into one of its integral divisors. The second flavor is of a conserved real quantity X partitioned into any real number $R < X$ such that $X/R = I$, where I is an integer, so that (R + R + R + … I times) = X. It is noted that the prefix '**equi**' is omitted in the term 'Singular Balanced Partition' since only one partition is possible here and therefore there is no need to give **equal** weights to some non-existing multiple partitions.

Figure 22 depicts Singular Balanced Partition of the integral quantity 33 into identical integral parts of 3. Here there are no bigs, no smalls, no mediums, but only one uniform size of 3. Certainly the motto 'small is beautiful' is irrelevant here; no quantitative comparison or analysis of any sort can be made; and it is impossible to establish any relationship between this partition and Benford's Law in any sense. Physical order of magnitude here is POM = Max/Min = 3/3 = 1, and order of magnitude is OOM = LOG(Max/Min) = LOG(3/3) = LOG(1) = 0, namely as low as they could be! There is no variation here at all.

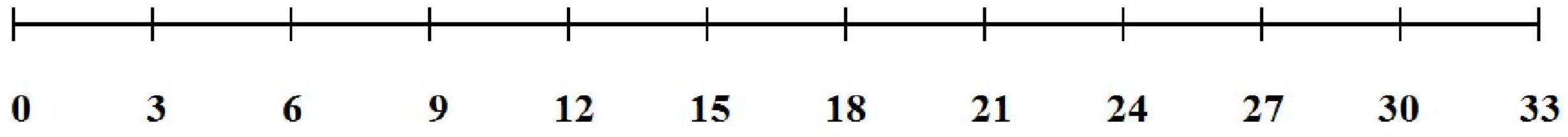

**Figure 22**: Singular Balanced Partition of 33 into Eleven Parts of 3's along the x-axis

Singular Balanced Partition is one extreme scenario where the allowed set of parts contains only one value Q, namely when the system is very inflexible, frugal, and totally unimaginative, not letting X be partitioned more liberally into all sorts of parts except for Q. The other extreme; the opposite pole of Singular Balanced Partition; is when the system is highly flexible, liberal and open to all sorts of values for the parts. So much so that the system allows all integral, fractional, rational, irrational, or whatsoever type of real numbers that exist between 0 and X! Here the allowed set of parts is {all real numbers on (0, X)}. Indeed, such a partition should be already very familiar with the reader, namely **Random Real Partition**! It is noted that the prefix '**equi**' is omitted in the term here since only a singular decisive random partition is needed and performed in this process - and which is sufficient to obtain the desired quantitative and near-Benford configurations in one fell swoop - given that plenty of parts are created. There is no need to give **equal** weights to some non-existing partitions.

Figure 23 depicts Random Real Partition of the quantity 33 into 32 parts of various real lengths, and which could be thought of as Random Pipe Breaking. Here there are very few big parts, some medium parts, and many small parts; all in the spirit of the small is beautiful principle. The relationship between this partition and Benford's Law is immediate and there is no need to repeat the experiment many times over and incorporate multiple partitions by constructing their aggregate set. This single set of resultant 32 parts in and of itself is not yet close enough to Benford because it does not contain enough parts, namely because the quantity (pipe) has not been broken thoroughly enough. To get directly to [nearly] Benford, all that is needed here is to break it more thoroughly into many more resultant parts (approximately over 5000 parts). Physical order of magnitude here is significantly larger as compared with the previous two examples of Figure 20 and Figure 22, coming at POM = Max/Min = 4.5/0.2 = 22.5, while order of magnitude is OOM = LOG(Max/Min) = LOG(4.5/0.2) = LOG(22.5) = 1.4.

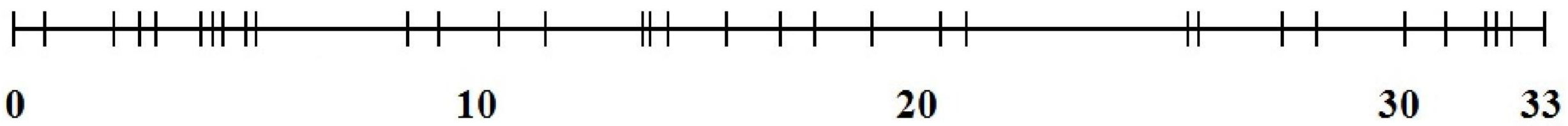

**Figure 23**: Random Real Partition of 33 into 32 real Parts along the x-axis

The choice to break the original quantity into 32 parts was somewhat deliberate, aiming to fit the number into the relationship $2^{INTEGER} = 32$ where INTEGER here is 5 and thus $2^5 = 32$ holds. This is so because of the possible [but erroneous] interpretation of Random Real Partition as Random Dependent Partition in the particular cases where [Number of Parts] = $2^{INTEGER}$.

Figure 24 depicts the supposed or imagined 5 partition stages where quantity 33 is gradually and randomly being broken into 32 much smaller real quantities in the spirit of Random Dependent Partition - resulting in the same exact quantitative configuration and having identical parts as in the Random Real Partition case of Figure 23. Yet the two processes of Random Real Partition and Random Dependent Partition are distinct in nature, and they almost always lead to distinct resultant set of parts, in spite of the apparent successful superimposition of the two into a singular partition shown in Figure 24.

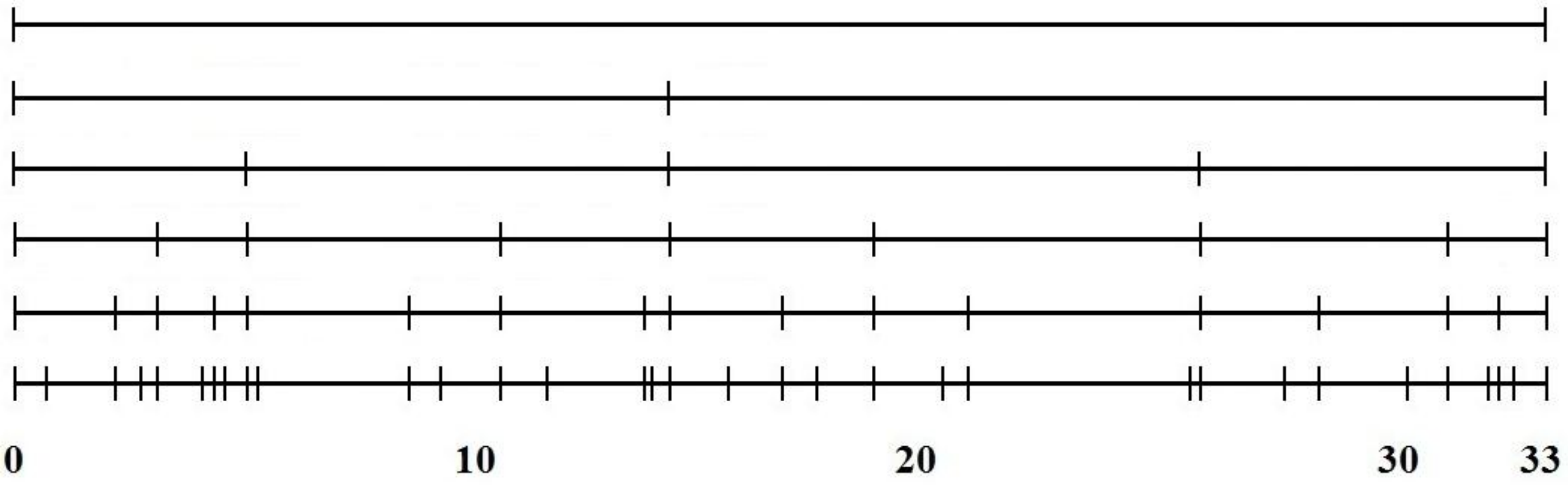

**Figure 24:** Erroneous Interpretation of Random Real Partition as Random Dependent Partition

**Random Dependent Partition** as in Random Rock Breaking involves taking a conserved real quantity X and breaking it into 2 real parts of random proportions chosen via the continuous Uniform(0, 1) - interpreted as the random ratio/percentage of the breakup. In the first stage, X is split into {X*Uniform(0, 1), X*(1 – Uniform(0, 1)). The process continues with the breaking up of each of these 2 parts separately but in the same random fashion, resulting in 4 parts altogether. Continuing breaking up the 4 into 8, the 8 into 16, the 16 into 32, and so forth, would result in the Benford configuration after about 10 or so such partition stages - depending on the desired level of compliance accuracy. Performing about 17 such stages would result in nearly perfect Benford configuration; and in the limit as the number of stages goes to infinity the logarithmic distribution is obtained exactly (in a mathematical rigorous way).

The complete success of arriving nearly perfectly (or exactly in the limit) at the Benford configuration for the process of Random Dependent Partition is contrasted with the slightly less successful story of Random Real Partition where Benford is found only approximately so, because the process is gradually metamorphosing into the Exponential Distribution which itself is only approximately Benford – assuming the creation of sufficiently large number of pieces/parts.

On the face of it, new marks in Random Dependent Partition are dependent on established older marks, because each new mark must to be placed between two older ones, while in Random Real Partition new marks are placed in a totally free and chaotic manner. Yet, a heuristic argument is made here, viewing the process of the Random Dependent Partition applying the Uniform(0, 1) and its resultant random x-axis placements of ($2^{STAGES}$ – 1) marks, as quite similar (yet not equivalent) to an independent generations of ($2^{STAGES}$ – 1) realizations from the Uniform(0, X) serving as marks on the x-axis – just as was seen in the corresponding cases of Figure 23 and Figure 24. Such a vista then lets us somehow unify in a limited sense the process of Random Real Partition with the process of Random Dependent Partition, because ultimately each scheme constitutes just a different description of events of the similar stochastic process which randomly places marks on (0, X) via the same distribution of the Uniform(0, 1)!

It is quite remarkable that these two random methods of placing marks on the x-axis via the same distribution of the Uniform(0, 1) would lead to somewhat distinct quantitative configurations for the resultant set of parts and therefore to distinct measures of compliance with Benford!

## [10] Flexibility in the Selection of Parts Leads to Benford and High POM

Let us summarize and compare three different partition schemes:

(1) **Singular Balanced Partition** – due to its extremely rigid rule of how parts are selected, restricting them exclusively to only one size Q, there exists neither big nor small, but only one unique size for all the parts, no quantitative or digital comparisons can be made, and certainly there exists no relationship to Benford's Law whatsoever. Here order of magnitude is as low as can possibly be, namely 0.

(2) **Complete Equipartition** and **Refined Equipartition** – due to their limited flexibility in the rule of how parts are selected, allowing a particular diversity of integral sizes only, while excluding fractional and irrational sizes, these processes yield a skewed quantitative configuration for the set of all possible partitions aggregated as one vast data set, where the small is numerous and the big is rare. Only Refined Equipartition could be approximately Benford (or exactly so in the limit) provided that N is a very large number, that n is much smaller than N, and assuming that the range of allowed sizes - namely {1, 2, 3, … n} - spans integral powers of ten, such as when n assumes the value of 10, 100, or 1000, and so forth. Here order of magnitude is well over 0, yet it is still a bit low.

(3) **Random Real Partition** – due to its extreme flexibility in the rule of how parts are selected, allowing for unaccountably infinite many real numbers on (0, X), namely any possible size whatsoever, this leads to an approximate Benford configuration in one fell swoop for just one single such random partition without any need to repeat the partition over and over again, and without the need to aggregate anything - assuming conserved quantity X is partitioned into very many real parts (approximately 5000 to 10000 pieces). Here order of magnitude is quite high, and especially so when numerous parts are created.

<u>Conclusion:</u> The more flexible a given partition process is in terms of how it selects its parts, the higher is resultant order of magnitude, the skewer is resultant set of parts, the more pronounced is the manifestation of the small is beautiful phenomenon, and possibly the closer it is to the Benford configuration.

## [11] Testing Compliance of Random Real Partition with Benford's Law

Another perspective on Random Real Partition is its description as in the following scheme:

Generate N realizations from the continuous Uniform(0, X).
Order them from low to high, add 0 on the very left, and add X on the very right.
The data set is: $\{0, U_1, U_2, U_3, \ \ldots \ , U_{N-2}, U_{N-1}, U_N, X\}$.
Generate the difference data set out of the one above.
This data set is: $\{(U_1 - 0), (U_2 - U_1), (U_3 - U_2), \ \ldots \ , (U_{N-1} - U_{N-2}), (U_N - U_{N-1}), (X - U_N)\}$.
This data set is conjectured to be nearly Benford as N gets large. In practical terms it might be sufficient for N to be about 5000 to 10000 for a reasonable fit to Benford, and that nothing or not much is gained by increasing N from this level, since beyond around this level saturation sets in.

It should be noted that Monte Carlo simulation run of N realizations from the continuous Uniform(0, X) yields the same quantitative configuration for the difference data set had it been based on the continuous Uniform(0, 1), or on the continuous Uniform(0, Y), and so forth. This is so since these processes differ only in the sense of having a different scale, but structurally they perfectly correspond. As far as digital configuration is concerned, for Benford data sets the Scale Invariance Principle guarantees that digits distribution are the same in any scale, although for data sets which are not perfectly Benford, scale does indeed matter. Since Random Real Partition is only approximately Benford, therefore scale does matter here, but only slightly so.

Let us perform several Monte Carlo empirical tests regarding Random Real Partition and its compliance with Benford, applying the Uniform as the distribution generating the marks.

Uniform(0, 10) Partitioned via 25,000 Marks - {27.6, 16.9, 13.1, 10.3, 9.1, 7.2, 6.3, 5.3, 4.4}
Benford's Law for First Significant Digits - {30.1, 17.6, 12.5, 9.7, 7.9, 6.7, 5.8, 5.1, 4.6}
SSD value is **9.2**, and such low value indicates that this partition is fairly close to Benford.
Here there are 25000**/**10 or 2500 marks per unit, and therefore this process corresponds to the Exponential Distributions with 2500 Lambda parameter value. In one computer simulation run for this Exponential Distribution, digits came as {27.2, 16.7, 13.3, 10.7, 9.2, 7.3, 6.3, 4.9, 4.4}.

Uniform(0, 800) Partitioned via 10,000 Marks - {32.2, 16.7, 11.3, 8.6, 7.5, 6.8, 6.0, 5.8, 5.1}
Benford's Law for First Significant Digits - {30.1, 17.6, 12.5, 9.7, 7.9, 6.7, 5.8, 5.1, 4.6}
SSD value is **8.2**, and such low value indicates that this partition is fairly close to Benford.
Here there are 10000**/**800 or 12.5 marks per unit, and therefore this process corresponds to the Exponential Distributions with 12.5 Lambda parameter value. In one computer simulation run for this Exponential Distribution, digits came as {32.3, 16.5, 11.4, 8.7, 7.5, 6.8, 6.3, 5.4, 5.1}.

Uniform(0, 38) Partitioned via 35,000 Marks - {32.7, 17.9, 11.5, 8.7, 7.4, 6.2, 5.5, 5.1, 5.0}
Benford's Law for First Significant Digits - {30.1, 17.6, 12.5, 9.7, 7.9, 6.7, 5.8, 5.1, 4.6}
SSD value is **9.7**, and such low value indicates that this partition is fairly close to Benford.
Here there are 35000**/**38 or 921.1 marks per unit, and therefore this process corresponds to the Exponential Distributions with 921.1 Lambda parameter value. In one computer simulation run for this Exponential Distribution, digits came as {32.8, 17.6, 11.8, 8.5, 7.4, 6.4, 5.6, 5.3, 4.7}.

Results are close to Benford, yet a more decisive result with lower SSD value (say less than 2) cannot be found here even if we partition an interval into many more parts. Saturation point is found somewhere around 5000, or 10000, or perhaps around 15000.

When Uniform(0, 38) for example is partitioned via only 1,000 marks, SSD values are somewhat higher. Six such Monte Carlo partition runs with 1,000 marks gave the following SSD values: 32.7, 52.6, 11.6, 26.8, 32.4, 21.1.

When Uniform(0, 38) for example is partitioned via only 100 marks, SSD values are significantly higher, since such partial and insufficient partition which yields only few parts is not effective enough and does not converge close enough to Benford and to the Exponential Distribution. Six such Monte Carlo partition runs with 100 marks gave the following SSD values: 114.8, 63.2, 77.1, 115.4, 43.5, 49.6.

## [12] Models of Random Real Partition Applying a Variety of Distributions

One wonders what happens if another statistical distribution is substituted for the continuous Uniform! Could Random Real Partition be generalized to other distributions?! Surely the correspondence to the Exponential Distribution would be ruined in such substitution of distribution, but our quest is to get ever closer to Benford with much lower SSD.

Let us perform several Monte Carlo empirical tests on other (non-Uniform) statistical distributions to examine the possibility of improving and generalizing this random process.

Normal(19, 4) Partitioned via 35,000 Marks - {28.5, 17.4, 13.2, 10.5, 8.3, 6.8, 6.0, 5.0, 4.3}
Benford's Law for First Significant Digits - {30.1, 17.6, 12.5, 9.7, 7.9, 6.7, 5.8, 5.1, 4.6}
SSD value is **4.0**, and this lower value (in comparison to the Uniforms) indicates that applying the Normal in partitions may be in general closer to Benford than the application of the Uniform.

Exponential(2.4) Partitioned via 20,000 Marks - {29.2, 17.4, 13.1, 9.7, 8.2, 7.0, 5.7, 5.4, 4.4}
Benford's Law for First Significant Digits - {30.1, 17.6, 12.5, 9.7, 7.9, 6.7, 5.8, 5.1, 4.6}
SSD value is **1.4**, and this extremely low value (in comparison to the Uniforms and the Normals) indicates that applying the Exponential in partitions may be in general closer to Benford than the applications of the Uniforms or the Normals.

Could the superior result of the partitioning via the Exponential be explained in terms of the fact that the Exponential distribution itself is quite close to Benford? Let us examine then random partitions with the applications of the highly logarithmic distributions of the Lognormal and k/x to get a clue.

k/x on (1, 10) Partitioned via 30,000 Marks - {30.0, 18.0, 12.2, 9.4, 7.8, 6.9, 5.7, 5.2, 4.7}
Benford's Law for First Significant Digits - {30.1, 17.6, 12.5, 9.7, 7.9, 6.7, 5.8, 5.1, 4.6}
SSD value is **0.4**, and such exceedingly low value indicates that applying the k/x distribution gets us extremely close to Benford!

Lognormal(9.3, 1.7) Partitioned via 35,000 M. - {30.0, 17.6, 12.4, 10.0, 8.0, 6.6, 6.0, 5.0, 4.4}
Benford's Law for First Significant Digits - {30.1, 17.6, 12.5, 9.7, 7.9, 6.7, 5.8, 5.1, 4.6}
SSD value is **0.2**, and such exceedingly low value indicates that applying the Lognormal gets us extremely close to Benford! [Note on Notation: Lognormal with shape = 1.7 and location = 9.3.]

Yet, intuitively it's very clear that the logarithmic behavior of the distribution being utilized to randomly partition the interval has nothing to do with the resultant logarithmic behavior of the parts. But if this is the case, then what could account for the correlation between logarithmic-ness of the partitioning distribution itself and the logarithmic-ness of the resultant set of parts as was observed in the simulations above? The straightforward answer to this dilemma is that it is not the logarithmic-ness of the partitioning distribution that induces better logarithmic results for the set of parts, but rather its skewness! Figure 25 depicts several random realizations from a highly (negatively) skewed distribution, resulting in many small parts on the right where density is high but only a few big parts on the left where density is low. Skewness implies that not all regions on the x-axis are equally likely, some regions have higher densities and some have lower densities. As a consequence, the regions with higher densities are those producing numerous marks which are (relatively speaking) narrowly crowded together and parenting many small parts. On the other hand, regions with lower densities, especially long tails of distributions, are those producing fewer marks which are (relatively speaking) widely spread apart from each other, and parenting only a few big parts. Clearly, skewness of the generating distribution induces more pronounced quantitative differentiation in the generated set of parts; and since quantitative differentiation is the driving force of digital logarithmic-ness, hence partitioning along skewed distributions induces more skewness and better result for the parts.

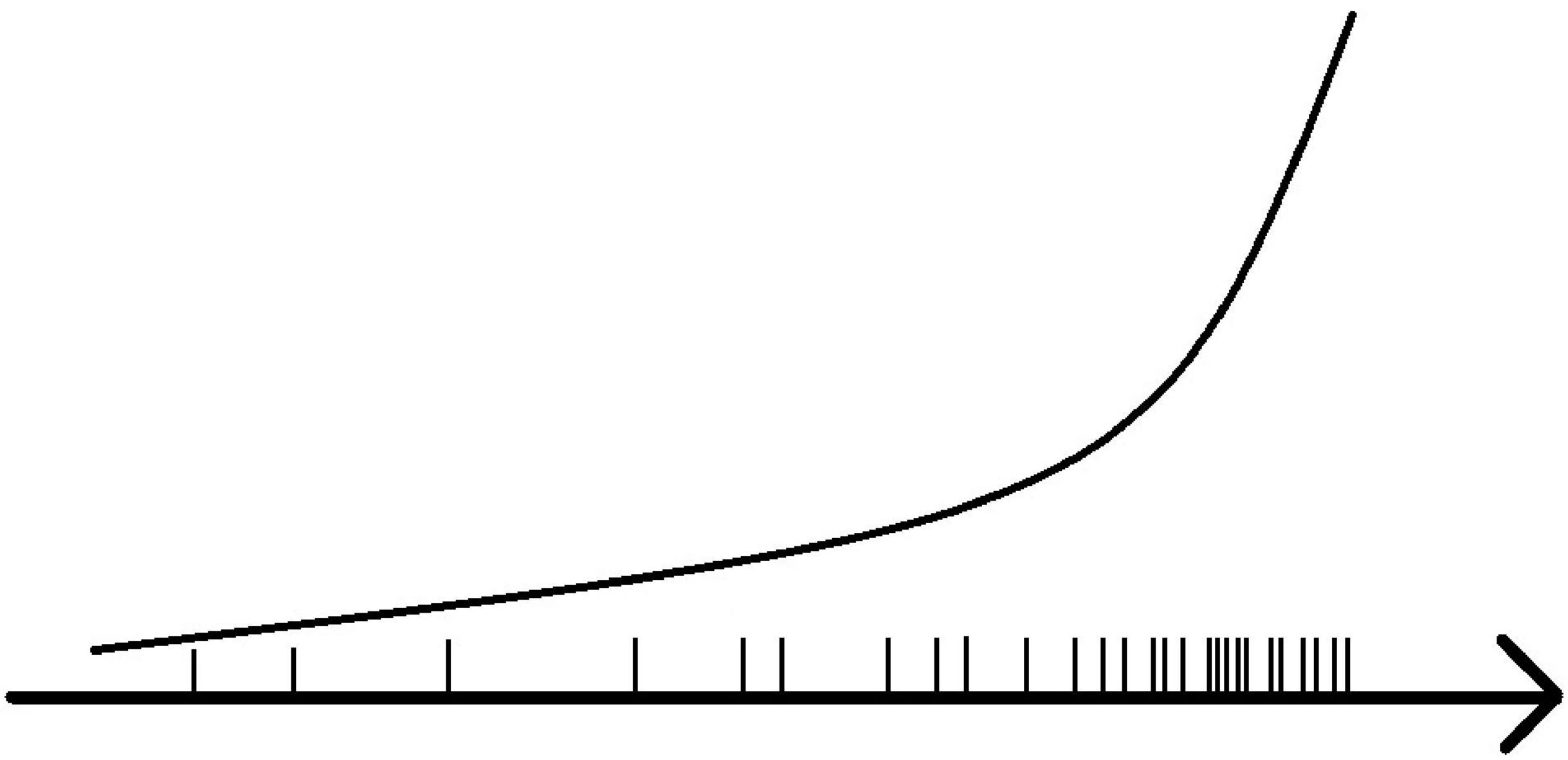

**Figure 25:** It is not Logarithmic-ness but Skew-ness which Improves Results in Real Partitions

Let us verify the above argument at least in one concrete case. The distribution PDF(x) = $k*x^3$ defined over (1, 50) [*where* $k = 4/(50^4 - 1) = 0.00000064$] is highly skewed, resembling a great deal the curve in Figure 25. Its digital configuration is totally non-Benford. This is so mostly because it has a rising histogram, where the small is rare and the big is numerous. Yet, according to the argument above, its highly skewed density implies that partitioning (1, 50) along it should lead to a strong Benford configuration for the parts! Let us then check this result empirically:

$k*x^3$ on (1, 50) Partitioned via 25,000 Marks - {30.8, 17.5, 12.0, 9.1, 8.0, 6.8, 5.6, 5.4, 4.8}
Benford's Law for First Significant Digits - {30.1, 17.6, 12.5, 9.7, 7.9, 6.7, 5.8, 5.1, 4.6}
SSD value is **1.3**, and this very low value indicates that partitioning along highly skewed distributions leads to stronger Benford results for the parts (in comparison to symmetrical ones).

Note: Here the Cumulative Function is $(k/4)*(x^4 - 1)$ over the range (1, 50), and it is set equal to the Uniform(0, 1) in Monte Carlo computer simulations in order to generate realizations.
The random number function in MS-Excel called RAND() generates values uniformly distributed on (0, 1). Hence, in the expression RAND() = $(k/4)*(x^4 - 1)$ we solve for x to obtain the relationship x = 4th root of [ RAND()*(4/k) + 1 ], x being a random realization from $k*x^3$ distribution.

Yet, there seems to be a stubborn refusal on the part of the Normal and especially on the part of the Uniform towards letting Random Real Partition achieving a near perfect convergence to the logarithmic configuration, and no matter how many marks are being placed on the x-axis, SSD never seems to manage to get below 6 or 5.

The failure to achieve complete Benfordness when the Uniform is chosen to serve as the distribution responsible for placing the marks on the x-axis can be neatly explained via the intimate connection between the Uniform and the Exponential distributions. This well-known result in Mathematical Statistics states that distances between [Uniformly] random markings along the length of one spacial dimension are distributed as in the Exponential Distribution. The Exponential Distribution is specified by the single parameter called Lambda. Lambda is the event rate, namely the average number of events per unit time, or the average number of spacial occurrences per unit length. In the context of Benford's Law, first digits of the Exponential Distribution are known to be close to LOG(1 + 1/d) but not close enough, no matter what value is assigned to parameter Lambda, with SSD fluctuating roughly between 5 and 13.

## [13] Deterministic Dependent Partition Applying Fixed Ratios

The submission to the publisher of the manuscript of Kossovsky's book in November 2013, and the publishing of an article about Random Dependent Partition in Dec 2013 by the mathematician Steven Miller, took place independently and at about the same time, both suggesting essentially the same process. Discussions about the process can be found in Kossovsky (2014) Chapter 92 titled "Breaking a Rock Repeatedly into Small Pieces is Logarithmic", and Miller et al (2013) "Benford's Law and Continuous Dependent Random Variables"; the latter containing a rigorous mathematical proof as well as the correct statement regarding fixed deterministic ratio of partitions.

In Kossovsky (2014) the description is of a piece of rock of a given weight being repeatedly broken into binary smaller pieces. In Miller Steven et al (2013) the description is of an original one-dimensional linear stick of length L being repeatedly broken into binary smaller segments. Surely, both descriptions, of rocks and sticks, representing weights and lengths, are simply particular manifestations of the same generic idea of repeated random divisions of a given conserved quantity into smaller and smaller ones.

Interestingly, Miller includes the case of a deterministic fixed p ratio [and its complement (1 - p)] breakup, such as in, say, 20% - 80%, or 40% - 60%, instead of utilizing random ratios via the Uniform(0, 1) distribution. The fixed ratio result is constrained to cases where LOG((1 - p)/p) cannot be expressed as a rational N/D value, N and D being integers. When LOG((1 - p)/p) is a rational number no convergence is found. Miller has provided rigorous mathematical proofs for both scenarios, for the random case, as well as for the deterministic case. There are two factors which render Miller's deterministic case less relevant to real-life physical data sets. The first factor is the extremely slow rate of convergence in the fixed deterministic case, which necessitates thousands if not tens of thousands of stages, in contrast to the random case which rapidly converges extremely close to the logarithmic after merely, say, 10 or 13 cycles! It may be that there exist some very long deterministic decomposition processes in nature which involve such huge number of stages, but one would conjecture that these must be quite rare in nature even if they exist at all, and that they are not the typical data sets that the scientist, engineer, or the statistician encounters. The second factor is the rarity with which decompositions in nature are conducted with such precise, fixed, and orderly ratio, and perhaps this never occurs at all. Mother Nature is known to behave erratically and chaotically when she feels weak and unable to hold her compounds intact anymore, passively letting them decompose slowly and gradually, and be partitioned in a random way, using totally random ratios. She is even more chaotic when she rages and in anger spectacularly explodes her constructs into bits and pieces rapidly in quick successions; and to expect her to deliberately, calmly, and steadily apply continuously the same fixed p ratio is unrealistic. Expecting to find in nature a decomposition process that (1) comes with an enormous number of stages, and (2) that it steadily keeps the same deterministic fixed p ratio throughout, is being doubly unrealistic.

As one simulation example, a rock weighing 33 kilograms is repeatedly broken in 13 stages into $2^{13}$ = 8192 pieces by randomly deciding on the breakup proportions via the Uniform(0, 1).

Breaking a 33-kilo Rock in 13 Stages - {29.9, 17.2, 12.6, 9.7, 8.1, 6.7, 5.9, 5.6, 4.3}
Benford's Law for First Order Digits - {30.1, 17.6, 12.5, 9.7, 7.9, 6.7, 5.8, 5.1, 4.6}

SSD value is 0.6, and such extremely low value indicates that Random Rock Breaking process is extremely close to Benford. The 3 smallest pieces and the 3 biggest pieces after the 13th stage are: {0.0000000000079, 0.0000000002123, 0.0000000004750, … , 0.42, 0.44, 1.45}.
Order of magnitude seems to be incredibly large, calculated as LOG(1.45/0.0000000000079) = LOG($1.83*10^{11}$) = 11.3, and (on the face of it) this guarantees a near perfect logarithmic behavior. But this is deceiving! The upper 1% and lower 1% whiskers (extreme outliers) are attempting to make us believe that the spread of the data is huge, but in fact it is not as dramatic as it seems. Avoiding the exaggeration of the whiskers and concentrating only on the core 98% of the data, we obtain: 1%-percentile = 0.0000000259, and 99%-percentile = 0.0645, hence a much more realistic order of magnitude is LOG(0.0645/0.0000000259) = LOG(2494364) = 6.4, and which is still considered rather unusually high in Benford's Law, guaranteeing an excellent fit to the logarithmic. CPOM = ($P_{90\%}$ / $P_{10\%}$) = (0.0068572/0.0000025) = 2697; it's quite high!
Figure 26 depicts the histogram of almost all the resultant pieces which lie on the short interval (0, 0.07). It shows severe skewness where the small is numerous and the big is rare. The choice of a logarithmic vertical scale enables us to see the overall data structure clearly, but unfortunately it masks the dramatic fall in the histogram, namely its severe skewness.

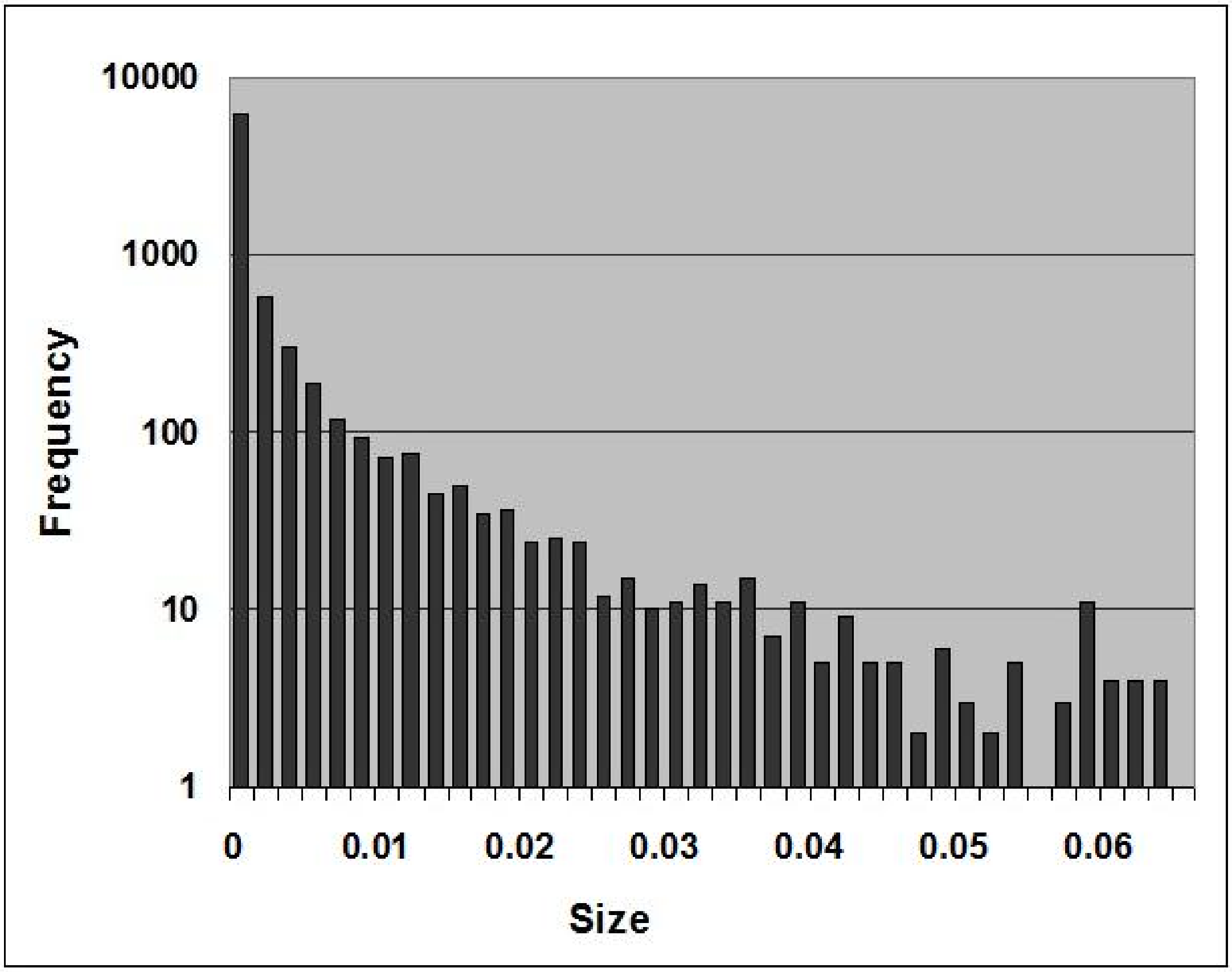

**Figure 26**: Histogram of the Pieces of 33-Kilogram Rock Broken Randomly in 13 Stages

Further scrutinizing the set of 8192 randomly obtained pieces after the 13th stage reveals that the original 33-kilogram rock has been thoroughly divided into totally distinct parts, so that the set of 8192 pieces does not contain any duplicated quantity! In other words, the resultant set of 8192 pieces contains 8192 distinct sizes! This fact should certainly be considered as a strong 'pro-partition' feature of the Random Rock Breaking model, explaining why it gets so fast and so close to Benford in one fell swoop, after only very few stages.

From numerous empirical experimentations (simulations), it is concluded that the rate of convergence for Random Rock Breaking is quite rapid, and after 10 stages, having merely $2^{10}$ or 1024 pieces, SSD is almost always below 10.0. After 14 stages, SSD is almost always below 1.0.

It is essential to visualize how quantities here evolve algebraically; and this is accomplished by writing the process carefully stage by stage. Assuming the original weight of the rock is 1 kilogram, $U_J$ being the Jth realization from the Uniform(0, 1) in the whole simulation scheme, and $(1 - U_J)$ being its complement:

{1}

-------------------

$\{U_1, (1 - U_1)\}$

---------------------------------------------------------------------

$\{U_2*(U_1),\ \ (1 - U_2)*U_1,\ \ U_3*(1 - U_1),\ \ (1 - U_3)*(1 - U_1)\}$

---------------------------------------------------------------------------------------------------------------

$\{U_4*(U_2*U_1),\ \ (1 - U_4)*(U_2*U_1),\ \ U_5*(1 - U_2)*U_1,\ \ (1 - U_5)*(1 - U_2)*U_1,$
$U_6*U_3*(1 - U_1),\ \ (1 - U_6)*U_3*(1 - U_1),\ \ U_7*(1 - U_3)*(1 - U_1),\ \ (1 - U_7)*(1 - U_3)*(1 - U_1)\}$

---------------------------------------------------------------------------------------------------------------

And so forth to higher stages.

Clearly, it would be exceedingly rare to find identical values (pieces) here. For example, it is possible in principle to have $U_4*(U_2*U_1) = U_6*U_3*(1 - U_1)$, but this would require a very particular set of choices of $U_J$'s and which is extremely unlikely and rare, having zero probability in a formal mathematical sense. The Uniform(0, 1) contains unaccountably infinite real numbers!

The above sequence of algebraic expressions demonstrates that Random Rock Breaking can be thought of also as a multiplicative process albeit with strong dependencies between the terms. Since multiplication processes in general are known to be highly skewed quantitatively and to have strong logarithmic digital tendencies, we expect (and get) skewness and Benford behavior here. This assertion relies on the fact that the arithmetical terms of the process involve numerous multiplicands, and this fact is more than sufficient to obtain a powerful logarithmic tendency and skewness for the products - given that multiplicands come with high order of magnitude. The use of the Uniform(0, 1) which ['supposedly'] possesses infinitely large order of magnitude guarantees that the process can easily overcome any possible challenge from the algebraic dependencies between the terms. Order of magnitude here is naively calculated as LOG(1/0) = LOG(Infinite) = Infinite. Yet, CPOM is (0.9)/(0.1) = 9, and not infinite! As discussed in Kossovsky (May 2016) "Arithmetical Tugs of War and Benford's Law", multiplication processes involving random variables with high order of magnitude lead to a strong and rapid convergence to the logarithmic. By its very nature, Random Rock Breaking model cannot use any other random variable with low order of magnitude such as for example Uniform(5, 7), because it needs to break a whole quantity of a particular rock weight into two fractions, and this can only be achieved via Uniform(0, 1) which is of high order of magnitude.

Let us now turn our attention to Miller's model of a deterministic fixed ratio of breakup.
It is essential to visualize how quantities evolve here algebraically; and this is accomplished by writing the process carefully stage by stage. Assuming the original weight of the rock is 1 kilogram; p is the fixed deterministic ratio; and s = (1 – p) is its complement:

{1}

--------

{p, s}

-------------------

{pp, ps, sp, ss}

--------------------------------------------

{ppp, pps, psp, pss, spp, sps, ssp, sss}

--------------------------------------------------------------------------------------------------------------

{pppp, ppps, ppsp, ppss, pspp, psps, pssp, psss, sppp, spps, spsp, spss, sspp, ssps, sssp, ssss}

--------------------------------------------------------------------------------------------------------------

{ppppp, pppps, pppsp, pppss, ppspp, ppsps, ppssp, ppsss, psppp, pspps, pspsp, pspss, psspp, pssps, psssp, pssss, spppp, sppps, sppsp, sppss, spspp, spsps, spssp, spsss, ssppp, sspps, sspsp, sspss, ssspp, sssps, ssssp, sssss}

--------------------------------------------------------------------------------------------------------------

And so forth to higher stages.

This relates to the Binomial Distribution. Clearly there are many repeating terms here.
In the 2nd stage, **ps = sp**. In the 3rd stage, **pps = psp = spp**, as there are 3 ways to order such a product, where s is on the right, or in the center, or on the left. Also: **pss = sps = ssp**. In the 4th stage, **ppps = ppsp = pspp = sppp**, as well as **ppss = psps = pssp = spps = spsp = sspp**.

In how many ways can $(P^{\mathbf{r}})(S^{\mathbf{p}})$ be arranged? First, let us define n = r + p, namely the number of stages. The number of ways they can be arranged is given by the Binomial Coefficient:

$$\binom{n}{r} = \frac{n!}{(r)!\,(n-r)!}$$

Applying the binomial coefficient for ppps = ppsp = pspp = sppp, namely $(P^{\mathbf{3}})(S^{\mathbf{1}})$:

$$\binom{4}{3} = \frac{4!}{(3)!\,(1)!} = \frac{4*3*2*1}{(3*2*1)(1)} = 4$$

Applying the binomial coefficient for ppss = psps = pssp = spps = spsp = sspp, namely $(P^{\mathbf{2}})(S^{\mathbf{2}})$:

$$\binom{4}{2} = \frac{4!}{(2)!\,(2)!} = \frac{4*3*2*1}{(2*1)(2*1)} = 6$$

Hence there are 4 ways to arrange ppps, ppsp, pspp, sppp.
Hence there are 6 ways to arrange ppss, psps, pssp, spps, spsp, sspp.

As a demonstration of the incredibly slow convergence rate in Miller's deterministic case, a rock weighing 33 kilograms is repeatedly broken in 13 stages into $2^{13} = 8192$ pieces by deterministically deciding on the breakup fixed ratio of p = 0.7 and s = (1 – p) = 0.3.

Breaking 33-kilo Rock, 13 Stages, 70% - 30% - {38.7, 4.4, 15.7, 15.7, 1.0, 3.5, 0.0, 20.9, 0.0}
Benford's Law First Order Significant Digits - {30.1, 17.6, 12.5, 9.7, 7.9, 6.7, 5.8, 5.1, 4.6}

Digit distribution is nowhere near Benford as yet after only 13 stages, and SSD value is 658.1.

Convergence is exceedingly slow! Deterministic Rock Breaking process requires a truly huge number of stages in order to converge to Benford.

Scrutinizing the above set of 8192 deterministically obtained pieces after the 13th stage reveals that the original 33-kilogram rock hasn't been divided into numerous sizes, and that the set of 8192 pieces contains numerous duplicated quantities. So much so, that there exists only 14 distinct sizes within the entire set of 8192 pieces! The details of these 14 sizes are as follow:

| **Size** | **Occurrences** |
|---|---|
| 0.0000053 | 1 |
| 0.0000123 | 13 |
| 0.0000286 | 78 |
| 0.0000668 | 286 |
| 0.0001560 | 715 |
| 0.0003639 | 1287 |
| 0.0008491 | 1716 |
| 0.0019812 | 1716 |
| 0.0046228 | 1287 |
| 0.0107865 | 715 |
| 0.0251685 | 286 |
| 0.0587266 | 78 |
| 0.1370287 | 13 |
| 0.3197337 | 1 |

For example, there are 1716 pieces of size 0.0019812. Each such piece springs from:

0.0019812 = 33*0.7*0.7*0.7*0.7*0.7*0.7*0.7*0.3*0.3*0.3*0.3*0.3*0.3, or from
0.0019812 = 33*0.3*0.7*0.7*0.7*0.7*0.7*0.7*0.3*0.3*0.7*0.3*0.3*0.3 or from
0.0019812 = 33*0.3*0.3*0.3*0.7*0.7*0.7*0.7*0.7*0.7*0.7*0.3*0.3*0.3
etc. etc.

and in general
$0.0019812 = 33*(0.7^{7})(0.3^{6})$.

In how many ways can $(0.7^{7})(0.3^{6})$ be arranged? The answer is 1716.
Here 7 + 6 = 13, signifying that there are 13 stages in this process.

The Binomial Coefficient here is:

$$\binom{13}{7} = \frac{13!}{(7)!\,(6)!} = \frac{6227020800}{5040 * 720} = \frac{6227020800}{3628800} = 1716$$

Such meager set of only 14 distinct sizes for the resultant set of 8192 pieces, with so many repetitions, does not bode well for Benford and skewness! Indeed the process hasn't even begun to converge here.

Such meager set of only 14 distinct sizes for the resultant set of 8192 pieces, with so many repetitions, reveals that the original 33-kilogram rock hasn't been thoroughly divided, and that this process of partition is incomplete.

Such meager set of only 14 distinct sizes for the resultant set of 8192 pieces, with so many repetitions, and the fact that the sizes are much smaller than 33, remind us so much of Refined Equipartition here, much more so than of Random Rock Breaking! [*well, only at this early 13th stage, but it's not so later*.]

Within the perspective of Refined Equipartition, these two facts also explain why the process hasn't even begun to arrive at the Benford configuration yet. In Refined Equipartition it was absolutely necessary to consider the set of all possible partitions to arrive at a Benford-like configuration, but here only a single partition is considered, and this explains why this process isn't yet Benford.

Interestingly, values within this set of 14 sizes increase steadily by a multiplicative factor of 2.33 (almost), pointing to a bit more similarity here in a sense with Refined Equipartition model which increases the sizes steadily with the additive value of 1.

# [14] Chaotic Rock Breaking

The connection or relevance of the Refined Equipartition model to physical real-life data sets is often questioned by those who know Mother Nature well and are familiar with the way she works, and it is posited that she would probably never bother to delicately break her quantities carefully only along sets of parts with exact integral relationships, and that she would definitely not restrict herself only to much smaller and refined pieces (namely n << N). Indeed, this line of thought suggests that also Random Real Partition and Random Dependent Partition are not the typical processes that she likes doing or even capable of performing.

Let us imagine a real-life assembly-line with workers and management as in typical large corporations, attempting to physically perform the Random Dependent Partition process on a very long metal pipe. First the workers decide on the first random location where the pipe is to be cut, followed by actual cutting, and then the designations of "1" and "2" tags for the newly created pieces are made. This is followed by the orderly cutting of piece "1" and then piece "2" using random location within each piece. Without such designations and tags, there exists the possibility that by mistake one piece is cut twice leaving the other piece intact. Next, the workers designate the newly created four pieces as "1", "2", "3", and "4". At this stage, piece designation becomes even more crucial to avoid confusion and mistakes, and to remember which pieces were already cut and which pieces are awaiting their turn. It is highly doubtful that Mother Nature is capable or even interested in such serious and rigid type of work.

Consideration of how Random Real Partition would take place in such real-life assembly-line setup also leads one to think that this is not the type of work that would interest Mother Nature much. In this case, hundreds or rather thousands of workers are set up at random points along the long metal pipe with saws, ceramic mills, diamond drills, and other such cutting tools, all ready to simultaneously cut the pipe when the order is given. Upon hearing the first ear-piercing whistle of the foreman, workers stand to attention ready to do the cutting. The moment the second ear-piercing whistle is heard, they all cut rapidly, forcefully, and simultaneously, while Mother Nature is looking at them benevolently from above, feeling pity for the workers for all their highly coordinated effort and forced concentration. This description is called Scenario A. It should be noted that there is no compelling reason to assume that in Random Real Partition the breakups should be 'performed' simultaneously. Indeed, there is no need whatsoever to introduce the time dimension here, although without timing and detailed description of how Random Real Partition is performed, the concept is not a partition 'process' per se, but rather mere abstraction of how a given quantity may be broken into many smaller parts. This is why one cannot ask Mother Nature to 'perform' such a partition in the abstract, unless she is told what to do precisely, stage by stage. What could be suggested here (to be called Scenario B) is the successive gentle markings one by one of random points along the long metal pipe, followed by the actual cutting in a random fashion all of these marks by a lone, underpaid, exploited, and stoic worker, who cuts the long pipe for hours on end, one mark at a time (chosen randomly).

Obviously, this discussion about Mother Nature is metaphorical, and surely there exist in nature some particular decomposition processes that perfectly match the mathematical models of Random Real Partition and Random Dependent Partition, and even perhaps Refined Equipartition approximately, but what is conjectured here is that the typical decomposition process in nature is highly chaotic, totally lacking structure.

How would temperamental Mother Nature go about breaking a rock or a pipe her way, leisurely, chaotically, and consistent with her free-spirit attitude and her strong dislike of regimentation?

Her first act in the process is the breaking of the original rock of weight X into two parts randomly via the continuous Uniform(0, 1) to decide on the proportions of the two fragments. There is no need whatsoever to designate any pieces with any tags thereafter, since order of breakups does not matter to her in the least. Her second act is the totally relaxed and random selection of any one of the two pieces, followed by its fragmentation into two parts randomly via the continuous Uniform(0, 1), resulting in 3 pieces. Her third act is the totally relaxed and random selection of any one of the three pieces, followed by its fragmentation into two parts randomly via the continuous Uniform(0, 1), resulting in 4 pieces. This continues on and on for sufficiently large number of stages to obtain the Benford configuration.

Obviously, after each stage, the number of existing pieces increases by one. Also, it is obvious that the total quantity of the entire system (overall sum - overall weight of all the pieces) is conserved throughout the entire process. The specific description of physical balls or rocks of uniform mass density being broken, and the focus on the weight variable, is an arbitrary one of course, and the generic model is of pure quantities and abstract numbers.

Schematically the process is described as follow:

1) Initial Set = {X}

2) Repeat C times:

Choose one value at random, remove it from the set of values, then split it as in {Uniform(0, 1), 1 – Uniform(0, 1)}, then place these two values back into the set.

3) Final Set of (C + 1) pieces is Benford, assuming C > 1000 approximately.

This is coined as ‘Chaotic Rock Breaking’. From the point of view of Mother Nature, this is the most natural and straightforward way to randomly break a rock into small fragments. She closes her eyes; picks up randomly one piece at a time; breaks it in a random fashion; throws the two pieces back into the pile; and leisurely repeats this procedure over and over again until she gets tired or bored, or until she notices that the Benford configuration has already been achieved almost perfectly – and which immediately extinguishes her motivation to do any further work.

The table in Figure 27 depicts digital results for 23 distinct Monte Carlo computer simulations of Chaotic Rock Breaking with varying number of resultant pieces (i.e. varying number of fragmentation acts), all starting with 100-kilogram rock, and using the random breakup ratio of the Uniform(0, 1).

In the first row of the table, 100-kilogam rock has been broken 50 times, resulting in 51 pieces, and here digital configuration is not Benford, as indicated by the very high 207.0 value for SSD. Figure 27 is not about a single process and 23 snapshots taken at different times, but rather about 23 independent and totally different Monte Carlo simulation runs (i.e. the computer simulation starts anew from the beginning with a single whole 100-kilogram rock for each of the 23 different rows.)

| # of Pieces | Digit 1 | Digit 2 | Digit 3 | Digit 4 | Digit 5 | Digit 6 | Digit 7 | Digit 8 | Digit 9 | SSD |
|---|---|---|---|---|---|---|---|---|---|---|
| **51** | 21.6 | 25.5 | 9.8 | 3.9 | 9.8 | 5.9 | 9.8 | 5.9 | 7.8 | **207.0** |
| **101** | 30.7 | 11.9 | 9.9 | 6.9 | 13.9 | 8.9 | 4.0 | 5.0 | 8.9 | **109.9** |
| **201** | 29.9 | 18.9 | 12.9 | 14.4 | 4.5 | 7.0 | 5.5 | 4.0 | 3.0 | **40.2** |
| **301** | 31.2 | 16.9 | 9.6 | 13.6 | 7.3 | 8.0 | 5.6 | 4.0 | 3.7 | **29.5** |
| **401** | 30.7 | 19.5 | 12.7 | 8.5 | 7.5 | 3.7 | 5.5 | 7.0 | 5.0 | **17.9** |
| **501** | 32.1 | 19.4 | 11.6 | 8.6 | 8.0 | 5.8 | 6.0 | 4.0 | 4.6 | **11.4** |
| **601** | 31.1 | 15.0 | 11.5 | 10.8 | 7.3 | 6.7 | 7.5 | 5.7 | 4.5 | **13.8** |
| **701** | 29.8 | 18.5 | 13.7 | 9.3 | 9.1 | 5.7 | 4.6 | 4.1 | 5.1 | **7.8** |
| **801** | 31.6 | 17.4 | 12.1 | 8.6 | 8.7 | 6.7 | 5.9 | 5.2 | 3.7 | **5.0** |
| **901** | 30.4 | 17.2 | 13.8 | 9.4 | 6.1 | 7.2 | 6.0 | 6.7 | 3.2 | **9.8** |
| **1001** | 29.3 | 18.1 | 11.3 | 10.1 | 9.5 | 6.7 | 5.5 | 5.9 | 3.7 | **6.5** |
| **1301** | 29.0 | 18.6 | 12.8 | 9.4 | 9.8 | 6.1 | 5.6 | 4.7 | 4.1 | **6.6** |
| **1501** | 32.1 | 17.6 | 12.1 | 9.6 | 7.5 | 5.6 | 5.4 | 4.9 | 5.3 | **6.4** |
| **1701** | 30.5 | 18.6 | 12.4 | 10.2 | 8.2 | 6.8 | 5.0 | 4.1 | 4.3 | **3.3** |
| **2001** | 30.1 | 16.4 | 13.4 | 9.8 | 7.4 | 6.4 | 5.6 | 5.6 | 5.3 | **3.4** |
| **2301** | 30.6 | 16.2 | 12.4 | 9.2 | 7.6 | 7.4 | 6.8 | 5.3 | 4.4 | **4.0** |
| **2501** | 29.9 | 17.0 | 12.8 | 8.9 | 9.0 | 6.9 | 5.6 | 5.2 | 4.8 | **2.6** |
| **3001** | 29.5 | 17.5 | 12.6 | 10.4 | 8.2 | 6.3 | 5.7 | 4.7 | 5.1 | **1.5** |
| **3501** | 30.2 | 18.1 | 11.5 | 9.9 | 7.8 | 6.8 | 5.8 | 5.7 | 4.3 | **1.7** |
| **4001** | 29.3 | 17.7 | 12.4 | 10.4 | 7.9 | 6.5 | 6.3 | 4.6 | 4.8 | **1.7** |
| **10001** | 30.3 | 18.0 | 12.7 | 9.3 | 7.7 | 6.6 | 5.9 | 4.9 | 4.6 | **0.5** |
| **15001** | 29.4 | 17.7 | 12.5 | 9.9 | 7.7 | 6.9 | 5.8 | 5.3 | 4.6 | **0.7** |
| **20001** | 30.0 | 17.2 | 12.4 | 9.9 | 8.0 | 6.8 | 6.0 | 5.1 | 4.8 | **0.3** |

**Figure 27**: Digital Results for 23 Chaotic Rock Breaking Processes Via Uniform(0, 1)

The chart in Figure 28 depicts the scatter plot of SSD versus the number of pieces for the simulation runs of the upper 20 rows of Figure 27. A logarithmic scale is used for the vertical axis for better visualization and clarity. Given that a value of SSD less than 10.0 is tolerated, and that any data set with SSD less than this arbitrary cutoff point of 10.0 is considered as Benford; then from these Monte Carlo simulation results one could conclude that Chaotic Rock Breaking with over 1000 fragmentations (pieces) is definitely Benford. When comparing this result with the Random Rock Breaking process where approximately 10 stages are needed at a minimum to arrive close enough to the Benford digital configuration with SSD below 10.0, and the implied $2^{10}$ or 1024 pieces, it seems that both, Chaotic and Random, have the same or very similar rates of convergence. Perhaps the adjective 'Chaotic' has stronger connotation with lack of order and total unpredictability than adjective 'Random' has, hence the motivation behind the coining of the terms 'Random Rock Breaking' and 'Chaotic Rock Breaking' reflecting the two distinct levels of randomness, yet they both seem to converge just as rapidly.

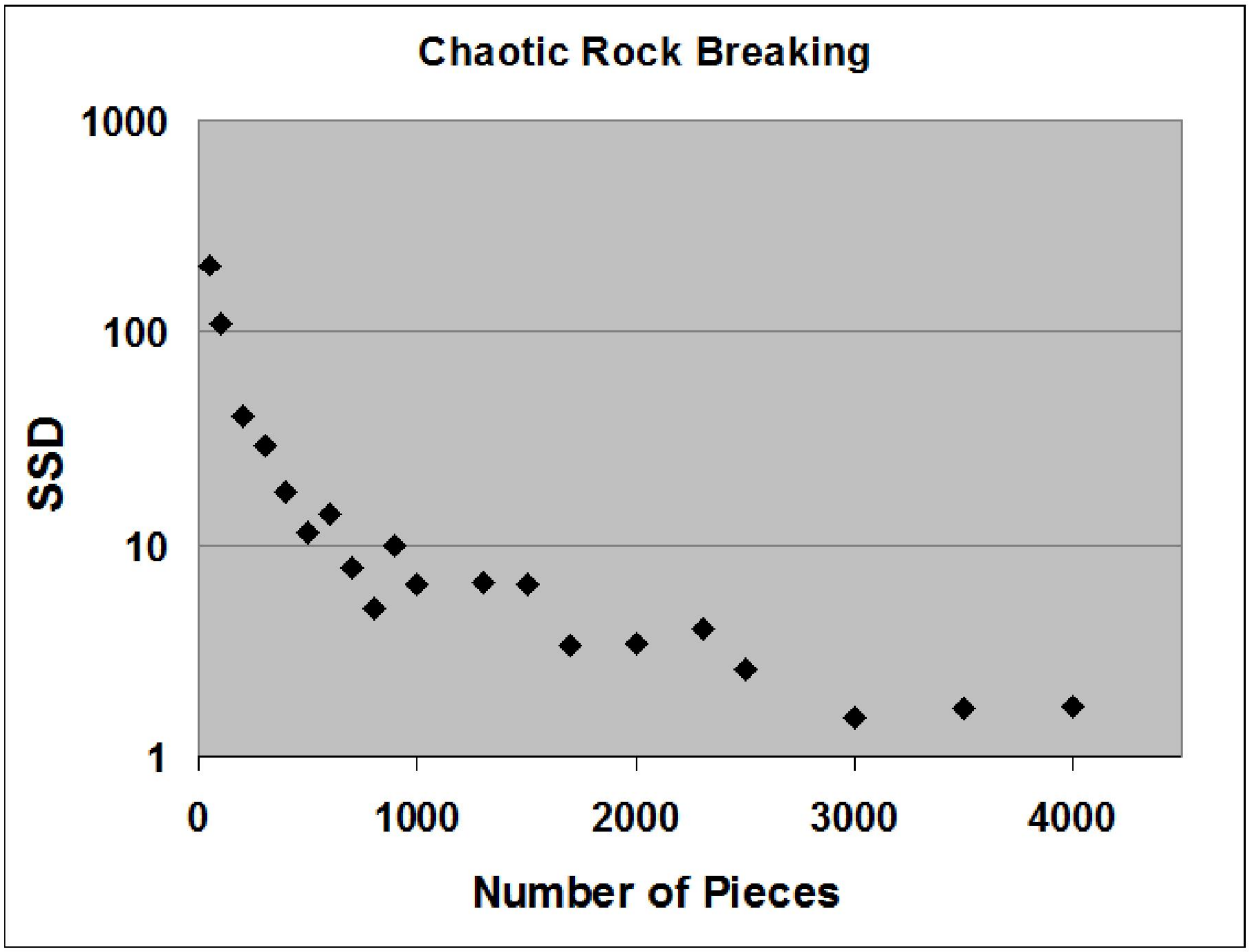

**Figure 28**: Plot of SSD Versus Number of Pieces for 20 Chaotic Rock Breaking Processes

It is beneficial to visualize how quantities here evolve algebraically; and this is accomplished by writing one possible random scenario stage by stage. Assuming the original weight of the rock is W kilogram, $U_J$ being the Jth realization from the Uniform(0, 1) in the whole simulation scheme, and $(1 - U_J)$ being its complement:

{W}     W is broken

---------------------------------------------------------------------------------------------------------------

$\{U_1*W,\ \ (1 - U_1)*W\}$      then only $U_1*W$ is broken

---------------------------------------------------------------------------------------------------------------

$\{U_2* U_1*W,\ \ (1 - U_2)* U_1*W,\ \ (1 - U_1)*W\}$      then only $(1 - U_1)*W$ is broken

---------------------------------------------------------------------------------------------------------------

$\{U_2* U_1*W,\ \ (1 - U_2)* U_1*W,\ \ U_3*(1 - U_1)*W,\ \ (1 - U_3)*(1 - U_1)*W\}$      then $U_2*U_1*W$

---------------------------------------------------------------------------------------------------------------

$\{U_4*U_2*U_1*W,\ \ (1 – U_4)*U_2*U_1*W,\ \ (1 - U_2)* U_1*W,\ \ U_3*(1 - U_1)*W,\ \ (1 - U_3)*(1 - U_1)*W\}$

---------------------------------------------------------------------------------------------------------------

And so forth, producing many more broken pieces.

Clearly, after sufficient number of such fragmentation stages (approximately 1000), on average each term (i.e. each piece) contains numerous multiplicands, and therefore the set of pieces can be thought of as emerging from a multiplication process with high order of magnitude, and thus leading to Benford. Roughly speaking, order of magnitude of the Uniform(0, 1) is calculated as LOG(1/0), and theoretically or potentially this is infinitely large. In reality it is finite and not as large, because realizations or simulations from the Uniform(0, 1) never truly get near 0; outliers are misleading and exaggerating variability; and Core Physical Order of Magnitude (CPOM) is by far the most appropriate measure here.

The common denominator in all three processes of Random Real Partition (Scenarios A and B), Random Rock Breaking, and Chaotic Rock Breaking, is that each process simply places marks randomly everywhere on the x-axis, not discriminating against any particular sub-sections within the relevant x-axis range. Yet, intriguingly, Random Rock Breaking and Chaotic Rock Breaking processes come with digital results that vary considerably from Random Real Partition [i.e. both converging fully to Benford in the limit]! If one attempts to visually follow 'in real time' the sequence of marks being formed onto the x-axis for all three processes, then it might be possible in some cases to distinguish between them and to know which process is actually being run. In Random Real Partition Scenario A, the observer sees all the marks being established simultaneously in one epic instance, although as discussed earlier, Random Real Partition is not clearly defined as a 'process' per se, involving the time dimension. In Random Rock Breaking the observer sees a whole new set of marks established between each older set of marks, in orderly and highly regimented stages. In Chaotic Rock Breaking as well as in Random Real Partition Scenario B, the observer sees new marks being established, one at a time, totally randomly and chaotically, without any apparent structure or regimentation, and without any regards to currently existing older marks, as if the process totally lacks memory of all the previous stages – yet these two processes lead to distinct quantitative and digital results!

The whole motivation in introducing the idea of Chaotic Rock Breaking is to allow Mother Nature to act naturally, totally lacking structure and order. Yet curiosity prompts us to investigate hybrid models of such chaotic fragmentation using the predictable, fixed, and deterministic ratio p. In other words, for the same process of Chaotic Rock Breaking, instead of random ratio of partition as in the Uniform(0, 1), a fixed ratio p is used throughout the entire process. Randomness is still preserved in these hybrid models because the piece to be broken is always being chosen randomly. Surely Mother Nature cannot remember nor concentrate hard enough to make sure that the same p ratio is applied in all partitions; and expecting this from her is really not realistic. In any case, let us explore empirical results from such hybrid models. The table in Figure 29 gives the results for the deterministic fixed ratio 85% - 15% of Chaotic 100-kilogram Rock Breaking. Figure 29 is not about a single process and various snapshots taken at different times, but rather about independent and totally different Monte Carlo simulation runs with varying lengths (i.e. the computer simulation starts anew from the beginning with a single whole 100-kilogram rock for each of the 12 different rows.)

Obviously, for this hybrid model with deterministic fixed 85% - 15% ratio, convergence to Benford is much slower as compared with the pure Chaotic Rock Breaking model. Moreover, it is not even known whether or not a much longer run well over 33092 pieces would manage to get SSD below 1 say. In other words, it is not even certain that full convergence in the hybrid 85% - 15% case can ever be achieved. Simulations well beyond 33000 prove somewhat time-consuming, and surely they should be explored in the future.

| # of Pieces | Digit 1 | Digit 2 | Digit 3 | Digit 4 | Digit 5 | Digit 6 | Digit 7 | Digit 8 | Digit 9 | SSD |
|---|---|---|---|---|---|---|---|---|---|---|
| 101 | 35.6 | 17.8 | 14.9 | 5.9 | 8.9 | 5.9 | 5.9 | 1.0 | 4.0 | 69.3 |
| 201 | 26.4 | 17.4 | 15.9 | 7.5 | 10.0 | 8.5 | 3.5 | 6.0 | 5.0 | 44.2 |
| 301 | 32.6 | 21.6 | 10.0 | 8.0 | 8.3 | 4.7 | 6.6 | 3.0 | 5.3 | 41.4 |
| 501 | 28.1 | 19.2 | 13.8 | 7.8 | 8.6 | 8.4 | 4.2 | 6.2 | 3.8 | 19.2 |
| 801 | 28.1 | 18.9 | 14.7 | 7.9 | 9.7 | 6.4 | 4.1 | 5.4 | 4.9 | 20.3 |
| 1001 | 27.7 | 18.7 | 15.5 | 10.2 | 8.0 | 7.3 | 4.9 | 4.8 | 3.0 | 20.0 |
| 2001 | 30.3 | 18.3 | 13.3 | 6.5 | 9.4 | 7.2 | 4.7 | 5.1 | 5.1 | 15.0 |
| 4001 | 28.8 | 18.0 | 14.5 | 7.8 | 9.6 | 7.6 | 3.9 | 5.5 | 4.1 | 17.1 |
| 7001 | 28.5 | 18.5 | 14.2 | 7.7 | 10.0 | 6.9 | 4.3 | 5.6 | 4.3 | 17.0 |
| 12001 | 28.2 | 17.4 | 14.6 | 8.0 | 9.9 | 7.1 | 4.4 | 6.3 | 4.1 | 18.5 |
| 25001 | 28.9 | 17.9 | 14.2 | 8.3 | 9.0 | 7.4 | 4.6 | 5.9 | 3.8 | 10.5 |
| 33092 | 28.8 | 17.7 | 14.2 | 9.0 | 8.9 | 7.0 | 5.0 | 5.5 | 3.8 | 7.1 |

**Figure 29**: Digital Results of Hybrid Chaotic Rock Breaking with Fixed 85% - 15% Ratio

It stands to reason that this Hybrid Chaotic Rock Breaking model with 85% - 15% fixed ratio converges to Benford by far faster than Miller's Deterministic Rock Breaking with 85% - 15% fixed ratio. As was indeed seen in the previous chapter, breaking 33-kilo rock in 13 stages, with fixed ratio of 70% - 30%, and yielding 8192 pieces, was not even close to Benford. This is so because Hybrid Chaotic Rock Breaking model still contains the random element within itself, in the sense that each piece about to be broken is chosen randomly, while Deterministic Rock Breaking on the other hand is a purely deterministic model, and there is nothing random about it.

The table in Figure 30 gives digital results for a variety of deterministic fixed ratios for the Hybrid Chaotic Rock Breaking model. All simulation runs start with 100-kilogram rock. There is a bit of a mystery here why some fixed p rates come out very close to Benford, while others show just the Benfordian tendency but without any obvious strong convergence. Certainly Miller's irrationality constraint of LOG((1 - p)/p) does not apply here at all since this is a very different mathematical model than Deterministic Rock Breaking, and even though the ratio is deterministic, yet the process selects the next piece about to be broken in a random manner. It is possible that ultimately in the limit when the number of pieces goes to infinity, or in a practical way as Monte Carlo simulations are run for a truly large number of pieces on a more powerful computer, full or partial convergence to Benford could be found here equally and uniformly for the entire variety of these deterministic fixed ratios.

| Fixed p Ratio | # Pieces | Digit 1 | Digit 2 | Digit 3 | Digit 4 | Digit 5 | Digit 6 | Digit 7 | Digit 8 | Digit 9 | SSD |
|---|---|---|---|---|---|---|---|---|---|---|---|
| 0.5000000 | 5001 | 30.2 | 9.7 | 20.7 | 8.0 | 2.2 | 8.0 | 12.6 | 0.0 | 8.7 | 256.3 |
| 0.2715402 | 5001 | 27.5 | 14.8 | 14.5 | 13.3 | 2.7 | 12.2 | 0.2 | 11.1 | 3.8 | 156.6 |
| 0.2777000 | 5001 | 31.4 | 18.2 | 12.1 | 7.4 | 8.8 | 7.2 | 6.1 | 3.2 | 5.5 | 13.5 |
| 0.3368796 | 2001 | 30.7 | 16.0 | 11.8 | 11.0 | 5.9 | 6.1 | 7.4 | 5.6 | 5.3 | 12.8 |
| 0.4254751 | 2001 | 27.0 | 22.8 | 10.7 | 6.5 | 8.7 | 7.0 | 4.3 | 3.1 | 9.7 | 83.1 |
| 0.0520000 | 2001 | 29.2 | 19.9 | 11.0 | 9.8 | 6.6 | 6.1 | 9.3 | 4.3 | 3.5 | 24.7 |
| 0.2524733 | 1001 | 29.9 | 18.8 | 14.5 | 8.6 | 7.2 | 5.9 | 5.2 | 5.7 | 4.3 | 8.5 |
| 0.2684584 | 1001 | 30.0 | 18.5 | 12.5 | 11.9 | 8.0 | 5.6 | 4.1 | 9.3 | 0.2 | 46.3 |

**Figure 30**: Digital Results of Hybrid Chaotic Rock Breaking with a Variety of Fixed Ratios

# [15] Random Minimum Breaking

Instead of randomly selecting the next piece to be partitioned as in Chaotic Rock Breaking, a process is envisioned where at each stage the minimum of all currently existing pieces is selected and partitioned randomly via the Uniform(0, 1). In other words, the minimum is constantly being partitioned in a random fashion, over and over again. Here, resultant data of this process rapidly approaches extreme quantitative skewness, and Benford digital behavior is confirmed. Surely, by selecting the minimum at each stage of partition, skewness is constantly being increased in the system. The fact that the smallest piece is partitioned into two pieces [thus increasing the number of the small in the system] while big pieces are left intact [leaving the number of the big in the system unchanged] implies that at each stage greater skewness is achieved.

In one Monte Carlo simulation run of 500-kilogram rock, with 400 stages of splitting up the minimum via the Uniform(0, 1), the following digital configuration was obtained:

Breaking Min of 500-kilo Rock in 400 Stages - {29.5, 17.3, 15.8, 10.3, 5.0, 5.0, 5.8, 6.5, 5.0}
Benford's Law First Order Significant Digits - {30.1, 17.6, 12.5, 9.7, 7.9, 6.7, 5.8, 5.1, 4.6}

Digit distribution is quite close to Benford, with SSD value of 24.9.

The biggest 18 values of the ordered resultant data set after the 400th stage are as follows:

{378.0, 63.6, 29.4, 18.9, 7.3, 2.0, 0.72, 0.08, 0.03, 0.00159, 0.00046, 0.000046, 0.00001125, 0.00000504, 0.000000518, 0.000000236, 0.000000125, 0.00000000258}

The final minimum after the 400th stage is the exceedingly low value of $8.28*10^{-306}$!

Log histogram for the resultant data set after the 400th stage is nearly uniform, implying that resultant data set is of the deterministic flavor in Benford's Law. The logarithm of successive elements of the ordered resultant data set decreases by 0.77 on average.

The logarithm of the biggest 18 values of the ordered resultant data set after the 400th stage are as follows:

{2.58, 1.80, 1.47, 1.28, 0.87, 0.29, -0.14, -1.07, -1.54, -2.80, -3.34, -4.34, -4.95, -5.30, -6.29, -6.63, -6.90, -8.59}

Surely, Mother Nature would adamantly refuse to perform this tedious process. She avoids at all cost such regimented and intense partition style where the next minimum must be carefully ascertained at each stage.

## [16]  Random Maximum Breaking

When the maximum is repeatedly selected and partitioned randomly via the Uniform(0, 1), resultant data set approaches the Uniform Distribution in the limit as the number of such partitions goes to infinity. Such total absence of quantitative skewness precludes Benford digital behavior. Here, the maximum is constantly being partitioned in a random fashion, over and over again. Surely, by selecting the maximum at each stage of partition, skewness is constantly being decreased in the system. The fact that the biggest piece is partitioned into two pieces [thus increasing the number of the big in the system] while small pieces are left intact [leaving the number of the small in the system unchanged] implies that at each stage greater uniformity and evenness is achieved.

In one Monte Carlo simulation run of 100-kilogram rock, with 4000 stages of splitting up the maximum via the Uniform(0, 1), the following digital configuration was obtained:

Breaking Max of 100-kilo Rock in 4000 Stages - {22.0, 23.0, 22.9, 21.5, 2.2, 2.3, 1.9, 2.0, 2.3}
Benford's Law First Order Significant Digits    - {30.1, 17.6, 12.5,  9.7, 7.9, 6.7, 5.8, 5.1, 4.6}

Digit distribution is not of the Benford configuration at all, as indicated by the very large SSD value of 424.9.

Quantitatively, resultant data set after the 4000th stage is uniformly distributed on the interval between 0.000007 and 0.049550. Out of 4001 pieces, 3225 pieces fall uniformly on the interval (0.01, 0.05), with only 776 pieces falling below 0.01. This explains why first digits are nearly evenly distributed between digits 1, 2, 3, and 4.

The smallest 7 values and the biggest 7 values, of the ordered resultant data set after the 4000th stage, are as follows:

{0.000007,  0.000010,  0.000012,  0.000032,  0.000045,  0.000051.  0.000053, . . .
. . .  0.049492,  0.049497,  0.049506,  0.049527,  0.049534,  0.049546,  0.049550}

Surely, Mother Nature would adamantly refuse to perform this tedious process. She avoids at all cost such regimented and intense partition style where the next maximum must be carefully ascertained at each stage.

# [17] Mathematical Model for Equipartition

Let us introduce standard notations for the Equipartition model. We begin with the Equipartition Parable of a group of 33 spies being separated into independent cells of maximum 5 spies each.

Quantity to be partitioned: X = 33
Allowed set of parts: $\{x_j\}$ = {1, 2, 3, 4, 5}   [ index j running from 1 to 5 ]

Writing down some particular partitions could prove quite tedious and too long at times. For example: **{1, 1, 1, 1, 1, 1, 1, 1, 1, 1, 1, 1, 1, 1, 1, 1, 1, 1, 1, 1, 2, 2, 2, 2, 2, 3}** is too long and ink-consuming. Since there are usually repeating terms in Equipartition, it would be much easier and more efficient to express them more concisely by simply counting the repetitions $n_j$ of each part $x_j$. For the partition above, there are 20 ones, 5 twos, 1 three, 0 fours, and 0 fives, hence $\mathbf{n_j}$ **= {20, 5, 1, 0, 0}**.

Using Dot Product notation, the fact that any given partition here must add up to 33 can be succinctly expressed as **33 = {20, 5, 1, 0, 0}*{1, 2, 3, 4, 5}**.

The other partitions

{5, 5, 5, 5, 5, 5, 2, 1}
{2, 4, 5, 5, 1, 3, 1, 1, 1, 5, 5}
{1, 1, 1, 1, 1, 1, 1, 1, 1, 1, 2, 2, 2, 2, 2, 3, 5, 5}
{4, 5, 4, 5, 4, 5, 2, 2, 2}
{3, 3, 3, 3, 3, 3, 3, 3, 3, 3, 3}

are written in the format X = $\{n_j\}$*$\{x_j\}$, where $\{x_j\}$ is a fixed vector, while $\{n_j\}$ varies:

33 = {1, 1, 0, 0, 6}*{1, 2, 3, 4, 5}
33 = {4, 1, 1, 1, 4}*{1, 2, 3, 4, 5}
33 = {10, 5, 1, 0, 2}*{1, 2, 3, 4, 5}
33 = {0, 3, 0, 3, 3}*{1, 2, 3, 4, 5}
33 = {0, 0, 11, 0, 0}*{1, 2, 3, 4, 5}

And which can be expanded and written explicitly as:

33 = 1*1 + 1*2 + 0*3 + 0*4 + 6*5
33 = 4*1 + 1*2 + 1*3 + 1*4 + 4*5
33 = 10*1 + 5*2 + 1*3 + 0*4 + 2*5
33 = 0*1 + 3*2 + 0*3 + 3*4 + 3*5
33 = 0*1 + 0*2 + 11*3 + 0*4 + 0*5

In general, a conserved positive integral quantity X is partitioned exclusively into positive integral pieces $\{x_j\}$.

These pieces are a fixed set of N positive integral quantities $\{x_j\}$ and are called the '**part set**'. The part set is fixed within any given Equipartition model, and all possible partitions are written with respect to this fixed set, but each Equipartition model has its own distinct part set.

$\{x_j\} = \{ x_1, x_2, x_3, \ldots, x_N \}$. Typically $\{x_j\}$ is a set of consecutive integers starting at 1, as:

$$\mathbf{\{x_j\} = \{ 1, 2, 3, \ldots, N \}}.$$

For any particular partition, each $x_j$ could be repeated T times and so $n_j = T$, or it may occur only once and so $n_j = 1$, or it may not occur at all and so $n_j = 0$.

For any particular partition, X is written as a linear combination of $\{x_j\}$ by way of $\{n_j\}$ as:

$$X = n_1*x_1 + n_2*x_2 + n_3*x_3 + \ldots + n_N*x_N$$

Or in Dot Product notations: $X = \{n_j\}*\{x_j\}$.

The '**number of parts set**', namely $\{n_j\}$, of N non-negative integers is of course not fixed, but rather it varies according to the particular partition chosen. Obviously the set $\{n_j\}$ is restricted to those linear combinations that add up to X.

The partition of X is then expressed as a linear combination of $\{x_j\}$ and $\{n_j\}$:

$$X = \sum_{j=1}^{N} n_j * x_j$$

The variable vector $\{n_j\}$ specifies a particular partition of X.

It is preferable that the part set $\{x_j\}$ starts at 1, namely that $x_1 = 1$. This is so in order to ensure that a partition will always exist for every X.

The part set $\{x_j\}$ should be conveniently written as monotonically increasing set, namely that $x_{j+1} > x_j$ for all j.

Obviously, the part set should consist only of distinct values, namely that $x_j \neq x_i$ for all $j \neq i$.

In the final description of this abstract model, all $\{n_j\}$ partitions are given equal weights. This is the reason for the prefix '**equi**' in the term 'Equipartition'.

The term 'configurational entropy' in Thermodynamics refers to the assumption that all possible system configurations are equally likely. In the same vein, here all possible partitions $\{n_j\}$ are considered as equally-likely, and this is consistent with the above assignment of equal weights to all partitions. Physics inspires us, so we borrow from physics the principle regarding equality of all possible configurations. Fortunately and perhaps surprisingly, imitating nature in this abstract mathematical model leads to Benford under certain conditions! Yet, this is **not** regarded as a statement of some physical law of nature about real-life actual partitions, and once this is stated mathematically it can be run and simulated successfully on the computer without direct reference to physical reality or any laws of nature.

Equipartitions fall into 3 distinct categories, Complete Equipartition, Refined Equipartition, and Irregular Equipartition.

**Complete Equipartition** is one in the style of Integer Partitions of Number Theory with a smooth and simple part set $\{x_j\} = \{1, 2, 3, \dots , X\}$. Here N = X and $x_N = X$. The part set $\{x_j\}$ increases monotonically and nicely by one integer at a time from 1 all the way to X. It is called 'complete' since it incorporates all possible breakups.

**Refined Equipartition** is one where the part set $\{x_j\}$ contains only integers that are much smaller than X thus ensuring that X is being broken only into much smaller and refined parts, resulting in a significant fragmentation of the original X quantity.
Formally stated: $N << X$, or as: $x_N << X$. Here the largest (last) element in the part set is much smaller than X. Also (for some results) it is thought of as a limiting process where $x_N$ (and by implication N) is fixed, while X tends to infinity. It should be noted that the part set $\{x_j\}$ in refined equipartition increases nicely and monotonically one integer at a time from 1, so that there are no gaps. Mathematical applications of general results from <u>restricted</u> Integer Partition in Number Theory are of course possible even for this limited/refined partition.

**Irregular Equipartition** is one in which the part set $\{x_j\}$ does not increase nicely by one integer at a time; so that it has some gaps where some integers between 1 and $x_N$ are 'missing'.
There exist a small number of applications for such irregular partitions, but usually there is nothing to gain by increasing the part set in such irregular manner. In any case, Irregular Equipartitions shall be omitted for the rest of the whole discussion. Letting all the integers between 1 and $x_N$ participate in the breakup of quantity X ensures a smooth and more thorough partition. Here, no mathematical results are offered for this messy irregular equipartition, although indeed some of the mathematical results and expressions given here may apply, or at least may constitute good approximations for Irregular Equipartitions as well.

Since Irregular Equipartitions are excluded from our discussion, the part set $\{x_j\}$ starts from 1, and increases monotonically by 1, until it reaches N, as it imitates the positive integers in a finite way; hence X can be expressed as:

$$X = n_1*1 + n_2*2 + n_3*3 + \ldots + n_N*N$$

Or equivalently, the partition of X can be expressed more concisely as the linear combination:

$$X = \sum_{j=1}^{N} n_j * j$$

The focus of the mathematical analysis here is on the configuration of the aggregated data set of all possible partitions, namely on the relative frequency or repetition of each $x_j$ within that vast data set. Indirectly, we can learn about the relative frequencies of $x_j$ by summing $n_j$ separately for each size j over all partitions, which is simply the total number of occurrences of $x_j$ within the entire equipartition model. An alternative measure is calculating AVG( $n_j$ ), namely the average number of parts $n_j$ of size $x_j$.

A single partition of X into part set $\{x_j\}$ is not Benford in the least. Only the aggregation of all possible partitions leads to Benford under certain conditions. For this aggregation, once the list of all possible partitions is available, one is interested in counting how many times $x_1$ occurs, and how many $x_2$ occurs, and so forth, namely the grand histogram of all $x_j$ incorporating all partitions as one vast data set. It should be noted that if the value of $n_1$ is 48 say for one particular partition, then this implies that $x_1$ occurs 48 times in this particular partition. In order to arrive at the aggregated $x_j$ histogram incorporating occurrences from all partitions, one needs to simply sum (over all partitions) those $n_1$ occurrences to obtain the grand count of $x_1$, and then sum (over all partitions) those $n_2$ occurrences to obtain the grand count of $x_2$, and so forth. Indeed, AVG( $n_j$ ) is actually this exact sum of $x_j$ re-scaled (i.e. divided) by the number of all possible partitions, hence AVG( $n_j$ ) conveys the relative proportions of $x_j$ as well.

P is defined as the number of all possible partitions for any particular equipartition scheme. $n_{kj}$ refers to $n_j$ of the k-th partition; with index k running from 1 to P. AVG ( $n_j$ ) is then:

$$\sum_{k=1}^{P}(n_{kj})/\,P \;=AVG(\,n_j)$$

The number of times each $x_j$ occurs within the <u>entire</u> equipartition scheme is:

$$\sum_{k=1}^{P} n_{kj} \;=[\#\;of\;occurrences\;of\;x_j\;in\;entire\;scheme]$$

$$AVG\big(n_j\big)*P=[\#\;of\;occurrences\;of\;x_j\;in\;scheme]$$

$$AVG\big(n_j\big)*[\#\;of\;Partitions]=[\#\;of\;occurrences\;of\;x_j\;in\;scheme]$$

For either Complete Equipartitions or Refined Equipartitions (but not for Irregular Equipartitions), mathematical analysis in Miller (2015) leads to an exponential-like discrete distribution for AVG( $n_j$ ) with the following expression:

$$\mathrm{AVG}(n_j) = \frac{1}{e^{\lambda x_j}-1}$$

Where the constant λ – called the Lagrange multiplier – is uniquely determine via the constraint:

$$X=\sum_{j=1}^{N}\frac{x_j}{e^{\lambda x_j}-1}$$

Let us derive this last equation:

$$AVG(Nj) = \sum_{k=1}^{P} (n_{kj}) / P$$

$$AVG(Nj) = \frac{1}{e^{\lambda xj} - 1}$$

Equating the two right hand sides we get:

$$\sum_{k=1}^{P} (n_{kj}) / P = \frac{1}{e^{\lambda xj} - 1}$$

Multiplying both sides by $x_j$ we get:

$$\sum_{k=1}^{P} x_j (n_{kj}) / P = \frac{x_j}{e^{\lambda xj} - 1}$$

Summing both sides over index j in order to aggregate all sizes, we get:

$$\sum_{j=1}^{N} \sum_{k=1}^{P} x_j (n_{kj}) / P = \sum_{j=1}^{N} \frac{x_j}{e^{\lambda xj} - 1}$$

Switching the order of the indices, we get:

$$\frac{1}{P} \sum_{k=1}^{P} \sum_{j=1}^{N} x_j n_{kj} = \sum_{j=1}^{N} \frac{x_j}{e^{\lambda xj} - 1}$$

Since original quantity X is being conserved in any given partition, the inner summation is X:

$$\sum_{j=1}^{N} x_j n_{kj} = \mathit{Dot\ Product\ of}\ \{x_j\} \text{ and } \{n_j\} = (\text{Total Quantity in a Partition}) = X$$

Hence: $$\frac{1}{P} \sum_{k=1}^{P} X = \sum_{j=1}^{N} \frac{x_j}{e^{\lambda xj} - 1}$$

Finally, evaluating the summation on the left side of the last equation, and applying the summation rule $\sum_{k=1}^{P} \text{Constant} = \text{P} * \text{Constant}$, we get:

$$\frac{1}{\text{P}}\sum_{k=1}^{P} \text{X} = \frac{1}{\text{P}}\,(P * X) = X = \sum_{j=1}^{N} \frac{\text{x}_\text{j}}{\text{e}^{\lambda \text{xj}} - 1}$$

Clearly, the above expression $\frac{\mathbf{1}}{e^{\lambda x_j} - \mathbf{1}}$ for AVG( $n_j$ ) is skewed in favor of the small (e.g. $n_1$), discriminating against the big (e.g. $n_N$). In other words, the expression is in a sense inversely proportional to $x_j$ in a complicated way (appearing within an exponent in the denominator).

The above expression for AVG( $n_j$ ) reminds us of the continuous Exponential Distribution $\lambda e^{-\lambda x}$ defined on the infinite real range of (0, + ∞), and which is known for being fairly close to Benford regardless of the value of its parameter λ, but the expression above is only for the integral values of the part set $\{x_j\}$, namely for the discrete (and very finite) set {1, 2, 3, … , N}, and therefore it is not possible at all to draw any conclusions whatsoever about possible logarithmic behavior here even in the approximate. Moreover, for $\{x_j\}$ = {1, 2, 3} say, digits 4 to 9 never get a chance to lead in base 10 (although the possibility exists that the frequency vector might fit Benford's Law base 4).

Yet, such result where quantities are skewed in favor of the small raises our hopes and induces us to explore the possibility that Benford's Law could be found here. Indeed, additional mathematical arguments in Miller (2015) for Refined Equipartitions (but not for Complete Equipartitions) lead to much superior result given that $x_N << X$ (or equivalently N << X), and formally in the limit as X goes to infinity while N (or equivalently $x_N$) remains fixed:

$$\text{AVG}(\text{n}_\text{j}) \approx \frac{(X/N)}{x_\text{j}}$$

This result implies a monotonically decreasing set of values for the average of $n_j$.
Here the expression for the average of $n_j$ is inversely proportional to $x_j$!

The above expression can also be written as:

$$(\text{x}_\text{j})*\text{AVG}(\text{n}_\text{j}) \approx \frac{X}{N}$$

Let us combine two results; the first result [definition rather] was discussed earlier:

$$AVG(n_j)*[\# \text{ of Partitions}] = AVG(n_j)*[P] = [\# \text{ of occurrences of } x_j \text{ in scheme}]$$

The second result is common sense and quite obvious:

$$[\text{total quantity for size } x_j \text{ in scheme}] = x_j*[\# \text{ of occurrences of } x_j \text{ in scheme}]$$

Together these relationships yield:

$$[\text{total quantity for size } x_j \text{ in scheme}] = x_j*[AVG(\ n_j\ )*[P]] = [x_j*AVG(\ n_j\ )]*[P]$$

From Miller's result above $[x_j*AVG(\ n_j\ )] \approx [X/N]$, hence we finally get:

$$[\text{total quantity for size } x_j \text{ in scheme}] \approx \frac{XP}{N}$$

Since P, X, N are all fixed values and independent of j for a given Refined Equipartition scheme, it follows that total quantity for each size $x_j$ is a constant, namely the same for all sizes.

Thus, an appealing interpretation of Miller's result is that Refined Equipartition endows the same quantitative portion from the entire quantity of the entire equipartition scheme of XP to each size of the N existing sizes, namely (XP)/N. This is so since there are N sizes of $x_j$, and each partition contains the quantity X within itself as its own sum, therefore XP is the quantity of all possible partitions in the entire scheme, namely the total quantity of that vast data set of the Refined Equipartition model. In other words: the entire quantity of the entire equipartition scheme XP is divided fairly and equally among the N sizes of $x_j$.

This result is gotten only if we thoroughly break up the conserved quantity X into very small and highly refined pieces – as is the case in Refined Equipartitions. Complete Equipartitions on the other hand do not lead to the above result because they also include 'crude', 'partial' or 'improper' partitions, where fairly large pieces are left intact.

Finally, the observation that variable AVG( $n_j$ ) is inversely proportional to $x_j$ implies that it is of the k/x distribution form, albeit in a discrete sense , not in a continuous sense. Since the continuous k/x density is known for its exact Benford behavior in cases when range falls exactly between 1 and an integral power of ten, namely $[1, 10^N)$, N being an integer greater than 0, such as for example on [1, 10 ), [1, 100 ), [1, 1000), and so forth, it follows that Refined Equipartition could possibly be Benford-like whenever part set $\{x_j\}$ consists of corresponding integral ranges, such as {1, 2, 3, 4, 5, 6, 7, 8, 9, 10}, or perhaps more appropriately on {1, 2, 3, 4, 5, 6, 7, 8, 9}, since real x-axis sub-range [9, 10) is 'covered' by integer 9.

To empirically test the first general equipartition result for AVG( $n_j$ ) in one concrete case, a computer program is run to produce Complete Equipartition for X = 25. Here the part set is { $x_j$ } = {1, 2, 3, … , 23, 24, 25}. The computer does not assume anything and does not apply any mathematical formulas, except that it gives equal weights to all possible partitions. It simply displays all possible partitions, one by one, converts them into {$n_j$} vector format, and then calculates AVG( $n_j$ ) directly. There are 1958 possible partitions here.

Empirical results - computer generated AVG( $n_j$ ):
{3.75, 1.65, 0.96, 0.63, 0.44, 0.31, 0.23, 0.17, 0.13, 0.09, 0.07, 0.05, 0.04, 0.03, 0.02, 0.015, 0.011, 0.008, 0.006, 0.004, 0.003, 0.002, 0.001, 0.001, 0.001}

Theoretical expression of $\mathbf{1/(e^{\lambda x_j} - 1)}$ for AVG( $n_j$ ):
{3.59, 1.57, 0.92, 0.60, 0.41, 0.30, 0.22, 0.16, 0.12, 0.09, 0.07, 0.06, 0.04, 0.03, 0.03, 0.020, 0.016, 0.012, 0.009, 0.007, 0.006, 0.004, 0.004, 0.003, 0.002}

This result shows a good fit of empirical to the theoretical expression of AVG( $n_j$ ). Here 0.24583 was the calculated value of the Lagrange multiplier $\lambda$. Figure 31 depicts AVG( $n_j$ ) for integers 1 to 12. There is nothing here resembling Benford's Law of course except for the conceptual observation that small is beautiful. One could examine the first digits distribution of these 25 integers, but nothing of significance would come out of such a study.

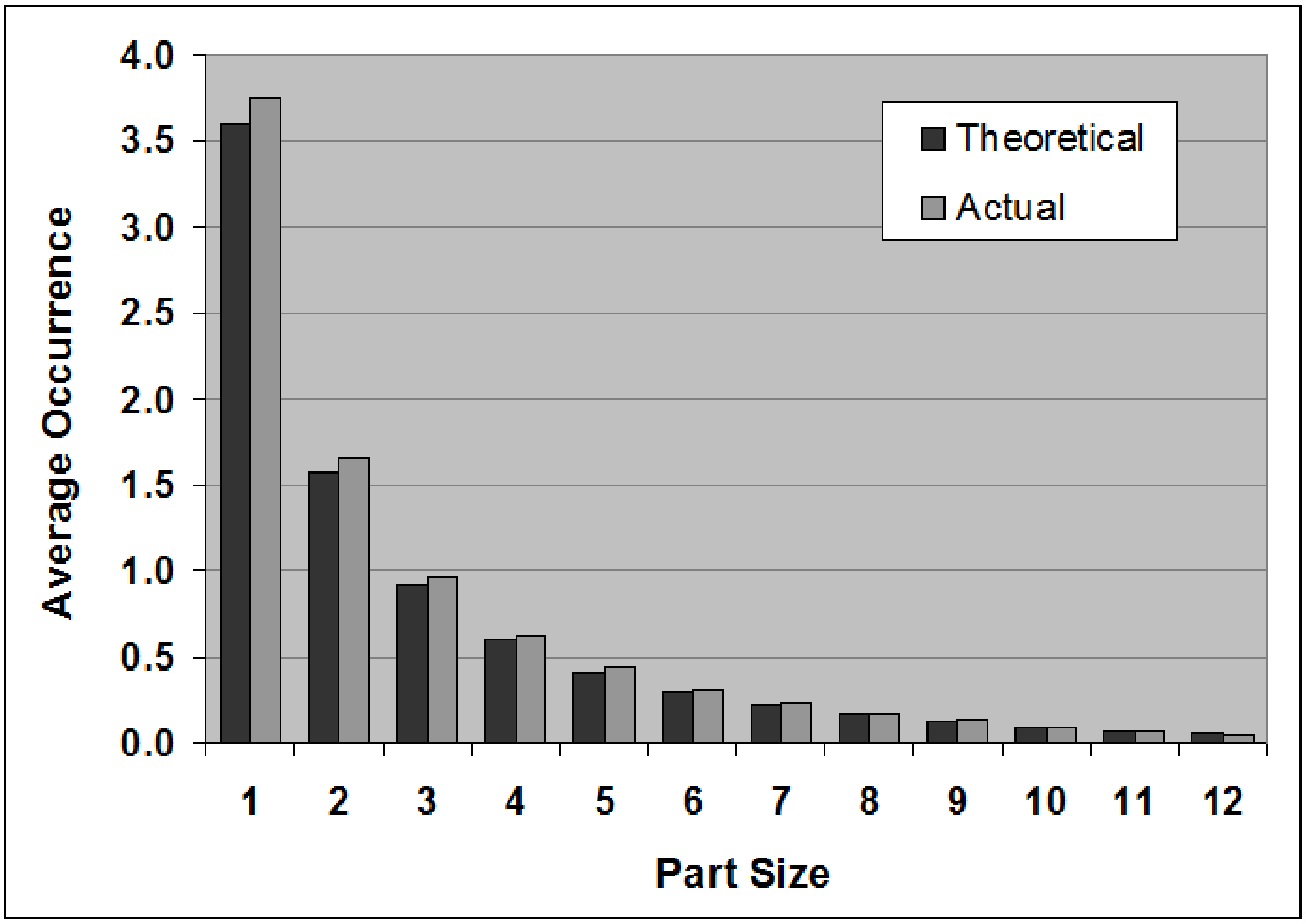

**Figure 31**: Good Fit between Empirical and Theoretical in Complete Equipartition of 25

Another empirical test is performed regarding the two expressions for AVG( $n_j$ ).
This computer program produces Refined Equipartition for X = 33 and part set { $x_j$ } = {1, 2, 3, 4, 5}, namely that of the Equipartition Parable. The computer does not assume anything and does not apply any mathematical formulas, except that it gives equal weights to all possible partitions. It simply displays all possible partitions, one by one, converts them into {$n_j$} vector format, and then calculates AVG( $n_j$ ) directly. There are 918 possible partitions here.

Empirical results - computer generated AVG( $n_j$ ) = {7.52, 3.52, 2.20, 1.53, 1.14}
Theoretical expression of **$1/(e^{\lambda x_j} - 1)$** for AVG( $n_j$ ) = {7.50, 3.52, 2.20, 1.54, 1.15}

This result shows an excellent fit between the empirical result and the theoretical expression. Here 0.1251 was the calculated value of the Lagrange multiplier λ. The 2nd theoretical expression (X/N)/$x_j$ assumes that N << X, and although this is principally Refined Equipartition, yet this condition is not completely satisfied here since 5 is not really much less than 33, consequently the fit of empirical to the theoretical is not as strong for this result.

Empirical results - computer generated AVG( $n_j$ ) = {7.52, 3.52, 2.20, 1.53, 1.14}
Theoretical expression of **(X/N)/$x_j$** for AVG( $n_j$ ) = {6.61, 3.30, 2.20, 1.65, 1.32}

Converting the above empirical results of AVG( $n_j$ ) into actual occurrences of $x_j$ is straightforward via the use of the expression [occurrences of $x_j$] = AVG( $n_j$ )*[# of partitions] = AVG( $n_j$ )*[918]. Hence for this Refined Equipartition: 1 occurs 6905 times, 2 occurs 3228 times, 3 occurs 2017 times, 4 occurs 1408 times, and 5 occurs only 1050 times.
The proportions of the occurrences of {1, 2, 3, 4, 5} are {47%, 22%, 14%, 10%, 7%}.
Total Quantity Per Size = [Occurrences of $x_j$ in scheme]*[$x_j$] = {6905, 6456, 6051, 5632, 5250}

There is nothing here resembling Benford's Law base 10 of course, except for the conceptual observation that small is beautiful. One could examine the first digits distribution of these 5 integers (which are just the integers themselves), but absolutely nothing of significance would come out of this study. Digits 6 to 9 never occur in any case. Yet, the nature of the part set here is approximately logarithmic-like because it is roughly of the k/x distribution form. A comparison to Benford's Law base 6 does not yield a very good fit, and it disappoints a bit:

Equipartition, X = 33, part set {1, 2, 3, 4, 5} - {47%, 22%, 14%, 10%, 7%}
Benford's Law Base 6 for the First Digits - {39%, 23%, 16%, 13%, 10%}

Conceptually explaining such skewed result of AVG( $n_j$ ) here is quite straightforward. It is simply the direct consequence of the model itself. Here total quantity is X = 33, and the part set is {$x_j$} = {1, 2, 3, 4, 5}. Since $n_1$ refers to $x_1$ which contains very little quantitative value (i.e. quantity 1) hence it often needs to be frequent and repetitive in order for 33 to be obtained; while $n_5$ refers to $x_5$ which contains much bigger quantitative value (i.e. quantity 5) hence it could often be quite infrequent and scarce and 33 may still be obtained.

Indeed, the underlying explanation why average $n_j$ is monotonically decreasing springs from a very profound, universal, and yet extremely simple principle regarding how a conserved quantity can be partitioned. The obvious principle or observation here is that: ‘One big quantity is composed of numerous small quantities’, or: ‘Numerous small quantities are needed to merge into one big quantity”. Hence partitioning a fixed conserved quantity into parts can be done roughly-speaking in two extreme styles, either via a breakup into many small parts, or via a breakup into few big parts. A more moderate style would be to have a mixture of all kinds of sizes, consisting of many small ones, some medium ones, and a few big ones. This conceptual outline explains why average $n_j$ is skewed quantitatively.

Since AVG($n_1$) represents the expected number of small things - it is big.
Since AVG($n_N$) represents the expected number of big things - it is small.

For an additional illustration of the ideas and results of this chapter, another concrete numerical example shall be given where total quantity X = 5, and the part set $\{x_j\} = \{1, 2, 3, 4, 5\}$.
This scenario is of the Complete Equipartition type, violating the constraint N << X.
In any case, the quantitative lesson this scenario teaches us is generic and universal in all partitions. Clearly $n_1$ can assume many values within its relatively large possible range of 0 to 5, given that the rest of $n_j$ values are such that it all adds up to 5. And clearly $n_5$ can assume only two values within its relatively smaller possible range of 0 to 1, given that the rest of $n_j$ values are such that it all adds up to 5. The exhaustive list of all 7 possible partitions written in Dot Product notations as $\{n_j\}*\{x_j\}$ is:

{5, 0, 0, 0, 0}*{1, 2, 3, 4, 5} = 5
{3, 1, 0, 0, 0}*{1, 2, 3, 4, 5} = 5
{2, 0, 1, 0, 0}*{1, 2, 3, 4, 5} = 5
{1, 2, 0, 0, 0}*{1, 2, 3, 4, 5} = 5
{1, 0, 0, 1, 0}*{1, 2, 3, 4, 5} = 5
{0, 1, 1, 0, 0}*{1, 2, 3, 4, 5} = 5
{0, 0, 0, 0, 1}*{1, 2, 3, 4, 5} = 5

AVG( $n_j$ ) = {12/7, 4/7, 2/7, 1/7, 1/7}
AVG( $n_j$ ) = {1.71, 0.57, 0.29, 0.14, 0.14}

The relationship (occurrences of $x_j$) = AVG( $n_j$ )*(# of partitions) is quite obvious here, namely (occurrences of $x_j$) = AVG( $n_j$ )*(7), thus AVG( $n_j$ ) = (occurrences of $x_j$)/(7).

It is essential to dispel any possible mistaken perception that in Complete Equipartitions each size possesses approximately the same portion of overall quantity, namely the false perception that $x_j$*(occurrences of $x_j$) for all j are roughly equal, they are not! Calculating this here we get: for size 1 overall quantity is 1*12; for size 2 overall quantity is 2*4; for size 3 overall quantity is 3*2; for size 4 overall quantity is 4*1; for size 5 overall quantity is 5*1; so that the vector regarding overall quantities for the five sizes is {12, 8, 6, 4, 5}, and which is decisively not equal, but rather uneven, as the small size of 1 earns 12 quantitative units while the big size of 5 earns only 5 quantitative units. This quantitative configuration is of extreme skewness, excessively favoring the small over the big, and this is indeed the case in all Complete Equipartitions. Benford digital configuration is not valid here.

The standard measuring rod (benchmark) with which all quantitative configurations should be compared to is the equitable configuration where all the sizes share equally and fairly the same portion of overall quantity, so that $x_j$*(occurrences of $x_j$) for each size j is ≈ (XP)/N as in Refined Equipartitions, or more generally stated for all types of partitions and equipartitions: when quantitative portions for all the sizes are the same.

NOTE: Since this is Complete Equipartition of 5, only $1/(e^{\lambda xj} - 1)$ expression for AVG( $n_j$ ) is valid, while $(X/N)/x_j$ expression is not appropriate here. The two comparisons are:

Empirical calculations for AVG( $n_j$ ) = {1.71, 0.57, 0.29, 0.14, 0.14}
Theoretical $1/(e^{\lambda xj} - 1)$ for AVG( $n_j$ ) = {1.63, 0.62, 0.31, 0.17, 0.10}
Theoretical $(X/N)/x_j$ for AVG( $n_j$ ) = {1.00, 0.50, 0.33, 0.25, 0.20}

As a concrete logarithmic example, total conserved quantity is X = 10,000,000; the part set is $\{x_j\}$ = {1, 2, 3, …, 997, 998, 999}. This is the Refined Equipartition case; hence both mathematical expressions for AVG($n_j$) can be applied. Lagrange multiplier $\lambda$ is calculated as 0.00009749151. Needless to say, this is not performed on the computer due to the incredibly large number of possible partitions here numbering in the trillions of trillions and more. Normal personal computers cannot even begin to handle such fantastically huge number of combinatorial possibilities. Hence ('empirical') computer corroboration of the theory is not possible in this case. Rather the theoretical expressions $1/(e^{\lambda x j} - 1)$ and $(X/N)/x_j$ are taken on faith and used with the aid of the computer to arrive at numerical results. Applying the expression $1/(e^{\lambda x j} - 1)$ we get:

AVG($n_1$) = **10257**
AVG($n_2$) = **5128**
AVG($n_3$) = **3419**

… etc.

AVG($n_{997}$) = **9.80**
AVG($n_{998}$) = **9.79**
AVG($n_{999}$) = **9.78**

The relationship [# of occurrences of $x_j$ in scheme] = AVG($n_j$)*[# of Partitions] leads to the relative percentages of $x_j$ without the need to find out the actual value of the # of Partitions.

1st digits of 999 $x_j$ values, $1/(e^{\lambda x j} - 1)$ - {32.4, 17.7, 12.2, 9.3, 7.5, 6.3, 5.5, 4.8, 4.3}
Benford's Law for First Digits order - {30.1, 17.6, 12.5, 9.7, 7.9, 6.7, 5.8, 5.1, 4.6}
SSD value is 6.1, and such low value indicates that this process is very close to Benford. The choice of the 1 to 999 range was deliberate in order to have it spanning integral powers of ten.

Here N << X, namely that 999 << 10,000,000, therefore the theoretical expression AVG( $n_j$ ) = $(X/N)/x_j$ can be applied. Results are as follows:

AVG($n_1$) = **10010**
AVG($n_2$) = **5005**
AVG($n_3$) = **3337**

… etc.

AVG($n_{997}$) = **10**
AVG($n_{998}$) = **10**
AVG($n_{999}$) = **10**

1st digits of 999 $x_j$ values, $(X/N)/x_j$ - {32.3, 17.6, 12.2, 9.3, 7.6, 6.4, 5.5, 4.8, 4.3}
Benford's Law for First Digits order - {30.1, 17.6, 12.5, 9.7, 7.9, 6.7, 5.8, 5.1, 4.6}
SSD value is 5.3, and such low value indicates that this process is very close to Benford.

The Achilles' heel of Refined Equipartition model is the highly unusual deterministic flavor of the main result AVG( $n_j$ ) ≈ (X/N)/$x_j$. It implies that resultant $x_j$ data is constantly and consistently logarithmic throughout its entire range when measured on mini sub-intervals between integral powers of ten such as (1, 10), (10, 100), and so forth, and that Digital Development Pattern does not exist. As mentioned in chapter VIII of Part I, practically all random data sets come with Digital Development Pattern, almost without any exception. Surely data derived from exponential growth series is of the k/x deterministic flavor, but this is rather rare, such as bank account frozen for decades or centuries steadily earning interest without any withdrawals or deposits, or growing bacteria in the laboratory with many weeks or months of hourly colony recordings. For this reason the applicability of Refined Equipartition to real-life data in the natural world is highly doubtful. Indeed, such probable lack of connection to real-life data is in perfect harmony with another lack of connection to the highly random nature of real-life partition processes which almost never occur along those neat integral lines of Refined Equipartition, and this can be nicely explained by those who intimately know Mother Nature and the way she works. She would probably never bother to carefully break her entities and quantities carefully only along part sets with exact integral relationships. She is simply too busy and is always rushing, building and destroying in a hurry, along whatsoever parts that seem the easiest or the fastest to use, including fractional and real messy part sets. Moreover, those statisticians expecting her to carefully abstain from breaking X along large pieces and to stick only to much smaller and refined pieces (i.e. the constraint of N << X for Refined Equipartition) are doubly naïve and do not really know how Mother Nature typically works. She does not like to be constrained in any way, and she is usually quite crude, rarely refined.

Random Real Partition, Random Dependent Partition, and Chaotic Rock Breaking, on the other hand, have that logarithmic flavor compatible with random distributions and real-life data sets, because these three processes come with a decisive Digital Development Pattern. Therefore these three random partition models are better candidates to explain data sets related to partitions in the natural world, more so than Refined Equipartition which lacks that nearly universal pattern of development. This is also highly consistent with the flexible and crude way these three random partition models are performed, allowing for any fractional and real value for the parts, as well as allowing for any size whatsoever, including very large and crude pieces. These three random partition models are thus much more compatible with the chaotic and crude manner Mother Nature does her daily work (especially Chaotic Rock Breaking as discussed in chapter 14) and could serve as more appropriate models for real-life partition-related data.

Summarizing the differences between these three partition processes and Refined Equipartition:

**Refined Equipartition:**
Not a realistic model, naïvely expecting Mother Nature to partition so carefully along integral parts, as well as so refinely by avoiding big pieces. Strangely not showing any Digital Development Pattern which accompanies almost all random data.

**Random Real Partition, Random Dependent Partition, and Chaotic Rock Breaking:**
Realistic models, allowing Mother Nature to behave more naturally and to partition along any integral, fractional, or real parts, including the permission to crudely break along big pieces. Showing a decisive Digital Development Pattern which is so typical in real-life random data.

## [18] Balls Distributed inside Boxes

**Oded Kafri** (2009) "Entropy Principle in Direct Derivation of Benford's Law".

Conceptually, we partition a big pile of balls into separate mini piles to reside inside boxes. Kafri's discussion and ideas lead to two distinct yet very similar descriptions:

**Model A:** All possible permutations of **L** undistinguishable balls inside **X** distinguishable boxes, aggregated as one large data set. In the same vein as the term 'configurational entropy' of Thermodynamics, here all possible permutations are considered as equally likely and thus are given equal weights. For example, for 5 undistinguishable balls inside 3 distinguishable boxes, the data set to be crowned as Benford is the number of balls in a box, namely the collection of all the (21*3 = 63) numbers in the left panel of Figure 32, zero included. This yields the skewed proportions of {28.6%, 23.8%, 19.0%, 14.3%, 9.5%, 4.8%} for the quantities {0, 1, 2, 3, 4, 5}, but which is not close enough to Benford Base 7. Kafri is suggesting further ignoring all the zeros, but a comparison to Benford Base 6 is still not satisfactory. Perhaps a scheme which averages numerous results with certain X and L parameters, together with some limiting process might yield Benford exactly. What starkly differentiates Kafri's model from Equipartition model is the insistence on considering the boxes as distinguishable, thus introducing order as a factor into the scheme. For Kafri, partitioning 5 into {3, 1, 1} or {1, 3, 1} or {1, 1, 3} signify three different post-partition states, whereas in Equipartition these three partitions are identical as the model disregards order altogether. That the small always outnumbers the big in Kafri's model can be seen clearly in Figure 32, and this is true for all L and X values whenever X > 2. The only exception is for X = 2 which yields equal proportions for all sizes.

| |
|---|
| { 5 , 0 , 0 } |
| { 0 , 5 , 0 } |
| { 0 , 0 , 5 } |
| { 4 , 1 , 0 } |
| { 4 , 0 , 1 } |
| { 1 , 4 , 0 } |
| { 0 , 4 , 1 } |
| { 1 , 0 , 4 } |
| { 0 , 1 , 4 } |
| { 3 , 1 , 1 } |
| { 1 , 3 , 1 } |
| { 1 , 1 , 3 } |
| { 3 , 2 , 0 } |
| { 3 , 0 , 2 } |
| { 2 , 3 , 0 } |
| { 0 , 3 , 2 } |
| { 2 , 0 , 3 } |
| { 0 , 2 , 3 } |
| { 2 , 2 , 1 } |
| { 2 , 1 , 2 } |
| { 1 , 2 , 2 } |

| | | | | | |
|---|---|---|---|---|---|
| 0 | 1 | 2 | 3 | 4 | 5 |
| 0 | 1 | 2 | 3 | 4 | 5 |
| 0 | 1 | 2 | 3 | 4 | 5 |
| 0 | 1 | 2 | 3 | 4 | |
| 0 | 1 | 2 | 3 | 4 | |
| 0 | 1 | 2 | 3 | 4 | |
| 0 | 1 | 2 | 3 | | |
| 0 | 1 | 2 | 3 | | |
| 0 | 1 | 2 | 3 | | |
| 0 | 1 | 2 | | | |
| 0 | 1 | 2 | | | |
| 0 | 1 | 2 | | | |
| 0 | 1 | | | | |
| 0 | 1 | | | | |
| 0 | 1 | | | | |
| 0 | | | | | |
| 0 | | | | | |
| 0 | | | | | |
| *18* | *15* | *12* | *9* | *6* | *3* |
| **28.6%** | **23.8%** | **19.0%** | **14.3%** | **9.5%** | **4.8%** |

**Figure 32**: All Possible Permutations of 5 Undistinguishable Balls in 3 Distinguishable Boxes

**Model B:** The physical random throw of L actual (and thus distinguishable) balls into X distinguishable boxes, and where the principle of entropy is not being applied in any way. Instead, this is thought of as a statistical/stochastic process where each ball is randomly landing in any one of the boxes, so that all the boxes have equal probability in attracting a thrown and flying ball. The variable to be crowned as Benford is the number of balls in a box as above. For example, for 5 balls thrown into 3 distinguishable boxes, Monte Carlo simulations yield the proportion {13.2%, 32.9%, 32.9%, 16.4%, 4.1%, 0.4%} for the occurrences of {0, 1, 2, 3, 4, 5}.

In Kossovsky (2014) it is shown these two descriptions of Kafri's scheme, namely model A and model B, regarding the existence of L balls inside X boxes, are truly two distinct scenarios yielding two distinct numerical and quantitative results.

Interestingly, Kafri's Balls and Boxes scheme could be viewed conceptually not only as partitions but also as consolidations. One vista is to consider an initial glued group of balls undergoing partitions, where many balls end up in potentially different boxes experiencing separations from each other. An alternative vista is to consider the throwing or placing of disconnected and dispersed individual balls into boxes as consolidations, where some balls end up together touching each other and sharing a box.

Modification of Kafri's model type A, fitting it into Refined Equipartition in a sense, should bring it around to the Benford configuration. In order to achieve that, it is necessary to add four new features (the first two features are actually part of Kafri's original suggestion):

To ensure that a genuine refined partition is occurring here, there must be some superficial restriction on the maximum number of balls residing within any given box, and such a limit must be made well below L. Such a limit on the number of balls in any given box would eliminate not only the possibility of one box containing all L balls, but would also eliminate the possibility of one box containing too many balls in general. This feature ensures that balls are well-broken and nicely spread out. The maximum number of balls per box is designated as B, and the constraint B << L is added to ensure that this corresponds to the Refined Equipartition constraint N << X.

The elimination of all empty boxes (zeros) from the calculations.

The condition X = L is postulated due to two arguments. (1) X < L - namely more balls than existing boxes - would not let us spread the balls in the most extreme way possible, namely having all the ball alone separated, and original L quantity totally broken up into 1's. We want to ensure that extreme fragmentation is possible. (2) X > L - namely more boxes than existing balls - would introduce redundant boxes into the system; boxes that are never going to host any balls, even if we break the L pile of balls into 1's. In conclusion: if the possibilities of X < L as well as X > L are eliminated, it then follows that X = L.

Order is eliminated from the model. All the balls are undistinguishable and all the boxes are undistinguishable. The configuration {1, 1, 3} signifies that two boxes are with one ball each, and that another box is with three balls, and this can also be equally written as {3, 1, 1}. No designations such as 'left box', 'center box', or 'right box' are made.

## [19] Division of Logarithmic Data along Small/Medium/Big Sizes

The main result of Refined Equipartition is that AVG( $n_j$ ) ≈ (X/N)/$x_j$. This implies that all the sizes obtain equal quantitative portions, namely that [ total quantity for size $x_j$ ] ≈ (XP)/N, so that all the sizes share the same portion fairly from the overall quantity of the system. The oval-shaped area in Figure 5 enables us to clearly visualize this property of the partition-resultant data where all sizes share fairly and equally in the bounty of the overall quantity of the system. The quest in this chapter is to examine real-life logarithmic data sets and abstract logarithmic distributions and to empirically determine whether or not the equitable and fair quantitative configuration of Figure 5 is indeed the norm for Benford-obeying data sets and distributions, as opposed to the alternative configurations of Figure 6 and Figure 7 where only the Small or only the Big dominate all other sizes.

Hence, each logarithmic data set or distribution under consideration shall be divided in its entirety into 3 size categories - designated as Small, Medium, and Big, and then quantitative portions for each of the three sizes shall be calculated. Hopefully, these 3 sizes would be found to share equally and fairly between them overall quantity of the entire system, namely that each of the 3 sizes should earn approximately 33% of the overall quantity in the entire system.

In order to divide a given data set into Small, Medium, and Big categories, it is necessary to decide on two border points, Border Point A and Border Point B, reasonably chosen within the range of the particular data under consideration. Consequently, all data points falling to the left of Border Point A are deemed as Small; all data points falling between Border Point A and Border Point B are deemed as Medium; and all data points falling to the right of Border Point B are deemed as Big. In other words, Border Points A and B are the threshold or cutoff points differentiating between the three sizes. This arrangement can be visualized in Figure 33 which depicts the classification of Small/Medium/Big via these two Border Points. The reason for the slightly lesser range allocated to the Medium shall be explained shortly. By convention, any data points falling exactly on Border Points A are deemed as Medium; and any data points falling exactly on Border Points B are deem as Big.

For smoothly spread data, it is only from the knowledge of the range of the data under consideration that we construct these two border points, and almost without utilizing any information regarding relative concentrations of values within the range provided by the data's density or histogram. For data with severely irregular spread, where the portions of the data on the margins are highly diluted so that the edges are considered outliers, these two border points are constructed in a way that attempts to ignore and omit these outliers from the calculations.

The most obvious way to choose Border Point A and Border Point B is by simply dividing the entire range, from its minimum to its maximum, equally into 3 sub-intervals, each having the same length, namely (Max – Min)/3. Yet, in order to avoid excessive influence from outliers on the very margins of data, a more reasonable approach here is to start at the 1st percentile point and to end at the 99th percentile point, utilizing these percentile points as the two edges of the core 98% range of data. Figure 34 depicts the arrangement of the two Border Points A and B.

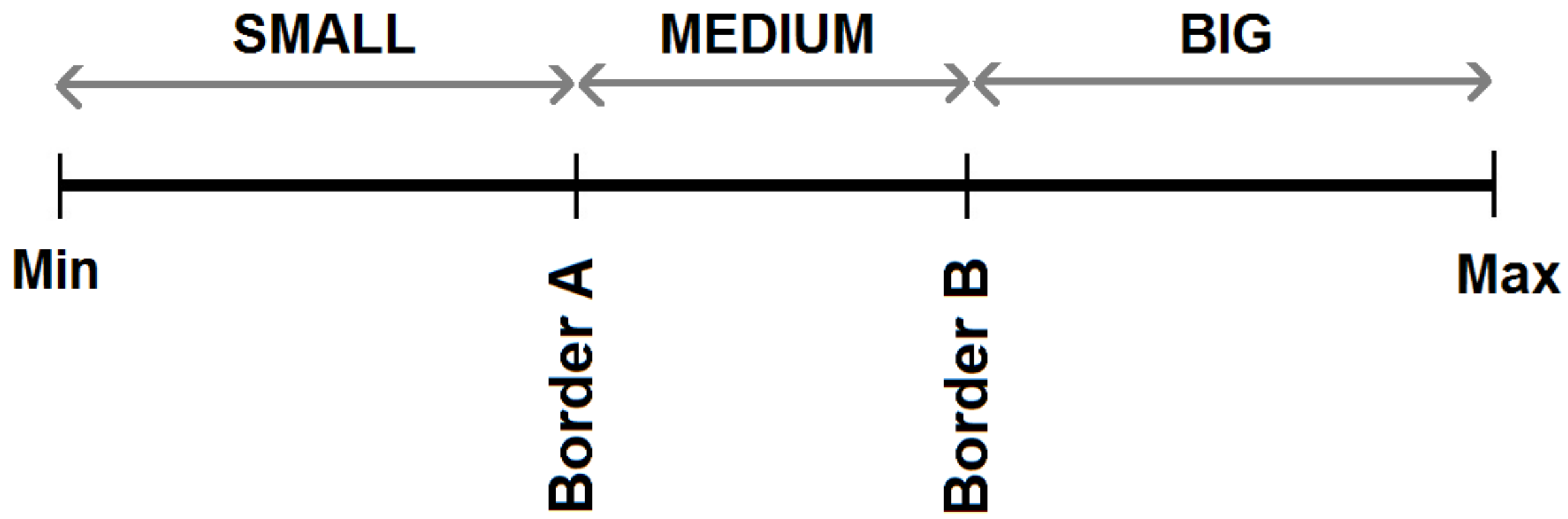

**Figure 33**: Classification of Small, Medium, and Big via Two Border Points

The 1% percentile point, or the 1st percentile point, denoted as $P_{1\%}$, is the value below which 1% of the ordered data may be found.

The 99% percentile point, or the 99th percentile point, denoted as $P_{99\%}$, is the value below which 99% of the ordered data may be found.

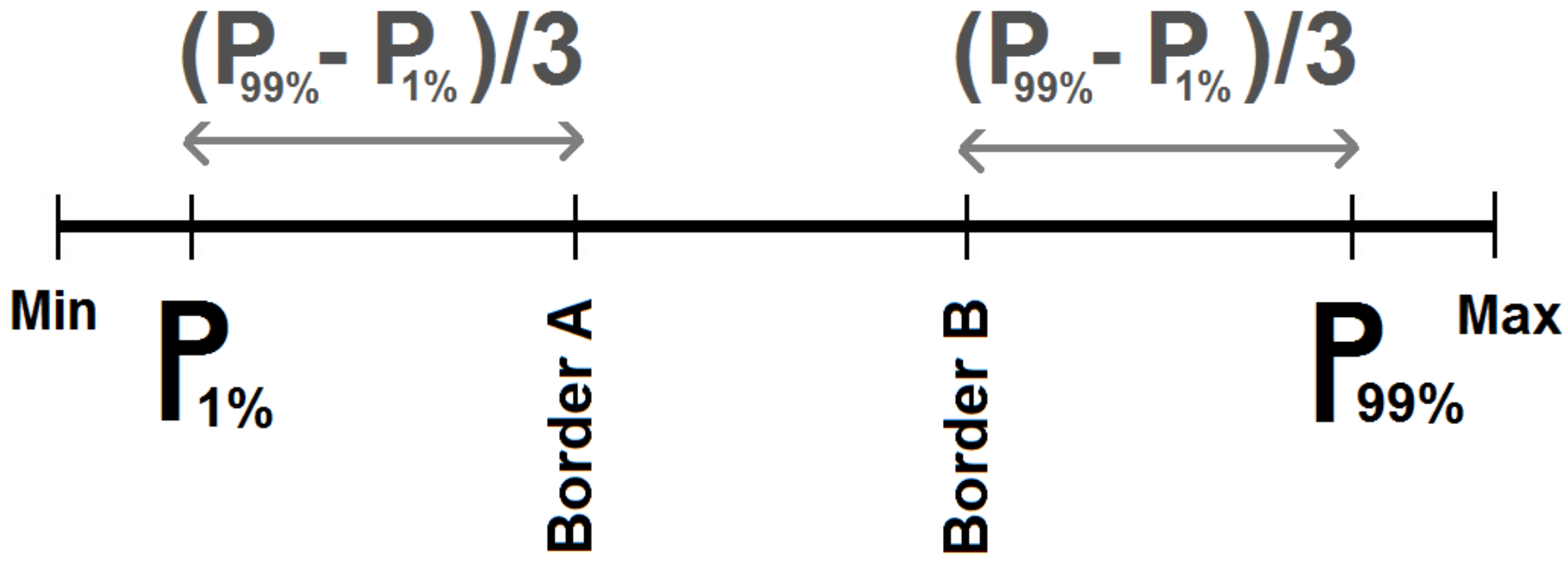

**Figure 34**: The Creation of Two Border Points via the Exclusion of Top 1% and Bottom 1%

In what might seem as a paradox, once Border Point A and Border Point B are chosen with the deliberate exclusion of the outliers, the entire set of numbers in the data are now incorporated into the designation Small/Medium/Big, including even those outliers on the margins. In other words, after excluding the margins from the determination of Border Points A and B, the points falling between Min and $P_{1\%}$ are to be included at the end and are allocated to Small, and the points falling between $P_{99\%}$ and Max are to be included at the end and are allocated to Big. This seemingly contradictory attitude towards outliers is motivated by the realization of the potential for significant adverse effects on the Border Points A and B due to outliers; as opposed to the insignificant and very mild effects of adding only very few extra data points (outliers) to Small and Big. By including the outliers in the classification scheme at the end, we ensure that nothing within the entire data set is omitted or neglected. The diagrams of Figure 33 and Figure 34 define the two border points as follows:

**Border Point A = $P_{1\%} + (P_{99\%} - P_{1\%})/3$**
**Border Point B = $P_{99\%} - (P_{99\%} - P_{1\%})/3$**

In one concrete numerical example, demonstrating the absolute necessity of avoiding the utilization of the maximum, minimum, and possible outliers in the construction of the two Border Points, a data set is imagined having 30,000 points which are approximately uniformly distributed on (5, 35), plus a single outlier value of 155. Allowing this outlier to influence decisions regarding Border Points A and B by partitioning the entire range from the minimum 5 to the maximum 155 into 3 equal sections of (155 – 5)/3 = (150)/3 = 50 width each, would cause the Small to be defined over (5, 55), and thus having the Small artificially earn nearly 100% of overall data portion. By using the 1st and the 99th percentiles, we arrive at a much more reasonable partitioning scheme along (5, 15.1) for the Small, [15.1, 24.9) for the Medium, and [24.9, 155) for the Big. It should be noted how Medium earns less range than either Big or Small.

For this data set, $P_{1\%} = 5 + 0.01*(35 - 5) = 5.3$, and $P_{99\%} = 5 + 0.99*(35 - 5) = 34.7$. It follows that Border Point A = $P_{1\%} + (P_{99\%} - P_{1\%})/3 = 5.3 + (34.7 - 5.3)/3 = 5.3 + 9.8 = 15.1$, and that Border Point B = $P_{99\%} - (P_{99\%} - P_{1\%})/3 = 34.7 - (34.7 - 5.3)/3 = 34.7 - 9.8 = 24.9$.

Medium is full of envy; it is bitterly complaining that its share on the entire range is less than that allocated to either Small or Big; and that it is being discriminated against. Subsequently, the algorithmist points out to Medium that differences are really tiny; that its loss is no more than 2% of the overall range, and finally Medium reluctantly accepts the arrangement so as not to appear as obstructionist, still whispering to itself all sorts of old grievances and ridiculous accusations against the Small and especially against the Big, such as in the alleged maltreatments and discriminations of middle children in large families.

By including the outliers at the end of the classification scheme we ensure that the entire data set is presented, and part of the motivation for this is the partition-vista of the data, which induces the desire that the partition would apply to all segments and parts of the data, even on the margins.

Perhaps the noun ‘Composition’ would be more appropriate for this vista than the noun ‘Partition’. Surely this vista is readily and nicely interpreted in the case of population data as the total population of the entire country. This vista is also readily and nicely interpreted in the case of county or province area data as the total area of the entire country. This vista is trivially interpreted in the case of data on the time intervals between earthquakes in a given year as simply one year time interval. But how could one possibly interpret for example the sum of all the distances from the Solar System to a large collection of stars?! And how could one possibly interpret for example the sum of the prices of all the items on sale in a big catalog of a large retail company?! In any case, this partition-vista of data shall be taken here, and therefore all outliers and data points on the margins are incorporated into the scheme. We wish to have the sum of all the parts equals exactly to the whole.

As an example of how all the quantitative results pertaining to this algorithm for a variety of logarithmic data sets and distributions are going to be presented below [namely the format and style of the presentation of the quantitative analysis], the imaginary data set supposedly generating the oval shape in Figure 5 will be presented as in Figure 35. Here the areas represent the quantitative portions for each of the 3 sizes, Small, Medium, and Big. The numbers within each area represent the number of parts/pieces/points. It should be emphasized that the numbers inside the areas do not represent quantitative portions, but rather the number of parts/pieces/points for a given size. In Figure 5 there are 220 small parts in the middle, 35 medium parts on the right, and only 7 big parts on the left. Since Figure 5 was drawn with the ‘artistic intention’ of endowing 1/3 of total oval area to each of the 3 sizes equally, this fact is reflected here a bit more clearly and exactly showing that each size obtains 1/3 of the total circular area.

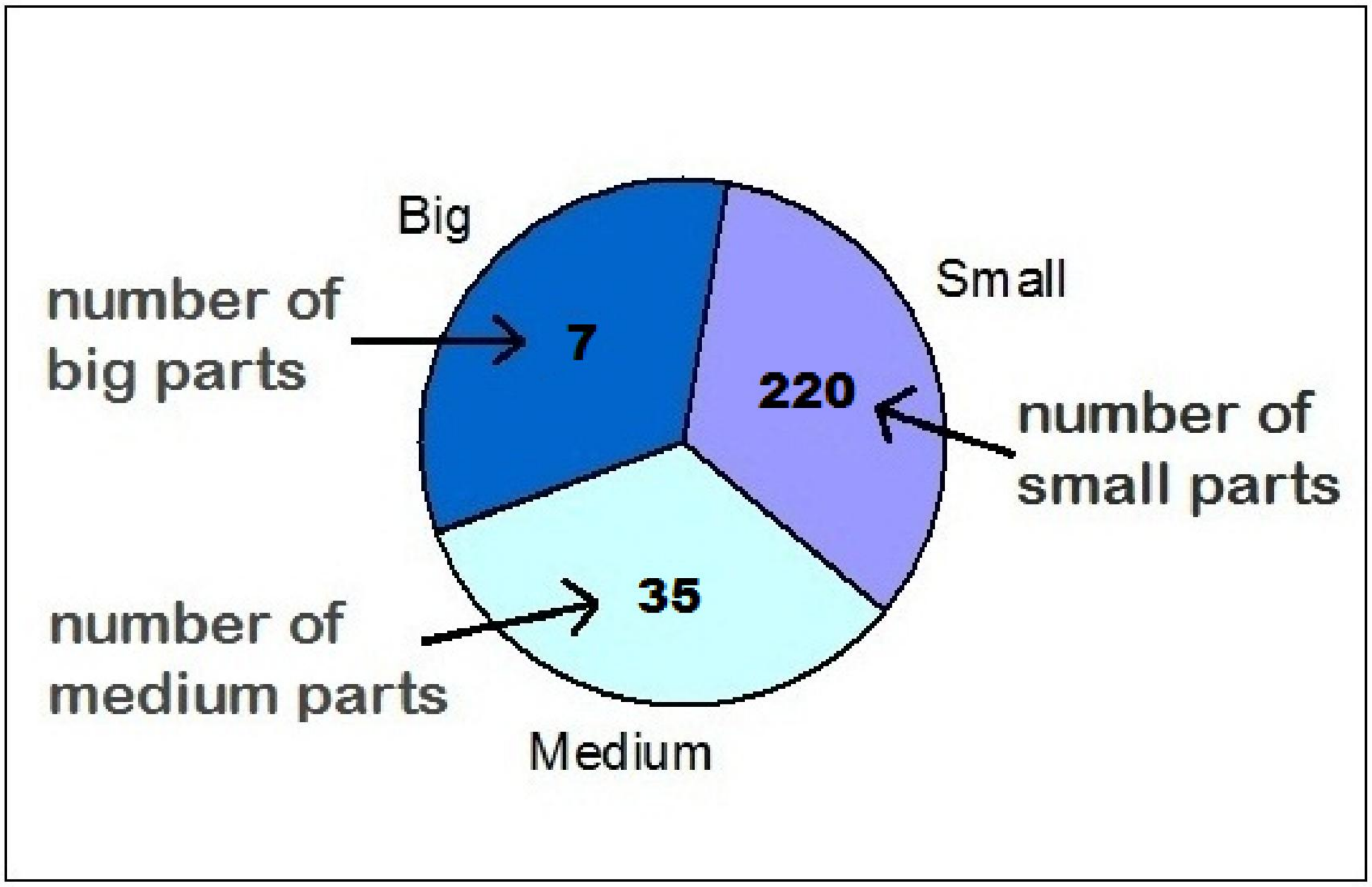

**Figure 35**: Quantitative Portions for 3 Sizes – Prototype Example for Figure 5

Figure 36 depicts quantitative results for Monte Carlo computer simulations obtaining 35,000 realizations from the Lognormal(9.3, 1.7). Quantitative portions are:
**Small = 50.8%** **Medium = 15.8%** **Big = 33.4%**

Figure 37 depicts quantitative results for the US Census Data on 2009 Population count of its 19,509 incorporated cities & towns. See Kossovsky (2014) Chapter 12. Quantitative portions are:
**Small = 36.7%** **Medium = 14.8%** **Big = 48.5%**

Figure 38 depicts quantitative results for the 19,452 Global Earthquakes occurring in 2012. The data here is of the time in the units of seconds between successive earthquakes. See Kossovsky (2014) Chapter 11 for more information. Quantitative portions are:
**Small = 46.0%** **Medium = 33.9%** **Big = 20.1%**

Figure 39 depicts quantitative results for the prices of 14,914 items on sale in the catalog of Canford Audio PLC. See Kossovsky (2014) Chapter 44. Quantitative portions are:
**Small = 53.3%** **Medium = 16.9%** **Big = 29.7%**

Figure 40 depicts quantitative results for the 8,192 Pieces in a 13-Stage Random Rock Breaking process in one Monte Carlo simulation run. Quantitative portions are:
**Small = 34.3%** **Medium = 15.3%** **Big = 50.4%**

Figure 41 depicts quantitative results for 30,000 values of the simulated Symmetrical Triangle on (1, 5) serving as exponents. The realizations from the simulations of the Triangle density are used as exponents of base 10; namely that the final data set is $10^{\text{Symmetrical Triangle}}$.
See Kossovsky (2014) Chapter 64 for discussions about the topic. Quantitative portions are:
**Small = 47.6%** **Medium = 24.0%** **Big = 28.4%**

Figure 42 depicts quantitative results for the US Market Capitalization as of Oct 9, 2016, encompassing 2,883 companies. The largest 6 companies were omitted in this analysis since they appear as extreme outliers here. Quantitative portions are:
**Small = 41.0%** **Medium = 12.7%** **Big = 46.2%**

Figure 43 depicts quantitative results for 20,000 values of the simulated chain of five Uniform(0, b) Distributions, where parameter b is chosen randomly from yet another such Uniform Distribution, while ultimate Uniform is with a fixed parameter b = 5666.
Schematically written as Uniform(0, Uniform(0, Uniform(0, Uniform(0, Uniform(0, 5666))))).
**Small = 48.9%** **Medium = 26.3%** **Bi g= 24.8%**

Figure 44 depicts quantitative results for the prices of 8,079 items on sale in the catalog of MD Helicopters Inc. for the year 2012. See link at http://www.mdhelicopters.com/v2/index.php.
MD Helicopters Inc. manufactures and retails helicopters and parts. Quantitative portions are:
**Small = 32.9%** **Medium = 15.8%** **Big = 51.3%**

Figure 45 depicts quantitative results for 30,000 simulated values of the perfectly and uniquely Benford k/x distribution defined on the interval (1, 10). Quantitative portions are:
**Small = 32.4%** **Medium = 32.0%** **Big = 35.5%**

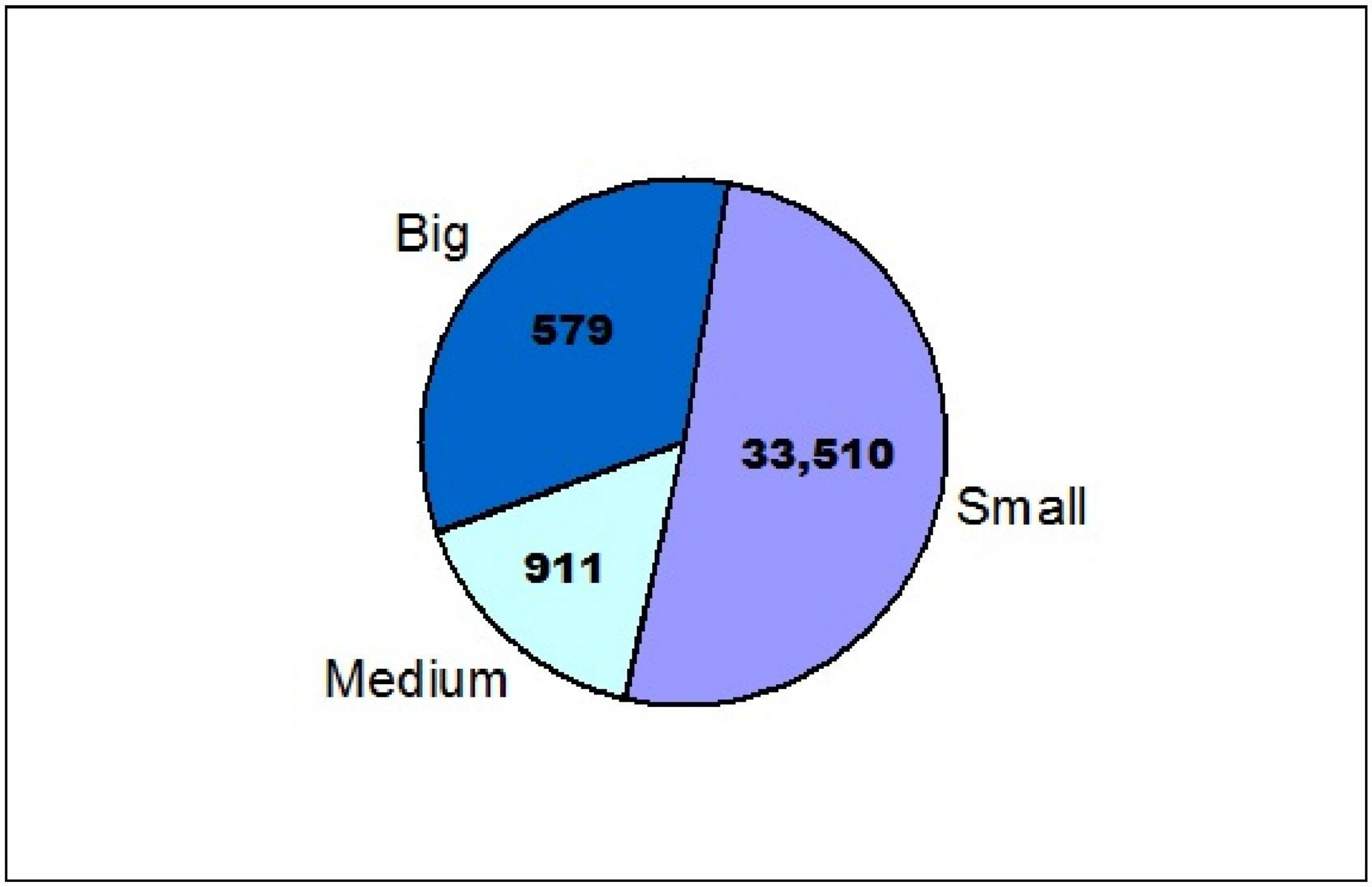

**Figure 36**: Quantitative Portions for 3 Sizes - Lognormal(9.3, 1.7)

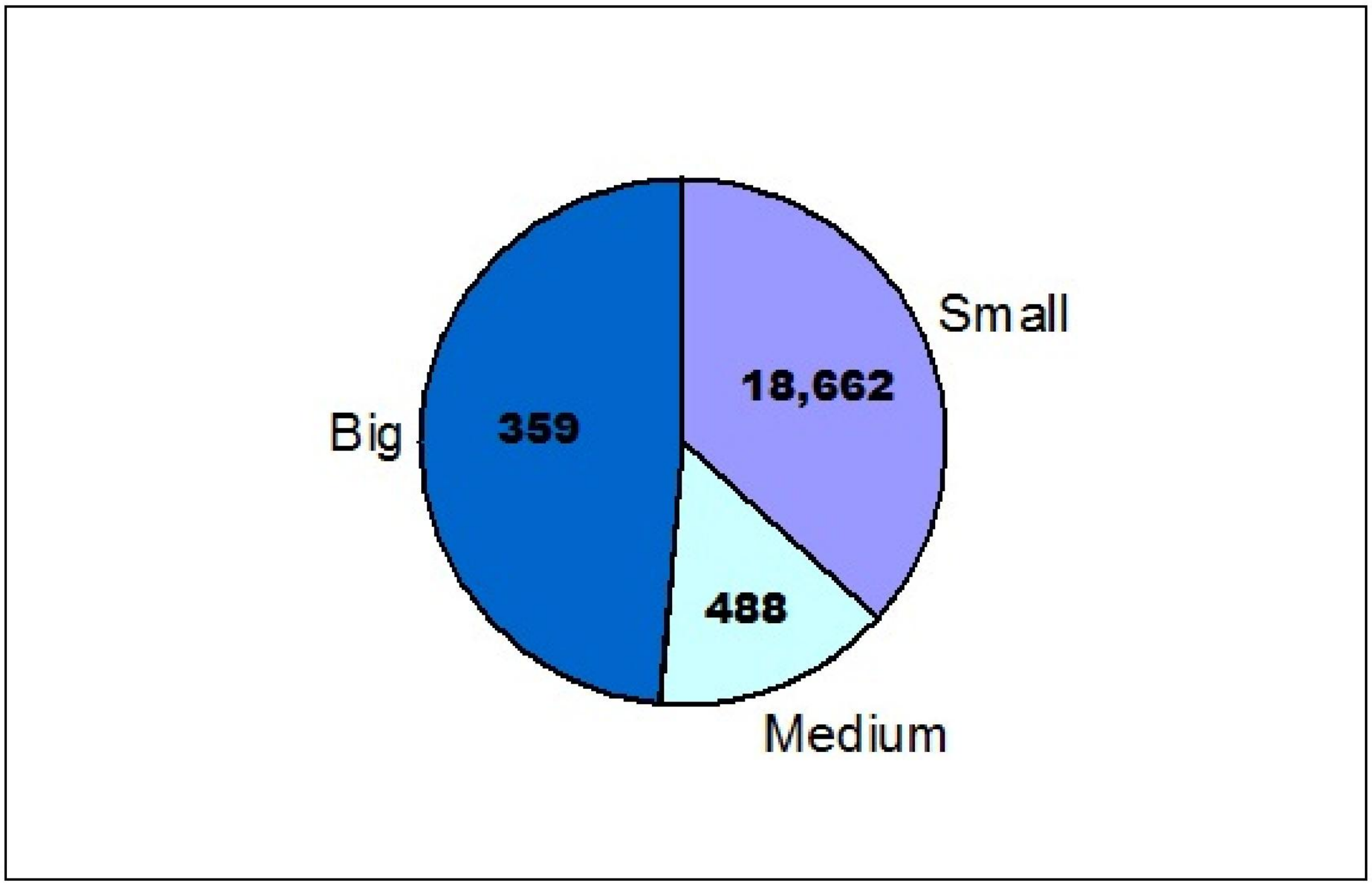

**Figure 37**: Quantitative Portions for 3 Sizes – US Population Census in 2009

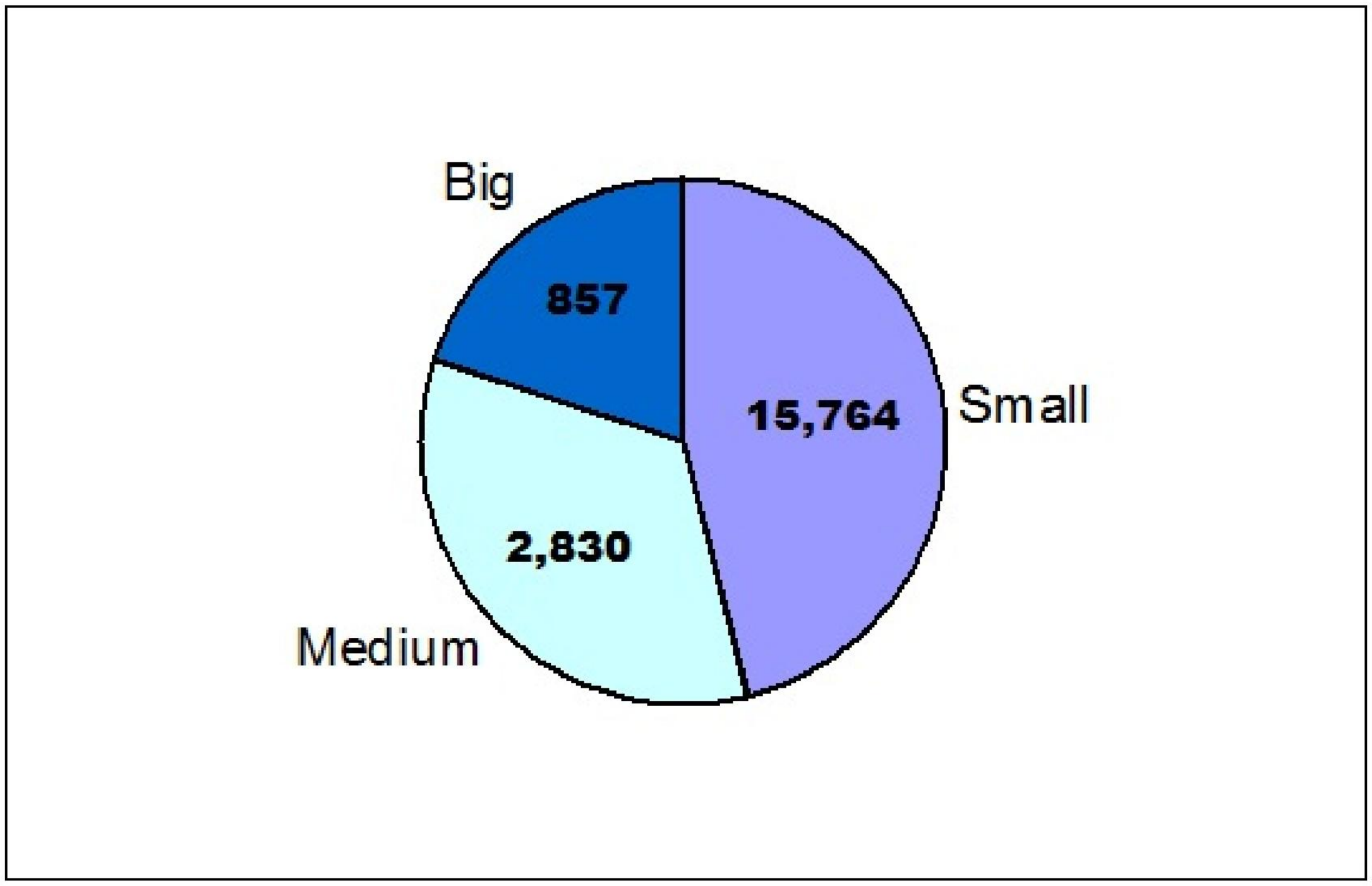

**Figure 38**: Quantitative Portions for 3 Sizes – Global Earthquakes in 2012

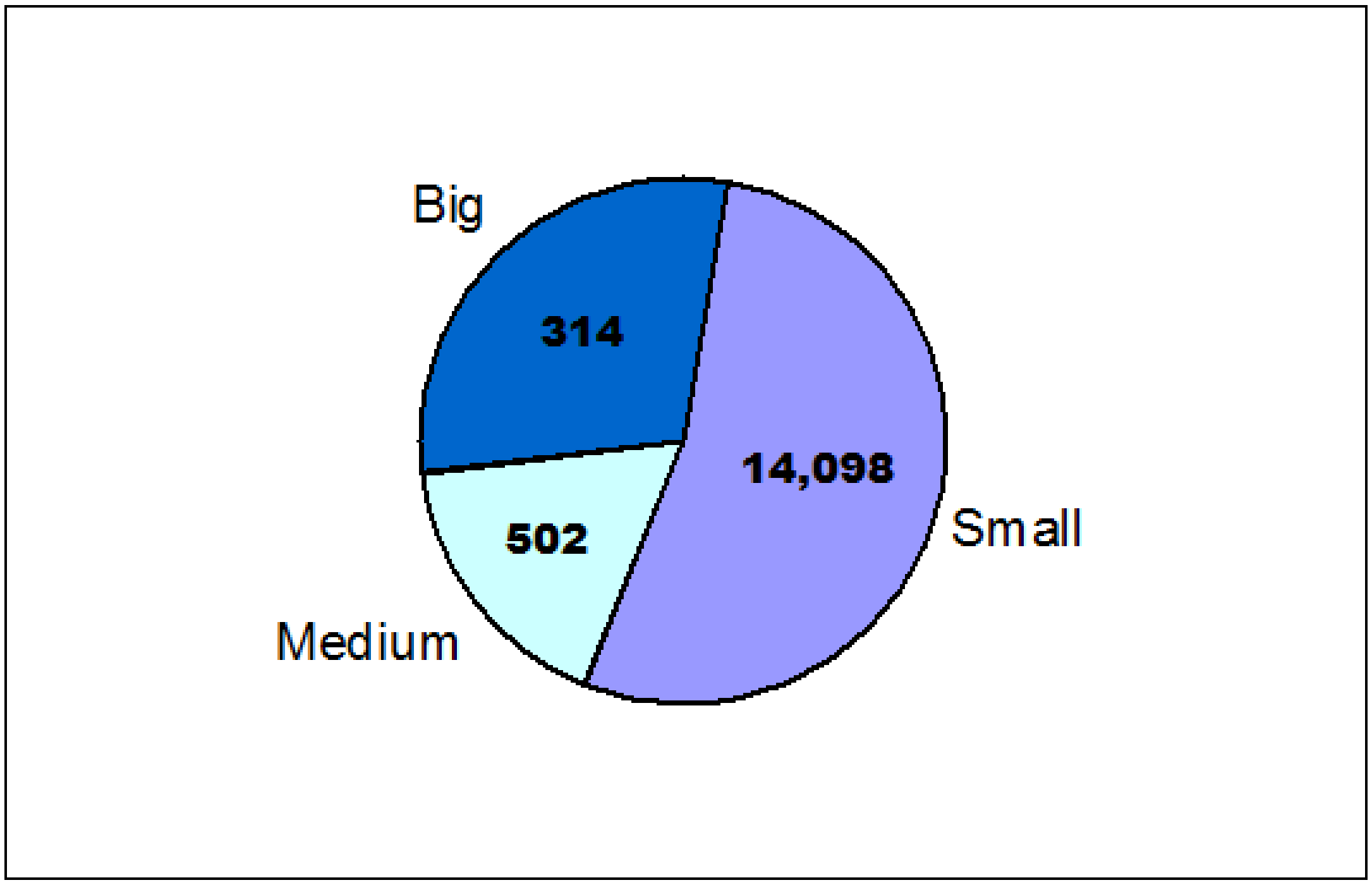

**Figure 39**: Quantitative Portions for 3 Sizes – Canford PLC Catalog

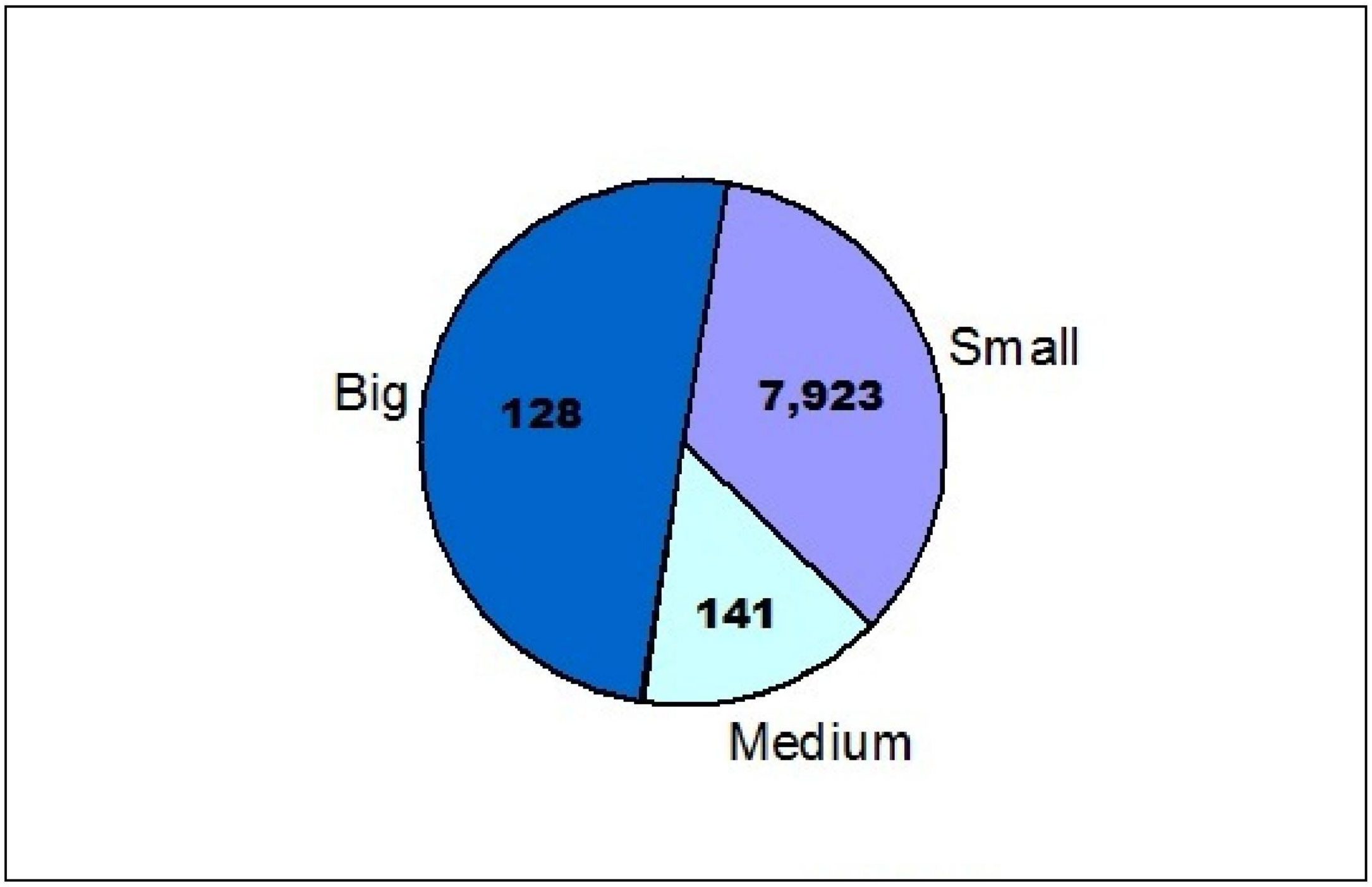

**Figure 40**: Quantitative Portions for 3 Sizes – Rock Breaking, 13 Stages, 8192 Pieces

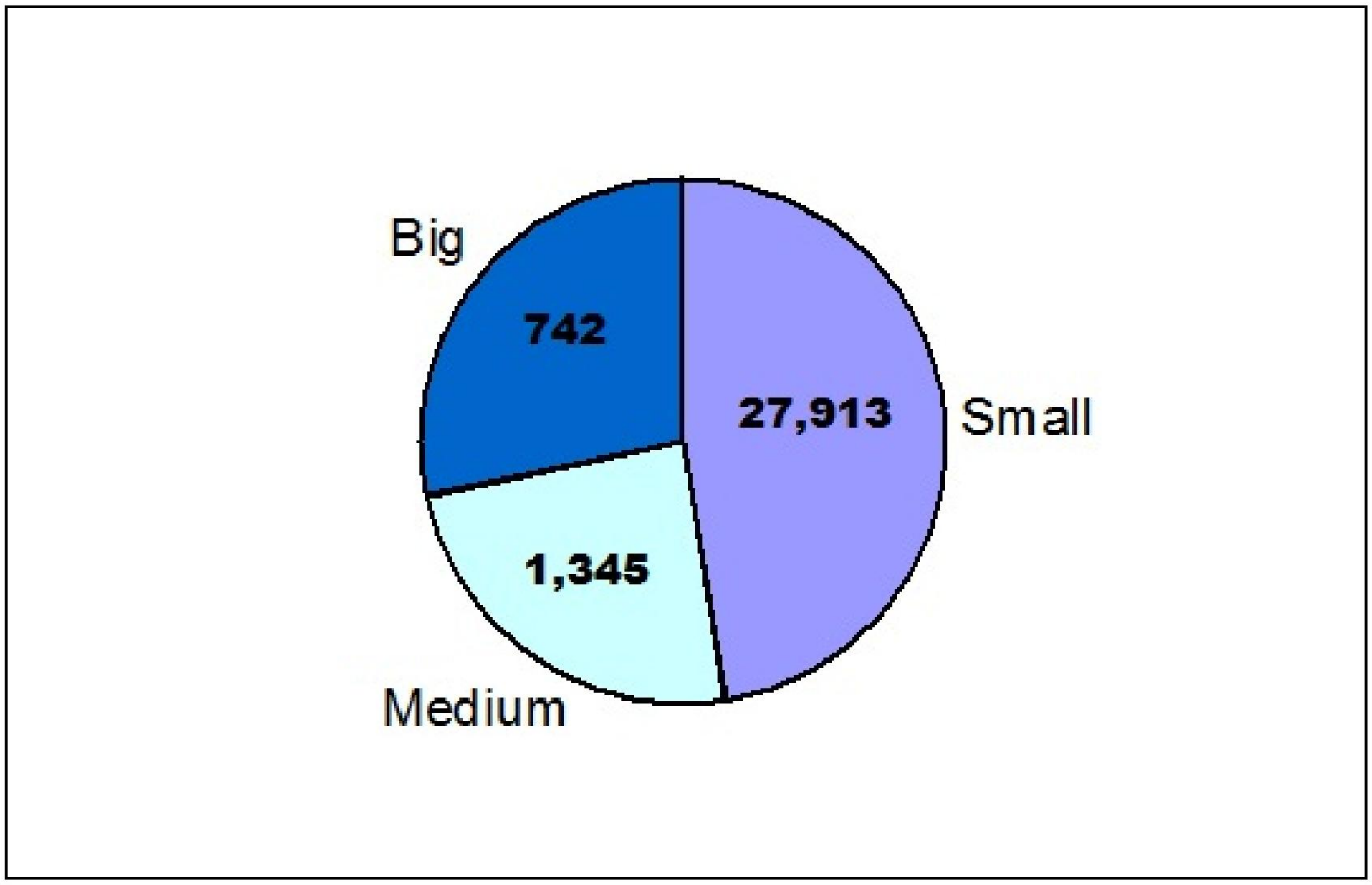

**Figure 41**: Quantitative Portions for 3 Sizes – Log is Symmetrical Triangle on (1, 5)

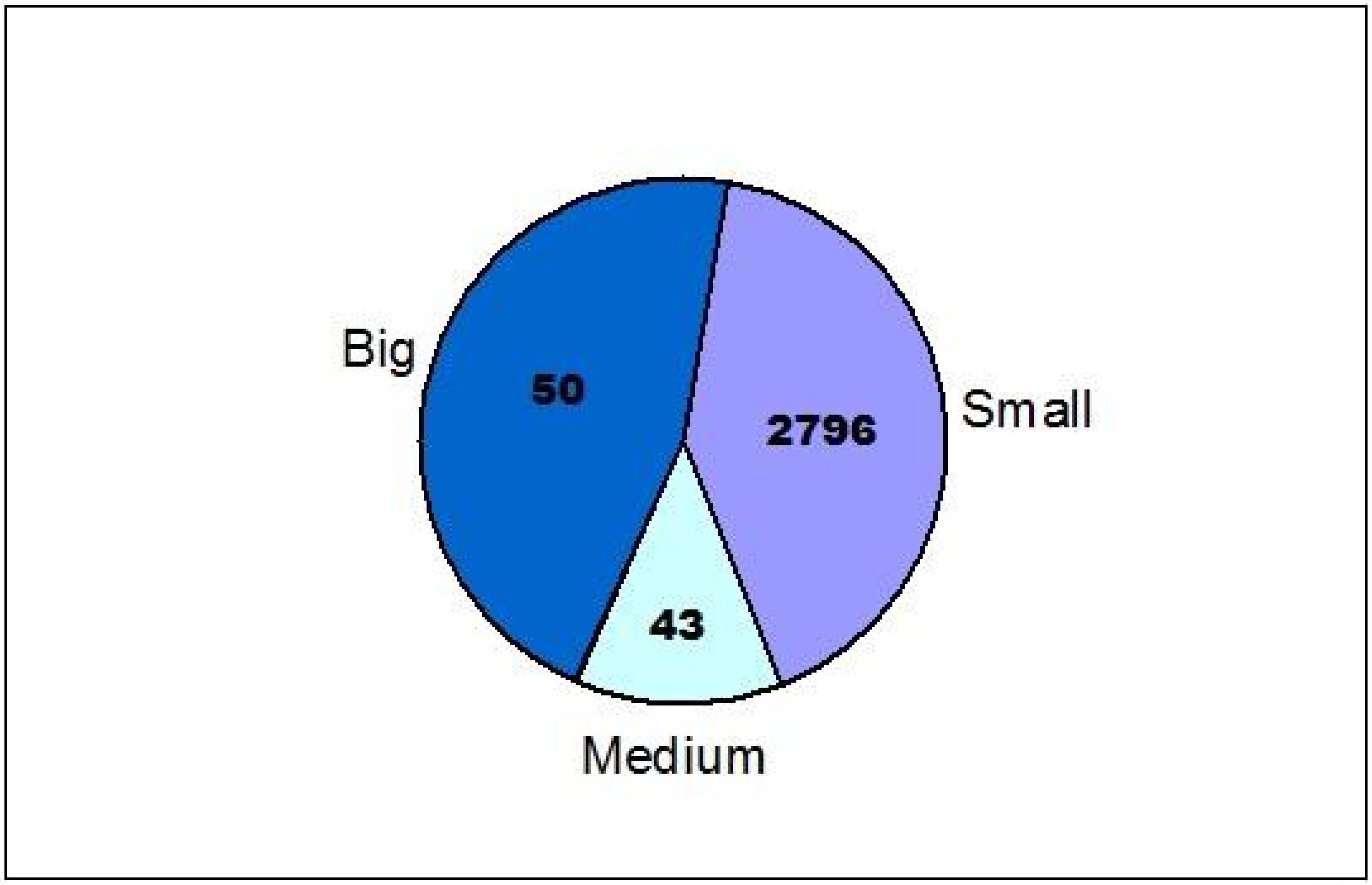

**Figure 42**: Quantitative Portions for 3 Sizes – US Market Capitalization - Oct 9, 2016

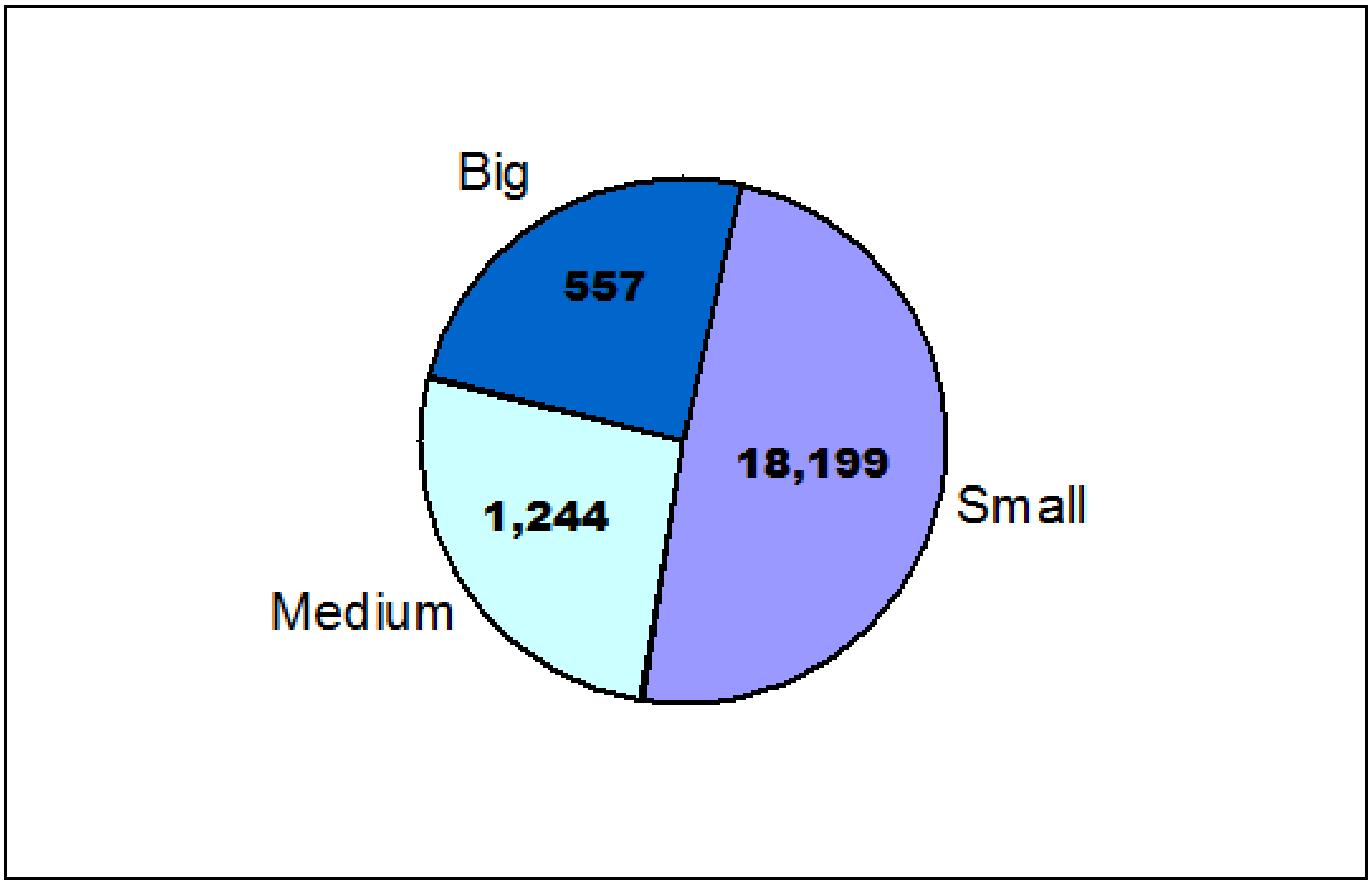

**Figure 43**: Quantitative Portions for 3 Sizes – Chain of 5 Uniform Distributions

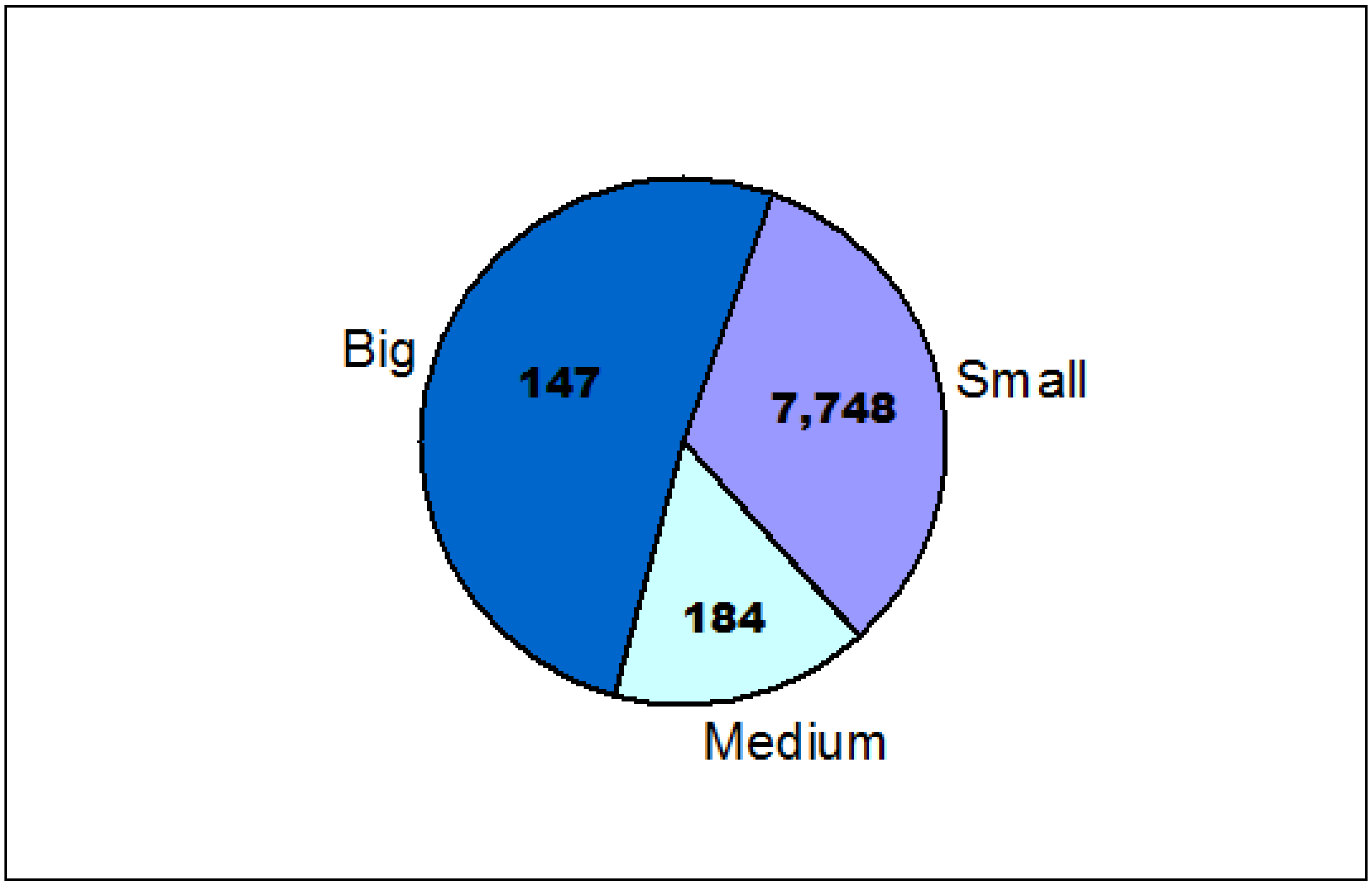

**Figure 44**: Quantitative Portions for 3 Sizes – MD Helicopters Inc. Catalog

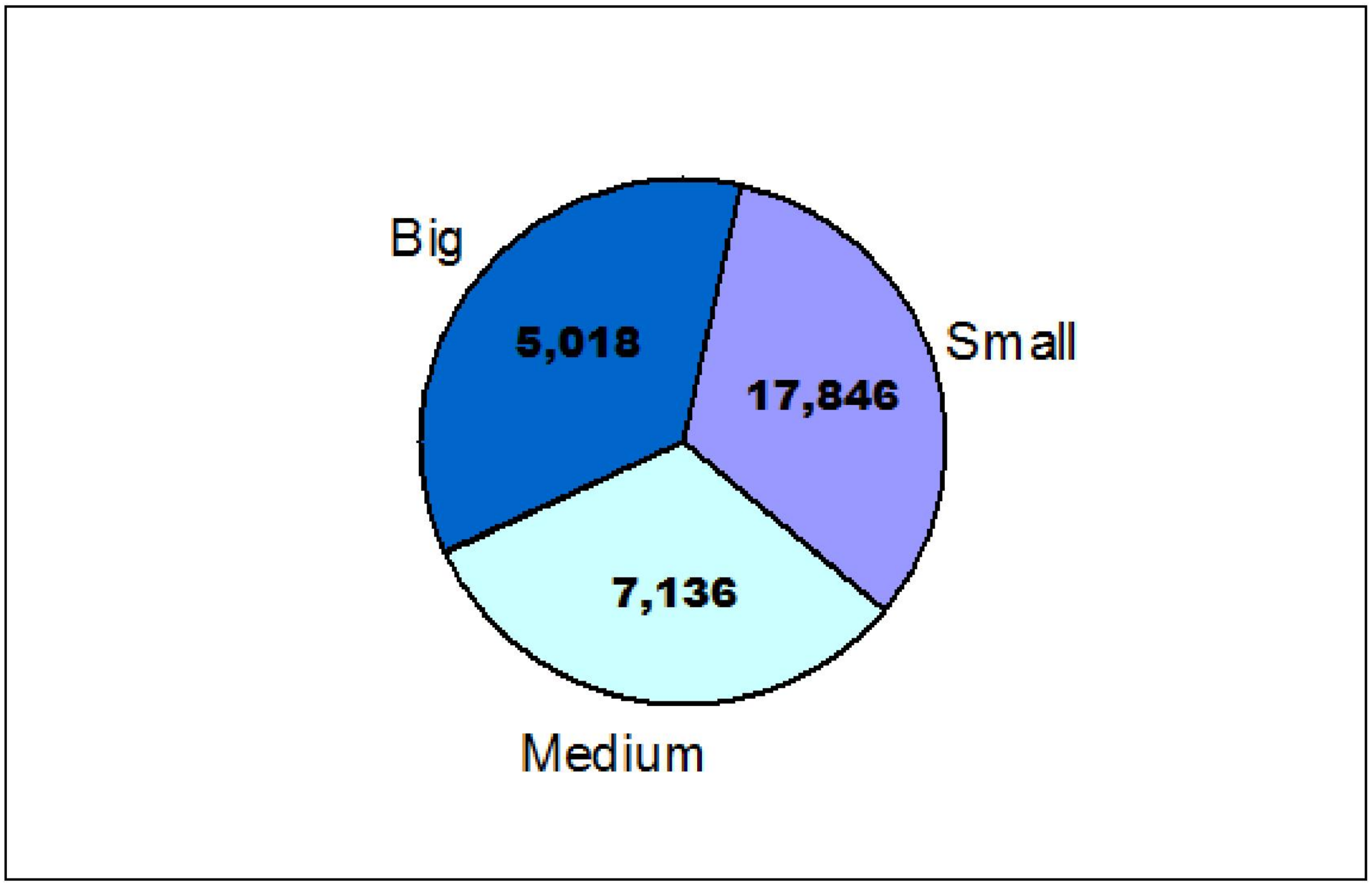

**Figure 45**: Quantitative Portions for 3 Sizes – The Perfectly Benford k/x over (1, 10)

The first common feature noted here in all of these ten logarithmic data sets and distributions, is the fact that (for the most part) no particular size dominate quantitative-wise, as if the three sizes always sincerely wish to share equally and fairly between them overall quantity of the entire system, namely 33% for each of the three sizes, although reality often intervenes and imposes upon them to give a bit to, or to take a bit from, one size or another, and even then it's never by much. The lowest quantitative portion for any size in the entire set of ten is for the most part 15% (except once in the case of US Market Capitalization which is 12.7%). The highest quantitative portion for any size in the entire set of ten is 53.3%. We never encounter here extreme situations where one size dominates with say 60% or 70% of quantitative portion, or where one size diminishes and obtains only say 10% or less.

The second common feature noted here in all of these ten logarithmic data sets and distributions, is the fact that the number of Small, Medium, and Big parts consistently reflects the small is beautiful phenomenon; and that in almost all of these ten data sets and distributions the Small is more numerous than the Medium, and the Medium is more numerous than the Big. There is almost no exception here to this strict and nearly universal rule [well, only for US Market Capitalization data set, the Big with 50 points has a very slight advantage over the Medium with 43 points, being a tiny bit more numerous, yet the Big and the Medium combined which earn in total 93 points pales in comparison to the Small which earns 2796 points!]

## [20] Division of Non-Logarithmic Data along Small/Medium/Big Sizes

Applying the above algorithm in the division of data into Small/Medium/Large sizes for a group of 5 non-logarithmic data sets and distributions shows decisive deviations not only from the small is beautiful principle, but also from the approximate or near equality in quantitative portions seen in all of the 10 logarithmic data sets and distributions of the previous chapter. Here for non-logarithmic data sets and distributions, one size often strongly dominates all other sizes, obtaining 60% or even 70% of overall quantity.

Here is a descriptive summary of the quantitative results for these 5 data sets and distributions:

Figure 46 depicts quantitative results for 25,000 realizations of Monte Carlo computer simulations from the Uniform Distribution on (5, 78,000). Surely, since the density here is flat, uniform, and horizontal, each size contains about the same number of points (a tiny bit less for Medium since it is defined as a bit shorter than either Small or Big). Hence, none is more frequent than the others; rather all three sizes occur with almost equal frequency. The ramification regarding quantitative portions is quite straightforward, as there are 8,342 Small points lying between the Minimum of 5 and Border Point A of 26,279; there are 8,220 Medium points lying between Border Point A of 26,279 and Border Point B of 51,737; and there are 8,438 Big points lying between Border Point B of 51,737 and the Maximum of 78,000, hence each size obtains substantially different quantitative portion depending on the qualitative territory its points lie in, and this produces the relative results where Big > Medium > Small. Quantitative portions are:
**Small = 11.2%** **Medium = 32.8%** **Big = 56.0%**

Figure 47 depicts quantitative results for 25,000 realizations of Monte Carlo computer simulations from the Normal(177, 40). Here Medium benefits substantially from the fact that most of the data points and also most of the quantities are around the center, namely around the mean/median/mode, as in all Normal distributions. Here Medium is by far the most frequent size in terms of having the largest number of data points. Medium also earns the most in terms of quantitative portions. Quantitative portions are:
**Small = 15.0%** **Medium = 56.6%** **Big = 28.5%**

Figure 48 depicts quantitative results for 25,000 realizations of Monte Carlo computer simulations from the $x^3$ distribution over (1, 50) discussed earlier in chapter 12. Here the curve rises sharply and constantly, hence not only the expression 'big is exceedingly beautiful' is true here, but there is also a consistent advantage here to bigger sizes over smaller sizes in both measures, namely in the number of data points, as well as in quantitative portions, where Big > Medium > Small. Quantitative portions are:
**Small = 4.7%** **Medium = 22.6%** **Big = 72.7%**

Figure 49 depicts quantitative results for the data set pertaining to areas of the 3,143 Counties in the US. See Kossovsky (2014) Chapter 45 for more information. Here Small is excessively more frequent than the other sizes, over and above its usual level of frequency according to Benford's Law, and especially regarding quantitative portions, which are:

**Small = 59.4%** **Medium = 13.2%** **Big = 27.3%**

Figure 50 depicts quantitative results for the State of Oklahoma payroll data of the Department of Human Services for the 1st quarter of 2012. This data can be found on their website https://data.ok.gov/Finance-and-Administration/State-of-Oklahoma-Payroll-Q1-2012/dqi7-zvab. Only those 2189 rows from the column 'Amount' pertaining to the Department of Human Services are considered. See Kossovsky (2014) Chapter 138, Page 592 for more information. Here the medium is the most significant in both measures, namely in the number of data points, as well as in quantitative portions, where Medium > Small and where Medium > Big. Here quantitative portions are:

**Small = 19.0%** **Medium = 66.7%** **Big = 14.4%**

In conclusion: a clear distinction is seen between the 10 logarithmic data sets and distributions and the 5 non-logarithmic ones. The non-logarithmic data sets strongly deviate from the small is beautiful principle regarding number of points per size, as well as allowing one size to strongly dominate (quantitative-portion-wise) all the other sizes, well above its supposed 33% allocation.

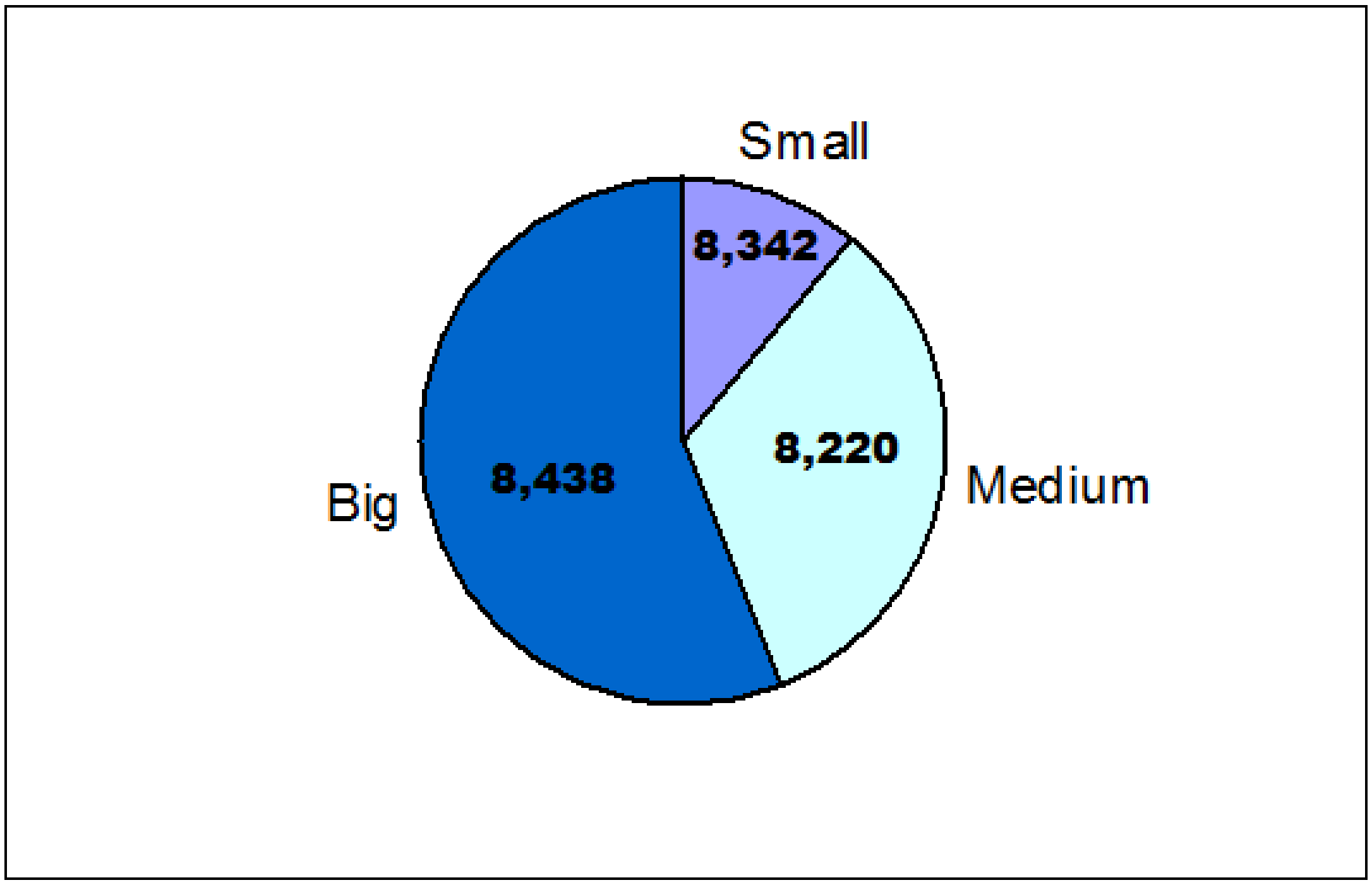

**Figure 46**: Quantitative Portions for 3 Sizes – Uniform Distribution on (5, 78000)

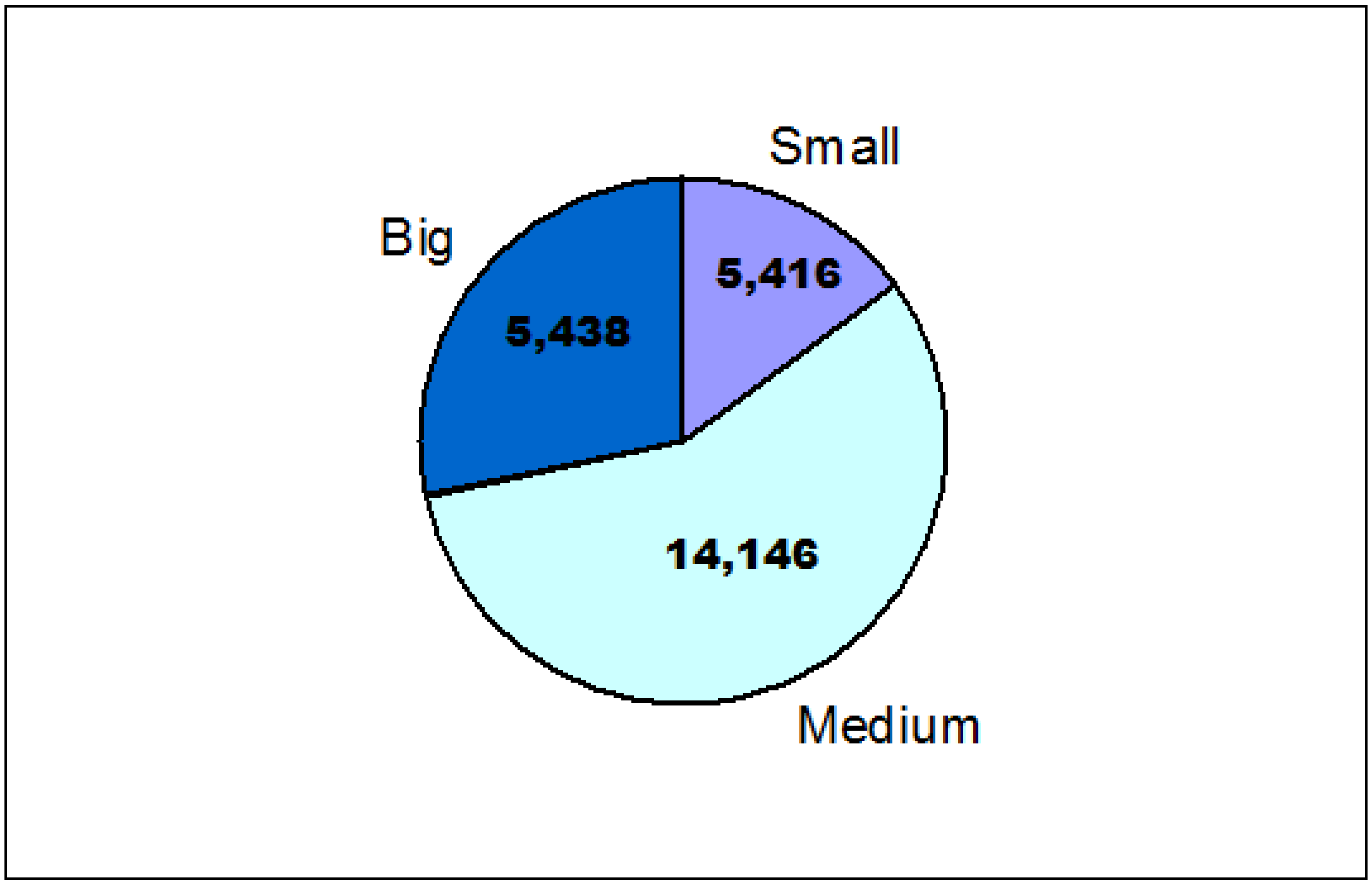

**Figure 47**: Quantitative Portions for 3 Sizes – Normal(177, 40)

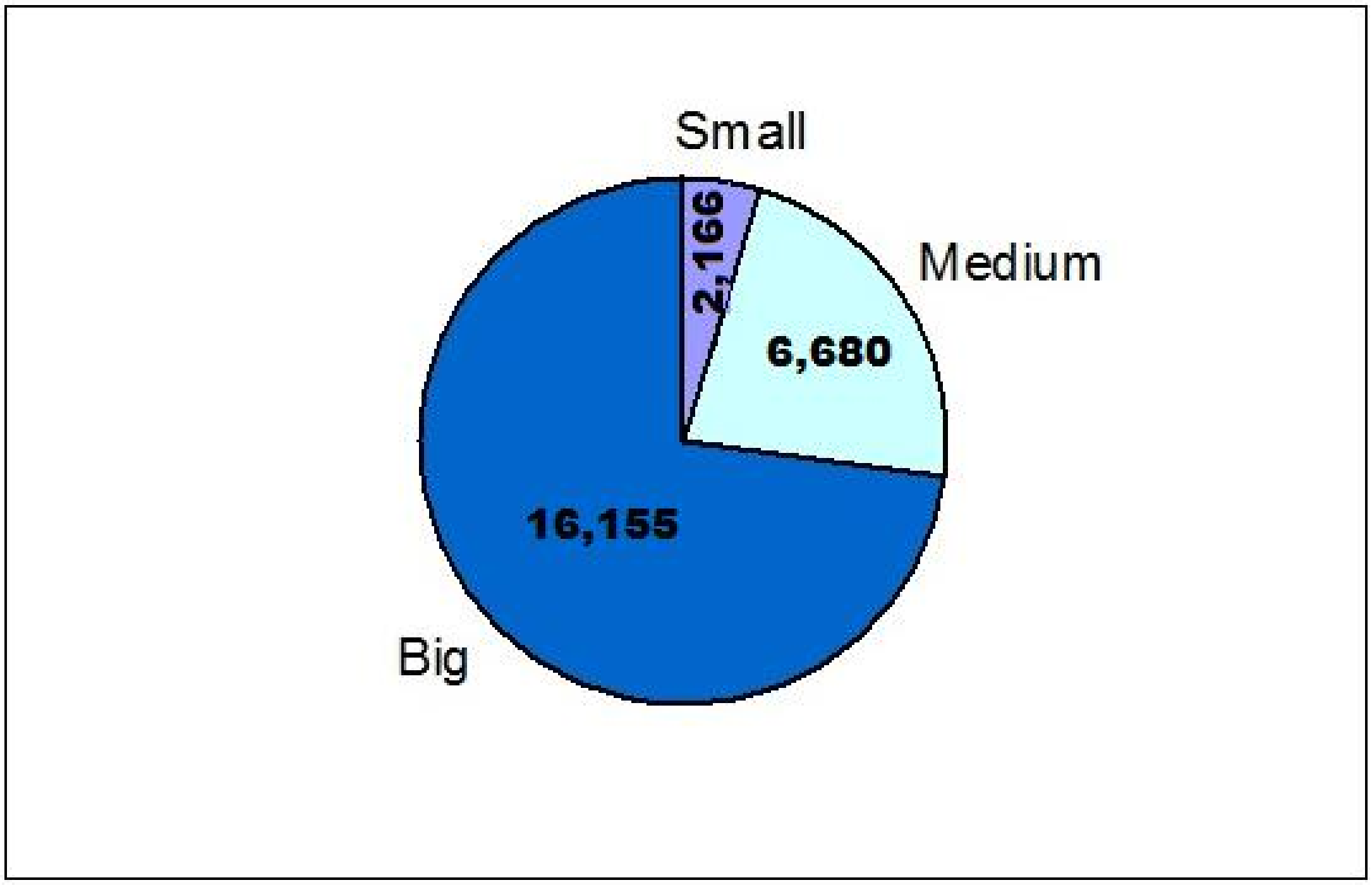

**Figure 48**: Quantitative Portions for 3 Sizes – $x^3$ Distribution over (1, 50)

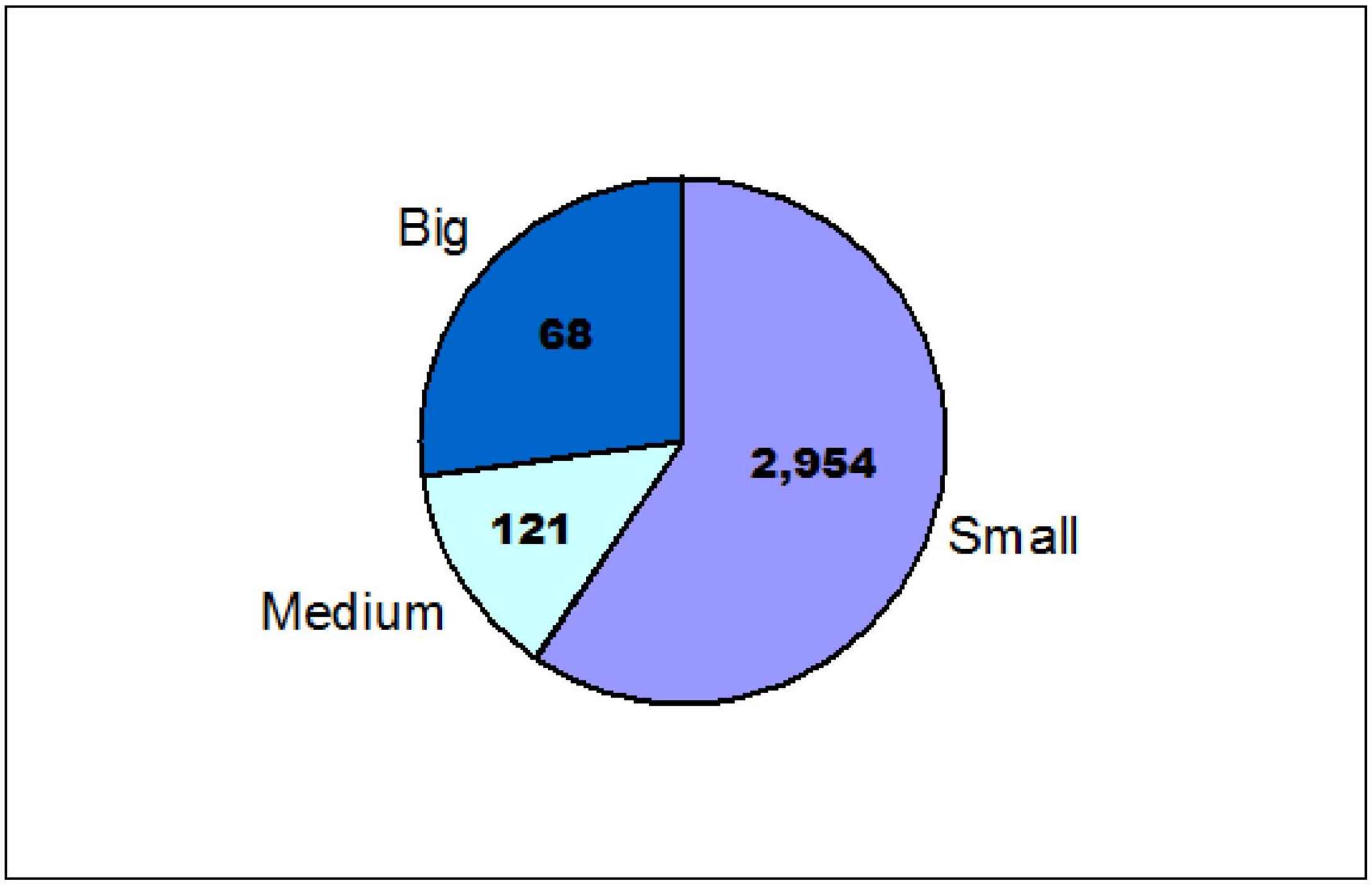

**Figure 49**: Quantitative Portions for 3 Sizes – US County Area Data

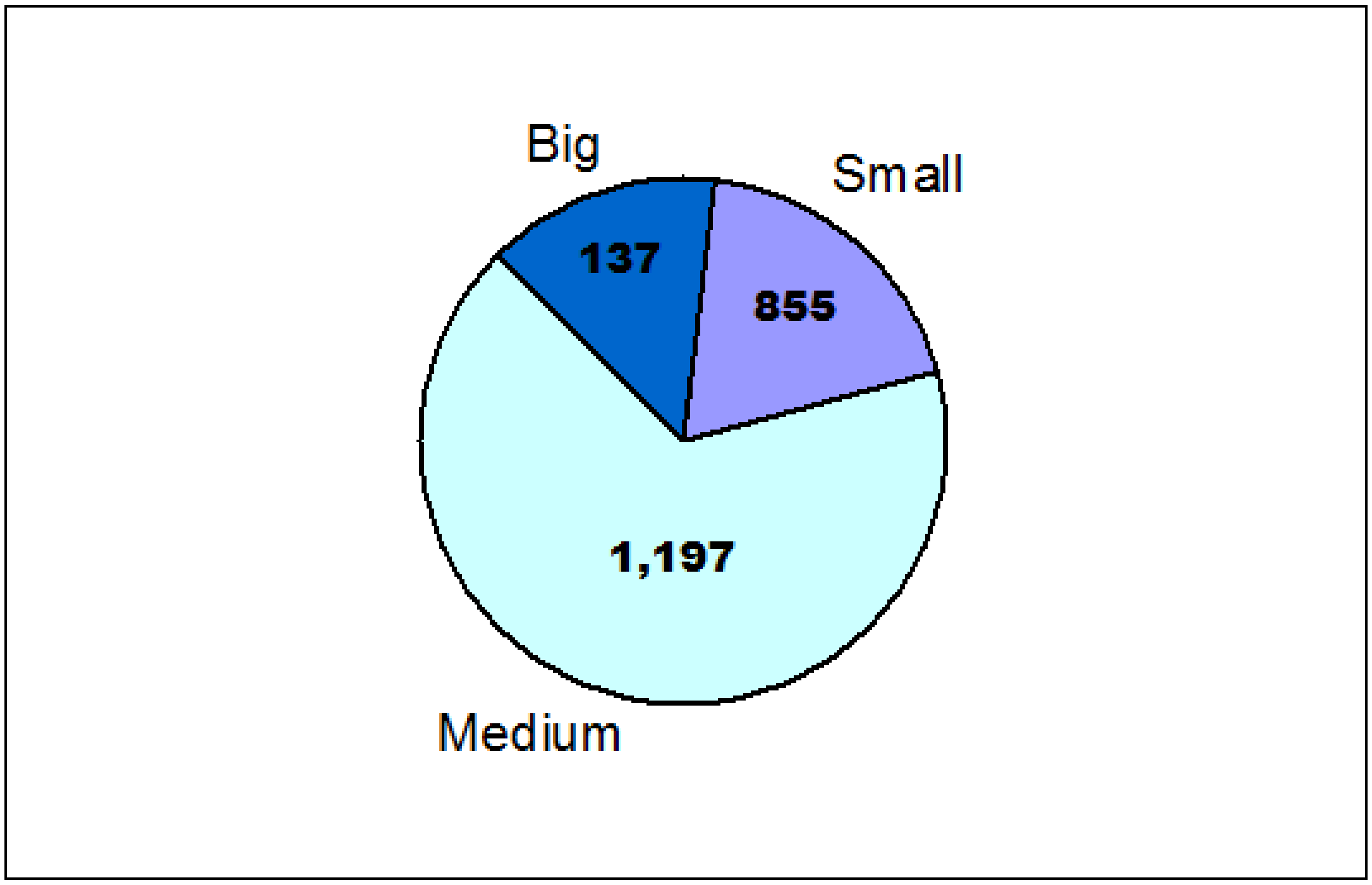

**Figure 50**: Quantitative Portions for 3 Sizes – Payroll Data Oklahoma State

## [21] The Constancy of Quantitative Portions in the k/x Case

Figure 45 of the perfectly Benford k/x distribution over (1, 10) comes the closest to that ideal, equitable, and fair 33.3% division for all the three sizes. Medium earns a tiny bit less than either Small or Big, but this is due only to its slightly lesser range as defined in the algorithm which endows the extra section of top 1% to Big and the extra section of bottom 1% to Small, and nothing extra to Medium. One must bear in mind though that k/x has that rare deterministic flavor in Benford's Law, totally lacking Digital Development Pattern. The distribution k/x is a unique case in the field Benford's Law. Indeed, from the definition of the k/x distribution we can deduce one of its essential properties, namely that by doubling quantity x we reduce its frequency (density height) by half. Surely for the transition such as: **x → 2x** the density of k/x always diminishes by half as in: **k/x → k/(2x)** = (k/x)/2. As a consequence, sums of quantities (approximately - rectangle area times some middle x value) are constant everywhere on the x-axis for all sub-intervals having an identical width on the x-axis.

Let us consider two adjacent regions under the k/x curve, a left region and a right region, both of equal length on the x-axis, as seen in Figure 51. Here, the left region has the advantage of having relatively higher density, but it also has the disadvantage of laying over lower x values. The right region has the disadvantage of having relatively lower density, but it has the advantage of laying over higher x values. As it happened, one factor offsets the other factor, and it all cancels out exactly, resulting in equal sums for the left and for the right regions. Let us prove this quantitative equality in the generic sketch of k/x distribution illustrated in Figure 51.

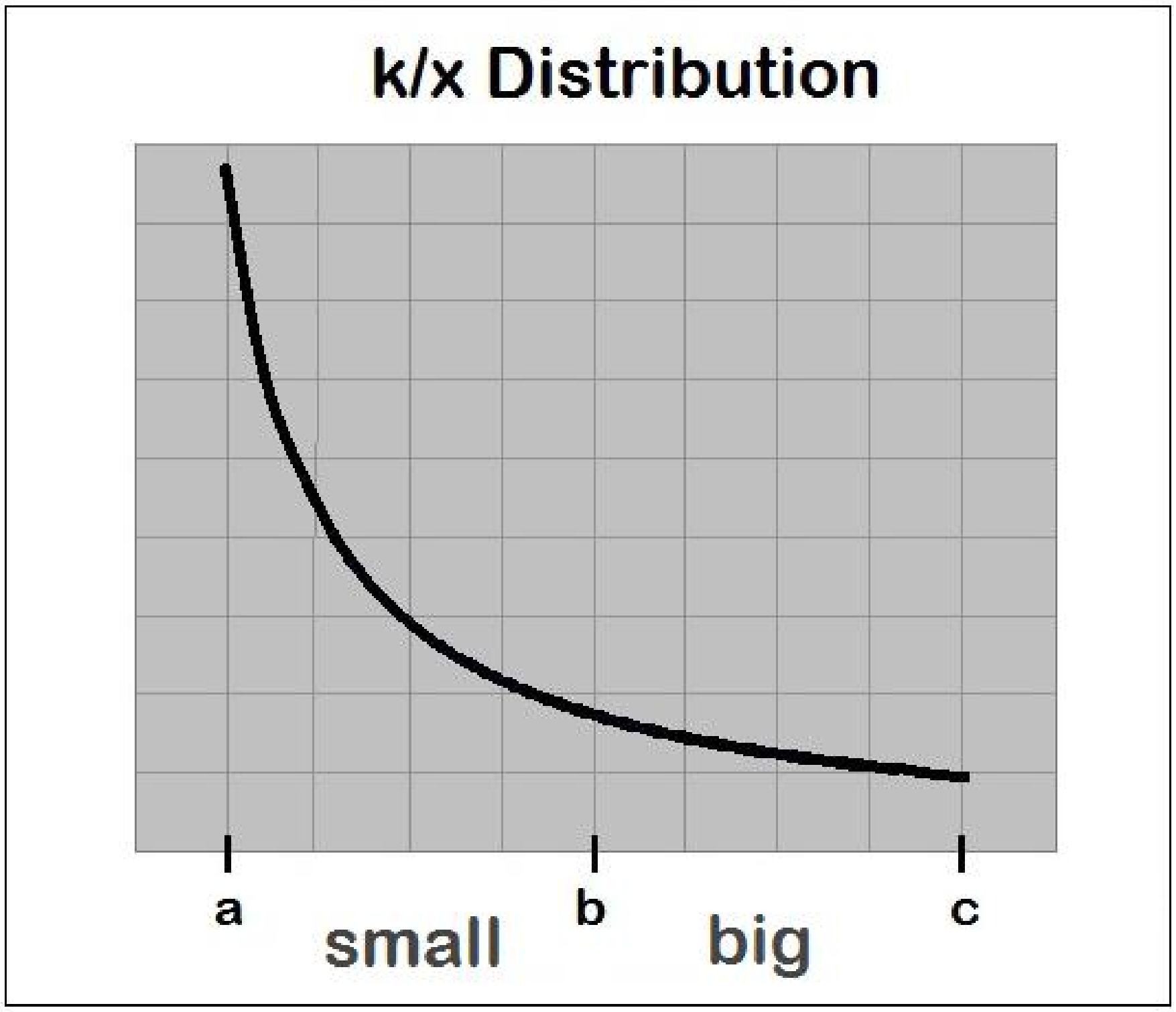

**Figure 51**: Equality of Quantitative Portions for k/x Distribution

When a continuous random distribution probability density is considered, it is not possible to sum 'amounts' in the usual manner by adding discrete values on the histogram; there are none here. To better illustrate the point, one has to trace back the path we always take going from a histogram of discrete values to an abstract continuous density probability curve (best fitting the actual data) by way of dividing the height of each rectangle of x-axis unity width by the total number of data points in the entire data set. Hence we define a fixed imaginary integer N representing the number of all the values within some imaginary data set perfectly fitting the density curve in question, namely fitting k/x defined over (a, c).

The product of an infinitesimal or tiny rectangular area under the k/x curve times N, should then represent the number of 'discrete values' falling within that infinitesimal sub-interval. If that product is then further multiplied by the value of the x-axis at the bottom of the infinitesimal rectangular, it yields the sum of all amounts within that tiny rectangular area. Algebraically:

Sum within infinitesimal sub-interval = xN[infinitesimal area]

It shall be assumed that (b – a) = (c – b), namely that the two Small and Big sub-intervals are of equal length. Let us first sum the quantitative portion of Small falling on the sub-interval (a, b). The overall sum of all the infinitesimal areas' sums within the sub-interval (a, b) is given by:

$$\text{Sum for (a, b)} = \sum \text{xN[mini area]} \quad \textit{[for all rectangles from a to b].}$$

Turning it into a definite integral, we get:

$$\text{Sum for (a, b)} = \int_a^b \text{xNf(x)dx}$$

$$\text{Sum for (a, b)} = \int_a^b \text{xN}(^{\text{k}}/_{\text{x}})\text{dx}$$

The x term cancels out, and we are left with:

$$\text{Sum for (a, b)} = \int_a^b \text{Nkdx}$$

$$\text{Sum for (a, b)} = \text{Nk(b} - \text{a)} = \text{Nk(Length of subinterval)}$$

Namely the same constant value of Nk[Length of sub-interval] for any sub-interval of comparable length. Hence the same result is obtained for Big falling on the sub-interval (b, c).

The setup of this proof surely resonates as something quite familiar in statistical methods. In the discrete case, the idea of summation is always employed in the definition of the average, as in [Avg] = [Sum]**/**[Number of Values]. Here too, the above expression ∫ xNf(x) dx corresponds to the generic definition of the average ∫ xf(x) dx *[over entire range]* in mathematical statistics, except that extra N term. But since N represents [Number of Values], it follows that the expression ∫ xNf(x) dx directly signifies sum.

## [22] Closer Quantitative Scrutiny of Random Logarithmic Data Sets

In light of the very consistent result in the rare case of the deterministic k/x distribution showing equality of quantitative portions by size, the critical question that arises here regards all the other much more common random data cases; namely what effect does Digital Development Pattern have on quantitative portions by size for logarithmic real-life data sets and distributions. Empirically, from the nine logarithmic random data sets and distributions of Figures 36 - 44, it seems that Small often obtains somewhat higher quantitative portions than the expected 33.3% of the purely Benford deterministic k/x case. The averages of quantitative portions by size for these nine random data sets and distributions are: **Small = 43%**, **Medium = 20%**, **Big = 37%**.

In order to gain better understanding about quantitative allocation by size for logarithmic data under the influence of Digital Development Pattern, it is necessary to let go of the parsimonious and simple division of data into merely 3 sizes of Small, Medium, and Big, and instead examine quantitative portions of numerous sizes by dividing the range of data into much smaller sub-intervals. Even dividing the range into only 9 categories say, 'very small', 'somewhat small', 'just small', 'almost medium, 'somewhat medium', 'just medium', 'almost big', 'somewhat big', 'very big' sizes, wouldn't do! It is necessary to examine a very large number of sizes, much larger than merely 9.

In Kossovsky (2014) Chapter 81, the concept of Leading Digits Inflection Point (LDIP) is discussed - exclusively in terms of its effects on digital behavior. LDIP is defined as the maximum or top point on the histogram of the logarithm-transformed data where it's temporary flat and horizontal. Also, $10^{LDIP}$ is the corresponding LDIP on the x-axis itself for the raw data. It is precisely at LDIP that the data or curve behaves purely logarithmically as in the k/x distribution case. LDIP is a digital turning point, where to the left of it digital configuration is milder than Benford, and to the right of it digital configuration is more extreme and skewer than Benford. Since Digital Development Pattern revolves around LDIP, the conjecture that arises naturally in the context of relative quantities is that LDIP also determines quantitative portions locally for mini sub-intervals on the x-axis; namely that LDIP is also a quantitative turning point.

Two factors lead to the more precise statement of the conjecture:

**(1)** The well-known fact that in almost all logarithmic (and even in some non-logarithmic) random data sets, the histogram 'very briefly' rises on the extreme left part of the x-axis for the lowest values in the beginning of the data set. The obvious implication regarding relative quantities is that quantitative portions on mini sub-intervals on the left-most part of the data increase as x increases, so that there the big contains more quantitative portion than the small. Surely, around the left-most part of the data, as x advances to higher values under taller heights of the local histogram, quantitative portions rise accordingly. All this occurs to the left of LDIP.

**(2)** Precisely at LDIP, the data behaves purely logarithmically as in the k/x distribution case. This fact implies that around LDIP quantitative portions on local mini sub-intervals are equal, as is the case for the k/x distribution.

Hence the conjecture extrapolates these two facts, and states that anywhere to the left of LDIP bigger sizes obtain higher quantitative portions than smaller sizes, while anywhere to the right of LDIP smaller sizes obtain higher quantitative portions than bigger sizes. Let us now turn to empirical observations in order to verify this theoretical conjecture. Indeed, closer empirical scrutiny applying numerous sizes (instead of merely 3) reveals a very consistent pattern in all logarithmic random data sets and distributions, where quantitative portions are dramatically different on the region to left of LDIP as compared with the region to right of LDIP. For the region to the right of LDIP, sub-intervals contain less and less quantitative portions as x gets larger. The exact opposite occurs for the region to the left of LDIP.

Figure 52 depicts the typical way quantitative portions occur in almost all random logarithmic data and distributions. Figure 52 utilizes 35,000 realizations from computer simulation runs of the Lognormal(9.3, 1.7). A total of 82 quantitative portions for 82 sizes are calculated here in steps of 1000 units of length on the x-axis, beginning from the origin 0 and ending at 82,000. Hence sums are calculated on [0, 1000), [1000, 2000), [2000, 3000), … , [80000, 81000), [81000, 82000). For example, the set of all the numbers falling within [0, 1000) are added (i.e. their sum is calculated) yielding the quantitative value of 1,554,816, and this is divided by the total quantitative sum of the entire 35,000 simulated Lognormal values which is 1,629,854,915, pointing to the ratio (1,554,816)**/**(1,629,854,915) or 0.095% as the quantitative portion.

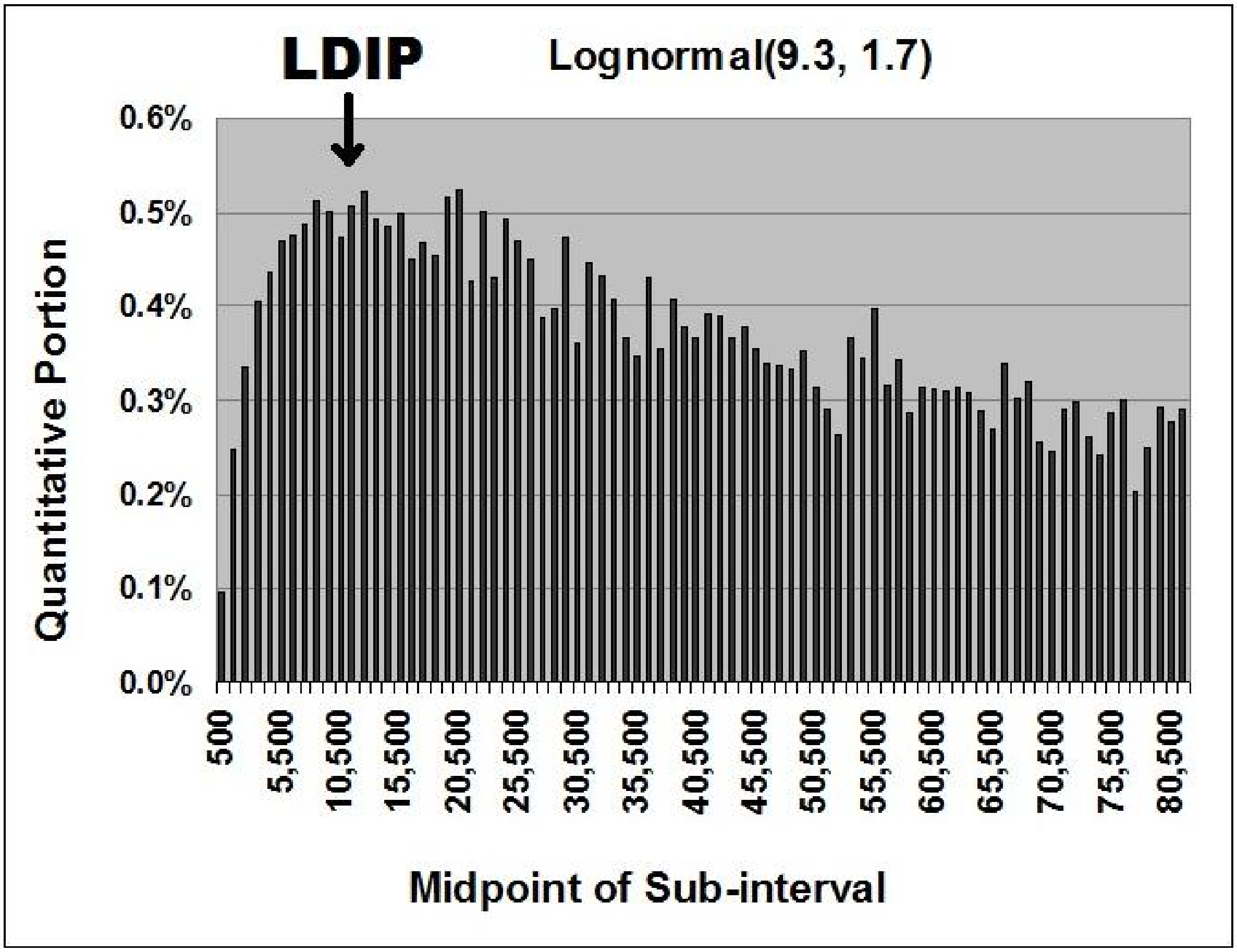

**Figure 52**: LDIP is the Turning Point in Quantitative Portions – Lognormal(9.3, 1.7)

In Kossovsky (2014) Chapter 81, Page 347, LDIP of the Lognormal is derived, yielding the generic expression $[e]^{\text{LOCATION PARAMETER}}$ for the LDIP on the x-axis. Applying this expression here we obtain $[e]^{\mathbf{9.3}} = [2.718281]^{\mathbf{9.3}} = 10{,}938$ as the LDIP on the x-axis. This calculation shows an excellent fit between the theoretical LDIP and the empirical observation of its location on the x-axis where a decisive quantitative reversal is occurring, as seen in Figure 52. It should be noted that the apparent falling long tail on the right in Figure 52 does indeed exist, and extended empirical examinations show that quantitative portions continue to fall further as additional mini sub-intervals are considered there for higher values of x well beyond 82,000. Hence in a sense, there is a strong resemblance (in overall-shape) between the typical histogram of random logarithmic data itself and its related chart of quantitative portions with the % unit for the vertical axis. Both appear to rapidly and impulsively rise strongly for a very short interval on the left in the beginning, only to then deeply regret this rushed act and to totally reverse themselves and fall gradually, slowly, and almost monotonically, on a much longer interval to the right.

A quantitative scheme with very few sizes applying very wide sub-intervals per size, typically encloses or ‘swallows’ LDIP within the smallest (first) size, and this totally masks the typical brief rise in quantitative portions around the left-most part of the data as was seen in Figure 52.

Figures 53 and 54 - which also pertain to the Lognormal(9.3, 1.7) - depict the typical way quantitative portions occur in almost all random logarithmic data sets and distributions when only very few sizes are created, applying very wide sub-intervals in the definition of sizes, and resulting in LDIP residing comfortably well within the smallest (first) size.

In the setup of Figures 53 and 54, the entire part of the range left of LDIP is being absorbed and easily swallowed within the smallest (first) size. The entire range of all 35,000 realizations of Lognormal(9.3, 1.7) is divided into much wider sub-intervals in long steps of 25,000 units of length on the x-axis, beginning from the origin 0 and ending at 375,000.

A total of 15 quantitative portions for 15 sizes are calculated here. Hence sums are calculated on [0, 25,000), [25,000, 50,000), … , [325,000, 350,000), [350,000, 375,000). For example, the set of the 23,987 numbers falling within [0, 25,000) are added (i.e. their sum is calculated) yielding the quantitative value of 182,794,046, and this is divided by the total quantitative sum of the entire 35,000 simulated Lognormal values which is 1,629,854,915, pointing to the ratio (182,794,046)**/**(1,629,854,915) or 11.2% as the quantitative portion. Also here, the set of the 4,457 numbers falling within [25,000, 50,000) are added (i.e. their sum is calculated) yielding the quantitative value of 158,330,171, and this is divided by the total quantitative sum of the entire 35,000 simulated Lognormal values which is 1,629,854,915, pointing to the ratio (158,330,171)**/**(1,629,854,915) or 9.7% as the quantitative portion.

Since the right edge of the first [0, 25,000) sub-interval is well beyond LDIP value of 10,938, no quantitative rise can be seen here whatsoever, and successive sub-intervals (i.e. sizes) consistently obtain lower and lower quantitative portions.

Note: Figure 54 depicts the relative [not the absolute/actual] quantitative portions for these 15 sizes. Total quantitative portion between 0 and 375,000 is only 64.5%, and the rest being 35.5% falls farther to the right. For example: area of 'Small 1' appears as (11.2%)/(64.5%) = 17.4%.

| Size Name | From | To | Quantitative Portion | Number of Data Points |
|---|---|---|---|---|
| Small 1 | 0 | 25,000 | 11.2% | 23,987 |
| Small 2 | 25,000 | 50,000 | 9.7% | 4,457 |
| Small 3 | 50,000 | 75,000 | 7.6% | 2,021 |
| Small 4 | 75,000 | 100,000 | 6.3% | 1,184 |
| Small 5 | 100,000 | 125,000 | 4.9% | 719 |
| Medium 1 | 125,000 | 150,000 | 4.3% | 513 |
| Medium 2 | 150,000 | 175,000 | 3.3% | 331 |
| Medium 3 | 175,000 | 200,000 | 2.9% | 251 |
| Medium 4 | 200,000 | 225,000 | 2.7% | 208 |
| Medium 5 | 225,000 | 250,000 | 2.5% | 171 |
| Big 1 | 250,000 | 275,000 | 2.4% | 149 |
| Big 2 | 275,000 | 300,000 | 1.9% | 109 |
| Big 3 | 300,000 | 325,000 | 2.0% | 107 |
| Big 4 | 325,000 | 350,000 | 1.4% | 67 |
| Big 5 | 350,000 | 375,000 | 1.4% | 61 |

**Figure 53**: Table of Quantitative Portions for 15 Sizes - Lognormal(9.3, 1.7)

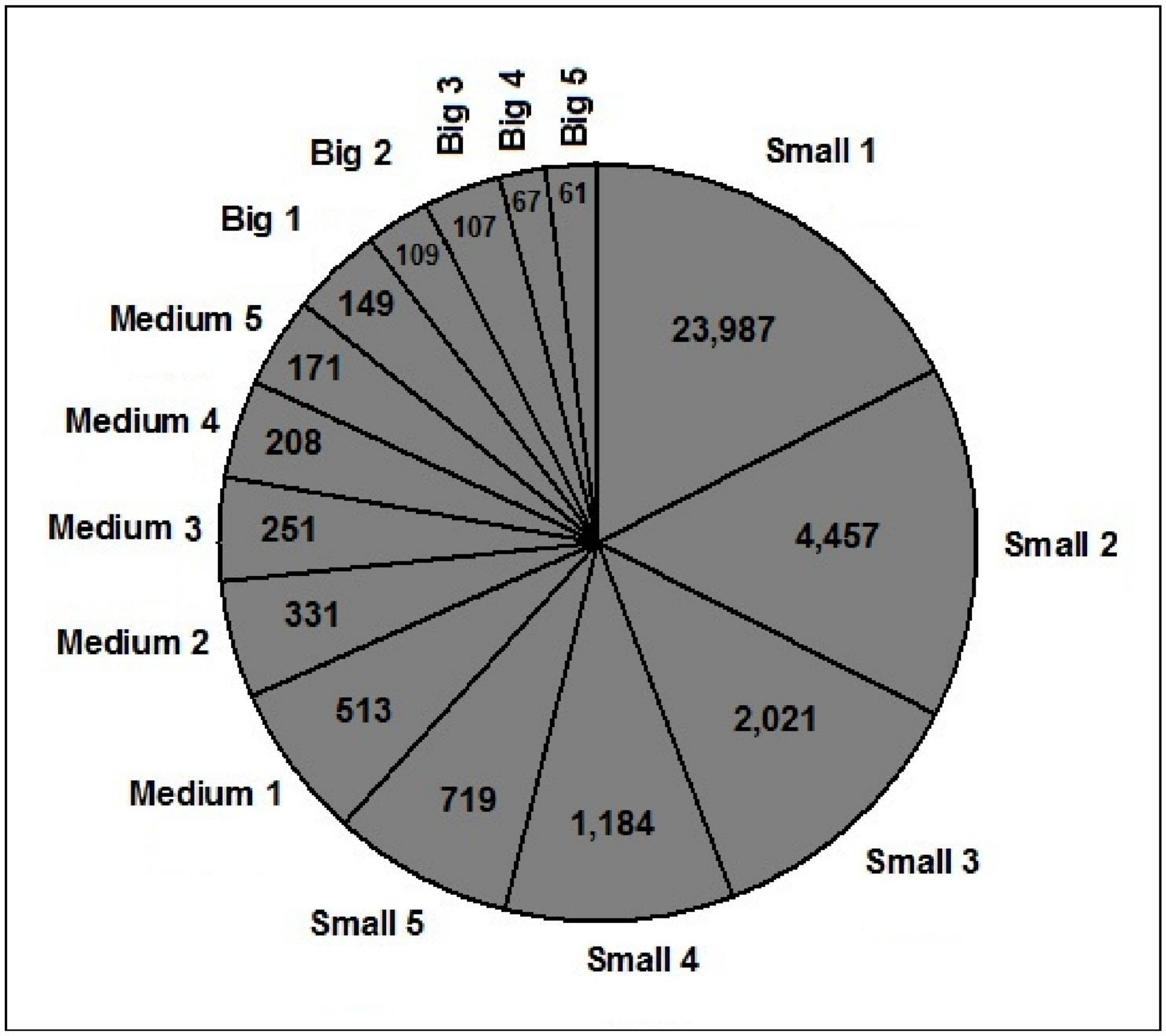

**Figure 54**: Chart of Quantitative Portions for 15 Sizes - Lognormal(9.3, 1.7)

Figure 55 is another demonstration of the typical way quantitative portions occur in almost all random logarithmic real-life and physical data sets. Figure 55 pertains to the data set of all 19,451 time intervals (in seconds) between successive global earthquakes in the year 2012. A total of 82 quantitative portions for 82 sizes are calculated in steps of 100 units of length on the x-axis, beginning from zero (which is very near the minimum value of 0.01) and ending at 8,200. Hence sums are calculated on [0, 100), [100, 200), [200, 300), … , [8000, 8100), [8100, 8200). For example, the set of all the numbers falling within [0, 100) are added (i.e. their sum is calculated) yielding the quantitative value of 71,101, and this is divided by the total quantitative sum of the entire 19,451 time intervals which is 31,617,664, pointing to the ratio (71,101)/(31,617,664) or 0.22% as the quantitative portion.

Note: The total quantitative sum of the entire 19,451 time intervals is 31,617,664. This value closely corresponds to the theoretical/calculated 60*60*24*366 = 31,622,400 seconds for the entire year, since 2012 was a leap year (given one extra day) with 366 days in total.

The approximate value of LDIP for the global 2012 earthquake data set [according to the conjecture] can be found by visually reading the highest point of the chart of Figure 55 where ‘curve’ is momentarily flat and horizontal. The value of 1550 approximately seems like a reasonable choice.

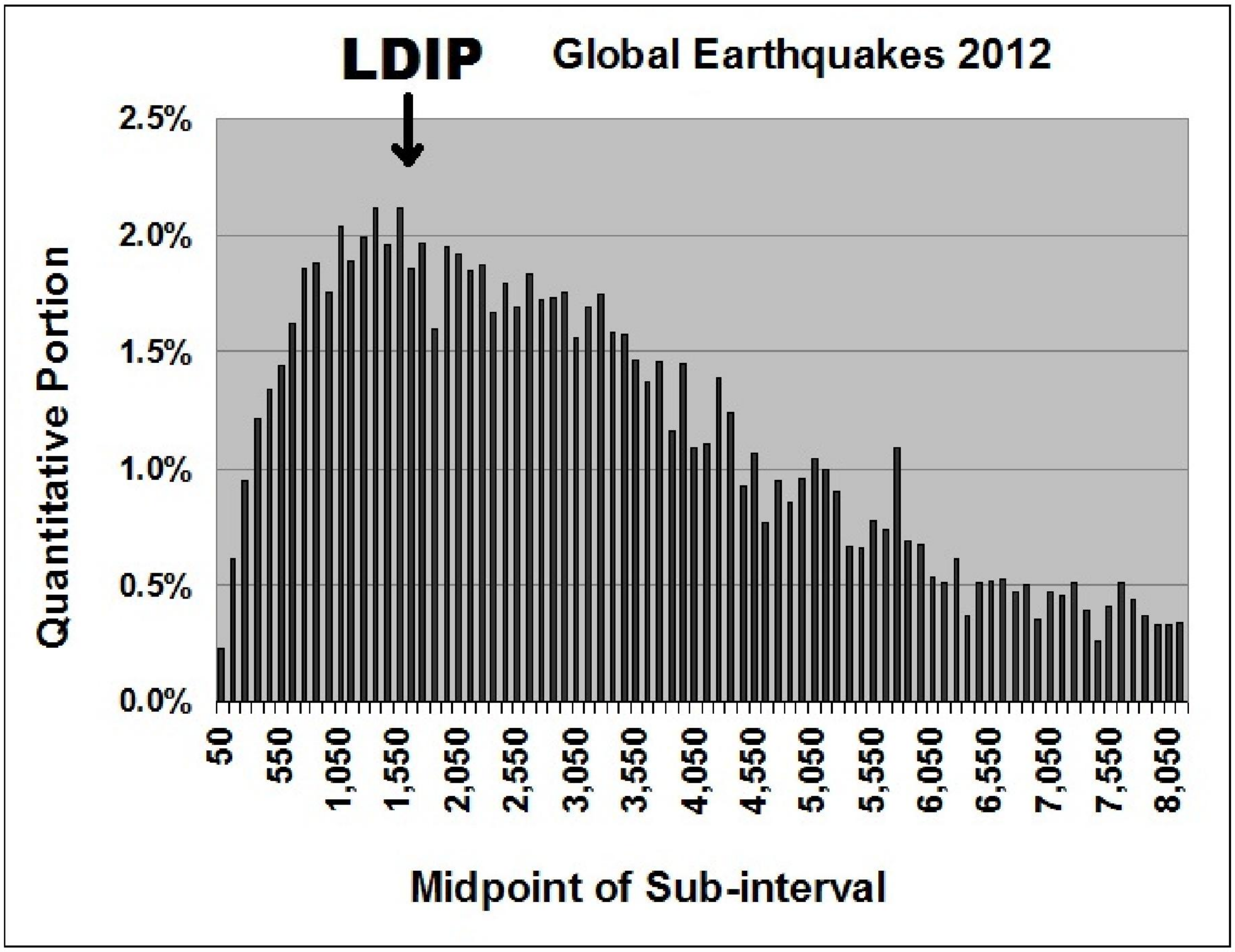

**Figure 55**: LDIP is the Turning Point in Quantitative Portions – Global Earthquake Data

Let us now figure out the approximate value of LDIP for the global 2012 earthquake data set by visually reading the highest point on its log histogram where 'curve' is momentarily flat and horizontal, as depicted in Figure 56. The choice of 3.2 log value seems like a reasonable one.

The LDIP on the x-axis itself should then be approximately $10^{3.2}$ = 1585, and this value corresponds nicely with the quantitatively-estimated LDIP shown in Figure 55.

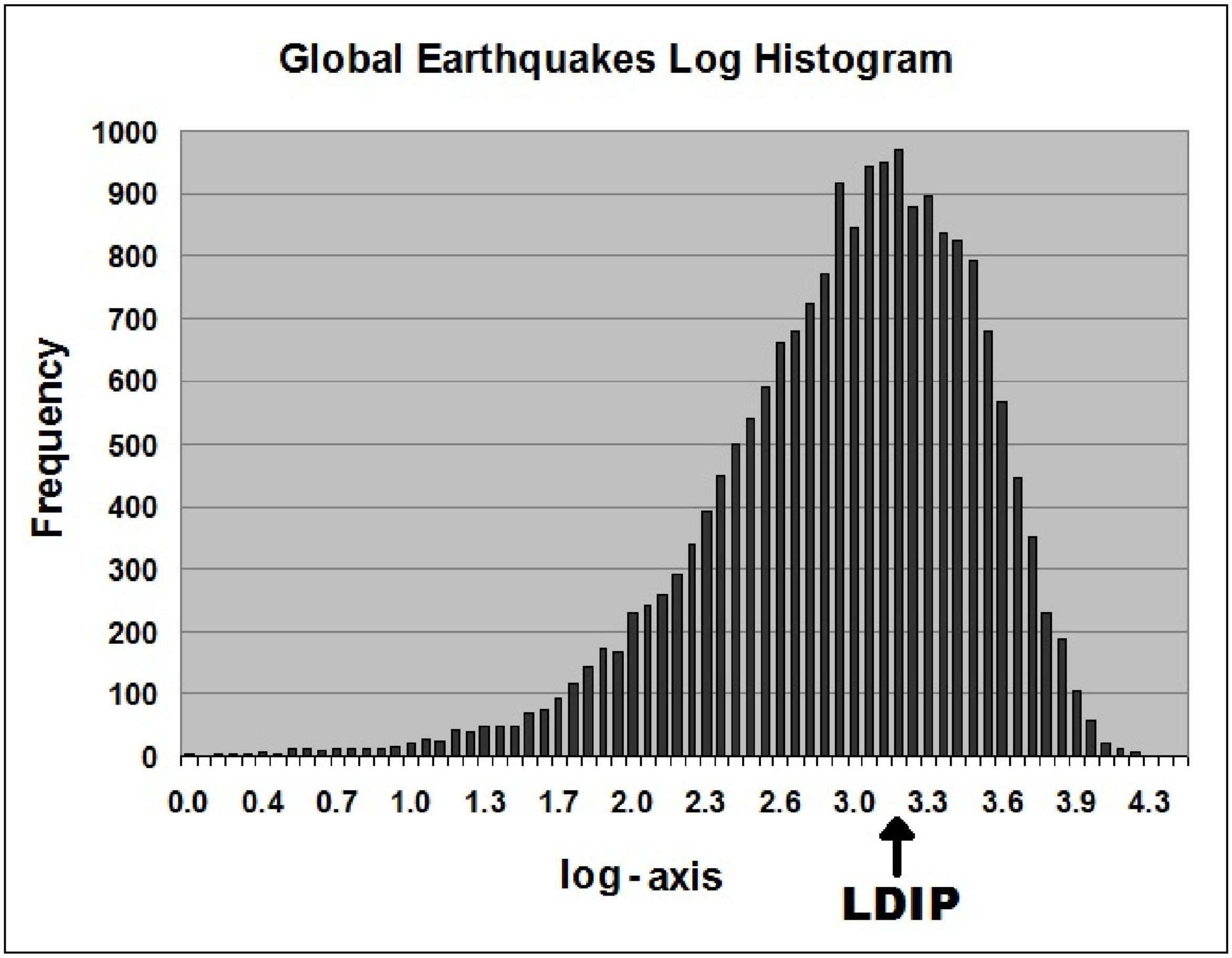

**Figure 56**: Visual Determination of LDIP as the Highest Log Point – Global Earthquake Data

Quantitative portions for the global 2012 earthquake data set applying only 3 sizes via the algorithm in chapter 19 yielded: Small = 46.0%, Medium = 33.9%, and Big = 20.1%. Figure 55 which applies 82 sizes provides a more detailed description of how quantitative portions are occurring here. According to the algorithm in chapter 19, Border Point A is 2764, and Border Point B is 5516. Therefore the definition of Small classifies those values which are less than 2764 as such. It follows that LDIP value of approximately 1585 is well within the range of the category of Small, and this fact gives rise to the expectation that quantitative proportions are such that Small > Medium > Big, as indeed shown by the result of the algorithm.

## [23] General Derivation of Log Density for Any Random Distribution

In order to rigorously prove the above conjecture, namely that Leading Digits Inflection Point is also the decisive quantitative turning point for any random distribution f(x), it is necessary first to derive the generic expression of the density distribution of the log-transformed g(y), namely g(log(x)), and then to demonstrate that the x-axis locations of its rise and fall corresponds exactly to the x-axis locations of the rise and fall of local mini quantitative portions.

Let f(x) be a continuous and differential function over the range R, serving as the probability density function for variable X.

Let g(y) be the probability density function for the base B logarithmically-transformed X variable, namely for $Y = LOG_B X$.

The Distribution Function Technique shall be applied using similar notations as in Freund's book "Mathematical Statistics", sixth edition, chapter 7.3.

The Distribution Function (Cumulative Distribution Function) of Y, denoted as G(y), is then:

G(y) = Probability[ Y < y ] = Probability[ $LOG_B X$ < y ].

Taking base B to the power of both sides of the last inequality we obtain:

G(y) = Probability[ $B^{LOG_B(X)} < B^y$ ] = Probability[ $X < B^y$ ], hence

$$G(y) = \int_{-\infty}^{B^y} f(x)dx$$

Applying the fundamental relationship in mathematical statistics between f(x) and F(x), namely:

$$f(x) = \frac{d(F(x))}{dx}$$

$$Probability\ Density\ Function(x) = \frac{d(Cumulative\ Function(x))}{dx}$$

We then differentiate G(y) with respect to **y**, and obtain:

$$g(y) = \frac{d}{dy}\int_{-\infty}^{B^y} f(x)dx$$

Letting **u(y) = B**$^y$, we then substitute **u** as a function of **y** for $B^y$ and obtain:

$$g(u(y)) = \frac{d}{dy}\int_{-\infty}^{u} f(x)dx$$

The Chain Rule states that:

If G(y) = f(u(y)), then G'(y) = f'(u(y))*u'(y)

and in Leibniz's notation:

$$\frac{dG}{dy} = \frac{dG}{du} * \frac{du}{dy}$$

[Taking Leibniz's notation to extreme, as if on faith, the **du** in the numerator 'cancels out' the **du** in the denominator, a procedure which highly annoys Calculus, provoking it to protest against the demeaning reduction of its rich complexity into simple Algebra.]

We obtain:

$$g(u(y)) = [\frac{d}{du}\int_{-\infty}^{u} f(x)dx\,] * [\frac{du}{dy}]$$

$$g(u(y)) = [\frac{d}{du}\int_{-\infty}^{u} f(x)dx\,] * [\frac{d}{dy}\,B^y\,]$$

The Second Fundamental Theorem of Calculus states that:

$$\frac{d}{dx}\int_{a}^{x} f(t)dt = f(x)$$

Hence g(y) - the Probability Density Function of variable y - is reduced to:

$$g(y) = f(B^y) * \frac{d}{dy}\,B^y$$

From the basic rules of derivatives of exponential functions we get:

$$g(y) = f(B^y) * \ln(B) * B^y$$

Using base 10 would imply that B = 10; recalling the fact that f stands for the probability density function for the original variable x itself; and recalling the fact that $Y = LOG_BX$, namely that $B^Y = x$; we finally obtain:

$$g(y) = f(B^y) * \ln(B) * B^y$$

$$g(y) = f(\mathrm{x}) * \ln(10) * \mathrm{x}$$

$$g(\log(x)) = \ln(10) * x * f(\mathrm{x})$$

Probability Density Function of Log (x) $= \ln(10) * x * f(\mathrm{x})$

## [24] Leading Digits Inflection Point is also Quantitative Turning Point

Leading Digits Inflection Point is defined as the location on the log-axis where the probability density function g(log(x)) curve is at its maximum, namely where g(log(x)) is momentarily flat, horizontal and uniform, neither rising nor falling. Extensive empirical evidences strongly indicate that the nature of random and statistical data (Benford and non-Benford) in extreme generality is characterized by log-gradualism; that log histogram almost never starts suddenly nor ends abruptly around some high initial or final value; rather it starts and ends very low near the log-axis itself; rising up to LDIP; then falling back towards the log-axis; mimicking an upside-down U curve; and that it nearly always shows a marked curvature, as it concaves down with negative 2nd derivative around the core of the data. This implies that there is typically only one unique maximum; namely only one LDIP.

In chapter 21 on the quantitative portions in the k/x distribution case, the generic expression for quantitative portion within any sub-interval (a, b) for <u>any</u> random distribution was given as:

$$\text{Sum for (a, b)} = \int_{a}^{b} N * x * f(x)\, dx$$

Clearly the two expressions, the one for quantitative portions as well as the one for log density, both revolve around the product **x*f(x)**, except for the irrelevant constants N and ln(10). The fact that the expression for quantitative portions involves a definite integral, while the expression for log density does not involve definite integrals is also irrelevant, since the former can be written locally simply as [b – a]*N*x*f(x) where b is only tiny bit greater than a. Hence, to the left of LDIP where x*f(x) is rising, quantitative portions are also rising, and to the right of LDIP where x*f(x) is falling, quantitative portions are also falling. Therefore the derivation concerning the generic expression of log density of the previous chapter constitutes a rigorous proof that LDIP is also a quantitative turning point, apart from serving also as a digital turning point.

## [25] Leading Digits Inflection Point is Number System Invariant

The definition of Leading Digits Inflection Point as the location on the log-axis where the curve is at its maximum clearly necessitates a positional number system as well as a particular base B. Without a positional number system no discussion about logarithms or digits and their distributions can take place. Remarkably, the position of LDIP on the x-axis is unchanged under a base transformation. Surely the log density of any given distribution appears differently in different bases, yet when LDIP on that B-base log-axis is translated into its related location on the x-axis, it always points to the same point regardless of the value of the base B. One doesn't need any rigorous mathematical proof or complex arguments in order to demonstrate this result, as it follows directly from the principles of the General Law of Relative Quantities as outlined in the 7th Section of Kossovsky (2014). Since it was shown earlier that LDIP is also the quantitative turning point, it then follows that it is base-invariant as well as number-system-invariant. Histograms and continuous density curves are invariant under a base transformation as well as being totally independent on the number system in use. The only changes that occur when a base is transformed or when a totally new number system is put into use (say Roman or Egyptian Numerals) are the symbols below the x-axis and to the left of the y-axis in the Cartesian Plane, but not the [quantitative meaning of the] actual set of tuples constituting the curve. LDIP is also the turning point in relative quantitative portions measured judiciously on tiny sub-intervals, and this quantitative portion is a primitive and basic concept, not necessitating any digits, base, or even a number system. Quantitative portion contribution from each data point in the data set is the confluence of the x-axis and the y-axis positional interplay; being the product [length of the x-value from origin]*[height of the y-axis].

## [26] Curved Two-Dimensional Random Partition Process

During a conversation with Don Lemons about the relationship between random partition processes and Benford's Law, the author proposed the most primitive perhaps or the most simplistic two-dimensional random partition of (say) an oval-shaped tablecloth, soft fabric, or a cardboard, where the cutting is performed in the freest possible way, without any restrictions. Figure 57 depicts an example of an original oval shape partitioned along random curved paths.

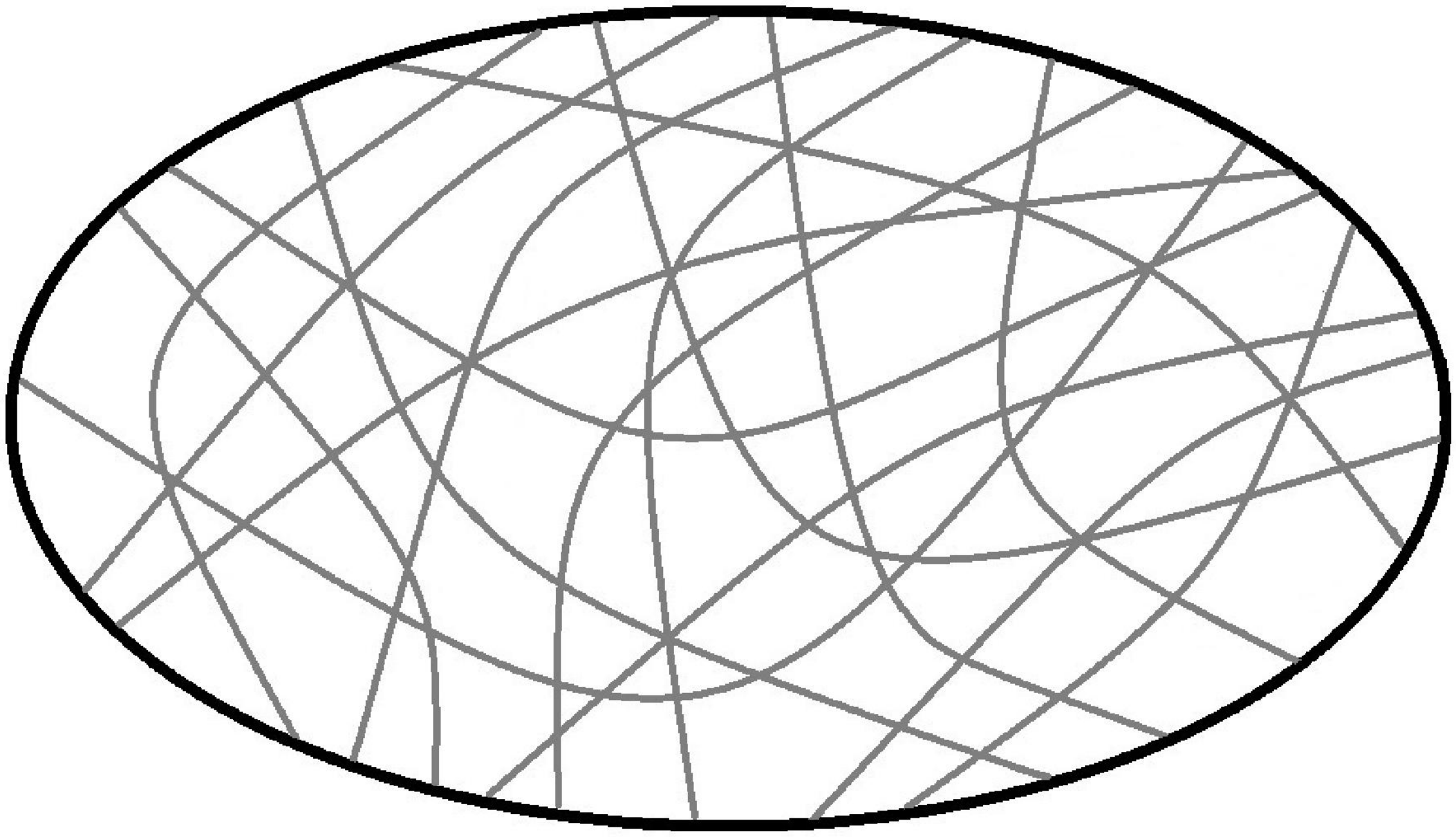

**Figure 57:** Oval-like Shaped Tablecloth is Partitioned via the Cutting along Curved Paths

A large and strong scissor is imagined, being allowed to curve around while cutting without constraints. This is done by cutting from any edge to any other edge, while turning around any which way one sees fit, and without stopping in the middle. The overall oval shape stays intact after each such cutting act, to let the next cutting act be performed also on an oval-shaped tablecloth, until the very end of the long cutting process, and then when no more cutting is desired, when the scissor is put away for good, the still-standing fragile oval with its physically cut pieces is dismantled, after which the oval shape is no longer intact, and its pieces are gathered into one big pile, followed by the recording of the values of the areas of all the pieces. The author conjectures that such restriction-less, natural, and spontaneous partition should lead to Benford even with only very few resultants pieces. Due to the complex two-dimensional curve-like nature of this partition model, the enormous challenge here is in setting up any Monte Carlo computer simulations in order to verify the conjecture, and especially in providing a rigorous mathematical proof.

## [27] Linear Two-Dimensional Random Partition Process

Don Lemons has suggested modifying the above oval-like curved partition process, performing it in a linear way instead, so as to obtain a process that could be more systematic, simpler, and by far easier to simulate and to provide a rigorous mathematical proof. This modified model is of strictly linear partitions, both vertically and horizontally, and with the original cardboard being a square or a rectangle, as opposed to a round oval shape.

This linear partition model performed on a square or on a rectangle, is done by creating random marks via the continuous Uniform on the y-axis vertically, in addition to creating random marks via the continuous Uniform on the x-axis horizontally, followed by the partitioning the original square or rectangle along these two sets of marks, resulting in the creation of numerous mini rectangles, and where the sum of all the mini areas is equal to the area of the original square or rectangle. The areas of these mini rectangles are measured and recorded to verify the Benford digital configuration for these values.

As an illustrative example of the linear model - visualized in Figure 58 - a square area of 7 meters width by 7 meters height is to be randomly partitioned via Monte Carlo simulations of 24 random horizontal marks and 24 random vertical marks using the continuous Uniform(0, 7) for each dimension to generate independent random realizations.

The actual cutting along those **24** random marks scattered throughout the x-axis horizontal dimension creates **25** horizontal width-compartments.

The actual cutting along those **24** random marks scattered throughout the y-axis vertical dimension creates **25** vertical height-compartments.

Together, these two-dimensional partitions create 25*25 or **625** random areas which are almost always rectangles. Figure 58 depicts the results of one Monte Carlo simulation of such partition.

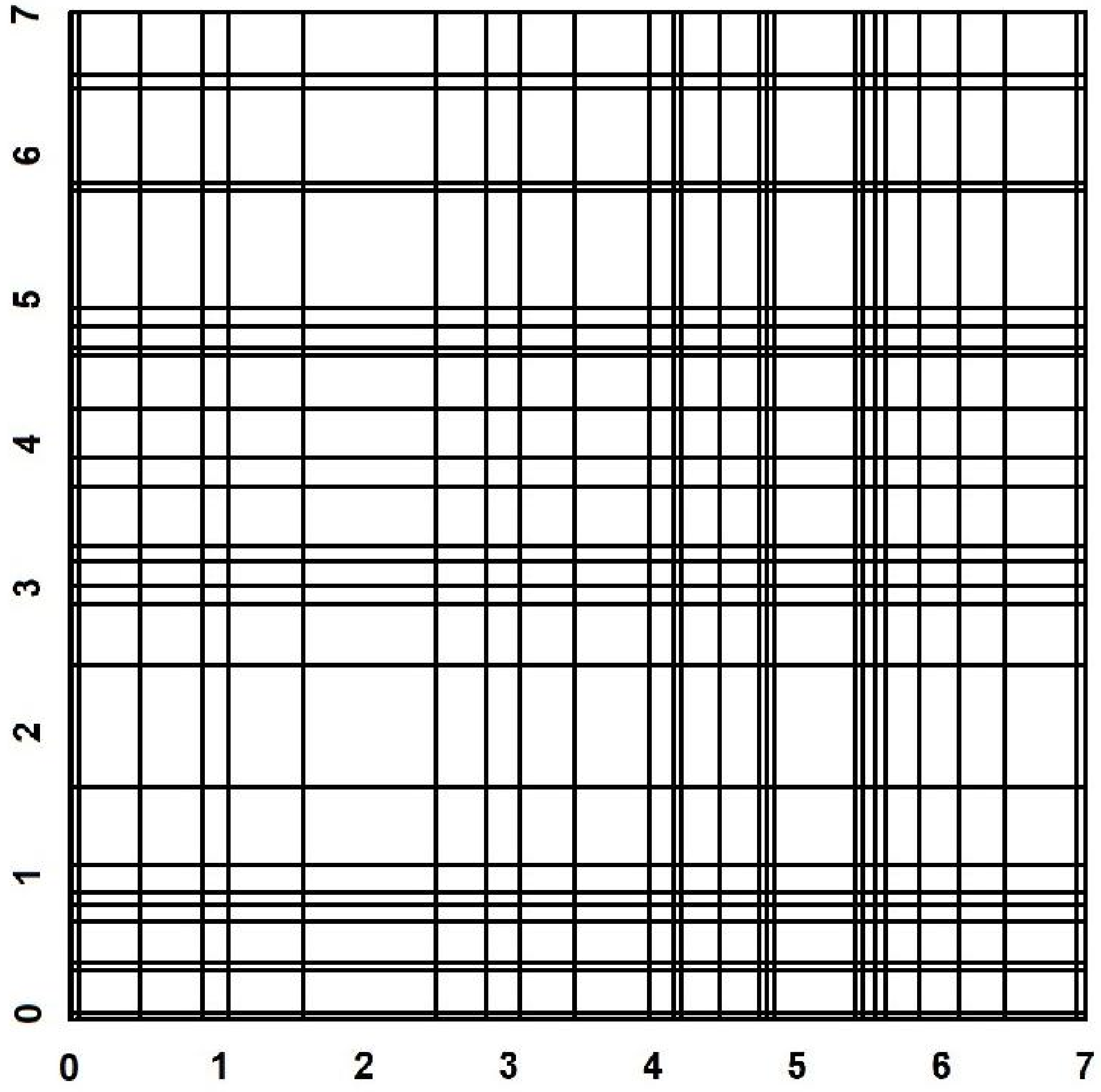

**Figure 58:** Simulated 24 Grids along the Horizontal Axis and 24 Grids along the Vertical Axis

The following is the set of these 24 simulated random marks along the **horizontal** dimension, sorted low to high, plus the two boundary values of 0 and 7:

{**0.000**, 0.023, 0.412, 0.862, 1.043, 1.660, 2.523, 2.826, 3.102, 3.447, 3.966, 4.130, 4.139, 4.366, 4.787, 4.857, 4.869, 5.416, 5.421, 5.489, 5.610, 5.821, 6.164, 6.450, 6.986, **7.000**}

The following is the set of the 25 differences in the above set, representing the 25 **width** values seen in Figure 58:

{0.023, 0.389, 0.450, 0.181, 0.617, 0.863, 0.303, 0.277, 0.345, 0.519, 0.164, 0.008, 0.227, 0.422, 0.070, 0.012, 0.546, 0.005, 0.068, 0.121, 0.211, 0.343, 0.286, 0.536, 0.014}

The following is the set of these 24 simulated random marks along the **vertical** dimension, sorted low to high, plus the two boundary values of 0 and 7:

{**0.000**, 0.032, 0.311, 0.346, 0.645, 0.794, 0.876, 1.065, 1.601, 2.450, 2.928, 3.006, 3.196, 3.323, 3.729, 3.903, 4.264, 4.664, 4.678, 4.848, 4.926, 5.755, 5.818, 6.528, 6.671, **7.000**}

The following is the set of the 25 differences in the above set, representing the 25 **height** values seen in Figure 58:

{0.032, 0.279, 0.035, 0.299, 0.148, 0.082, 0.189, 0.536, 0.849, 0.478, 0.078, 0.190, 0.127, 0.406, 0.174, 0.361, 0.400, 0.014, 0.170, 0.078, 0.829, 0.063, 0.710, 0.143, 0.329}

These two sets of differences are multiplied [i.e. each element in one set is multiplied by each element of the other set], yielding the set of 635 products representing the areas of the 635 mini rectangles created by the partition process.

First digits proportions are fairly close to Benford, even for this limited 25 by 25 partition:

Random 625 Rectangles - {30.1%, 18.2%, 11.2%, 11.2%, 7.7%, 5.9%, 5.9%, 4.2%, 5.6%}
Benford's Law 1st Digits - {30.1%, 17.6%, 12.5%, 9.7%, 7.9%, 6.7%, 5.8%, 5.1%, 4.6%}

Figure 59 depicts the bar chart of the digital comparison to Benford's Law. The low Sum of Squares Deviation (SSD) value of 7.0 for the set of resultant 625 rectangles indicates that this digit configuration is in good conformity with Benford's Law. Figure 60 depicts the histogram of the areas of resultant 625 rectangles, confirming the general principle in random partition processes that predicts decisive quantitative skewness where the relatively small is much more numerous than the relatively big.

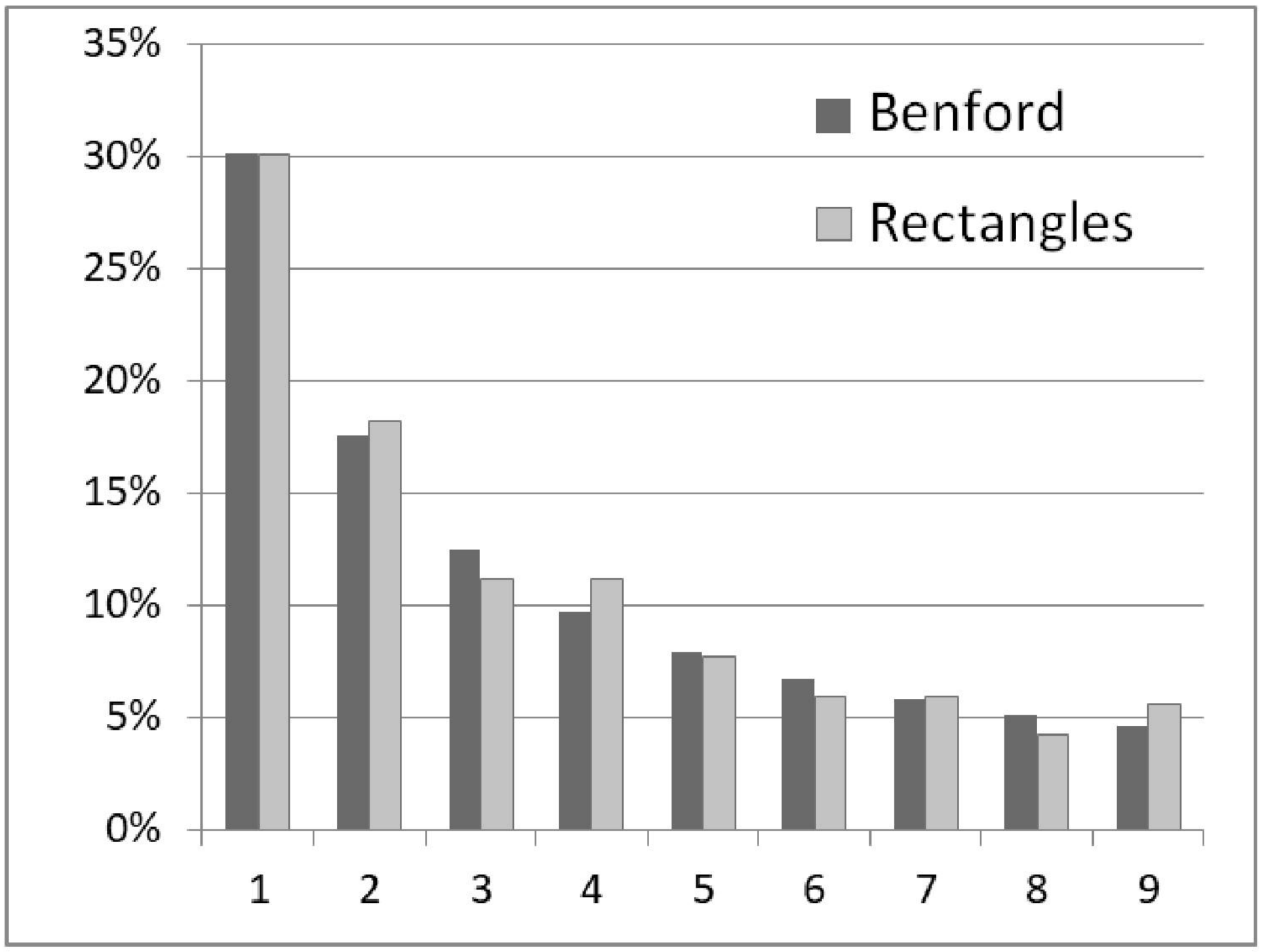

**Figure 59:** Good Fit to Benford - Simulated 625 Random Rectangles

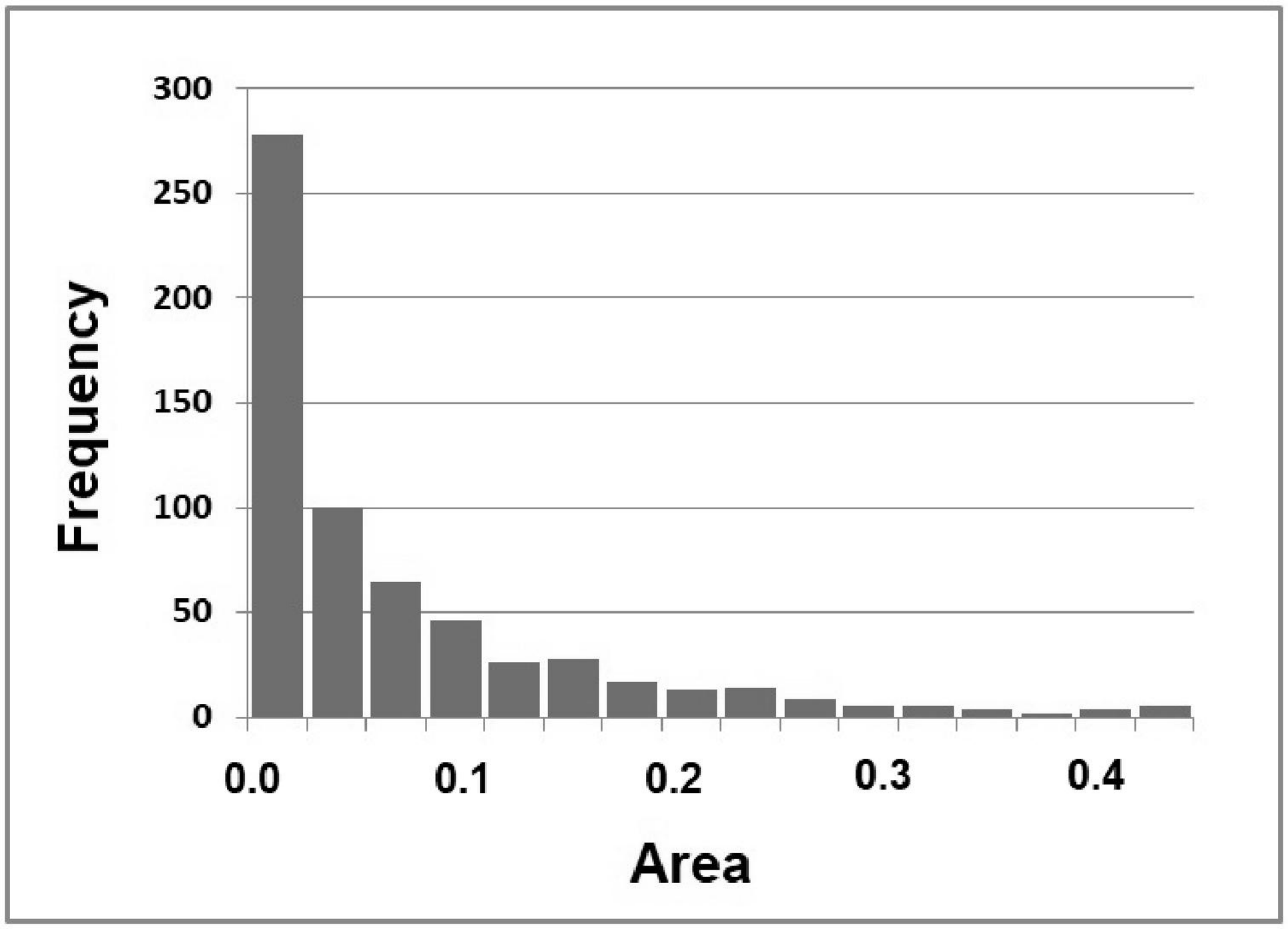

**Figure 60:** Skewed Histogram with a Tail to the Right - Simulated 625 Rectangles

Performing 8 additional Monte Carlo simulations of the above 25 by 25 linear partition scheme yielded a variety of results, mostly with slightly higher SSD values. Yet, quite surprisingly though, the average digit configuration of these 8 simulations came out extremely close to Benford, with the very low SSD value of 1.9! And this is so in spite of the fact that this modest 25 by 25 partition process involves only 48 actual cutting acts, and yields only 625 parts! Figure 61 depicts these 8 simulation results, including the digit-by-digit average of these 8 results at the bottom line (in bold font). Figure 62 depicts the excellent fit to Benford of the average of these 8 simulations.

| Digit 1 | Digit 2 | Digit 3 | Digit 4 | Digit 5 | Digit 6 | Digit 7 | Digit 8 | Digit 9 | SSD |
|---|---|---|---|---|---|---|---|---|---|
| 33.3% | 16.5% | 11.0% | 8.6% | 8.3% | 7.4% | 5.0% | 5.8% | 4.2% | **16.5** |
| 28.8% | 16.3% | 11.5% | 9.6% | 7.7% | 7.5% | 7.8% | 5.6% | 5.1% | **9.8** |
| 30.9% | 18.2% | 10.4% | 10.4% | 8.8% | 6.9% | 5.0% | 4.8% | 4.6% | **7.5** |
| 28.3% | 17.8% | 13.4% | 10.7% | 8.5% | 6.6% | 4.5% | 5.9% | 4.3% | **7.9** |
| 28.2% | 19.4% | 12.2% | 9.8% | 9.8% | 5.0% | 4.5% | 6.9% | 4.5% | **18.2** |
| 29.8% | 17.9% | 10.2% | 9.8% | 9.3% | 5.6% | 5.9% | 6.4% | 5.1% | **10.3** |
| 31.8% | 16.2% | 12.2% | 9.9% | 7.8% | 6.2% | 6.7% | 5.1% | 4.0% | **6.7** |
| 27.7% | 18.4% | 11.8% | 9.9% | 9.1% | 6.6% | 7.2% | 5.3% | 4.0% | **10.8** |
| **29.8%** | **17.6%** | **11.6%** | **9.8%** | **8.7%** | **6.5%** | **5.8%** | **5.7%** | **4.5%** | **1.9** |

**Figure 61:** 8 Distinct Simulations of 25 by 25 Partitions – Average is Nearly Benford

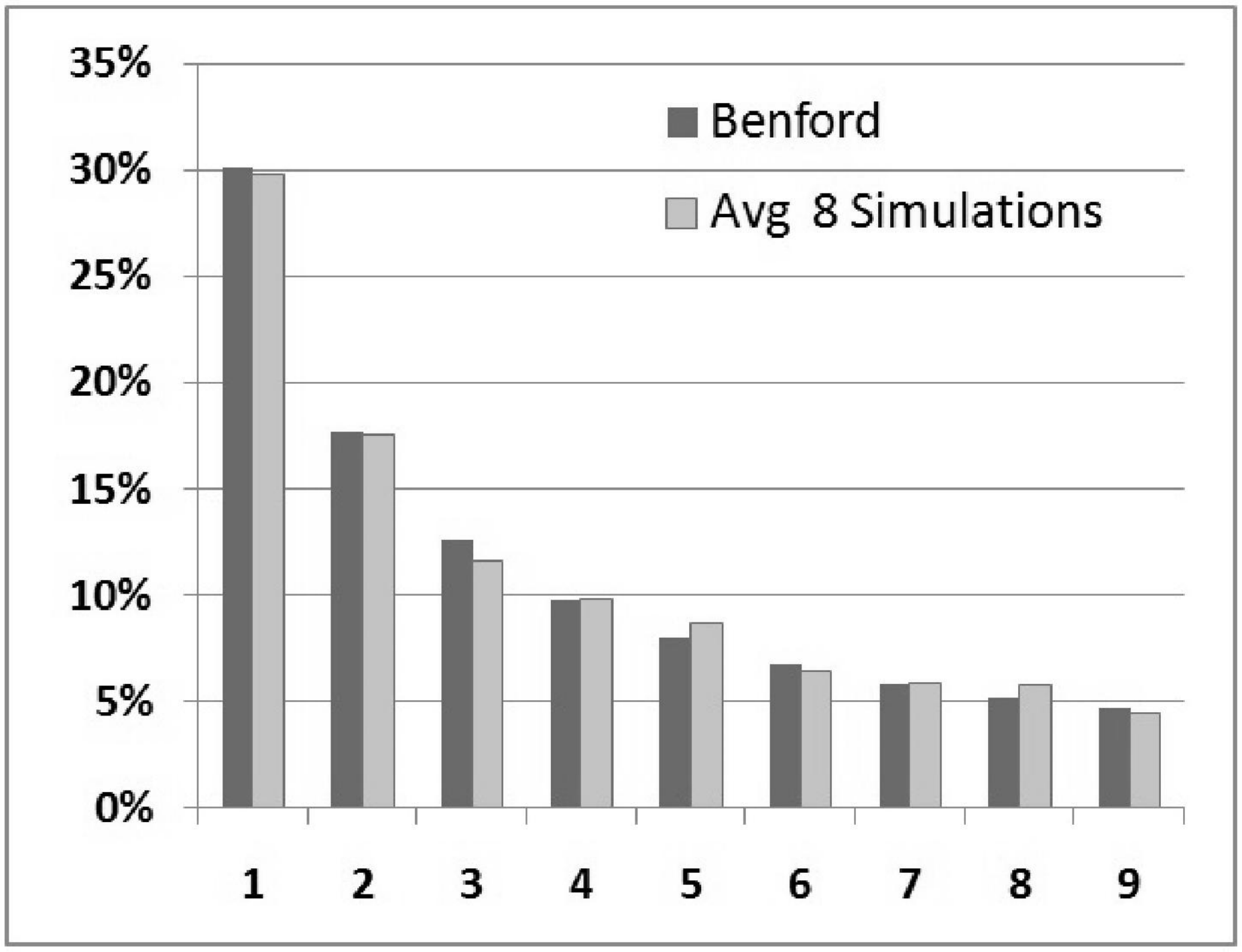

**Figure 62:** Excellent Fit to Benford - Average of 8 Simulations of 25x25 Partitions

Yet, this result actually should not be surprising. Indeed, these 8 distinct simulations could be imagined or thought of as a single complex partitioning of a long 56 by 7 original rectangle made of cardboard or tablecloth, containing within itself 8 squares (of dimension 7 by 7 each), and where each square undergoes 25 by 25 linear partition process independently of its adjacent squares. For such a vista, the satisfactory result with the very low value of 1.9 for SSD is very much expected due to the thorough and very refine partition process of this long rectangle, yielding numerous mini rectangular parts, namely yielding 8*625 or 5000 mini rectangles, all of which naturally should be very close to Benford. All this is in the spirit of the generic principle learnt in this entire article, namely that almost any partition process that is performed in a random fashion without too many constraints and yielding numerous pieces or parts converges to the Benford digital configuration and shows a marked preference for the small at the expense of the big.

Superior and practically perfect results are gotten by increasing the number of simulated grids to 499 by 499 resulting in 500*500 or 250,000 random rectangles, and where compliance with Benford's Law is as perfect as can be gotten for any random data type.

Random 250,000 Rectangles - {29.9%, 17.7%, 12.5%, 9.7%, 8.0%, 6.7%, 5.8%, 5.1%, 4.6%}
Benford's Law First Digits - {30.1%, 17.6%, 12.5%, 9.7%, 7.9%, 6.7%, 5.8%, 5.1%, 4.6%}

Figure 63 depicts the bar chart of the digital comparison to Benford's Law. The exceedingly low Sum of Squares Deviation (SSD) value of 0.1 for the set of resultant 250,000 rectangles indicates that this digit configuration is practically in complete conformity with Benford's Law!

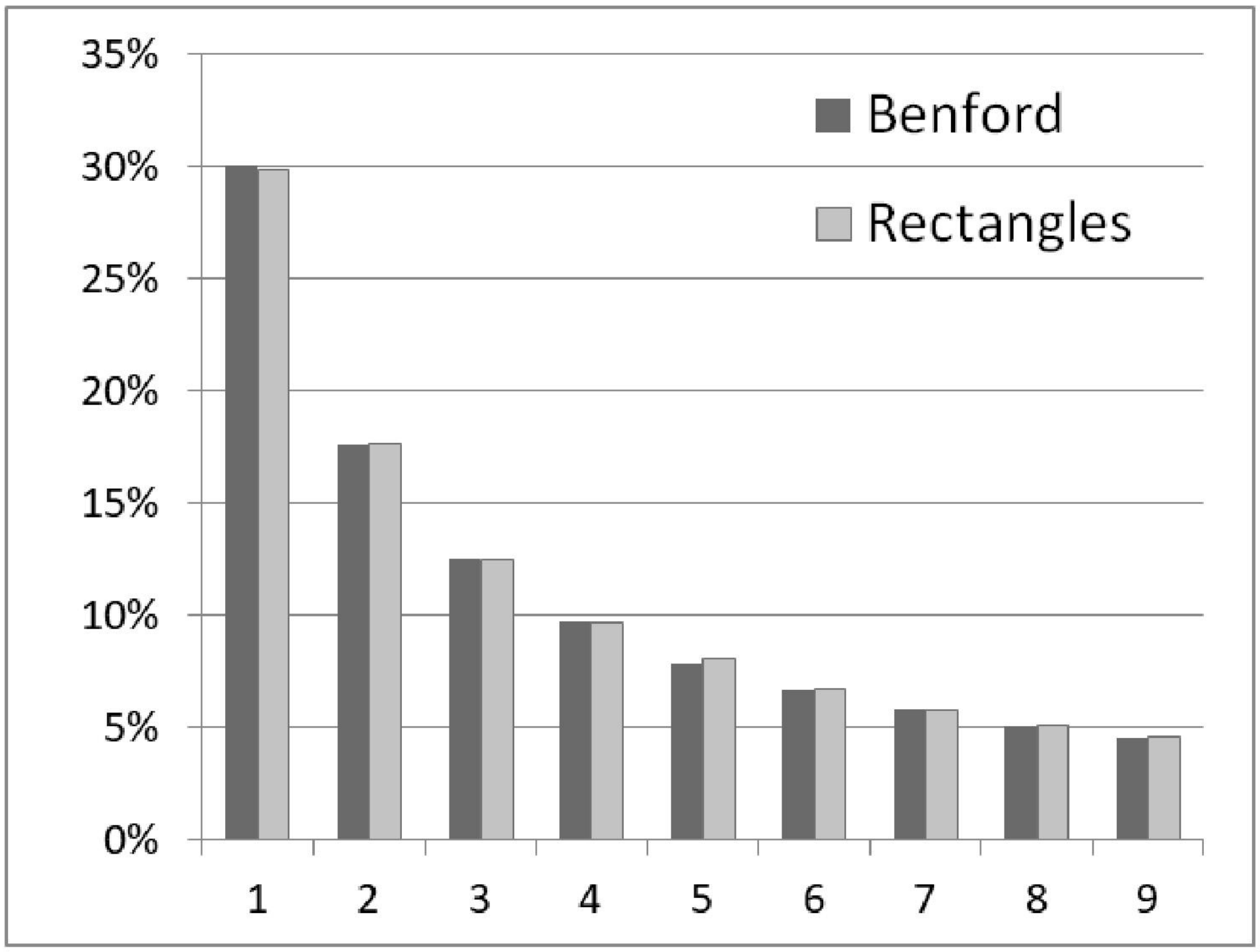

**Figure 63:** Perfect Fit to Benford's Law - Simulated 250,000 Rectangles

## [28] Two Vistas of the Linear Random Partition Process

**The mathematical-arithmetical vista** of the linear partition model is of two distinct one-dimensional random markings placements. Hence, assuming plenty of markings on the x-axis along the horizontal dimension, the horizontal width-compartments (i.e. differences between the markings) are distributed as in the Exponential Distribution [as discussed earlier in this article in chapters 5 and 11]. The same can be said about the vertical y-axis dimension. In addition, since the value of each mini rectangular area is obtained via the multiplication of its width by its height, therefore the model can be viewed as involving the random multiplicative process of two Exponential Distributions [known for its superb Benford convergence.]

The model is said to stand on two legs: (1) two distinct one-dimensional markings placement processes leading to two Exponentials, and (2) a multiplicative process. Both legs are known to tend to the Benford digital phenomenon in and of themselves. Therefore, the confluence of these two processes reinforces the effects of each on digits configuration leading to full convergence to Benford. Chapter 95 titled 'Hybrid Causes Leading to Logarithmic Convergence' in Kossovsky (2014) explores this frequently occurring manifestation of the Benford phenomenon.

**The physical partition vista** of the linear partition model is of the singular partitioning of a rectangular-shaped tablecloth, soft fabric, or a cardboard, where the cuttings are performed strictly in a linear way along the horizontal and vertical dimensions – and without any reference to multiplicative involvement.

Conceptually, for the physical partition vista, this partition process is much more effective in its 2nd stage as compared with its 1st stage. In the 1st stage, random horizontal cuts are made, signifying a normal one-dimensional partition process which can only lead to the Exponential configuration at most. But in the 2nd stage where random vertical cuts are made, each such single vertical line from end-to-end performs multiple partitions because it cuts across many already-existing horizontal ones! This is why the process of partition accelerates in the 2nd stage, and this may explain why we are able to fully converge to Benford after the 2nd stage, instead of obtaining only the approximate Benford digital configuration of the Exponential.

The superb digital results obtained via the 500 by 500 partition, with its resultant 250,000 random rectangles, and where compliance with Benford's Law was practically perfect, can never be matched by simply partitioning one-dimensional linear length, say a pipe, via 250,000 partitions, since such one-dimensional process leads only to the Exponential Distribution and its approximate and not exact Benford digital configuration (due to the saturation that always sets in beyond 5000 to 10000 partitions, where a near perfect fit to the Exponential is gotten, and where further improvement in digital configuration is not possible, as results continue to be Exponential no matter how many more cuts are made).

The comparison of the one-dimensional model with the two-dimensional model, for much fewer partitions, yielding only say 30*30 rectangles as compared with only say 900 linear pipelets, does not show any significant differentiation in the rate of convergence, as both models appear to yield the same degree of digital closeness to the LOG(1 + 1/d) configuration.

## [29] Order of Magnitude Considerations for the Convergence of the Model

As a different perspective on the various partition models, the reason one-dimensional pipe partition model does not lead to a full convergence to the Benford configuration is due to the insufficient order of magnitude (OOM) of the Exponential Distribution itself which hovers around the 2.7 level, while within the field of Benford's Law it is well known that order of magnitude should be approximately at least 3 for full compliance. The intimate connection between Benford's Law and order of magnitude of data is practically universal across all random data, and practically the only exception is the k/x distribution defined over integral powers of ten, such as (1, 10). Even though the support of the Exponential Distribution is the infinite range (0, ∞), in practical terms, real-life Exponential data is never exactly 0, nor infinite. Order of magnitude here is best calculated via the 1% percentile and the 99% percentile, namely as LOG(99% percentile**/**1% percentile), not via the minimum and the maximum as in LOG(max**/**min), so as to avoid the potential exaggeration and distortion of order of magnitude due to outliers, and this method of calculation always leads to resultant OOM of approximately 2.7, regardless of the value of parameter Lambda λ.

For the linear two-dimensional partition model on the other hand, resultant order of magnitude of the areas of the random rectangles (i.e. the product of two Exponentials) is theoretically the sum of the individual orders of magnitude of each Exponential, namely 2.7 + 2.7, or 5.4. In reality, it's somewhat less, yet definitely greater than 3. Therefore the linear two-dimensional model fully converges to the Benford digital configuration.

As outlined in chapter 7 of Kossovsky (May 2016) arxiv.org article titled "Arithmetical Tugs of War and Benford's Law", resultant theoretical order of magnitude of the product of any two variables X and Y is simply the addition of the individual orders of magnitude, namely $OOM_{X*Y} = OOM_X + OOM_Y$. Yet in reality, for any finite number of computer simulated multiplications of two random variables, resultant OOM of multiplied variables is somewhat lower in actual simulations then the theoretical expectation of $OOM_X + OOM_Y$. In any case, the net effect of the multiplicative process is certainly of increasing OOM and greater variability.

For the actual Monte Carlo simulation mentioned earlier of 499 by 499 grids resulting in 250,000 random rectangles, the OOM of the horizontal set of differences came out as only 2.35, and the OOM of the vertical set of differences came out as only 2.59, while the OOM of their product (i.e. the areas) came out as 3.62, and which is comfortably well above the OOM threshold of approximately 3, and therefore the partition process is practically in perfect agreement with Benford's Law. The above-mentioned OOM values were robustly calculated via the 1% percentile and the 99% percentile, and not via the minimum and the maximum, so as to avoid the potential distortions from outliers, and to prevent them from exaggerating OOM.

## [30] Two-Dimensional Linear Random Partition Process in Stages

In the linear partition process, all the marks on the axes are drawn simultaneously in a sense, so that no mark can claim seniority or extra importance as being drawn before any other mark.

Another random partition process is suggested here, one with carefully executed stages, and which also converges to Benford fully, and even more rapidly, as the number of stages grows. As shown in Figure 64, the process begins with a single square. Two points are chosen randomly and independently along the horizontal x-axis and the vertical y-axis, leading to two lines drawn within the square as in an asymmetrical cross, and resulting in four rectangles. Figure 65 depicts the 2nd stage, which is the continuation of the partition process after the 1st stage of Figure 64, and where within each of the four rectangles two random points are chosen in order to draw asymmetrical crosses within them, resulting in sixteen rectangles.

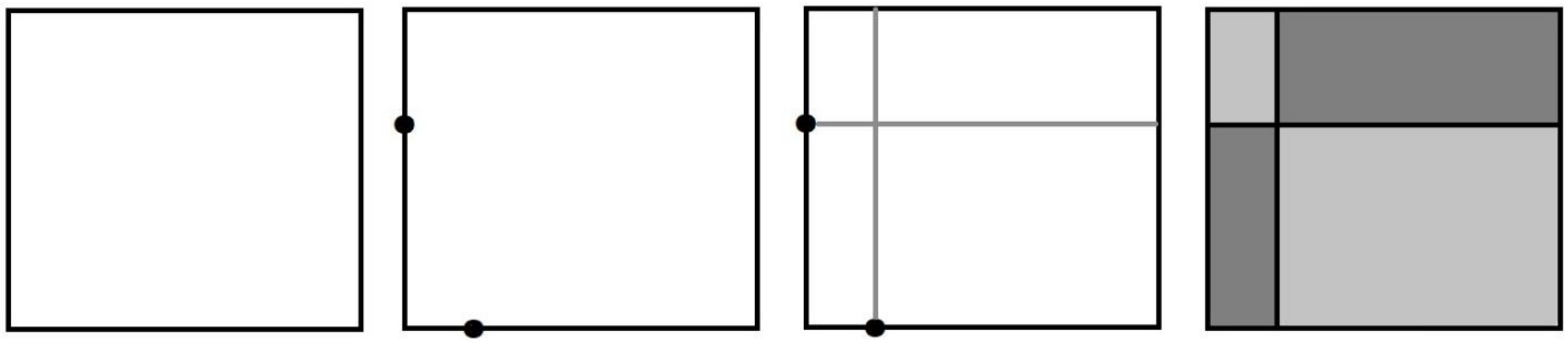

**Figure 64:** Two-Dimensional Partition Process after 1 Stage Yielding 4 Random Rectangles

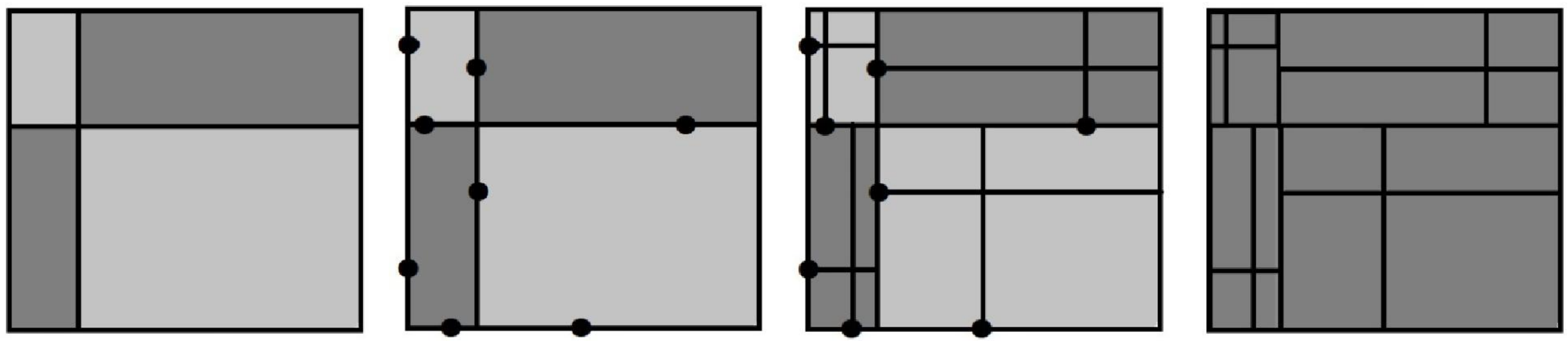

**Figure 65:** Two-Dimensional Partition Process after 2 Stages Yielding 16 Random Rectangles

Figure 66 depicts the 3rd stage, which is the continuation of the partition process after the 2nd stage of Figure 65, and where asymmetrical crosses are drawn within each of the 16 rectangles, resulting in a total of 64 random rectangles. Surely, Figures 64, 65, and 66 constitute just one unique scenario of the 3-stage partition process, and in principle there are infinitely many other possible scenarios of this 3-stage process.

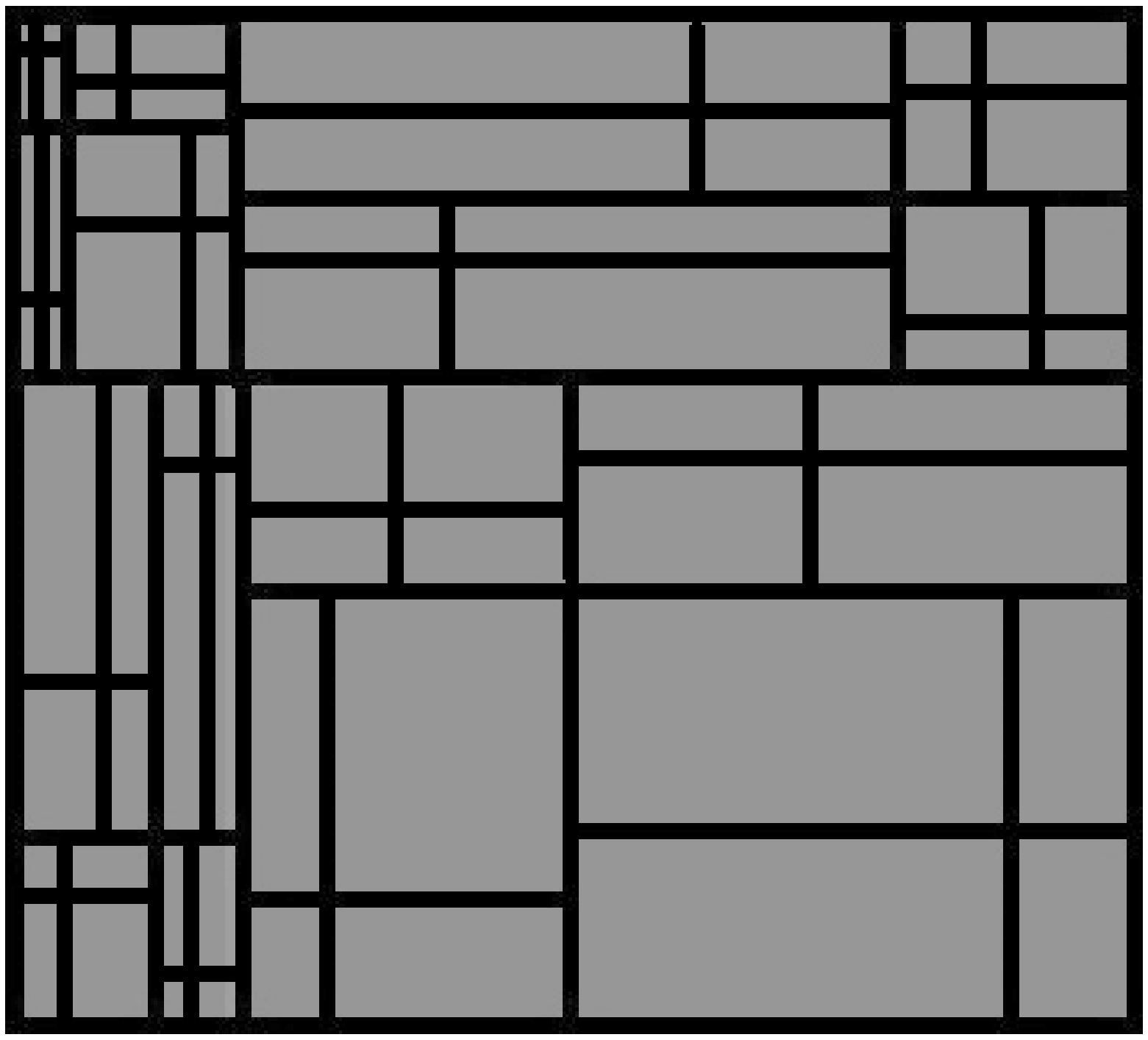

**Figure 66:** Partition Process after 3 Stages Yielding 64 Rectangles

Figure 67 depicts 21 distinct Monte Carlo simulations results of the 3-stage random partition process performed on an original 7 by 7 square. Each such process yields 64 random rectangles. Digit configuration recorded for these 21 simulations varies widely as can be seen in Figure 67. The last row at the bottom is the calculated average – digit-by-digit – of these 21 simulations. The last column on the right depicts the Sum Squares Deviations (SSD) measure for each simulation.

| Digit 1 | Digit 2 | Digit 3 | Digit 4 | Digit 5 | Digit 6 | Digit 7 | Digit 8 | Digit 9 | SSD |
|---|---|---|---|---|---|---|---|---|---|
| 26.6% | 17.2% | 10.9% | 9.4% | 3.1% | 7.8% | 10.9% | 4.7% | 9.4% | **89.1** |
| 32.8% | 12.5% | 7.8% | 15.6% | 3.1% | 6.3% | 4.7% | 9.4% | 7.8% | **143.6** |
| 26.6% | 21.9% | 10.9% | 9.4% | 14.1% | 9.4% | 1.6% | 6.3% | 0.0% | **118.4** |
| 39.1% | 14.1% | 18.8% | 4.7% | 9.4% | 3.1% | 4.7% | 4.7% | 1.6% | **182.4** |
| 23.4% | 10.9% | 23.4% | 15.6% | 6.3% | 6.3% | 7.8% | 3.1% | 3.1% | **257.0** |
| 26.6% | 10.9% | 18.8% | 14.1% | 3.1% | 15.6% | 6.3% | 3.1% | 1.6% | **231.3** |
| 26.6% | 12.5% | 20.3% | 6.3% | 14.1% | 1.6% | 4.7% | 6.3% | 7.8% | **188.7** |
| 25.0% | 26.6% | 15.6% | 6.3% | 7.8% | 4.7% | 6.3% | 1.6% | 6.3% | **147.5** |
| 26.6% | 15.6% | 12.5% | 9.4% | 6.3% | 6.3% | 9.4% | 9.4% | 4.7% | **50.5** |
| 43.8% | 14.1% | 10.9% | 6.3% | 7.8% | 7.8% | 3.1% | 4.7% | 1.6% | **230.8** |
| 26.6% | 10.9% | 10.9% | 6.3% | 7.8% | 14.1% | 9.4% | 9.4% | 4.7% | **156.5** |
| 29.7% | 17.2% | 15.6% | 9.4% | 14.1% | 6.3% | 4.7% | 1.6% | 1.6% | **71.1** |
| 37.5% | 15.6% | 10.9% | 6.3% | 10.9% | 6.3% | 4.7% | 3.1% | 4.7% | **87.4** |
| 37.5% | 18.8% | 17.2% | 6.3% | 4.7% | 1.6% | 3.1% | 6.3% | 4.7% | **135.1** |
| 28.1% | 15.6% | 12.5% | 7.8% | 6.3% | 7.8% | 6.3% | 4.7% | 10.9% | **56.3** |
| 28.1% | 17.2% | 17.2% | 4.7% | 4.7% | 4.7% | 7.8% | 12.5% | 3.1% | **126.3** |
| 28.1% | 21.9% | 14.1% | 10.9% | 10.9% | 3.1% | 4.7% | 1.6% | 4.7% | **61.9** |
| 29.7% | 15.6% | 15.6% | 6.3% | 7.8% | 10.9% | 4.7% | 6.3% | 3.1% | **48.4** |
| 21.9% | 23.4% | 12.5% | 12.5% | 9.4% | 9.4% | 6.3% | 0.0% | 4.7% | **145.2** |
| 32.8% | 12.5% | 15.6% | 10.9% | 7.8% | 7.8% | 3.1% | 4.7% | 4.7% | **53.4** |
| 21.9% | 21.9% | 18.8% | 12.5% | 6.3% | 6.3% | 6.3% | 3.1% | 3.1% | **142.2** |
| **29.5%** | **16.5%** | **14.8%** | **9.1%** | **7.9%** | **7.0%** | **5.7%** | **5.1%** | **4.5%** | **7.4** |

**Figure 67:** 21 Distinct Simulations of 3-Stage Partition - Digit Configuration Varies Widely

On the face of it, nothing of the exact or nearly-exact Benford configuration can be seen in Figure 67. Indeed, digit 1 fluctuates from 21.9% to 43.8%, while digit 9 fluctuates even more wildly, from 0% to 10.9%, and SSD values are as high as 50 to 150 approximately. And surely it can easily be argued that although the staged-partition process ultimately leads to Benford as the number of stages (and the number of rectangles) increases, yet, expecting to obtain the Benford configuration so early after merely 3 stages with only 64 generated rectangles is nothing but an illusion and wishful thinking on the part of Benford aficionados and digit enthusiasts (such as this author). Staring at the random structure of the assorted rectangles in Figure 66 representing one particular simulation trajectory of the 3-stage random partition process with only 64 resultant rectangles, and employing whatever intuition or personal instinct, one would never expect to get anywhere near the LOG(1 + 1/d) configuration!

Yet, even at this early stage, Benford does indeed manifest itself almost exactly! Albeit in a hidden and indirect way, namely in the calculation of the average – digit-by-digit – of these 21 simulations! Calculated SSD for this average is the fairly low value of 7.4!

Average Digit-By-Digit - {29.5%, 16.5%, 14.8%, 9.1%, 7.9%, 7.0%, 5.7%, 5.1%, 4.5%}
Benford's Law 1st Digits - {30.1%, 17.6%, 12.5%, 9.7%, 7.9%, 6.7%, 5.8%, 5.1%, 4.6%}

Figure 68 depicts the good fit to Benford of the average of the 21 simulations.

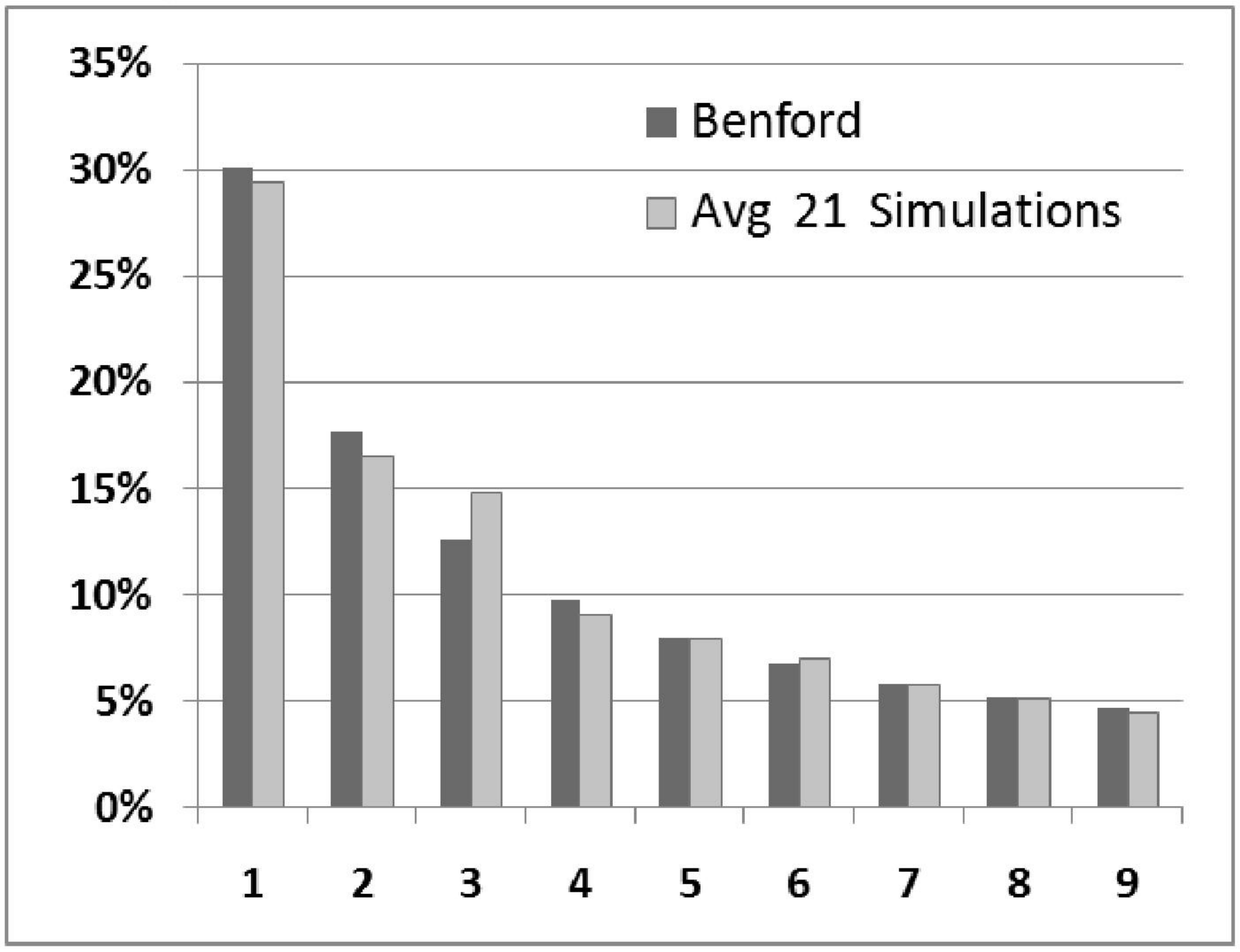

**Figure 68:** Good Fit to Benford after the 3rd Stage for Average of 21 Simulations

Yet, this result actually should not be surprising. Indeed, these 21 distinct simulations could be imagined or thought of as a single complex partitioning of a long 147 by 7 original rectangle made of cardboard or tablecloth, containing within itself 21 squares (of dimension 7 by 7 each), and where each square undergoes three-stage random partition independently of its adjacent squares. For such a vista, the satisfactory result with the low value of 7.4 for SSD is very much expected due to the thorough and very refine partition process of this long rectangle, yielding numerous mini rectangular parts, namely yielding 21*64 or 1344 mini rectangles, all of which naturally should be very close to Benford.

Figure 69 follows the long trajectory of one scenario of the staged partition process, from the original 7 by 7 square to the huge set of 65,536 rectangles after the 8th stage. Figure 69 starts recording from the 3rd stage the snapshots of the number of the rectangles, the digit configuration, sum of squares deviations (SSD), and order of magnitude (OOM), for each stage. Order of magnitude is calculated after the elimination of the top 1% of values and the bottom 1% of values, so as not to let extreme computer simulation values influence or distort order of magnitude. This is so since even a single outlier here as a minimum or as a maximum can exaggerate order of magnitude a great deal.

Clearly, the 8th stage yields digital results that are practically in perfect agreement with Benford's Law, resulting in having the exceedingly low SSD value of 0.1, and nothing really is going to be gained by continuing the process to the 9th stage. We have managed to obtain practically perfect Benford configuration with only 65,536 rectangles, as compared with the linear model of 500 by 500 partitions with its numerous 250,000 rectangles.

| Stage | # Rec. | Digit 1 | Digit 2 | Digit 3 | Digit 4 | Digit 5 | Digit 6 | Digit 7 | Digit 8 | Digit 9 | SSD | OOM |
|---|---|---|---|---|---|---|---|---|---|---|---|---|
| 3 | 64 | 35.9% | 15.6% | 12.5% | 14.1% | 7.8% | 7.8% | 1.6% | 1.6% | 3.1% | 91.0 | 4.1 |
| 4 | 256 | 28.9% | 20.7% | 11.3% | 9.4% | 7.4% | 7.4% | 3.1% | 5.1% | 6.6% | 24.7 | 4.8 |
| 5 | 1024 | 29.6% | 16.9% | 11.6% | 10.2% | 9.9% | 6.2% | 5.6% | 4.9% | 5.3% | 6.4 | 5.8 |
| 6 | 4096 | 29.4% | 17.7% | 12.3% | 9.7% | 8.4% | 6.3% | 6.3% | 5.2% | 4.7% | 1.3 | 6.7 |
| 7 | 16384 | 29.7% | 18.1% | 12.8% | 9.5% | 7.5% | 6.7% | 6.0% | 5.0% | 4.8% | 0.8 | 7.2 |
| 8 | 65536 | 30.2% | 17.6% | 12.4% | 9.6% | 7.9% | 6.8% | 5.7% | 5.1% | 4.7% | 0.1 | 7.8 |
| Benford's Law: | | 30.1% | 17.6% | 12.5% | 9.7% | 7.9% | 6.7% | 5.8% | 5.1% | 4.6% | 0.0 | ---- |

**Figure 69:** Snapshots of # Rectangles, Digit Configuration, SSD, OOM - Stage 3 to Stage 8

Figure 70 depicts the histogram of the areas of resultant rectangles after the 8th stage of that particular random trajectory of Figure 69, confirming the general principle in random partition processes that predicts decisive quantitative skewness, where the relatively small is much more numerous than the relatively big. A logarithmic scale is used for the vertical frequency axis.

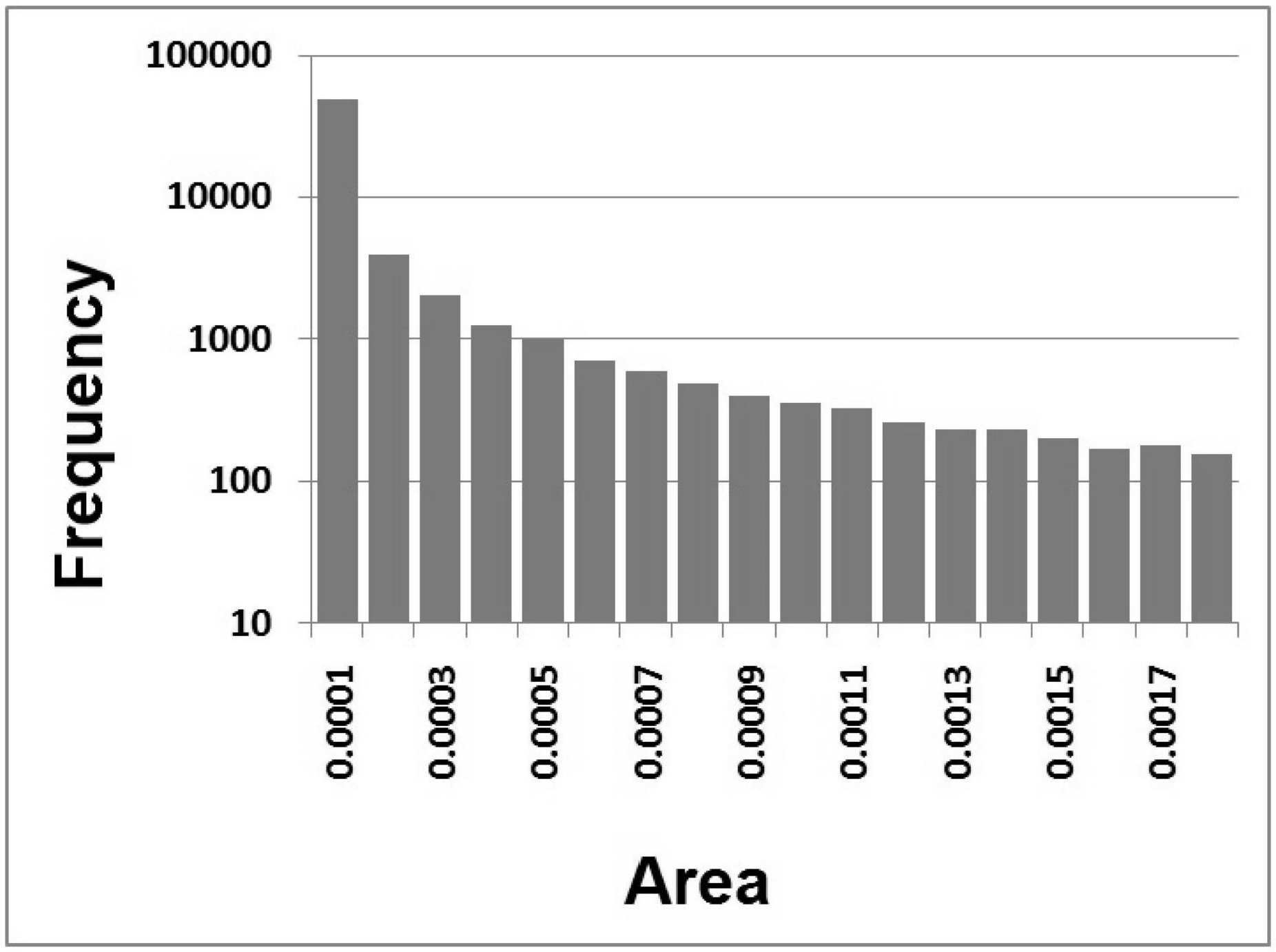

**Figure 70:** Skewed Histogram with a Tail to the Right - 8 Stages – 65,536 Rectangles

## [31] Three-Dimensional Linear Random Partition Process

An extension of the linear two-dimensional partition model is a linear partition performed on three-dimensional cubes or rectangular prisms, leading to the multiplication of three Exponential Distributions. Here the focus is on the volume of solid objects, which is the product of three dimensions, the length, the width, and the height. Figure 71 depicts one example of such three-dimensional linear random partitions where the x-axis is randomly partitioned into 4 compartments, the y-axis is randomly partitioned into 6 compartments, and the z-axis is randomly partitioned into 4 compartments, resulting in 4*6*4 or 96 random rectangular prisms.

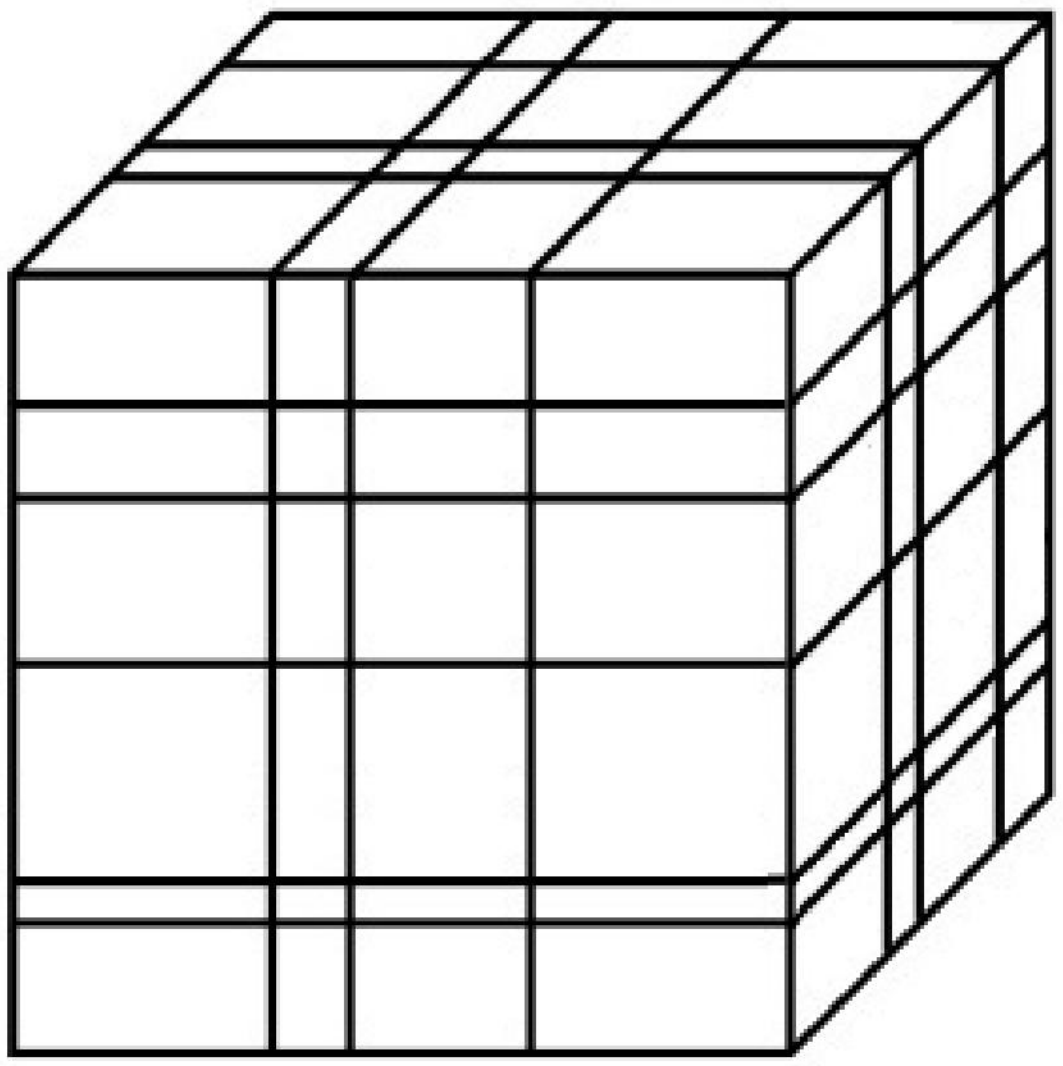

**Figure 71:** Three-Dimensional Random Partition - 4 by 6 by 4

Since three dimensions are multiplied by each other, namely three Exponential Distributions, resultant volume earns an even stronger multiplicative push towards the Benford digital configuration as compared with the two-dimensional random area, although this also requires sufficient number of partitions for each dimension. Order of magnitude here is very large, namely the sum of each order of magnitude of the three Exponentials which theoretically is around 2.7 + 2.7 + 2.7, or 8.1, but in reality only around 4.5 to 5.0 or so, and which easily surpasses the OOM threshold of 3, thus the Benford configuration can be readily obtained here even with a lower number of partitions for each dimension.

Figure 72 depicts 9 distinct Monte Carlo simulations results of three-dimensional 9 by 9 by 9 random partition processes performed on an original 7 by 7 by 7 cube. Each such process yields 9*9*9 or 729 random rectangular prisms. The last row at the bottom is the calculated average – digit-by-digit – of these 9 simulations, and which is extremely closed to Benford with the very low SSD value of 1.0. Thus we have managed to obtain in a sense quickly and easily truly excellent Benford configuration with merely nine such processes, each involving 9 + 9 + 9 or 27 actual cutting actions across the cube! Figure 73 depicts the excellent fit to Benford of the average of the 9 simulations.

| Digit 1 | Digit 2 | Digit 3 | Digit 4 | Digit 5 | Digit 6 | Digit 7 | Digit 8 | Digit 9 | SSD |
|---|---|---|---|---|---|---|---|---|---|
| 32.2% | 19.3% | 11.8% | 8.6% | 7.7% | 5.6% | 6.0% | 4.8% | 3.8% | **11.0** |
| 29.9% | 18.1% | 13.3% | 9.3% | 8.4% | 6.0% | 6.0% | 4.1% | 4.8% | **2.8** |
| 29.9% | 18.2% | 12.9% | 9.3% | 7.1% | 7.4% | 6.4% | 4.7% | 4.0% | **2.8** |
| 31.0% | 16.5% | 12.8% | 10.4% | 7.7% | 5.2% | 7.5% | 4.5% | 4.4% | **8.4** |
| 32.8% | 17.8% | 10.6% | 8.8% | 8.1% | 7.3% | 5.1% | 6.3% | 3.3% | **15.8** |
| 30.2% | 18.1% | 12.9% | 8.1% | 9.2% | 6.4% | 4.8% | 4.5% | 5.8% | **7.4** |
| 30.5% | 18.5% | 12.5% | 8.6% | 8.0% | 6.9% | 6.4% | 5.3% | 3.3% | **4.2** |
| 30.9% | 17.1% | 12.9% | 9.7% | 7.3% | 7.3% | 5.9% | 4.3% | 4.7% | **2.5** |
| 29.4% | 18.0% | 12.8% | 8.9% | 8.4% | 7.4% | 4.9% | 5.8% | 4.5% | **3.2** |
| **30.7%** | **18.0%** | **12.5%** | **9.1%** | **8.0%** | **6.6%** | **5.9%** | **4.9%** | **4.3%** | **1.0** |

**Figure 72:** Nine Distinct Simulations of 9 by 9 by 9 Partition – Average is Practically Benford

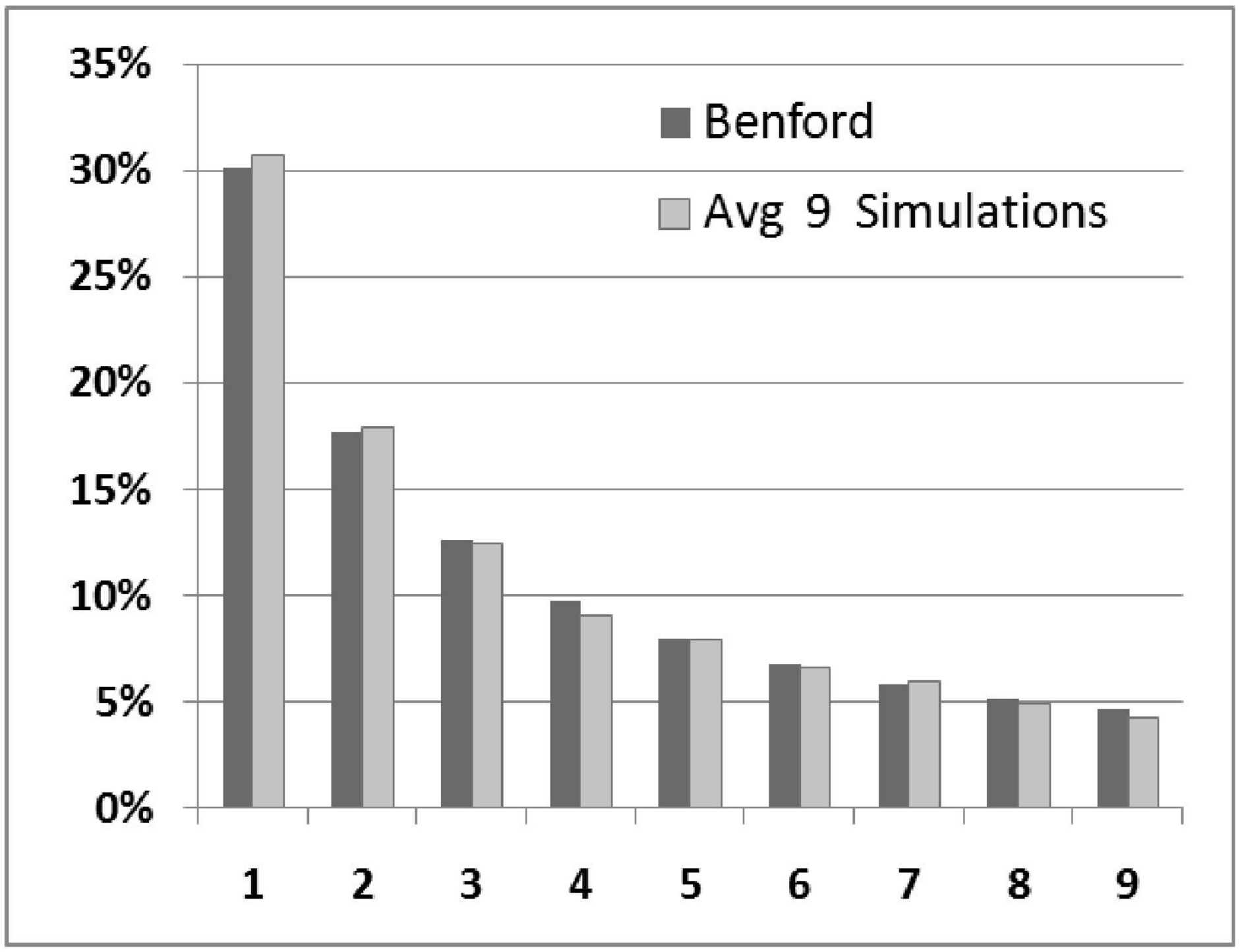

**Figure 73:** Excellent Fit to Benford for Average of 9 Simulations of 9 by 9 by 9

## Acknowledgement:

In part, this article also analyzes and expands on the works of Lemons, Kafri, and Miller.

A groundbreaking article titled "On the Number of Things and the Distribution of First Digits" is published by **Don Lemons** in 1986. It explores an averaging scheme on the set of all possible partitions of a real quantity X into smaller real parts {Δx, 2Δx, 3Δx, …} without explicitly applying anything from Physics. Lemon arrives at a distribution proportional to 1/x which is known to be Benford when defined over ranges with an integral exponent difference between the minimum and the maximum [*namely that LOG(maximum) – LOG(minimum) = Integer*].

**Oded Kafri** publishes an innovative article in 2009 regarding a particular 'balls and boxes' scheme, after exploring during the previous years possible connections between the principles of Information Theory, Entropy in Thermodynamics, and Benford's Law. The distribution of moveable or flexible balls inside fixed and rigid boxes can be interpreted as the partitioning of a large pile of balls into much smaller piles about to reside inside the boxes.

**Steven Miller** publishes an article in 2015 presenting a mathematically rigorous partition model termed 'equipartition', successfully leading to the Benford configuration. His model is a variation on Lemons' original insight, substituting integral quantities for real ones; applying results from Integer Partition and Number Theory; as well as 'borrowing' concepts from Physics.

The author would like to thank Don Lemons, Steven Miller, and Oded Kafri for their helpful comments and correspondence regarding the details of their models.

# REFERENCES

**Andrews, George** (1976). "The Theory of Partitions." Cambridge University Press.
**Benford, Frank** (1938). "The Law of Anomalous Numbers". Proceedings of the American Philosophical Society, 78, 1938, p. 551.
**Buck Brian, Merchant A., Perez S.** (1992). "An Illustration of Benford's First Digit Law Using Alpha Decay Half Lives". European Journal of Physics, 1993, 14, 59-63.
**Carslaw, Charles** (1988). "Anomalies in Income Numbers: Evidence of Goal Oriented Behavior". The Accounting Review, Apr. 1988, 321–327.
**Deckert Joseph, Myagkov Mikhail, Ordeshook Peter** (2011). "Benford's Law and the Detection of Election Fraud". Political Analysis 19(3), 245–268.
**Durtschi Cindy, Hillison William, Pacini Carl** (2004). "The Effective Use of Benford's Law to Assist in Detecting Fraud in Accounting Data". Auditing: A Journal of Forensic Accounting, 1524-5586/Vol. V( 2004), 17–34.
**Freund, John** (1999). "Mathematical Statistics". Sixth Edition. Pearson Education & Prentice Hall International, Inc.
**Gaines J. Brian, Cho K. Wendy** (2007). "Breaking the (Benford) Law: Statistical Fraud Detection in Campaign Finance". The American Statistician, Vol. 61, No. 3, pages 218-223.
**Hamming, Richard** (1970). "On the Distribution of Numbers". Bell System Technical Journal 49(8): 1609-25.
**Kafri, Oded** (2009). "Entropy Principle in Direct Derivation of Benford's Law". March 2009, http://arxiv.org/abs/0901.3047.
**Kafri, Oded & Hava** (2013). "Entropy, God's Dice Game".
**Kossovsky, Alex Ely** (2012). "Towards A Better Understanding of the Leading Digits Phenomena". City University of New York. http://arxiv.org/abs/math/0612627
**Kossovsky, Alex Ely** (Dec 2012). "Statistician's New Role as a Detective - Testing Data for Fraud". http://revistas.ucr.ac.cr/index.php/economicas/article/view/8015
**Kossovsky, Alex Ely** (2013). "On the Relative Quantities Occurring within Physical Data Sets". http://arxiv.org/ftp/arxiv/papers/1305/1305.1893.pdf
**Kossovsky, Alex Ely** (2014). "Benford's Law: Theory, the General Law of Relative Quantities, and Forensic Fraud Detection Applications". World Scientific Publishing Company. August 2014. ISBN: 978-981-4583-68-8.
**Kossovsky, Alex Ely** (2015). "Random Consolidations and Fragmentations Cycles Lead to Benford's Law". http://arxiv.org/ftp/arxiv/papers/1505/1505.05235.pdf
**Kossovsky, Alex Ely** (March 2016). "Prime Numbers, Dirichlet Density, and Benford's Law". http://arxiv.org/ftp/arxiv/papers/1603/1603.08501.pdf
**Kossovsky, Alex Ely** (May 2016). "Arithmetical Tugs of War and Benford's Law". http://arxiv.org/abs/1410.2174
**Kossovsky, Alex Ely** (June 2016). "Exponential Growth Series and Benford's Law" . http://arxiv.org/abs/1606.04425
**Leemis Lawrence, Schmeiser Bruce, Evans Diane** (2000). "Survival Distributions Satisfying Benford's Law". The American Statistician, Volume 54, Number 4, November 2000, 236-241.

**Lemons, S. Don** (1986) “On the Number of Things and the Distribution of First Digits”. The American Association of Physics Teachers. Vol 54, No. 9, September 1986.
**Leuenberger Christoph, Engel Hans-Andreas** (2003). “Benford’s Law for the Exponential Random Variables”. Statistics and Probability Letters, 2003, 63(4), 361-365.
**Ley, Eduardo** (1996). "On the Peculiar Distribution of the U.S. Stock Indices Digits". The American Statistician, Volume 50, No. November 4, 1996, pages 311-313.
**Miller, Steven** (2008). “Chains of Distributions, Hierarchical Bayesian Models and Benford's Law”. Jun 2008, http://arxiv.org/abs/0805.4226.
**Miller Steven, Joseph Iafrate, Frederick Strauch** (2015). “Equipartitions and a Distribution for Numbers: A Statistical Model for Benford's Law”. Williams College, August 2013.
**Miller Steven, Joseph Iafrate, Frederick Strauch** (2013). “When Life Gives You Lemons - A Statistical Model for Benford’s Law”. Williams College, August 2013.
http://web.williams.edu/Mathematics/sjmiller/public_html/math/talks/small2013/williams/Iafrate_SummerPoster2013.pdf
**Newcomb, Simon** (1881). “Note on the Frequency of Use of the Different Digits in Natural Numbers”. American Journal of Mathematics, 4, 1881, 39-40.
**Pinkham, Roger** (1961). “On the Distribution of First Significant Digits”. The Annals of Mathematical Statistics, 1961, Vol.32, No. 4 , 1223-1230.
**Raimi, A. Ralph** (1969). “The Peculiar Distribution of First Digit”. Scientific America, Sep 1969: 109-115.
**Raimi, A. Ralph** (1976). “The First Digit Problem”. American Mathematical Monthly, Aug-Sep 1976.
**Raimi, A. Ralph** (1985). “The First Digit Phenomena Again”. Proceedings of the American Philosophical Society, Vol. 129, No 2, June, 1985, 211-219.
**Ross, A. Kenneth** (2011). “Benford’s Law, A Growth Industry”. The American Mathematical Monthly, Vol. 118, No. 7, Pg. 571–583.
**Sambridge Malcolm, Tkalcic Hrvoje, Arroucau Pierre** (2011). “Benford’s Law of First Digits: From Mathematical Curiosity to Change Detector”. Asia Pacific Mathematics Newsletter October 2011.
**Sambridge Malcolm, Tkalcic Hrvoje, Jackson Andrew** (2010). “Benford’s Law in the Natural Sciences”. Geophysical Research Letters, Volume 37, 2011, Issue 22, L22301.
**Saville, Adrian** (2006). ”Using Benford’s Law to detect data error and fraud: An examination of companies listed on the Johannesburg Stock Exchange”. Gordon Institute of Business Science, University of Pretoria, South African Journal of Economics and Management Sciences, 9(3), 341-354. http://repository.up.ac.za/handle/2263/3283.
**Shao Lijing, Ma Bo-Qiang** (2010a). “The Significant Digit Law in Statistical Physics”. http://arxiv.org/abs/1005.0660, 6 May 2010.
**Shao Lijing, Ma Bo-Qiang** (2010b). “Empirical Mantissa Distributions of Pulsars”. http://arxiv.org/abs/1005.1702, 12 May 2010. Astroparticle Physics, 33 (2010) 255–262.
**Varian, Hal** (1972). “Benford's Law”. The American Statistician, Vol. 26, No. 3.

**Patent:** The U.S. Patent Office # 9,058,285. Inventor: Alex Ely Kossovsky.
Date Granted: June 16, 2015. http://www.google.com/patents/US20140006468
Titled: “Method and system for Forensic Data Analysis in fraud detection employing a digital pattern more prevalent than Benford s Law”.

January 19, 2020
Alex Ely Kossovsky
akossovsky@gmail.com